\documentclass[showpacs,superscriptaddress,floatfix,prl,notitlepage,reprint]{revtex4-2}
\usepackage{graphicx,amsfonts,amssymb,amsmath,hyperref,hypcap,enumerate}
\usepackage{braket}
\usepackage{lipsum} 
\usepackage{amsthm}
\usepackage{multirow}
\usepackage{graphicx}
\usepackage{dcolumn}   
\usepackage{bm}        
\usepackage{amssymb}   
\usepackage{amsmath}
\usepackage{braket}
\usepackage{epstopdf}
\usepackage{mathtools}
\usepackage{ulem}
\usepackage{color}
\usepackage{subfigure}
\usepackage{diagbox}
\usepackage{longtable}
\usepackage{makecell}
\usepackage{bbold}

\begin{document}
\title{Temporal Berry Phase and the emergence of Bose-Glass-Analog Phase in a Clean U(1) Superfluid}
\author{Ryuichi Shindou}
\email{rshindou@pku.edu.cn} 
\affiliation{International Center for Quantum Materials, Peking University, Beijing 100871, China}
\author{Pengwei Zhao}
\affiliation{International Center for Quantum Materials, Peking University, Beijing 100871, China}
\author{Xiaonuo Fang}
\affiliation{School of Physics, Peking University, Beijing 100871, China}
\date{\today}
\begin{abstract} 
The (2+1)-dimensional U(1) sigma model with temporal Berry phase term captures zero-temperature phase-fluctuation-driven superfluid (SF) transitions. From renormalization group (RG) analysis of its dual representation -- vortex loop gas model --, we clarify that the Berry phase leads to space-time anisotropic interference in vortex-loop proliferation, resulting in a quasi-disordered phase with short-ranged spatial yet persistent temporal phase coherence. The phase shares physical properties of the Bose glass phase known from disordered boson systems, suggesting a unified topological origin for the emergence of the glassy phase in phase-fluctuation-driven superfluid transitions. 
\end{abstract}
\maketitle

\vspace{1em} 

\textit{Introduction---}
  Superfluid transition is one of the most foundational macroscopic quantum phenomena in physics. Conventionally, the transition is described by the Ginzburg-Landau $\phi^4$ theory for the superfluid order parameter $\phi \equiv|\phi|e^{i\theta}$, where superfluid amplitude $|\phi|$ condenses at the critical point and establishes simultaneously the long-range coherence of the superfluid phase $\theta$. This picture works well for systems with large boson density, where the amplitude fluctuation dominates the critical behavior. However, in a broad class of physical systems, including underdoped high-temperature superconductors~\cite{loeser1996,ding1996,renner1998,corson1999}, ultrathin superconducting films~\cite{hebard1990,goldman1998,larkin2005}, Josephson junction arrays~\cite{resnick1981,newrock2000}, and strongly correlated boson systems~\cite{hadzibabic2006,desbuquois2012,choi2013}, the boson density is suppressed to a small but finite value by certain physical mechanisms such as system geometries~\cite{bezryadin2000} and correlation effect~\cite{emery1995}. Thereby, the phase fluctuation emerges as the dominant driver of the superfluid transition.

    In this phase-fluctuation-dominated regime, the critical physics is captured by a U(1) sigma model $Z \equiv \int d\theta \exp[-S]$ with 
\begin{align}
S \equiv \frac{1}{2g}\int d^{3}{\bm x} |{\bm \nabla} \theta|^2 + i\chi \int d^{3}{\bm x} \partial_{\tau} \theta, \label{eq1}
\end{align} 
where the superfluid (SF) phase $\theta$ fluctuates in space ${\bm r}$ and (imaginary) time ${\tau}$ with ${\bm x}\equiv (\tau,{\bm r})$. The Berry phase term along the temporal direction originates from the canonical commutation relation between boson density and phase, where $\chi$ is given by the mean boson density $\overline{\rho}=|\phi|^2$~\cite{wiegel1978,jacobs1984,fisher1989b,herbut2007}. For two-dimensional (2D) classical systems, the phase-fluctuation-driven transition is fully described by the Berezinskii-Kosterlitz-Thouless (BKT) theory~\cite{berezinskii71,berezinskii72,kosterlitz1973,kosterlitz1974,kosterlitz2017}, where the  unbinding of vortex-antivortex pairs drives the transition, a mechanism exactly mapped to the 2D Coulomb gas problem. Extending the BKT theory to the zero-temperature quantum transition in the two spatial dimensions elevates vortex-antivortex pairs into closed vortex loops in (2+1)-dimensional space-time. An early work by Popov established a duality mapping between the three-dimensional (3D) vortex-loop-driven superfluid transition~in the sigma model~\cite{onsager1949,feynman1955} and 3D classical magnetostatics of quantized electric current loops~\cite{popov1973,wiegel1973}. Subsequent works developed renormalization group (RG) theories of the vortex-loop-driven transition~\cite{kotsubo1986,williams1987,shenoy1989,janke1990,williams1993,chattopadhyay1994,williams1999,williams2004,goldbart2009}, drawing parallels with the 2D BKT physics where short vortex loops screen the $1/r$ Coulomb interactions. 

\begin{figure}[b]
\centering
\includegraphics[width=0.9\linewidth]{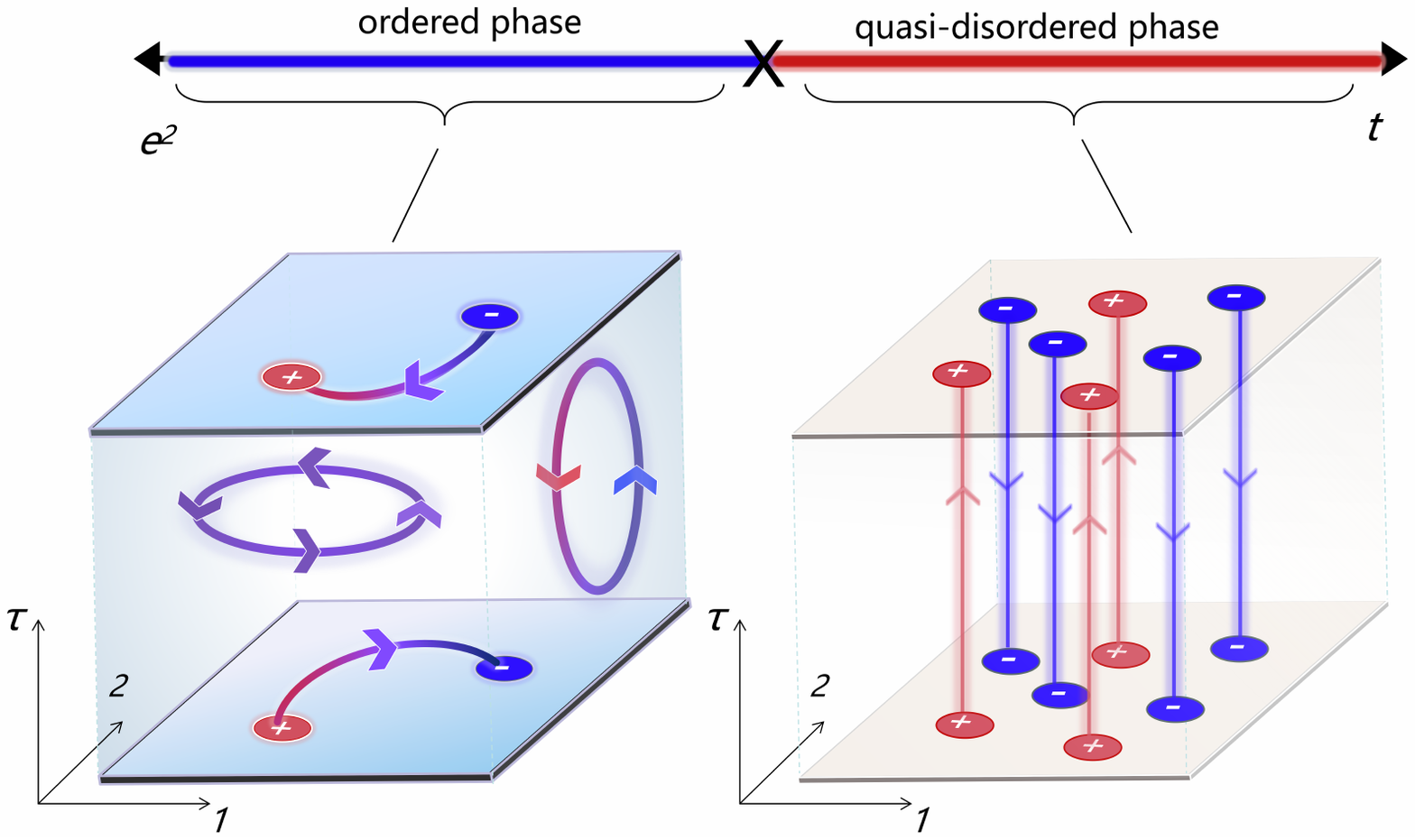}
\caption{A schematic picture of quasi-disordered phases in (2+1) dimensional U(1) sigma with temporal Berry phase. }
\label{Fig:1}
\end{figure}
    
     The role of the temporal Berry phase has been studied extensively in clean correlated lattice boson models~\cite{fisher1989a,read1989,read1990,tanaka2015}. However, its precise impact on the space-time anisotropy of superfluid phase correlations remains poorly understood~\cite{ma1986,fisher1989a,fisher1989b}. Conventionally, adding the temporal Berry phase to the U(1) sigma model has been viewed as simply changing the dynamical exponent from $z=1$ to $z=2$ while still describing a direct continuous transition between ordered and disordered phases~\cite{herbut2007}. However, whether the Berry phase can qualitatively alter the transition scenario -- for instance, by inducing an intermediate glassy phase -- remains an open question. A compelling hint comes from a duality transformation~\cite{banks1977,einhorn1978,savit1978,peskin1978,savit1980,fisher1989a,kiometzis95,herbut2007} that maps the sigma model to a 3D type-II superconductor in an external magnetic field~\cite{blatter1994,huse1992,fisher1989c,shindou2025,zhao2025}. In this dual description, a vortex line lattice (VLL) phase~\cite{abrikosov1957,larkin1970,safar1992,hetzel1992,pastoriza1994,doyle1995,zeldov1995,schilling1996,nguyen1996,nguyen1998a,nguyen1998b,ryu1998,nguyen1999} emerges adjacent to the normal  phase~\cite{nelson1989}. The VLL phase exhibits an anisotropic magnetic-monopole-field (MMF) correlation: long-ranged along the field direction, but short-ranged in perpendicular directions~\cite{fisher1989a,shindou2025,zhao2025}. Under duality, this anisotropic MMF correlation maps to long-range temporal phase coherence coexisting with short-range spatial coherence in the sigma model. The duality thus predicts an intermediate glassy phase characterized by extreme space-time anisotropy in the U(1) phase coherence~\cite{shindou2025,zhao2025}. Notably, such glassy behavior in boson systems has traditionally been attributed to extrinsic quenched disorder~\cite{fisher1989b,giamarchi1987,giamarchi1988}. However, recent quanutm Monte Carlo studies on disorder-free quasilattices have reported a superfluid-to-Bose-glass transition at low boson density~\cite{ciardi2023,zhu2023}, demonstrating that glassiness can emerge from intrinsic mechanisms alone. This lends indirect numerical support to the possibility of the Berry-phase-induced glassy phase in clean systems.

         In this Letter, we apply renormalization group theory to the (2+1)-dimensional U(1) sigma model with the temporal Berry phase, formulated through the duality mapping. We clarify that the Berry phase fundamentally alters the short-vortex-loop screening physics of the 3D magnetostatics, yielding a divergent space-time anisotropy of the magnetic permeability in an intermediate coupling regime. The divergent anisotropy polarizes vortex lines along the imaginary time direction, reducing the 3D magnetostatics problem to the 2D electrostatics problem. Thereby, spatial proliferation of time-polarized vortex lines destroys spatial coherence while preserving temporal coherence [Fig.~\ref{Fig:1}]. The resulting quasi-disordered phase shares essential U(1) phase coherence properties of the Bose glass, establishing topological interference as a unified origin for the glassy phase.

\textit{vortex loop gas and topological interference}
    The 3D sigma model with the Berry phase term can be studied by a partition function of the magnetostatics of vortex loops (`electric current loops'),
\begin{align}
Z = &\int {\cal D}{\sf a} \exp\Big[-\int d^{3}{\bm x} 
\Big(\frac{(\nabla\times {\sf a})^2_{\tau}}{2\mu_{\tau}}+\frac{|(\nabla\times {\sf a})_{\bm r}|^2}{2\mu_{\bm r}}\Big)\Big]  \nonumber \\
& \int {\cal D}\Gamma \!\ y^{L_{\Gamma}} \!\  
\exp\Big[i2\pi e \oint_{\Gamma} d{\bm l} \cdot {\sf a}
- i2\pi \chi S_{\bm r}[\Gamma] \Big]. \label{eq2}
\end{align}
The action with the isotropic magnetic permeability $[\mu_{\tau}=\mu_{\bm r}=1]$ can be legitimately obtained from the U(1) sigma model by the duality mapping~\cite{popov1973,lee1991}. Thereby, the phase gradient term is decoupled in terms of Stratonovich-Hubbard (SH)
vector field ${\bm \varphi}$, $\frac{1}{2g} |{\bm \nabla} \theta|^2 = \frac{1}{2} 
|{\bm \varphi}|^2 + i e {\bm \varphi}\cdot {\bm \nabla} \theta$ with $e=1/\sqrt{g}$. The gradient vector ${\bm \nabla}\theta$ is further decomposed into divergence free part ${\bm b}$ (vortex part) and a rotation free part ${\bm \nabla}\psi$ (spin-wave part). An integration over $\psi$ leads to a divergence-free constraint upon the SH field, yielding ${\bm \varphi}={\bm \nabla}\times {\sf a}$, and the isotropic Maxwell action from $|{\bm \varphi}|^2$. A rotation of the divergence free part is given by quantized flux line along closed vortex loops  $\Gamma$, ${\bm \nabla} \times {\bm b}({\bm x})=2\pi \oint_{\Gamma} d{\bm l} \delta^3({\bm x}-{\bm l})$, so that ${\bm \varphi}\cdot{\bm b}$ becomes the `magnetic' coupling between the vector potential ${\sf a}$ and closed vortex loops (`electric current loops'), $\int d^3{\bm x}\!\ {\bm \varphi}\cdot{\bm b} = 2\pi \oint_{\Gamma}d{\bm l}\cdot {\sf a}$. Meanwhile, the temporal Berry phase confers a finite phase factor upon each closed vortex loop~\cite{tanaka2015,zhao2024}, and the phase factor is given by a projected area enclosed by each loop onto the spatial plane ($\bm r$-plane). The Berry phase term becomes a sum of the projected areas $S_{\bm r}[\Gamma]$ over all vortex loops. $\int {\cal D}\Gamma$ in eq.~(\ref{eq2}) is a path integral over vortex loops configuration. $y$ is fugacity for vortex loop element with a unit length, and $L_{\Gamma}$ is total length of vortex loops. The action with $y=1$ can also be directly obtained from a correlated boson model with a finite mean boson density $\overline{\rho}$, where the magnetic reluctivity $1/\mu_{\tau}$ along the time direction is given by the repulsive boson interaction, the reluctivity $1/\mu_{\bm r}$ along the spatial direction is given by the boson mass normalized by the boson density $\overline{\rho}$~\cite{lee1991,supplemental}.

        In Eq.~(\ref{eq2}), the vector potential ${\sf a}$ mediates the $e^2/r$ Coulomb interaction between vortex loop elements. In the absence of the Berry phase term $[\chi=0]$, the phase transition in the model is driven by space-time proliferation of vortex loops: ordered phase is a weak coupling phase with vanishing fugacity $y$ and divergent $e$, and disordered phase is a strong coupling phase with divergent fugacity $y$ and vanishing $e$. The temporal Berry phase $\chi$ interferes destructively with the proliferation process of those loops with finite projected areas on the spatial plane, by imparting the phase factors to them~\cite{zhao2024,shindou2025,zhao2025}. Due to the destructive interference, one may expect that the phase transition with $\chi\ne 0$ is generally driven by vortex lines polarized along the time direction [Fig.~\ref{Fig:1}]: the polarized vortex lines are free from the phase factor. In the correlated boson model, bosons experience the time-polarized vortex lines as static magnetic fluxes with quenched dynamics. Thus, upon their spatial proliferation, the bosons strongly scatter off this intrinsic, self-generated quenched disorder, resulting in their Anderson localization -- genuine Bose-glass characteristics~\cite{fisher1989b}. To probe this scenario, we first clarify the screening to magnetic permeability and vortex fugacity renormalization with $\chi\ne 0$.

\begin{figure}[t]
\centering
\includegraphics[width=\linewidth]{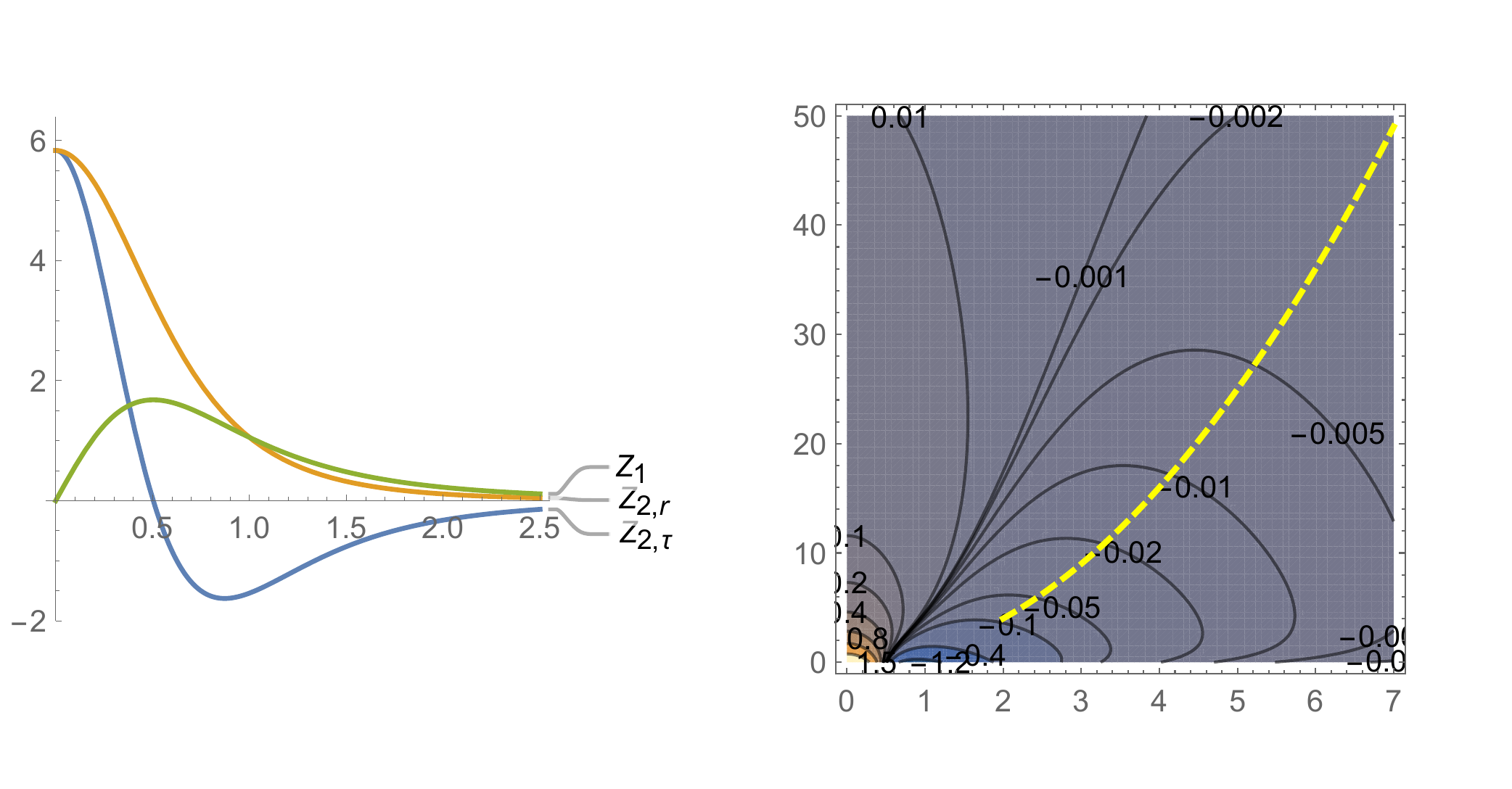}
\caption{(left) $Z_{2,\tau}$ [blue], $Z_{2,{\bm r}}$ [yellow], and $Z_1$[green] as a function of $\nu$ from Eqs.~(\ref{eq6},\ref{eq7}) with $y=1$. (Right) Contour plot of $Z_{2,\tau}$ of Eq.~(\ref{eq11}) with $t_{+}=0$ as a function of $\nu$ [horizontal axis] and $t_{-}\equiv t_{\tau}-t_{\bm r}$ [vertical axis]. A Dashed yellow line [$t_{-}=\nu^2$] depicts a `gorge' of negative value of $Z_{2,\tau}$.}
\label{Fig:2}
\end{figure}

\textit{space-time anisotropic screening}
    The magnetic permeability is screened by shorter vortex loops. To calculate this with $\chi\ne 0$, we decompose the sum over vortex loops into sums over larger vortex loops $\Gamma_{>}$ and a short vortex loop $\Gamma_{<}$, 
\begin{align}
\int {\cal D}\Gamma \!\ y^{L_{\Gamma}}  =\int {\cal D}\Gamma_{>} \!\  y^{L_{\Gamma_{>}}} \Big( 1 +   \int{\cal D}\Gamma_{<} \!\ y^{L_{\Gamma_{<}}} \Big). \label{eq3}
\end{align}
A typical size of the short vortex loop is  a lattice constant $a_0$, while the vector potential varies slowly in this scale. Thus, we use a gradient expansion in the integral associated with $\Gamma_{<}$, $\oint_{\Gamma_{<}} d{\bm l}\cdot {\sf a} = S_0 ({\bm \nabla}\times {\sf a})\cdot {\bm n}$ . Here $S_0$ is an area enclosed by the short vortex loop, and ${\bm n}\equiv (n_{\tau},{\bm n}_{\bm r})$ is a unit vector normal to a coplanar plane in which the short loop resides. We then expand $\exp[i2\pi e S_0 ({\bm \nabla}\times {\sf a})\cdot {\bm n}]$ in power of the electric charge $e$, while split the Berry phase term into a longer-loops contribution and the short-loop contribution, $S_{\bm r}[\Gamma_{>}+\Gamma_{<}]=S_{\bm r}[\Gamma_{>}] + S_0 n_{\tau}$. The expansion in $e$ can be justified by a systematic $\epsilon$ expansion, where the entire argument is generalized to $d$ spatial dimension with $d+1=2+\epsilon$~\cite{supplemental}. The first and second expansion terms then take the form of the Maxwell action in the presence of an applied magnetic field along $\tau$,  
\begin{align}
&\int {\cal D} \Gamma \!\ y^{L_{\Gamma}} \!\ e^{ie \oint_{\Gamma} d{\bm l}\cdot{\sf a} - i2\pi \chi S_{\bm r}[\Gamma]} \nonumber \\
& \ = \int {\cal D}\Gamma_{>} \!\ y^{L_{{\Gamma}_{>}}} \!\  e^{ie \oint_{\Gamma_{>}} d{\bm l}\cdot{\sf a} - i2\pi \chi S_{\bm r}[\Gamma_{>}]} \!\ \bigg[1 + \int d^3{\bm X} \Big\{ \!\ \delta f + \nonumber \\
& \delta \chi_{\tau} ({\bm \nabla}\times {\sf a})_{\tau} 
- \frac{\delta \nu_{\tau}}{2} 
({\bm \nabla}\times {\sf a})^2_{\tau}   -\frac{\delta\nu_{\bm r}}{2} |({\bm \nabla}\times {\sf a})_{\bm r}|^2 +\cdots \Big\} \bigg].  \label{eq4}
\end{align}
Here $\int {\cal D}\Gamma_{<}$ comprises of an integral $\int d^3{\bm X}$ over a center of mass coordinate ${\bm X}$ of the short vortex loop, and an integral over size and direction ${\bm n}$ of the short loop. In Eq.~(\ref{eq4}), the integral over size and direction,  and $y^{L_{<}}$ are included in $\delta \chi_{\tau}$, and $\delta \nu_{a}$ $(a=\tau,{\bm r})$ [see Eqs.~(\ref{eq6p},\ref{eq6},\ref{eq7})]. By re-exponentiating the expanded terms and combining them into a bare Maxwell term, we obtain the screening $\delta \nu_{a}$ on the magnetic reluctivity $(a=\tau,{\bm r})$,  
\begin{align}
Z &= \int {\cal D} {\sf a} e^{-{\cal S}_{0}} \int {\cal D}\Gamma_{>} y^{L_{\Gamma_{>}}} 
e^{i2\pi e \oint_{\Gamma_{>}}d{\bm l}\cdot {\bm a} - i 2\pi \chi S_{\bm r}[\Gamma_{>}]} \nonumber \\
{\cal S}_{0} &= \int d^3{\bm x} \Big(\frac{(({\bm \nabla}\times {\sf a})_{\tau}-\mu_{\tau} \delta \chi_{\tau})^2}{2(\mu_{\tau}-\mu^2_{\tau}\delta \nu_{\tau})} + \frac{|({\bm \nabla}\times {\sf a})_{\bm r}|^2}{2(\mu_{\bm r}-\mu^2_{\bm r}\delta \nu_{\bm r})}  \nonumber \\
& \hspace{2cm} + \cdots \Big).  \label{eq5}
\end{align}
The induced magnetic field $\delta \chi_{\tau}$ can be translated into the screening on the temporal Berry phase by a gauge transformation, $({\bm \nabla} \times {\sf a})_{\tau} \rightarrow ({\bm \nabla} \times {\sf a})_{\tau} +\mu_{\tau}\delta \chi_{\tau}$, $\chi \rightarrow \chi +  e \mu_{\tau}\delta \chi_{\tau}$. In the renormalization group theory, we softly constrain the size of the short vortex loop within a slice of $[a_0, a_0 b]$ with the renormalization scale parameter $b\gtrsim 1$. After the integration over the size and the direction ${\bm n}$ of $\Gamma_{<}$, we obtain the recursive screening on the Berry phase term, and temporal and spatial components of the reluctivity,
\begin{align}
\delta \chi_{\tau} = \pi^3 & e a^{-1}_0  Z_1 \ln b, \,\ \,\  \delta \nu_{a}=\frac{\pi^5}{2} e^2 a_0 Z_{2,a} \ln b, \label{eq6p} \\ 
Z_{1} &= \frac{5}{2} \int^{1}_{-1}d n_{\tau} \frac{in_{\tau} \!\ y^{a_0 \pi}}{(1+i\nu n_{\tau})^{\frac{7}{2}}},   \label{eq6} \\
\left(\begin{array}{c} 
Z_{2,\tau} \\
Z_{2,{\bm r}} \\
\end{array}\right)
&= \frac{35}{4}  \int^{1}_{-1} dn_{\tau} \!\  
\left(\begin{array}{c} 
n^2_{\tau} \\ 
\frac{1-n^2_{\tau}}{2} \\
\end{array}\right)  
\frac{y^{a_0 \pi}}{(1+i\nu n_{\tau})^{\frac{9}{2}}}.  \label{eq7}
\end{align}
Here, $\nu \equiv 2\pi S_0 \chi =\frac{\pi^2 a^2_0}{2} \chi$ is a normalized Berry phase parameter, and we assume that the short vortex loop takes a coplanar circular loop with its diameter $a_0$. Notice that a relation $e\partial_{\chi}\delta \chi_{\tau}=\delta \nu_{\tau}$ [or equivalently $\partial_{\nu} Z_{1} =Z_{2,\tau}$] is consistent with the magnetostatics analogy of $\chi/e$, $\delta \chi_{\tau}$ and $\delta \nu_{\tau}$ being magnetic flux density, an applied magnetic field and the magnetic reluctivity, respectively. Eq.~(\ref{eq7}) shows that the screening of the short vortex loop in the spatial plane $[n_{\tau} = \pm 1]$ increases $\nu_{\tau}$, and the screening of the vortex loop parallel to the temporal axis $[n_{\tau}= 0]$ increases $\nu_{\bm r}$. Importantly, the finite Berry phase suppresses the former screening more effectively than the latter through the Berry phase factor $\nu$, i.e. $Z_{2,\tau}<Z_{2,{\bm r}}$ [see Fig.~\ref{Fig:2}]. Thus, the magnetic screening in the presence of the temporal Berry phase term $\chi$ leads to an enhancement of the space-time anisotropy $\mu_{\tau}/\mu_{\bm r} \equiv \nu_{\bm r}/\nu_{\tau}$ of the magnetic permeability. As shown below, it is this enhanced space-time anisotropy that polarizes vortex lines along the time direction.

\textit{vortex fugacity renormalization} 
    The vortex fugacity is renormalized by a short-ranged part of the $e^2/r$ Coulomb interaction. In the space-time isotropic case, it is given by $E_{v}=\frac{\pi e^2}{2} \oint\oint_{a_0<|{\bm l}-{\bm l}^{\prime}|<a_0 b} d{\bm l}\cdot d{\bm l}^{\prime}/|{\bm l}-{\bm l}^{\prime}|$ with the lattice constant $a_0$ and the RG scale $b$. An integration over one of the coordinates gives the total loop length $L_{\Gamma}$, while the integration over the relative coordinate  $\Delta l = |{\bm l}-{\bm l}^{\prime}|$  gives a logarithmic dependence on the renormalization scale parameter $b$, $E_{v}=\pi e^2 L_{\Gamma} \int^{a_0 b}_{a_0} d(\Delta l)/\Delta l=\pi e^2 L_{\Gamma} \ln b$. Incorporating this Coulomb energy with the Boltzmann factor $e^{-E_v}$ into the action shifts the logarithmic fugacity, $\ln y \rightarrow \ln y - \pi e^2 \ln b$.  After a length rescale $[L^{\prime}_{\Gamma}=L_{\Gamma}b^{-1}]$, this gives an RG equation for the logarithmic fugacity  $t=\ln y$ as $dt/d\ln b = t - \pi e^2$~\cite{williams1987,shindou2025}.

        In the space-time anisotropic case $[\mu_{\tau} \ne \mu_{\bm r}]$, we decompose the magnetic vector potential ${\sf a}$ into fast mode ${\sf a}^{>}$ and slow mode ${\sf a}^{<}$. An integration over the fast mode ${\sf a}^{>}$ renormalizes the fugacity of a vortex loop $\Gamma$ into the space-time anisotropic form,   
\begin{align}
        y^{L_{\Gamma}} \rightarrow  \exp \bigg[\oint_{\Gamma} ds 
    \Big\{ t_{\tau} \Big(\frac{dl_{\tau}(s)}{ds}\Big)^2 + 
    t_{\bm r} \Big|\frac{d{\bm l}_{\bm r}(s)}{ds}\Big|^2\Big\}\bigg], \label{eq8}
\end{align}
with $t_{\tau} \ne t_{\bm r}$. Here, ${\bm l}(s)\equiv (l_{\tau}(s),{\bm l}_{\bm r}(s))$ is the space-time coordinate of a vortex loop element parameterized by one-dimensional parameter $s$. $d{\bm l}(s)/ds$ is a tangential vector along the loop, where $s$ has a length scale, $\oint_{\Gamma} ds = L_{\Gamma}$. The renormalization group equation for logarithmic fugacity $t_{\tau}$ for vortex loop element along time direction and $t_{\bm r}$ along spatial direction are calculated from the integration over ${\sf a}_{>}$, 
\begin{align}
\frac{dt_{\tau}}{d\ln b} = t_{\tau} - \pi e^2 \mu_{\bm r}, \,\  \frac{dt_{\bm r}}{d\ln b} = t_{\tau} - \pi e^2 \sqrt{\mu_{\tau}\mu_{\bm r}}. \label{eq9}
\end{align}
The first term of the equation reflects that the tree-level scaling dimension of $t_{\tau}$ and $t_{\bm r}$ is 1: they are defined per unit length. The second term of the equation can be understood from the short-ranged part of the Coulomb interaction $E_{v}$ in two limiting cases;  (i) the vortex line of length $L_{\Gamma}$ polarized along the time direction, and (ii) the vortex line of length $L_{\Gamma}$ polarized along the space directions. Thereby, we may use a scale transformation, $\tau^{\prime}=\tau$, ${\bm r}^{\prime}={\bm r} \sqrt{\mu_{\tau}/\mu_{\bm r}}$, ${\sf a}^{\prime}_{\tau}={\sf a}_{\tau}/\sqrt{\mu_{\bm r}}$, ${\sf a}^{\prime}_{\bm r}={\sf a}_{\bm r}/\sqrt{\mu_{\tau}}$. The transformation brings the anisotropic permeability into the isotropic one, $(\mu_{\tau},\mu_{\bm r},e)\rightarrow (1,1,e\sqrt{\mu}_{\bm r})$, while it stretches the vortex line along space by a factor of $\sqrt{\mu_{\tau}/\mu_{\bm r}}$ , and keeps unchanged the vortex line polarized along time.  Thus, from the isotropic argument, we obtain $E_{v}=\pi e^2 \mu_{\bm r}  L_{\Gamma} \ln b$  for the case (i), and $E_{v}=\pi e^2 \mu_{\bm r} L_{\Gamma} \sqrt{\mu_{\tau}/\mu_{\bm r}} \ln b = \pi e^2 \sqrt{\mu_{\tau}\mu_{\bm r}} \!\ L_{\Gamma}\ln b$ for the case (ii). Incorporating these two into $t_{\tau} L_{\Gamma}$ and $t_{\bm r}L_{\Gamma}$, respectively,  we obtain the second term of Eq.~(\ref{eq9}).  As the fugacity of a vortex loop for the anisotropic case is generalized into Eq.~(\ref{eq8}), we also replace $y^{L_{\Gamma}}$ in Eq.~(\ref{eq2}) by Eq.~(\ref{eq8}), and replace $y^{a_0\pi}$ in Eqs.~(\ref{eq6},\ref{eq7}) by $\exp[\frac{\pi a_0}{2} \{(1-n^2_{\tau}) t_{\tau} + (1+n^2_{\tau}) t_{\bm r}\}]$ [see Eq.~(\ref{eq11})].

       Eq.~(\ref{eq9}) indicates that the enhancement of $\mu_{\tau}/\mu_{\bm r}$ suppresses the vortex loop element along the spatial direction more significantly than the vortex loop element along the temporal direction, polarizing the vortex loop along time. The polarization of the vortex loop element further enhances the space-time anisotropy $\mu_{\tau}/\mu_{\bm r}=\nu_{\bm r}/\nu_{\tau}$, as the screening of the short vortex loops parallel to the temporal axis $[n_{\tau}=0]$ increases $\nu_{\bm r}$. This leads to a generic runaway divergence of the space-time anisotropy parameter.  In fact, the RG phase diagram at $\chi\ne 0$ exhibits an intermediate coupling regime between weak and strong coupling phases, where $\mu_{\tau}/\mu_{\bm r}$ runs away and diverges at a finite RG scale $b$.   

\textit{RG analysis}
     The RG equation is obtained by the recursive integrations of the short-vortex loop $\Gamma_{<}$ and the fast mode ${\sf a}_{>}$. At each recursive step, we rescale the length scale $[{\bm x} \equiv (\tau,{\bm r})\rightarrow {\bm x}^{\prime}={\bm x} b^{-1}]$, and normalize the slow mode ${\sf a}_{<}$ such that the spatial component ${\mu}_{\bm r}$ of the magnetic permeability is kept to the unit. With this convention, we obtain the RG equations for the electric charge square $e^2$, the normalized Berry phase parameter $\nu\equiv \frac{\pi^2 a^2_0}{2}\chi$, the space-time anisotropy parameter $\gamma_{\tau}\equiv \mu_{\bm r}/\mu_{\tau}=1/\mu_{\tau}$, and the logarithmic fugacities $t_{\tau}$ and $t_{\bm r}$:      
\begin{align}
\left\{\begin{array}{l}
\frac{de^2}{d\ln  b} = e^2 - \frac{\pi^5}{2} e^4 Z_{2,{\bm r}},  \\
\frac{d\nu}{d\ln b} = 2\nu  - \frac{\pi^5}{2} e^2 \gamma^{-1}_{\tau} Z_{1}, \\
\frac{d\gamma_{\tau}}{d\ln b} = \frac{\pi^5}{2} e^2 \big( Z_{2,{\tau}} - \gamma_{\tau} Z_{2,{\bm r}}\big),  \\
\frac{dt_{\tau}}{d\ln b} = t_{\tau} - \pi e^2, \,\ 
\frac{dt_{\bm r}}{d\ln b}= t_{\bm r} - \pi \frac{e^2}{\sqrt{\gamma_{\tau}}}, \\
\end{array}\right. \label{eq10}
\end{align}
and  
\begin{align}
\left\{\begin{array}{l} 
Z_{1} 
= \frac{5}{2} 
\int^{1}_{-1}dn_{\tau}  \!\   e^{\frac{\pi}{2}(t_{+}-n^2_{\tau} t_{-})} \!\ 
\frac{in_{\tau} }{(1+i\nu n_{\tau})^{\frac{7}{2}}},  \\
Z_{2,\tau} = \frac{35}{4}  \int^{1}_{-1} dn_{\tau} \!\  e^{\frac{\pi}{2}(t_{+}-n^2_{\tau} t_{-})} \!\ 
\frac{n^2_{\tau}}{(1+i\nu n_{\tau})^{\frac{9}{2}}}, \\ 
Z_{2,{\bm r}} = \frac{35}{8} \int^{1}_{-1} dn_{\tau} 
\!\ e^{\frac{\pi}{2}(t_{+}-n^2_{\tau} t_{-})} \!\ 
\frac{1-n^2_{\tau}}{(1+i\nu n_{\tau})^{\frac{9}{2}}},  \\ 
\end{array}\right. \label{eq11}
\end{align}
with $t_{\pm}\equiv t_{\tau}\pm t_{\bm r}$. Here, $t_{\tau}$ and $t_{\bm r}$ as well as $e^2$ are normalized by $1/a_0$. The tree level scaling dimension of 
$e^2$, $\nu$ and $\gamma_{\tau}$ are $1$, $2$ and $0$, respectively. As $\mu_{\bm r}$ is set to the unit by the normalization, the recursive screening onto $\mu_{\bm r}$ appears in the equations for $e^2$ and $\gamma_{\tau}$.

\begin{figure}[t]
\centering
\includegraphics[width=0.9\linewidth]{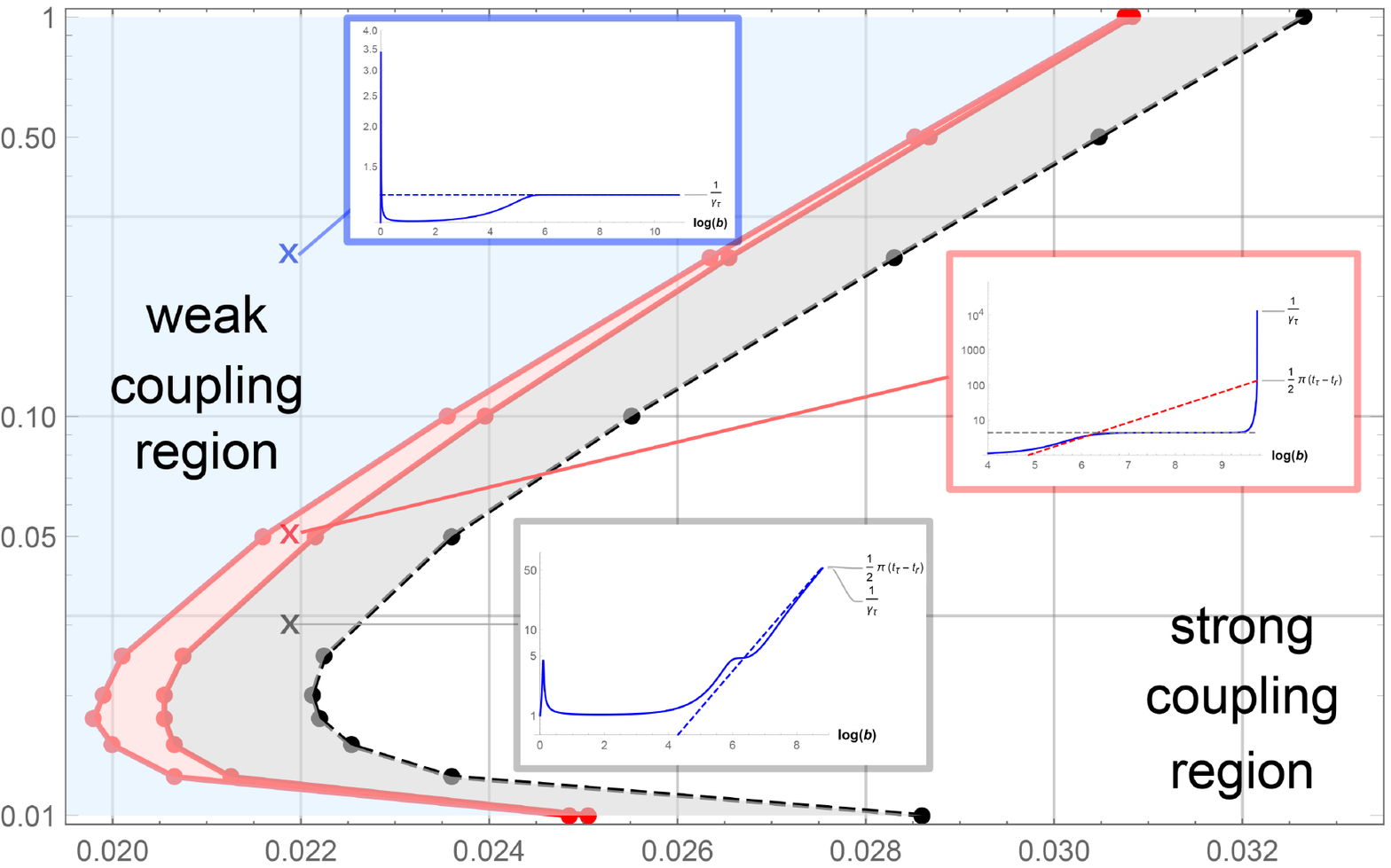}
\caption{Phase diagram numerically determined from the RG equations with initial values of $e^2$ [vertical axis], $t_{\tau}=t_{\bm r} \equiv t$ [horizontal axis], $\gamma_{\tau}=1$ and $\nu=1.2$. The light blue (gray or white) region is the weak-coupling (strong-coupling) region with divergent (vanishing) $e^2$ and $\nu$, and negatively (positively) divergent $t_{+}\equiv t_{\tau}+t_{\bm r}$. The light red region is the intermediate coupling region, where $\gamma_{\tau}$ vanishes at a finite RG scale $b$. Three inset figures show $\gamma^{-1}_{\tau}$ as a function of the RG scale $b$ in these three regions.}
\label{Fig:4}
\end{figure}

        The equations are solved numerically for a set of initial coupling constants. Their renormalized values in the large-$b$ limit determine the RG phase diagram. The diagram features three distinct regimes: a weak-coupling region where the fugacity vanishes while $e^2$, $\nu$ diverge and $\gamma_{\tau}$ remains constant in the large-$b$ limit; an intermediate-coupling regime where $\gamma_{\tau}$ vanishes at {\it finite} RG scale $b$ with positive $t_{\tau}$; and a strong-coupling region where the fugacity diverges while $e^2$ and $\nu$ vanish in the large-$b$ limit.  Importantly, a transformation of initial couplings from weak to strong coupling regions necessitates passage through the intermediate coupling region, where $\gamma_{\tau}$ goes to zero at a finite $b$. Such intermediate region arises because of  a `gorge' of a negative value of $Z_{2,t}$ at a finite $\nu$ region [Fig.~\ref{Fig:2}]. Namely, upon any transition from the weak-coupling to the strong-coupling regions, the decrease of the Berry phase parameter $\nu$  is driven by a large positive $Z_1$ in Eq.~(\ref{eq10}). Thus, by the time $\nu$ enters the gorge of the negative value $Z_{2,t}$ [Fig.~\ref{Fig:2}], $Z_{1}$ has already grown significantly large, and the large negative $Z_{2,t}=\partial_{\nu}Z_{1}$ reduces $\gamma_{\tau}$ to zero at a finite RG scale $b$ in Eq.~(\ref{eq10}). Since the large value of $Z_1$ comes from large positive $t_{+}=t_{\tau}+t_{\bm r}$, $t_{\tau}$ is inevitably positive when $\gamma_{\tau}\rightarrow 0+$ at the finite $b$: $t_{\tau}>t_+/2>0$.   

      The onset of $\gamma_{\tau}=0+$ with $t_{\tau}>0$ and $e^2 \ne 0$ at finite $b$ signals a Landau-pole-type instability in the intermediate coupling region. Namely, the anisotropic Coulomb energy with $\gamma_{\tau}\rightarrow 0+$ and finite $e^2$ polarizes all the vortex loop elements along $\tau$,  leaving only the temporal component ${\sf a}_{\tau}(\tau,{\bm r})$ of the gauge field being coupled with the vortex degrees of freedom in Eq.~(\ref{eq2}). As a result, the (2+1) dimensional partition function reduces to the classical partition function for the 2D isotropic Coulomb gas. In this reduction, the temporally-uniform component ${\sf a}_{\tau}(\omega=0,{\bm r}) \equiv \int^{\beta_{\tau}}_{0}  
{\sf a}_{\tau}(\tau,{\bm r}) d\tau$ plays the role of the 2D electrostatic potential, and the fugacity $e^{t_{\tau}\beta_{\tau}}$ for the polarized vortex line becomes the fugacity of electric charge in the 2D Coulomb gas. Thus, the zero-temperature limit $[\beta_{\tau}\rightarrow \infty]$ with positive $t_{\tau}$ always brings the system into the 2D plasma phase, where a spatial proliferation of the time-polarized vortex lines destroys the spatial correlation of the U(1) phase, while keeping intact the temporal U(1) phase coherence.  In fact, upon their spatial proliferation, a partition function of the boson model becomes an averaged quantum partition function of free boson systems in a background of dense, randomly distributed static magnetic fluxes~\cite{supplemental}. As all single-particle states in such 2D unitary-class systems become localized, the Anderson localization of the boson endows the resulting quasi-disordered phase with all the Bose-glass characteristics  -- spatial short-range superfluid correlation, finite compressibility, boson insulating transport, algebraic decay of the superfluid correlation in time~\cite{supplemental,fisher1989b}.

 \textit{Summary}—  Topology serves as a fundamental element in the formation of emergent phases and phase transitions, manifesting itself as topological terms in partition functions for quantum many-body systems~\cite{zhang1989,gogolin1998,schnyder2008,kitaev2009,altland2010,qi2011,chiu2016,khmelnitskii1983,pruisken1984,pruisken1987,altland2014,haldane1983,affleck1987,haldane1988,read1989,read1990}. This Letter proposes a topological mechanism for the emergence of the quasi-disordered phase in (2+1) dimensional U(1) sigma model with the temporal Berry phase term, that shares the essential physical properties with the Bose glass phase~\cite{fisher1989b,supplemental}.

\textit{Acknowledgements}-- RS thank Gary A. Williams for communications. The work was supported by the National Basic Research Programs of China (No. 2024YFA1409000) and the National Natural Science Foundation of China (No. 12074008 and No. 12474150). 

\bibliographystyle{unsrt}
\bibliography{ref}

\clearpage
\begin{widetext}
\numberwithin{equation}{section}
\section*{Supplemental Material for ``Temporal Berry Phase and the emergence of Bose-Glass-Analog Phase in a Clean U(1) Superfluid"} 

 U(1) nonlinear sigma model (NLSM) with a Berry phase term in $D=d+1$ dimension describes a zero-temperature partition function for $d$-dim.  phase-fluctuation-driven superfluid (SF) transitions, where SF phase fluctuates in space $r$ and time $\tau$, while SF amplitude is constrained around a finite value by a certain physical mechanism. The Berry phase term along the time direction  originates from a quantum-mechanical commutation relation between SF amplitude and phase. An ordered-disordered phase transition of the model is driven by a spatial proliferation of vortex excitations, where the Berry phase term confers a complex phase factor upon the partition function with the vortex excitations, possibly altering a nature of the disordered phase through an interference effect caused by the complex phase factor. In this paper, we map the U(1) NLSM into a $D$-dimensional generalization of the Maxwell action, whose electromagnetic field couples with vortex excitations, mediating the long-ranged $1/|x|^{D-2}$ Coulomb interaction between vortex hypersurface elements. Based on this dual model, we derive a renormalization group (RG) equation of the phase-fluctuation-driven SF transition, where the screening of smaller vortex hypersurfaces renormalizes the electromagnetic constitutive constant in the Maxwell action, and the short-range part of the Maxwell action renormalizes the vortex fugacity. By analyzing the scaling property of the RG equation around a saddle-fixed point using the authentic $\epsilon \equiv D-2$ expansion, we obtain the critical exponent $\nu$ and scaling dimension $y_{\chi}$ of the Berry phase parameter $\chi$ around the phase transition at $\chi=0$. Using the $D=3$ RG equation, we clarify that the screening effect in the presence of the Berry phase term yields a space-time anisotropy in the constitutive constant of the Maxwell action, while the space-time anisotropy polarizes vortex lines locally along the $\tau$ direction. Our numerical solutions of the $D=3$ RG equation demonstrate that the phase transition at $\chi\ne 0$ is driven by the proliferation of the vortex lines along $\tau$, indicating the emergence of the quasi-disordered phase next to the ordered phase. In the quasi-disordered phase, the SF phase correlation time is divergent, while correlation length is finite.

\vspace{1em} 
\section{\label{secI}U(1) NLSM and its dual form}
The U(1) NLSM with the 1-dimensional Berry phase term is given by 
\begin{align}
Z = \int {\cal D}\theta \exp \Big[-\frac{1}{2g}\int d^D{\bm x} \!\ \!\  
{\bm \nabla} \theta \cdot {\bm \nabla} \theta  - i\chi \int d^D{\bm x} \!\ \!\ \partial_\tau \theta \!\ \!\ \Big]. 
\end{align}with ${\bm x}=(\tau,{\bm r})$, ${\bm \nabla} \equiv (\partial_{\tau},\partial_{\bm r})$, and U(1) phase $\theta({\bm x})$. The Berry phase term $\chi$ originates from a quantum-mechanical commutation relation between the amplitude and phase of the superfluid (SF) order parameter, while the SF amplitude is constrained to a finite value by a specific physical mechanism.  Quantum fluctuation parameter $g$ plays the role of a coupling constant in the model. The perturbative $\beta$ function for the coupling constant has no disordering effect in general $D$, indicating that the phase transition to its disordered phase is entirely driven by the vortex proliferation associated with the U(1) phase. To analyze this topological phase transition for the general $D$ dimension, we can decompose the phase gradient vector ${\bm \nabla} \theta$ into a gradient field ${\bm \nabla}\phi$ and a solenoidal field ${\bm b}$ with ${\bm \nabla} \cdot {\bm b}=0$ and $\oint {\bm \nabla}\phi \cdot d{\bm l}=0$,
\begin{align}
{\bm \nabla} \theta = {\bm b} + {\bm \nabla}\phi. 
\end{align}
The gradient and solenoidal fields constitute the spin-wave part $Z_{\rm sw}$ and the vortex part $Z_{\rm v}$ of the partition function, respectively. 
\begin{align}
Z &= Z_{\rm sw}\!\ Z_{\rm v}, \nonumber \\
Z_{\rm sw} &= \int {\cal D}\phi \exp\Big[-\frac{1}{2g}\int d^D{\bm x} \!\ \!\  
{\bm \nabla} \phi \cdot {\bm \nabla} \phi\!\ \Big], \nonumber \\
Z_{\rm v} & = \int {\cal D}{\bm b} \exp \Big[ -\frac{1}{2g}\int d^D{\bm x} \!\ \!\  
{\bm b} \cdot {\bm b}\!\
- i\chi \int d^D{\bm x} \!\ \!\ b_{\tau} \!\ \!\ \Big]. 
\end{align}
In the case of $D=2$, the vortex excitation forms pairs of a vortex and an antivortex, and the space-time rotation of the solenoidal field is characterized by a dipole vector, 
\begin{align}
\partial_{\tau} b_{1}({\bm x}) - \partial_{1} b_{\tau}({\bm x}) = 2\pi \sum_{j} 
\big( \delta^2({\bm x}-{\bm R}_{j,{\rm v}}) - \delta^2({\bm x}-{\bm R}_{j,{\rm av}})\big). 
\end{align}
Here ${\bm R}_{j,{\rm v}}$ and ${\bm R}_{j,{\rm av}}$ are the two-dimensional coordinates of the vortex and antivortex in the $j$th dipole vector.  For $D=3$, the vortex excitation forms a closed loop, and the rotation of the solenoidal field, $v_{\mu\nu}({\bm x}) \equiv \partial_{\mu}b_{\nu}(\bm x)-\partial_{\nu}b_{\mu}({\bm x})$, becomes nothing but a vortex field with a quantized flux line along the loop, 
\begin{align}
v_{\mu\nu}({\bm x}) &\equiv \epsilon_{\mu\nu\lambda} J_{\lambda}({\bm x}), \nonumber \\
J_{\lambda}({\bm x}) &= 2\pi \sum_j \int_{\partial \Gamma_j} ds \frac{\partial R_{j,\lambda}(s)}{ds} \delta^3({\bm x}-{\bm R}_j(s)).  
\end{align}
Here $\partial \Gamma_j$ is the $j$-th closed vortex loop, whose element is parameterized by the one-dimensional parameter $s$ as ${\bm R}_j(s)$.  The vortex excitation for $D\ge 4$ generally forms a $(D-2)$-dimensional closed hyper-surface, and  $\partial \Gamma_j$ becomes a $j$-th closed vortex hyper-surface, whose space-time coordinate ${\bm R}_j({\bm s})$ is parameterized by $(D-2)$ variables ${\bm s}\equiv (s_1,s_2,\cdots,s_{D-2})$. Thereby, the exterior derivative of the solenoidal field is given by an $(D-2)$-dimensional ${\bm s}$-integral of the quantized flux over the vortex hyper-surface, 
\begin{align}
v_{\mu\nu}({\bm x})& \equiv 
\varepsilon_{\mu\nu\lambda\rho\cdots\sigma} 
\frac{J_{\lambda\rho\cdots\sigma}({\bm x})}{(D-2)!}, \label{def-V} \\ 
J_{\lambda\rho\cdots\sigma}({\bm x}) & \equiv 2\pi \sum_{j}\int_{\partial \Gamma_j} d^{D-2}{\bm s} \,\ \varepsilon_{ab\cdots c}
\Big(\frac{\partial R_{j,\lambda}({\bm s})}{\partial s_a}\frac{\partial R_{j,\rho}({\bm s})}{\partial s_b}
\cdots \frac{\partial R_{j,\sigma}({\bm s})}{\partial s_c}\Big) \, \ \delta^{D}({\bm x}-{\bm R}_j({\bm s})). 
\label{def-J}
\end{align}
Here $\varepsilon_{\mu\nu\cdots}$ and $\varepsilon_{ab\cdots}$ are antisymmetric tensors in the $D$-dimensional and $(D-2)$-dimensional spaces, respectively. The quantized flux for $D\ge 4$ takes the form of a $D$-dimensional antisymmetric tensor, and it is given by  a product between the delta function, $\delta^{D}({\bm x}-{\bm R}_j({\bm s}))$, and an oriented volume ${\sf g}_{j}({\bm s})$ of a parallelepiped at ${\bm R}_j({\bm s})$,  
\begin{align}
{\sf g}_{j,\lambda\rho\cdots\sigma}({\bm s}) \equiv  \sum_{a,b,\cdots,c=1,2,\cdots,D-2}\varepsilon_{ab\cdots c}
\Big(\frac{\partial R_{j,\lambda}({\bm s})}{\partial s_a}\frac{\partial R_{j,\rho}({\bm s})}{\partial s_b}
\cdots \frac{\partial R_{j,\sigma}({\bm s})}{\partial s_c}\Big). \label{Eq:oriented} 
\end{align}
Here, the parallelepiped is formed by the $(D-2)$ linearly independent vectors, $\partial {\bm R}_j({\bm s})/\partial s_{a}$ ($a=1,\cdots,D-2$), that are tangential to the vortex hyper-surface at ${\bm R}_j({\bm s})$. In this paper, we consider only vortex excitations with the $2\pi$ vorticity, as vortex excitations with higher vorticity can be expressed as a sum of vortex excitations with the $2\pi$ vorticity.

    The vortex part of the partition function is given by a sum over possible space-time configurations of multiple vortex hyper-spheres. To this end, the ${\bm b}^2$ term is decomposed by a dual field ${\bm \varphi}$,  
\begin{align}
Z_{\rm v} = \int {\cal D}{\bm b} \int {\cal D}{\bm \varphi} \exp \Big[-\int d^D{\bm x} \!\  
\big(\frac{1}{2}
\varphi^2_{\mu} + \frac{i}{\sqrt{g}} \varphi_{\mu} b_{\mu} + i\chi b_{\tau} \big)\Big].
\end{align}
The solenoidal field is divergence free; so is the auxiliary field ${\bm \varphi}$. Such a one-form field can be derived from a two-form field by the Hodge star. Namely, the auxiliary field can be represented by an antisymmetric tensor $u_{\mu\nu}=-u_{\nu\mu}$ as $\varphi_{\nu} = \partial_{\mu}u_{\mu\nu}$.   
\begin{align}
Z_{\rm v} = \int {\cal D}{\bm b} \int {\cal D}{\bm u} \exp \Big[-\int d^D{\bm r} \!\  
\big[\frac{1}{2}
(\partial_{\mu} u_{\mu\nu})(\partial_{\lambda} u_{\lambda\nu}) 
-\frac{i}{2\sqrt{g}}  u_{\mu\nu} (\partial_{\mu}b_{\nu}-\partial_{\nu}b_{\mu}) + i\chi b_{\tau} \big]\Big].
\label{Eq:Zv-1}
\end{align}
For $D=2$,  the $u_{\mu\nu}$ reduces to a scalar potential, $u_{\tau 1}= -u_{1\tau}=\phi_E$, where the first term becomes an electric part $(\nabla \phi_E)^2$ of the Maxwell action, and the coupling between $u_{\mu\nu}$ and  $v_{\mu\nu}$ becomes the electric coupling between the scalar potential and electric charge. For $D=3$, the antisymmetric tensor reduces to a magnetic vector potential $u_{\mu\nu}=\epsilon_{\mu\nu\lambda} {\sf a}_{\lambda}$, where the $(\partial  u)^2$ term becomes a magnetic part of the Maxwell action, and the $u_{\mu\nu}v_{\mu\nu}$ term becomes the magnetic coupling between the vector potential and the electric current. These suggest that the antisymmetric tensor $u_{\mu\nu}$, the $(\partial u)^2$ term, and the coupling constant $e \equiv 1/\sqrt{g}$ can be regarded as higher-dimensional $(D\ge 4)$ generalizations of the dual vector potentials, the Maxwell term, and the electric charge in electromagnetism, respectively.

      In the presence of the vortex excitations, the Berry phase term endows the partition function with a complex phase factor. Namely, the integral over the imaginary time $\tau$ can be taken in $\int d^{D}{\bm x}\,\ \partial_{\tau}\theta(\bm x)$, yielding the phase factor proportional to a $(D-1)$-dimensional volume ${\cal A}_{\tau,j}$ inside the vortex hyper-surface {\it projected} to a $\tau=$constant plane,  
\begin{align}
{\cal A}_{\tau,j} \equiv \int_{(\tau,{\bm r})\in \Gamma_j} d^{D-1}{\bm r}, 
\end{align}
with $\int d^D{\bm x} b_{\tau}({\bm x}) = 2\pi \sum_j {\cal A}_{\tau,j}$.

     As the right hand side of Eq.~(\ref{Eq:Zv-1}) is given only by the space-time coordinates of the $(D-2)$-dimensional vortex hyper-surface, the integral over ${\bm b}({\bm x})$ can be rewritten into a sum over all possible space-time configurations of vortex hyper-surfaces, 
\begin{align}
Z_{\rm v} =& \int {\cal D}{\bm u} \, \ \exp \Big[-\frac{1}{2}\int d^D{\bm x} \!\  
(\partial_{\mu} u_{\mu\nu})(\partial_{\lambda} u_{\lambda\nu}) \Big]\nonumber \\
& \Bigg\{ 1 + \sum^{\infty}_{N=1} \frac{1}{N!} \prod^N_{j=1} \bigg(\int {\cal D}^{D} {\bm X}_j \int {\cal D}^D {\bm Y}_j \int {\cal D}{\sf g}_j({\bm s}) \exp\big[E_j\big] 
\bigg) \nonumber \\
&\exp\Big[-\frac{i}{2} e  \int d^D{\bm x}\,\ u_{\mu\nu}({\bm x}) v_{\mu\nu}({\bm x}) - i2\pi \chi \sum_{j} {\cal A}_{\tau,j} \Big]\Bigg\}.  \label{Eq:Zv-1a}
\end{align}
Here $\exp[E_j]$ and ${\bm X}_j$ denote the fugacity and the center of mass of the $j$th vortex hyper-surface, respectively, while ${\bm Y}_j/2$ is a vector connecting the center of mass and one of vortex hyper-surface elements, ${\bm R}_j({\bm s}=0)$.  The integral over the oriented volume ${\sf g}_j({\bm s})$ of the parallelepiped element is taken at every ${\bm s}$ on the $j$th hyper-surface. The integral over ${\sf g}_{j}({\bm s})$ enumerates all shapes of the hyper-surface. We choose the surface element at ${\bm s}=0$ to be the closest to the center of mass, $|{\bm R}_j({\bm s})-{\bm X}_j|\ge |{\bm R}_j({\bm s}=0)-{\bm X}_j|=|{\bm Y}_j|/2$.  For typical vortex hyper-surfaces, $|{\bm Y}_j|$ can be regarded as a linear dimension of the hyper-surface size. Note that the integral over $ {\sf g}_{j,\lambda\rho\cdots\sigma}({\bm s})$ has no scaling dimension, as both ${\bm s}$ and ${\bm R}_j({\bm s})$ have the length-scale dimension in Eq.~(\ref{Eq:oriented}). For $D=2$, ${\bm X}_j\pm {\bm Y}_j/2$  are the coordinates of vortex and antivortex in the $j$th pair, respectively, so that the configurational integral for $D=2$ is given only by $\prod^N_{j=1}(\int {\cal D}^2{\bm X}_j \int {\cal D}^2{\bm Y}_j)$. In order to make the right hand side to be dimensionless, we normalize  ${\bm X}_j$ and ${\bm Y}_j$ by an ultraviolet (UV) cutoff length scale $a_0$, 
\begin{align}
\int {\cal D}^D {\bm X}_j \equiv \frac{1}{a^{D}_0} \int d^D {\bm X}_j, \,\ \int {\cal D}^{D}{\bm Y}_j \equiv \frac{1}{a^D_0} \int d^D{\bm Y}_j. 
\end{align} 
The symmetry factor of $1/N!$  in Eq.~(\ref{Eq:Zv-1a}) accounts for the multiple counting of an identical configuration that comes from $\prod^N_{j=1}(\int {\cal D}^D{\bm X}_j \int {\cal D}^D{\bm Y}_j \int {\cal D}{\sf g}_{j}({\bm s}))$.

      In the case of $D=2$, a Gaussian integration over $u_{\mu\nu}$ leads to the 2D logarithmic Coulomb interaction between vortices. For $D\ge 3$, a Gaussian integration over the dual vector potential yields the $1/|{\bm x}-{\bm x}^{\prime}|^{D-2}$ Coulomb interaction between two parallelepiped elements at ${\bm x}$ and ${\bm x}^{\prime}$ on a closed vortex hyper-surface. When ${\bm x}^{\prime}$ is further integrated over the $(D-2)$-dimensional surface, the interaction leads to a logarithmic divergence from ${\bm x}^{\prime} \simeq {\bm x}$. To regularize these logarithmic ultraviolet (UV) divergences in $D\ge 2$, we introduce a UV cutoff in the momentum-energy integral for the Maxwell action and introduce a UV cutoff for the size of the $j$th vortex hyper-surface $|{\bm Y}_j|$: 
\begin{align} 
\int d^D {\bm x} \, \ \big(\partial_{\mu}u_{\mu\nu}({\bm x})\big) 
\big(\partial_{\lambda} u_{\lambda\nu}({\bm x})\big) = & \int_{|{\bm k}|<\Lambda} \frac{d^{D}{\bm k}}{(2\pi)^D}\,\ k_{\mu}k_{\lambda} u_{\mu\nu}({\bm k}) u_{\lambda\nu}(-{\bm k}), \nonumber \\
\int d^D {\bm x} \, \ u_{\mu\nu}({\bm x}) v_{\mu\nu}({\bm x}) = & \int_{|{\bm k}|<\Lambda} \frac{d^{D}{\bm k}}{(2\pi)^D}\,\  u_{\mu\nu}({\bm k}) v_{\mu\nu}(-{\bm k}), \nonumber \\
\int {\cal D}^D {\bm Y}_j =& \int_{|{\bm Y}_j|>a_0} {\cal D}^D {\bm Y}_j.
\end{align}
The Coulomb interaction mediated by the vector potential with $|{\bm k}|>\Lambda$ is considered to be included in the vortex core energy $E_j$ [see section~\ref{secIIb}].

     The vortex excitations for $D=2$ are point defects, where the vortex fugacity is parameterized by a scalar constant: $E_j=w$.  For $D\ge 3$, the vortex excitation has a finite $(D-2)$-dimensional volume. Thereby, in addition to the constant $w$, the vortex core energy $E_j$ must also have a term that depends on the volume and local geometry of the vortex hyper-surface. In the case of the space-time isotropic system without the Berry phase term $[\chi=0]$, the additional fugacity term is proportional to the ($D-2$)-dimensional volume of the vortex hyper-surface,  i.e. 
\begin{align}
E_j = w + t \int_{\partial\Gamma_j} d^{D-2}{\bm s} \sqrt{\det \hat{\bf g}_j({\bm s})} \ \ \ \ \ \ {\rm for} \, \ \chi=0.  \label{Eq:fugacity-1}
\end{align}
Here, $(D-2)\times(D-2)$ real symmetric matrix $\hat{\bf g}_j({\bm s})$ is the first fundamental form, 
\begin{align}
[\hat{\bf g}_{j}({\bm s})]_{ab} \equiv \sum_{\mu=\tau,1,2,\cdots,D-1} \frac{\partial R_{j,\mu}({\bm s})}{\partial s_a} \frac{\partial R_{j,\mu}({\bm s})}{\partial s_b}, \label{Eq:1st-f}
\end{align}
for $a,b=1,\cdots,D-2$. Note that the volume of a hyper-surface is generally given by the ${\bm s}$-integral of the square-root of the determinant of its first fundamental form~\cite{goldbart2009,frankel2004}. In the case of the space-time anisotropic case with finite $\chi$, $E_j$ is parameterized by a vortex core energy density, which is a function of time component $\hat{\bf g}_{j,\tau}$ and spatial component $\hat{\bf g}_{j,{\bm r}}$ of the metric tensor, 
\begin{align}
E_j  \equiv w + \int_{\partial \Gamma_j} d^{D-2}{\bm s} \,\ t\big(\hat{\bf g}_{j,\tau}({\bm s}), \hat{\bf g}_{j,\bm r}({\bm s})\big), \ \ \ \ {\rm for} \,\ \chi\ne 0. 
\label{Eq:fugacity-2}
\end{align}
These metric tensors take the $(D-2)\times (D-2)$ real symmetric matrices,  
\begin{align}
[\hat{\bf g}_{j,\tau}({\bm s})]_{ab} \equiv \frac{\partial R_{j,\tau}({\bm s})}{\partial s_a} \frac{\partial R_{j,\tau}({\bm s})}{\partial s_b}, \ \ \ 
[\hat{\bf g}_{j,{\bm r}}({\bm s})]_{ab} \equiv \sum_{\mu=1,2,\cdots,D-1} \frac{\partial R_{j,\mu}({\bm s})}{\partial s_a} \frac{\partial R_{j,\mu}({\bm s})}{\partial s_b}, \label{Eq:1st-f-2}
\end{align}
which are invariant under arbitrary $O(D-1)$ rotation of the spatial coordinate ${\bm r}$ [$(\tau,{\bm r}) \rightarrow (\tau,\tilde{\bm r})$ with $\tilde{\bm r}=O(D-1)\!\ \cdot {\bm r}$]  as well as under the mirror with respect to the time $[(\tau,{\bm r})\rightarrow (-\tau,{\bm r})]$. Thus, the fugacity parameter $E_j$ is also invariant under the spatial rotation and/or the mirror operation of the vortex hyper-surface. In Sec.~\ref{secIIb}, we will justify Eqs.~(\ref{Eq:fugacity-1},\ref{Eq:fugacity-2}) by deriving the fugacity renormalization for these cases.   

     In the presence of the Berry phase term $\chi$, the electromagnetic constitutive constant in the Maxwell action acquires the space-time anisotropy through the screening effect of smaller vortex excitations [see Sec.\ref{secIIa}]. Thus, we include the space-time anisotropy parameter $\gamma_{\tau}$ into the Maxwell action,
\begin{align}
Z_{\rm v} =& \int {\cal D}{\bm u} \, \  \exp \bigg[-\frac{1}{2}\int_{|\partial u|<\Lambda|u|} d^D{\bm x} \!\  \Big\{ {\gamma}_{\tau} 
(\partial_{\mu} u_{\mu\tau})(\partial_{\lambda} u_{\lambda\tau}) +  \sum^{D-1}_{\nu=1}
(\partial_{\mu} u_{\mu\nu})(\partial_{\lambda} u_{\lambda\nu})\Big\}\bigg]\nonumber \\
& \Bigg\{ 1 + \sum^{\infty}_{N=1} \frac{1}{N!} \prod^N_{j=1} \bigg(\int {\cal D}^D {\bm X}_j \int_{|{\bm Y}_j|>a_0} {\cal D}^D {\bm Y}_j \int {\cal D}{\sf g}_j({\bm s}) \, \  \exp\big[ E_{j}\big]\bigg) \nonumber \\
&\exp\Big[-\frac{i}{2}  e \int_{|\partial u|<\Lambda |u|} d^D{\bm x}\,\ u_{\mu\nu}({\bm x}) v_{\mu\nu}({\bm x}) - i2\pi \chi \sum_{j} {\cal A}_{\tau,j} \Big]\Bigg\}, \label{Eq:Zv-2}
\end{align}
with 
\begin{align}
\int_{|\partial u|<\Lambda|u|} d^D{\bm x} \!\ (\partial_{\mu} u_{\mu\nu}({\bm x}))(\partial_{\lambda} u_{\lambda\nu}({\bm x})) \equiv & \int_{|{\bm k}|<\Lambda} \frac{d^D{\bm k}}{(2\pi)^D} \!\ k_{\mu} u_{\mu\nu}({\bm k}) \!\ k_{\lambda}u_{\lambda\nu}(-{\bm k}), \nonumber \\
\int_{|\partial u|<\Lambda|u|} d^D{\bm x} \!\  u_{\mu\nu}({\bm x}) v_{\mu\nu}({\bm x})  \equiv & \int_{|{\bm k}|<\Lambda} \frac{d^D{\bm k}}{(2\pi)^D} \!\ u_{\mu\nu}({\bm k}) \!\ v_{\mu\nu}(-{\bm k}).   
\end{align}
\section{\label{secII}Renormalization Group Theory}
     The partition function Eq.~(\ref{Eq:Zv-2}) can be studied using renormalization group (RG) analyses, where  short-distance degrees of freedom are recursively integrated out. Since the tree-level scaling dimension of the electric constant $e$ is $(D-2)/2$, and the RG equation for $\chi=0$ has a stable fixed point at $e^2 \simeq {\cal O}(\epsilon)$ for $D=2+\epsilon$ [see section~\ref{secIIIa}], we regard $e$ as a small quantity and treat the coupling constant perturbatively in the renormalization group analysis. The perturbative treatment is justified for $D\gtrsim 2$.     

     \subsection{\label{secIIa}one-loop renormalization to the constitutive constants and Berry phase term}
     In Eq.~(\ref{Eq:Zv-2}), the UV cutoff length appears in (i) the integral over the vortex surface size $|{\bm Y}_j|$, and in (ii) the momentum integral in the Maxwell term. Following the previous work~\cite{kosterlitz1973,kosterlitz1974,williams1987,shenoy1989,shindou2025},  we first decompose the configuration sum over each vortex surface into a smallest vortex-surface case and larger vortex-surface case, 
\begin{align}
     \int_{|{\bm Y}_j|> a_0} {\cal D}^D {\bm Y}_j = \int_{|{\bm Y}_j|>a_0 b} {\cal D}^D {\bm Y}_j + \int_{a_0 b > |{\bm Y}_j|>a_0 } {\cal D}^D {\bm Y}_j. \label{Eq:Zv-3}
\end{align}
Here $b$ stands for a renormalization scale parameter with $\ln b \gtrsim 1$. In the hyper-cubic lattice with the lattice constant $a_0$, the vortex surface in the smallest vortex surface case becomes a $(D-2)$-dimensional unit hyper-cubic plaquette. To mimic such a hyper-cubic plaquette in the continuum model, we consider a symmetric $(D-2)$-sphere with diameter $a_0$ as the smallest surface. Suppose that ${\bm e}_0$ is a unit vector normal to a plane in which the $(D-2)$-sphere lies. Then, the configuration sum of the smallest vortex surface is given by an integral of its center of mass, and an integral of $\bm e_0$ over the $(D-1)$-sphere $S_{D-1}$,  
\begin{align}
\int {\cal D}^D {\bm X}_ j \int_{a_0<|{\bm Y}_j|<a_0 b} {\cal D}^D {\bm Y}_j \int {\cal D}{\sf g}_{j}({\bm s})  \simeq   a^{-2D}_0 \int d^D {\bm X}_j \int^{a_0 b}_{a_0} Y^{D-1}_j dY_j \int_{S_{D-1}} {\cal D}^{D-1} {\bm e}_0. 
\label{Eq:Zv-4}
\end{align}
By substituting Eq.~(\ref{Eq:Zv-3},\ref{Eq:Zv-4}) into Eq.~(\ref{Eq:Zv-2}) and keeping only the first order in $\ln b$, we obtain 
\begin{align}
Z_{\rm v} =& \int {\cal D}{\bm u} \, \  \exp\bigg[-\frac{1}{2}\int_{|\partial u|<\Lambda|u|} d^D{\bm x} \!\  \Big\{ \gamma_{\tau} 
(\partial_{\mu} u_{\mu\tau})(\partial_{\lambda} u_{\lambda\tau}) + \sum^{D-1}_{\nu=1}
(\partial_{\mu} u_{\mu\nu})(\partial_{\lambda} u_{\lambda\nu})\Big\}\bigg]\nonumber \\
& \Bigg\{ 1 + \sum^{\infty}_{N=1} \frac{1}{N!} \prod^N_{j=1} \bigg(\int {\cal D}^{D} {\bm X}_j \int_{|{\bm Y}_j|>a_0 b} {\cal D}^D {\bm Y}_j \int {\cal D}{\sf g}_j({\bm s}) \, \  \exp\big[ E_{j}\big]\bigg) \nonumber \\
&\exp\Big[-\frac{i}{2} e \int_{|\partial u|<\Lambda |u|} d^D{\bm x}\,\ u_{\mu\nu}({\bm x}) v_{\mu\nu}({\bm x}) - i2\pi \chi \sum_{j} {\cal A}_{\tau,j}  \Big] \Bigg\} + \ln b\,\  Z^{\prime}_{\rm v}. \label{Eq:Zv-5}
\end{align}
Here $Z^{\prime}_{\rm v}$ contributes to the one-loop renormalization to $Z_{\rm v}$, and it is given by  
\begin{align}
Z^{\prime}_{\rm v} &= \int {\cal D}{\bm u} \, \  \exp\bigg[-\frac{1}{2}\int_{|\partial u|<\Lambda|u|} d^D{\bm x} \!\  \Big\{ \gamma_{\tau} 
(\partial_{\mu} u_{\mu\tau})(\partial_{\lambda} u_{\lambda\tau}) + \sum^{D-1}_{\nu=1}
(\partial_{\mu} u_{\mu\nu})(\partial_{\lambda} u_{\lambda\nu})\Big\}\bigg]\nonumber \\
&  \int {\cal D}^D{\bm X}_0 \int_{S_{D-1}} {\cal D}^{D-1} {\bm e}_0   \,\ \exp\big[E_0-i2\pi \chi {\cal A}_{\tau,0}\big]\exp\Big[-\frac{i e}{2}  \int_{|\partial u|<\Lambda|u|} d^D{\bm x}\,\ u_{\mu\nu}({\bm x}) v^{>}_{\mu\nu}({\bm x})\Big]\nonumber \\
&\Bigg\{ 1 + \sum^{\infty}_{N=2} \frac{1}{(N-1)!} \prod^{N-1}_{j=1} \bigg(\int {\cal D}^{D} {\bm X}_j \int_{|{\bm Y}_j|>a_0 b} {\cal D}^D {\bm Y}_j \int {\cal D}{\sf g}_j({\bm s}) \, \  \exp\big[ E_{j}\big]\bigg) \nonumber \\
&\exp\Big[-\frac{i}{2} e \int_{|\partial u|<\Lambda|u|} d^D{\bm x}\,\ u_{\mu\nu}({\bm x}) v^{<}_{\mu\nu}({\bm x}) - i2\pi \chi \sum^{N-1}_{j=1} {\cal A}_{\tau,j}  \Big] \Bigg\}. \label{Eq:Zv-6}
\end{align}
${\bm X}_0$ and $E_0$ are the center-of-mass coordinate and fugacity parameter of the smallest $(D-2)$-sphere, respectively,
\begin{align}
E_0 =  \begin{cases}
w & {\rm for} \,\ \,\ D=2, \\
w + \int_{|{\bm R}_0({\bm s})|=a_0/2} d^{D-2}{\bm s}\,\ t\big({\bf g}_{0,\tau}({\bm s}),{\bf g}_{0,{\bm r}}({\bm s})\big) & {\rm for} \,\ \,\  D>2. 
\end{cases}
\label{Eq:E0} 
\end{align}
${\cal A}_{\tau,0}$ is the $(D-1)$-dimensional volume inside the smallest sphere projected onto a $\tau=$constant plane. In Eq.~(\ref{Eq:Zv-6}), the vortex field $v_{\mu\nu}({\bm x})$ is decomposed into a vortex field $v^{>}_{\mu\nu}({\bm x})$ of the smallest vortex surface with $a_0<|{\bm Y}_0|<a_0 b$ and a vortex field $v^{<}_{\mu\nu}({\bm x})$ of the others with $a_0b<|{\bm Y}_j|$. For $D=2$, they are given by 
\begin{align} 
v^{>}_{\tau 1}({\bm x}) &= 2\pi \Big(\delta^2\big({\bm x}-{\bm X}_0-\frac{a_0}{2} {\bm e}_1\big)
- \delta^2\big({\bm x}-{\bm X}_0+\frac{a_0}{2} {\bm e}_1\big)\Big), \label{Eq:def-J2-2D} \\
v^{<}_{\tau 1}({\bm x}) &= 2\pi \sum^{N-1}_{j=1}\Big(\delta^2\big({\bm x}-{\bm X}_j-\frac{{\bm Y}_j}{2}\big)
- \delta^2\big({\bm x}-{\bm X}_j+\frac{{\bm Y}_j}{2}\big) \Big), \label{Eq:def-J3-2D}
\end{align}
respectively. Here, ${\bm e_1}$ in Eq.~({\ref{Eq:def-J2-2D}}) is the unit vector perpendicular to ${\bm e}_0$, and vortex and antivortex in the smallest vortex-antivortex pair are coordinated by ${\bm X}_0 \pm \frac{a_0}{2} {\bm e}_1$, respectively [$\{{\bm e}_0,{\bm e}_1\}$ forms the right-handed orthonormal basis in the 2-dimensional system]. For $D>2$,  $v^{>}_{\mu\nu}({\bm x})$ and $v^{<}_{\mu\nu}({\bm x})$ are given by, 
\begin{align}
v^{>}_{\mu\nu}({\bm x})& \equiv  
\frac{2\pi \varepsilon_{\mu\nu\lambda\rho\cdots\sigma} }{(D-2)!} \,\
\int_{|{\bm R}_0({\bm s})|=a_0/2} d^{D-2}{\bm s} \,\ \varepsilon_{a\cdots b}
\Big(\frac{\partial R_{0,\lambda}({\bm s})}{\partial s_a}
\cdots \frac{\partial R_{0,\sigma}({\bm s})}{\partial s_b}\Big) \, \ \delta^{D}({\bm x}-{\bm X}_0-{\bm R}_0({\bm s})), \label{def-J2} \\
v^{<}_{\mu\nu}({\bm x})& \equiv \frac{2\pi \varepsilon_{\mu\nu\lambda\rho\cdots\sigma} }{(D-2)!} \,\
\sum^{N-1}_{j=1}\int_{\partial \Gamma_j} d^{D-2}{\bm s} \,\ \varepsilon_{a\cdots b}
\Big(\frac{\partial R_{j,\lambda}({\bm s})}{\partial s_a}
\cdots \frac{\partial R_{j,\sigma}({\bm s})}{\partial s_b}\Big) \, \ \delta^{D}({\bm x}-{\bm R}_j({\bm s})), 
\label{def-J3}
\end{align}
where $\partial \Gamma_j$ is a larger vortex hyper-surface with $|{\bm Y}_j|>a_0 b$.

    Suppose that the unit normal vector ${\bm e}_0$ has an angle $\theta$ with respect to the time $\tau$ direction,
\begin{align}
{\bm e}_0 \equiv (\cos\theta,\sin\theta\cos\theta^{\prime},\cdots,\sin\theta\sin\theta^{\prime}\cdots\sin\theta^{\prime\prime}). \label{Eq:e0-angles}
\end{align}
Then, the fugacity parameter $E_0$ of the smallest vortex sphere in Eq.~(\ref{Eq:E0}) is a function only of $\theta$, and it is symmetric around $\theta=\pi/2$: $E_0=E_0(\theta)=E_0(\pi-\theta)$. This is because both temporal and spatial components of the metric tensor, $\hat{\bf g}_{j,r}$ and $\hat{\bf g}_{j,\tau}$ are generally invariant under the spatial rotations around the $\tau$ axis as well as under the inversion of $\tau$. The projected volume ${\cal A}_{\tau,0}$ inside the smallest $(D-2)$-sphere is also given by a function of $\theta$, while it is odd under $\theta \rightarrow \pi-\theta$, 
\begin{align}
{\cal A}_{\tau,0} = \cos\theta \int^{\frac{a_0}{2}}_{0} r^{D-2} dr \int^{\pi}_{0} 
\big(\sin\lambda_1\big)^{D-3}d\lambda_1 \cdots \int^{2\pi}_{0} d\lambda_{D-2} =
2A \Big(\frac{a_0}{2}\Big)^{D-1} \cos\theta,  \label{Eq:projected-volume}
\end{align}
with 
\begin{align}
A \equiv \frac{1}{D-1} \frac{\pi^{\frac{D-1}{2}}}{\Gamma\big(\frac{D-1}{2}\big)} = \frac{1}{2} \frac{\pi^{\frac{D-1}{2}}}{\Gamma\big(\frac{D+1}{2}\big)}.  \label{Eq:A-coefficient}
\end{align}
Note that Eq.~(\ref{Eq:projected-volume}) is also valid for $D=2$.

   Now that the vector potential $u_{\mu\nu}({\bm x})$ changes slowly in space and time on the scale of $a_0$, we use a gradient expansion to evaluate $\int_{|\partial u|<\Lambda |u|} d^D{\bm x}\,\ u_{\mu\nu}({\bm x}) v^{>}_{\mu\nu}({\bm x})$. In the following, we evaluate the expansion for $D>2$ and $D=2$ separately. In the end, we observe that the final results [Eqs.~(\ref{Eq:rg3},\ref{Eq:rg4},\ref{Eq:rg5},\ref{Eq:rg6}, \ref{Eq:Zv-8})] for general $D$ are consistent with the results for the $D=2$ case. 

\subsubsection{\label{secIIa1}gradient expansion for general $D$}
   To exercise the expansion for general $D>2$, we first introduce the spherical coordinate system for the symmetric $(D-2)$-sphere. Suppose that the unit vector ${\bm e}_0$ is normal to the plane where the sphere lies, and $\{{\bm e}_0,{\bm e}_1,\cdots,{\bm e}_{D-1}\}$ forms the right-handed orthonormal basis in the $D$-dimensional system. The parallelepiped element of the symmetric sphere can be parameterized by $(D-2)$ angle variables, ${\bm \lambda} \equiv (\lambda_1, \lambda_2, \cdots,\lambda_{D-3},\lambda_{D-2})$, 
\begin{align}
{\bm R}_0({\bm \lambda}) =& \frac{a_0}{2}\Big(\cos\lambda_1 {\bm e}_1 + \sin\lambda_1 \cos\lambda_2 {\bm e}_2 + 
\sin\lambda_1 \sin\lambda_2 \cos\lambda_3 {\bm e}_3 +\cdots \nonumber \\
& + \sin\lambda_1 \cdots \sin\lambda_{D-3} \cos\lambda_{D-2} {\bm e}_{D-2} + \sin\lambda_1 \cdots \sin\lambda_{D-3} \sin\lambda_{D-2} {\bm e}_{D-1}\Big). \label{Eq:scs}
\end{align}
We set the integral domain as $\lambda_j \in [0,\pi]$ $(j=1,\cdots,D-3)$, and  $\lambda_{D-2} \in [0,2\pi]$.  By replacing ${\bm s}$ in Eq.~(\ref{def-J2}) by $\frac{a_0}{2}{\bm \lambda}$, we obtain the vortex field $v^{>}_{\mu\nu}({\bm r})$ from the smallest sphere as follows, 
\begin{align}
v^{>}_{\mu\nu}({\bm x}) = & 2\pi \Big(\frac{a_0}{2}\Big)^{D-3} \int^{\pi}_{0} (\sin\lambda_1)^{D-3}  d\lambda_1 
\int^{\pi}_{0} (\sin \lambda_2)^{D-4} d\lambda_2 \cdots \int^{\pi}_{0} \sin\lambda_{D-3} d\lambda_{D-3} \nonumber \\
& \int^{2\pi}_{0}  d\lambda_{D-2} \big({\bm e}_{0,\mu} {\bm R}_{0,\nu}({\bm \lambda}) - {\bm e}_{0,\nu} {\bm R}_{0,\mu}({\bm \lambda}) \big) \,\ \delta^{D}({\bm x}-{\bm X}_0-{\bm R}_0({\bm \lambda})). 
\end{align}
The integral region of ${\bm \lambda}$ can be decomposed into a pair of two domains, which are transformed into each other by $\lambda_j \rightarrow \pi-\lambda_j$ $(j=1,\cdots,D-3)$ and $\lambda_{D-2}\rightarrow \lambda_{D-2}+\pi$. Namely, we can rewrite also this in the following way, 
\begin{align}
v^{>}_{\mu\nu}({\bm x}) = & 2\pi   \Big(\frac{a_0}{2}\Big)^{D-3} \int^{\pi}_{0} (\sin\lambda_1)^{D-3}  d\lambda_1 
\int^{\pi}_{0} (\sin \lambda_2)^{D-4} d\lambda_2 \cdots \int^{\pi}_{0} \sin\lambda_{D-3} d\lambda_{D-3} \int^{\pi}_{0}  d\lambda_{D-2} \nonumber \\
& \big({\bm e}_{0,\mu} {\bm R}_{0,\nu}({\bm \lambda}) - {\bm e}_{0,\nu} {\bm R}_{0,\mu}({\bm \lambda}) \big) \,\ \Big( \delta^{D}({\bm x}-{\bm X}_0-{\bm R}_0({\bm \lambda})) 
- \delta^{D}({\bm x}-{\bm X}_0+{\bm R}_0({\bm \lambda}))\Big). \label{Eq:gradient-expansion} 
\end{align}
Eq.~(\ref{Eq:gradient-expansion}) facilitates an expansion of $\int_{|\partial u|<|u|} d^D{\bm x}\,\ u_{\mu\nu}({\bm x}) v^{>}_{\mu\nu}({\bm x})$ in the power of the UV length scale $a_0$,
\begin{align}
\int_{|\partial u|<\Lambda|u|} d^D{\bm x}\,\ u_{\mu\nu}({\bm x}) v^{>}_{\mu\nu}({\bm x}) &= 2\pi  \Big(\frac{a_0}{2}\Big)^{D-3} \int^{\pi}_{0} (\sin\lambda_1)^{D-3}  d\lambda_1 
 \cdots  \int^{\pi}_{0}  d\lambda_{D-2} \nonumber \\
& \big({\bm e}_{0,\mu} {\bm R}_{0,\nu}({\bm \lambda}) - {\bm e}_{0,\nu} {\bm R}_{0,\mu}({\bm \lambda}) \big) \,\ \Big( u_{\mu\nu}({\bm X}_0+{\bm R}_0({\bm \lambda})) 
- u_{\mu\nu}({\bm X}_0-{\bm R}_0({\bm \lambda}))\Big), \nonumber \\
&= 4\pi \Big(\frac{a_0}{2}\Big)^{D-3} \int^{\pi}_{0} (\sin\lambda_1)^{D-3}  d\lambda_1 
 \cdots  \int^{\pi}_{0}  d\lambda_{D-2} \nonumber \\ 
& \big({\bm e}_{0,\mu} {\bm R}_{0,\nu}({\bm \lambda}) - {\bm e}_{0,\nu} {\bm R}_{0,\mu}({\bm   \lambda}) \big) \!\ \Big({\bm R}_{0,\phi}({\bm \lambda}) \! \ \partial_{\phi} u_{\mu\nu}({\bm X}_0)  + {\cal O}(\partial^3 u)
\Big) \nonumber \\
& = 8\pi   \Big(\frac{a_0}{2}\Big)^{D-1} A \!\ {\bm e}_{0,\mu} \partial_{\nu} u_{\mu\nu}({\bm X}_0) + {\cal O}(\partial^3 u). \label{Eq:gradient-expansion2}
\end{align}
From the second to third lines, we used  
\begin{align}
\int^{\pi}_{0} (\sin\lambda_1)^{D-3}  d\lambda_1 \cdots \int^{\pi}_{0}  d\lambda_{D-2}\!\ {\bm R}_{0,\nu}({\bm \lambda}) 
{\bm R}_{0,\phi}({\bm \lambda}) = A \Big(\frac{a_0}{2}\Big)^{2} (\delta_{\nu,\phi} - {\bm e}_{0,\nu}{\bm e}_{0,\phi}),
\end{align}
with Eq.~(\ref{Eq:A-coefficient}). Substituting Eq.~(\ref{Eq:gradient-expansion2}) into Eq.~(\ref{Eq:Zv-6}), and expanding the smallest vortex hyper-sphere term in the power of the electric coupling constant $e$, we obtain 
\begin{align}
Z^{\prime}_{\rm v} =& \int {\cal D}{\bm u} \, \  \exp \bigg[-\frac{1}{2}\int_{|\partial u|<\Lambda|u|} d^D{\bm x} \!\  \Big\{ \gamma_{\tau} 
(\partial_{\mu} u_{\mu\tau})(\partial_{\lambda} u_{\lambda\tau}) + \sum^{D-1}_{\nu=1}
(\partial_{\mu} u_{\mu\nu})(\partial_{\lambda} u_{\lambda\nu})\Big\}\bigg]\nonumber \\
&  \int {\cal D}^D {\bm X}_0 \int_{S_{D-1}} {\cal D}^{D-1} {\bm e}_0   \,\ \exp\big[E_0(\theta)-i4\pi \Big(\frac{a_0}{2}\Big)^{D-1}  A \chi \cos \theta \big]\nonumber \\
&\Big[ 1 - 4\pi i e   \Big(\frac{a_0}{2}\Big)^{D-1} A \!\ {\bm e}_{0,\mu} \partial_{\nu} u_{\mu\nu}({\bm X}_0)  - 8\pi^2 e^2 \Big(\frac{a_0}{2}\Big)^{2(D-1)} A^2 \big({\bm e}_{0,\mu} \partial_{\nu} u_{\mu\nu}({\bm X}_0) \big) \big({\bm e}_{0,\phi} \partial_{\lambda} u_{\phi\lambda}({\bm X}_0) \big)  + {\cal O}(\partial^3 u, \partial u \partial^3 u)\Big]\nonumber \\
&\Bigg\{ 1 + \sum^{\infty}_{N=1} \frac{1}{N!} \prod^{N}_{j=1} \bigg(\int {\cal D}^D {\bm X}_j \int_{|{\bm Y}_j|>a_0 b}{\cal D}^D {\bm Y}_j \int {\cal D}{\sf g}_j({\bm s}) \, \  \exp\big[ E_{j}\big]\bigg) \nonumber \\
&\exp\Big[-\frac{i}{2} e \int_{|\partial u|<\Lambda|u|} d^D{\bm x}\,\ u_{\mu\nu}({\bm x}) v^{<}_{\mu\nu}({\bm x}) - i2\pi \chi \sum^{N}_{j=1} {\cal A}_{\tau,j}  \Big] \Bigg\}. \label{Eq:Zv-7}
\end{align}
From Eq.~(\ref{Eq:Zv-6}) to Eq.~(\ref{Eq:Zv-7}), we have changed $N$ by one, $N_{\rm new}=N_{\rm old}-1$, and replaced Eq.(~\ref{def-J3}) by  
\begin{align}
v^{<}_{\mu\nu}({\bm x})&  = \frac{2\pi \varepsilon_{\mu\nu\lambda\rho\cdots\sigma} }{(D-2)!} \,\
\sum^{N}_{j=1}\int_{\partial \Gamma_j} d^{D-2}{\bm s} \,\ \varepsilon_{a\cdots b}
\Big(\frac{\partial R_{j,\lambda}({\bm s})}{\partial s_a}
\cdots \frac{\partial R_{j,\sigma}({\bm s})}{\partial s_b}\Big) \, \ \delta^{D}({\bm x}-{\bm R}_j({\bm s})). \nonumber 
\end{align}
From Eq.~(\ref{Eq:e0-angles}), the integral over ${\bm e}_0$ is given by, 
\begin{align}
\int_{S_{D-1}} {\cal D}^{D-1}{\bm e}_0 \equiv \int^{\pi}_{0} (\sin\theta)^{D-2} d\theta \int^{\pi}_{0} (\sin \theta^{\prime})^{D-3} d\theta^{\prime} \cdots \int^{2\pi}_{0} d\theta^{\prime\prime}.  \label{Eq:e_0-integral} 
\end{align}
The integral over ${\bm e}_0$ is calculated up to the second order in $e$, 
\begin{align}
&\int {\cal D}^D{\bm X}_0 \int {\cal D}^{D-1}{\bm e}_0 \!\ e^{E_0(\theta) - i4\pi \chi  \big(\frac{a_0}{2}\big)^{D-1} A \cos\theta} = \frac{V}{a^D_0} \delta f, \label{Eq:rgf0} \\
&\int {\cal D}^D{\bm X}_0 \int_{S_{D-1}} {\cal D}^{D-1}{\bm e}_0 \!\  e^{E_0(\theta) - i4\pi \chi \big(\frac{a_0}{2}\big)^{D-1}  A \cos\theta} (-4\pi i e)\!\  \Big(\frac{a_0}{2}\Big)^{D-1} A \,\ {\bm e}_{0,\mu} \partial_{\nu}u_{\mu\nu}({\bm X}_0)  = -\delta \chi \int d^D{\bm x} \,\ \partial_{\nu} u_{\tau\nu}({\bm x}), \label{Eq:rg1} \\
&\int {\cal D}^D{\bm X}_0 \int {\cal D}^{D-1}{\bm e}_0 \!\ e^{E_0(\theta) - i4\pi \chi  \big(\frac{a_0}{2}\big)^{D-1} A \cos\theta}\!\ (8\pi^2 e^2)  \Big(\frac{a_0}{2}\Big)^{2(D-1)} A^2 \,\ {\bm e}_{0,\mu} {\bm e}_{0,\phi} \partial_{\nu}u_{\mu\nu}({\bm X}_0) \partial_{\lambda}u_{\phi\lambda}({\bm X}_0) \nonumber \\ 
& = \delta \gamma_{\tau} \int d^D{\bm x}\,\ \partial_{\nu} u_{\nu\tau}({\bm x})\partial_{\lambda} u_{\lambda\tau}({\bm x}) +  \delta \gamma_{\bm r} \int d^D{\bm x}\,\ \sum_{\mu=1,\cdots,D-1} \partial_{\nu} u_{\nu\mu}({\bm x})\partial_{\lambda} u_{\lambda\mu}({\bm x}), \label{Eq:rg2} 
\end{align}
with 
\begin{align}
\delta f & \equiv \frac{D-1}{2} A\int^{\pi}_{0} d\theta \,\ e^{E_0(\theta)} 
\cos \Big[4\pi \chi  \big(\frac{a_0}{2}\big)^{D-1} A\cos\theta\Big], \label{Eq:rgf0a}  \\
\delta \chi &\equiv 4\pi e \frac{a^{-1}_0}{2^{D-1}} A \int_{S_{D-1}} {\cal D}^{D-1} {\bm e}_0 \,\ e^{E_0(\theta)} 
\sin \Big[4\pi \chi  \big(\frac{a_0}{2}\big)^{D-1} A\cos\theta\Big] \cos\theta,  \label{Eq:rg3} \\
\delta \gamma_{\tau} &\equiv 8\pi^2 e^2  \frac{a^{D-2}_0}{2^{2D-2}} A^2 \int_{S_{D-1}} {\cal D}^{D-1} {\bm e}_0 \,\ e^{E_0(\theta)} 
\cos \Big[4\pi \chi  \big(\frac{a_0}{2}\big)^{D-1} A\cos\theta\Big] \cos^2\theta, \label{Eq:rg4} \\
\delta \gamma_{\bm r} &\equiv 8\pi^2 e^2 \frac{a^{D-2}_0}{2^{2D-2}} A^2 \frac{1}{D-1} 
\int_{S_{D-1}} {\cal D}^{D-1} {\bm e}_0 \,\ e^{E_0(\theta)} 
\cos \Big[4\pi \chi \big(\frac{a_0}{2}\big)^{D-1} A\cos\theta\Big] \sin^2\theta. \label{Eq:rg5} 
\end{align}
Here $E_{0}(\theta)$ is given by Eq.~(\ref{Eq:E0}) for $D\ge 3$, 
\begin{align}
E_0(\theta) = 
w + \int_{|{\bm R}_0({\bm s})|=a_0/2} d^{D-2}{\bm s}\,\ t\big({\bf g}_{0,\tau}({\bm s}),{\bf g}_{0,{\bm r}}({\bm s})\big). \nonumber 
\end{align}
By substituting Eqs. (\ref{Eq:rg1},\ref{Eq:rg2}) into Eq. (\ref{Eq:Zv-6},\ref{Eq:Zv-5}), we finally obtain the one-loop  renormalization of the electromagnetic constitutive constants, 
\begin{align}
Z_{\rm v} =&  e^{V\delta f} \int {\cal D}{\bm u} \, \  \exp \bigg[-\frac{1}{2}\int_{|\partial u|<\Lambda|u|} d^D{\bm x} \!\  \Big\{ \overline{\gamma}_{\tau} 
(\partial_{\mu} u_{\mu\tau})(\partial_{\lambda} u_{\lambda\tau}) 
+ \overline{\gamma}_{\bm r}\sum^{D-1}_{\nu=1}
(\partial_{\mu} u_{\mu\nu})(\partial_{\lambda} u_{\lambda\nu}) - 2   \delta \chi \ln b \!\ \partial_{\mu} u_{\mu\tau} \Big\}\bigg] \nonumber \\
&\Bigg\{ 1 + \sum^{\infty}_{N=1} \frac{1}{N!} \prod^{N}_{j=1} \bigg(\int {\cal D}^{D} {\bm X}_j \int_{|{\bm Y}_j|>a_0 b} {\cal D}^D {\bm Y}_j \int {\cal D}{\sf g}_j({\bm s}) \, \  e^{E_{j}}\bigg) 
e^{-\frac{ie}{2}  \int_{|\partial u|<\Lambda|u|} d^D{\bm x}\,\ u_{\mu\nu}({\bm x}) v_{\mu\nu}({\bm x}) - i2\pi \chi \sum^{N}_{j=1} {\cal A}_{\tau,j} } \Bigg\}, \label{Eq:Zv-8}
\end{align}
with 
\begin{align}
\overline{\gamma}_{\tau} = \gamma_{\tau} + 2\delta \gamma_{\tau}  \ln b, \,\ \overline{\gamma}_{\bm r} = 1 +2 \delta \gamma_{\bm r}    \ln b. \label{Eq:rg6}
\end{align}
$\delta f$ contributes to an inhomogeneous scaling relation of the free energy density. As we are primarily interested in renormalization equations for the coupling constants, we will not delve into the free energy renormalization in this paper. 

\subsubsection{\label{secIIa2}gradient expansion for $D=2$}
    For $D=2$, Eq.~(\ref{Eq:def-J2-2D}) directly yields the gradient expansion of $\int_{|\partial u|<\Lambda |u|} d^D{\bm x}\,\ u_{\mu\nu}({\bm x}) v^{>}_{\mu\nu}({\bm x})$, 
\begin{align}
\int_{|\partial u|<\Lambda |u|} d^D{\bm x}\,\ u_{\mu\nu}({\bm x}) v^{>}_{\mu\nu}({\bm x}) 
&= 4\pi \int d^2{\bm x}\,\ u_{\tau 1}({\bm x}) \Big(\delta^2\big({\bm x}-{\bm X}_0 - \frac{a_0}{2} {\bm e}_1\big) - \delta^2 \big({\bm x}-{\bm X}_0 + \frac{a_0}{2} {\bm e}_1\big) \Big), \nonumber \\
&= 4\pi a_0 {\bm e}_{1,\lambda} \partial_{\lambda} u_{\tau 1}({\bm X}_0) + {\cal O}(\partial^3 u) =  4\pi a_0 {\bm e}_{0,\mu}\partial_{\nu} u_{\mu\nu} 
+ {\cal O}(\partial^3 u). \label{Eq:gradient-2D-1}
\end{align}
In fact, this is consistent with Eq.~(\ref{Eq:gradient-expansion2}): $A$ in Eq.~(\ref{Eq:A-coefficient}) is equal to one for $D=2$. Thus, Eqs.~(\ref{Eq:rg3},\ref{Eq:rg4},\ref{Eq:rg5}, \ref{Eq:Zv-8}, \ref{Eq:rg6}) are also valid for $D=2$, where $E_0(\theta)=w$ and $\int D{\bm e}_0 = \int^{2\pi}_{0} d\theta$.   

\subsubsection{\label{secIIa3}renormalization to Berry phase term}
    $\delta \chi \partial_{\mu} u_{\mu\tau}$ in Eq.~(\ref{Eq:Zv-8}) can be absorbed into the antisymmetric integral variable $u_{\mu\tau}=-u_{\tau\mu}$ by the following gauge transformation,
    \begin{align}
\partial_{\mu}u_{\mu\tau}-\frac{\delta \chi  \ln b}{\overline{\gamma}_{\tau}} &\rightarrow
\partial_{\mu}u_{\mu\tau}, \nonumber \\ 
\partial_{\mu}u_{\mu\lambda}&\rightarrow  \partial_{\mu}u_{\mu\lambda}, 
\end{align}
with $\lambda=1,\cdots,D-1$. Here, the summation over $\mu=\tau ,1,\cdots,D-1$ is assumed. For example,  one can consider the following variable changes, 
\begin{align}
    u_{\mu\tau}({\bm x}) \rightarrow u_{\mu\tau}({\bm x}) + \frac{\delta \chi \ln b}{\overline{\gamma}_{\bm r}} \delta_{\mu 1} r_1, \ \ 
    u_{\tau\mu}({\bm x}) \rightarrow u_{\tau\mu}({\bm x}) - \frac{\delta \chi \ln b}{\overline{\gamma}_{\bm r}} \delta_{\mu 1} r_1.
\end{align}
This yields a one-loop renormalization of the Berry phase term, 
\begin{align}
\frac{ie}{2}  \int d^D{\bm x}\,\ u_{\mu\nu}({\bm x}) v_{\mu\nu}({\bm x}) 
&\rightarrow 
 \frac{ie}{2}  \int d^D{\bm x}\,\ u_{\mu\nu}({\bm x}) v_{\mu\nu}({\bm x}) +  \frac{i e \delta\chi\ln b}{\overline{\gamma}_\tau} \int d^D{\bm x}\,\ r_1 v_{1\tau}({\bm x}), \nonumber \\
& = \frac{ie}{2}  \int d^D{\bm x}\,\ u_{\mu\nu}({\bm x}) v_{\mu\nu}({\bm x}) -  \frac{i e \delta\chi\ln b}{\overline{\gamma}_\tau} \int d^D{\bm x}\,\ b_{\tau}({\bm x}),
\label{Eq:total-derivative}
 \end{align}
From the first line to the second line, we took partial integrals over $r_1$ and $\tau$ with $v_{1\tau}=\partial_{r_1} b_{\tau} - \partial_{\tau} b_{1}$, dropping the associated surface terms. The surface terms can be neglected because the boundary of the system is sufficiently far from any vortex excitations, and vortex excitations always take the form of closed hyper-surface (For $D=2$, a pair of vortex and antivortex).  To summarize, we reach the following partition function after the integration over the smallest vortex hyper-sphere degrees of freedom;
\begin{align}
Z_{\rm v} =& \int {\cal D}{\bm u} \exp \bigg[-\frac{1}{2}\int_{|\partial u|<\Lambda|u|} d^D{\bm x} \!\  \Big\{ \overline{\gamma}_{\tau} 
(\partial_{\mu} u_{\mu\tau})(\partial_{\lambda} u_{\lambda\tau})  + \overline{\gamma}_{\bm r}\sum^{D-1}_{\nu=1}
(\partial_{\mu} u_{\mu\nu})(\partial_{\lambda} u_{\lambda\nu})  \Big\}\bigg] \nonumber \\
&\Bigg\{ 1 + \sum^{\infty}_{N=1} \frac{1}{N!} \prod^{N}_{j=1} \bigg(\int {\cal D}^{D} {\bm X}_j \int_{|{\bm Y}_j|>a_0 b} {\cal D}^D {\bm Y}_j \int {\cal D}{\sf g}_j({\bm s}) \, \  e^{E_{j}}\bigg) 
e^{-\frac{ie}{2} \int_{|\partial u|<\Lambda|u|} d^D{\bm x}\,\ u_{\mu\nu}({\bm x}) v_{\mu\nu}({\bm x}) - i2\pi \overline{\chi} \sum^{N}_{j=1} {\cal A}_{\tau,j} } \Bigg\}. \label{Eq:Zv-9}
\end{align}
with the renormalization to the Berry phase term, 
 \begin{align}
 \overline{\chi} \equiv \chi - \frac{e\delta\chi}{\gamma_{\tau}} 
 \ln b. \label{Eq:rg7}
 \end{align}

 \subsection{\label{secIIb}one-loop renormalization to vortex fugacity}

     The Maxwell term and the coupling between the vector potential fields and vortex excitations are given by the following action, 
\begin{align}
\frac{1}{2}\int_{|\partial u|<\Lambda|u|} d^D{\bm x} \!\  \Big\{ \gamma_{\tau} 
(\partial_{\mu} u_{\mu\tau})(\partial_{\lambda} u_{\lambda\tau})  + \sum^{D-1}_{\nu=1}
(\partial_{\mu} u_{\mu\nu})(\partial_{\lambda} u_{\lambda\nu})  \Big\} + \frac{ie}{2} \int_{|\partial u|<\Lambda|u|} d^D{\bm x}\,\ u_{\mu\nu}({\bm x}) v_{\mu\nu}({\bm x}). \nonumber 
\end{align}
The integration over the fast component of the vector potential field yields the short-ranged part of the Coulomb interaction between vortex hyper-surface elements. The short-ranged part of the Coulomb interaction can be included as the renormalization to the vortex core energy density. For general $D$,  the renormalization to the vortex core energy density is given by a function of the temporal and spatial components of the metric tensors defined in Eqs.~(\ref{Eq:1st-f},\ref{Eq:1st-f-2}), justifying Eq.~(\ref{Eq:fugacity-2},\ref{Eq:fugacity-1}) a posteriori.    
     
     \subsubsection{\label{secIIb1}vortex fugacity renormalization for general $D$}
     To determine the form of the short-ranged part of the Coulomb interaction for general $D\!\ (>2)$ and $\gamma_{\tau} \ne 1$ cases, let us take the Fourier transform of the relevant parts of the action, 
     \begin{align} 
    &-\sum^N_{j=1} 2\delta E_j \equiv \nonumber \\
     &\int_{b^{-1}\Lambda<|{\bm k}|<\Lambda} \frac{d^D{\bm k}}{(2\pi)^D} 
     \Big\{\gamma_{\tau} k_{\mu}u_{\mu\tau}({\bm k}) k_{\lambda} u_{\lambda\tau}(-{\bm k}) +  \sum_{\phi=1,\cdots,D-1}k_{\mu}u_{\mu\phi}({\bm k}) k_{\lambda} u_{\lambda\phi}(-{\bm k})  + \frac{i e}{2} \big( u_{\mu\nu}({\bm k}) v_{\mu\nu}(-{\bm k}) +  u_{\mu\nu}(-{\bm k}) v_{\mu\nu}({\bm k}) \big)
     \Big\} \nonumber \\
&     =  \int_{b^{-1}\Lambda<|{\bm k}|<\Lambda} \frac{d^D{\bm k}}{(2\pi)^D} 
     \Big\{\gamma_{\tau} \varphi_{\tau}(\bm k)\varphi_{\tau}(-{\bm k}) 
     +  \sum_{\phi=1,\cdots,D-1} \varphi_{\phi}(\bm k)\varphi_{\phi}(-{\bm k})  + ie \big( \frac{k_{\mu}}{k^2} \varphi_{\nu}({\bm k}) v_{\mu\nu}(-{\bm k}) + \frac{k_{\mu}}{k^2} \varphi_{\nu}(-{\bm k}) v_{\mu\nu}({\bm k})\big) \Big\} \nonumber \\
     & = e^2 \int_{b^{-1}\Lambda<|{\bm k}|<\Lambda} \frac{d^D{\bm k}}{(2\pi)^D} 
     \frac{k_{\mu}k_{\lambda}}{k^4} \!\ v_{\mu\nu}({\bm k}) v_{\lambda\nu}(-{\bm k})\!\ [{\bm C}^{-1}]_{\nu\nu} + {\rm const}.  \label{Eq:fugacity-rg1}
     \end{align}
To facilitate the Gaussian integration over the vector potential field for general dimension $D$, we chose a specific gauge in the right hand side [see also Sec.~\ref{secIV} for the alternative choice], and express the vector potential in terms of the divergence-free auxiliary field $\varphi$, 
\begin{align}
 u_{\mu\nu}({\bm k})=\frac{1}{k^2} \big(k_{\mu} \varphi_{\nu}({\bm k}) 
 - k_{\nu} \varphi_{\mu}({\bm k})\big). \label{Eq:gauge1}
 \end{align}
 $[{\bm C}]$ and its inverse are the $D\times D$ diagonal matrices, 
 \begin{align}
 [{\bm C}] = \left(\begin{array}{cc} 
 \gamma_{\tau} & \\
 &  \mathbf{1}_{(D-1)\times (D-1)}
 \end{array}\right), \,\   [{\bm C}^{-1}] = 
\alpha_0 \mathbf{1}_{D\times D} + \alpha_1 {\bm \eta},
 \end{align}
 with 
 \begin{align}
 \alpha_0 \equiv \frac{\gamma^{-1}_{\tau}+(D-1)}{D}, \,\ \alpha_1 \equiv \frac{-\gamma^{-1}_{\tau}+1}{D}, \,\ {\bm \eta} \equiv  \left(\begin{array}{cc} 
 -(D-1) & \\
 & \mathbf{1}_{(D-1)\times (D-1)} \label{Eq:a0a1eta}
 \end{array}\right). 
 \end{align}
Then, take the sum over $\mu,\lambda,\nu=0,\cdots,D-1$ in Eq.~(\ref{Eq:fugacity-rg1}),
\begin{align}
k_{\mu}k_{\lambda} v_{\mu\nu}v_{\lambda\nu} \big(\alpha_0 \delta_{\nu\nu} + \alpha_1 {\bm \eta}_{\nu\nu}\big) =& \Big(\frac{1}{(D-2)!}\Big)^2 (\alpha_0 \delta_{\nu\nu}+\alpha_1{\bm \eta}_{\nu\nu}) k_{\mu}k_{\lambda}\varepsilon_{\mu\nu\sigma\phi\cdots\psi} 
\varepsilon_{\lambda\nu\xi\rho\cdots\eta} J_{\sigma\phi\cdots\psi} J_{\xi\rho\cdots\eta},
\nonumber \\
 =&  \sum_{\sigma<\phi<\cdots<\psi} J_{\sigma\phi\cdots\psi} J_{\sigma\phi\cdots\psi} \Big(\alpha_0 k^2 - \alpha_1 \big(\sum_{\mu} {\bm \eta}_{\mu\mu} k^2_{\mu}\big) - \alpha_1 k^2 \big({\bm \eta}_{\sigma\sigma}+{\bm \eta}_{\phi\phi}
+ \cdots  + {\bm \eta}_{\psi\psi}\big) \Big). \label{Eq:fugacity-rg2}
\end{align}
In the first line, the summations of the repeated indices over $\tau,1,\cdots,D-1$ are implicitly assumed. In the second line, the summations of $\sigma$, $\phi,\cdots,\psi$  over $\tau,1,\cdots,D-1$ are taken in such a way that $\sigma<\phi<\cdots<\psi$ is satisfied $[\tau=0<1,\cdots,D-1]$, while the summation of $\mu$ over $0,1,\cdots,D-1$ is uncorrelated with $\sigma,\phi,\cdots,\psi$.  To derive Eq.~(\ref{Eq:fugacity-rg2}), we used the following formula,
\begin{align}
\sum_{\nu}\varepsilon_{\mu\nu\sigma\phi\cdots\psi} \varepsilon_{\lambda\nu\xi\rho\cdots\eta} &= \left(\begin{array}{ccccc}
\mu & \sigma & \phi & \cdots & \psi \\
\lambda& \xi & \rho & \cdots & \eta \\
\end{array}\right)_{\bm 1} \equiv \sum_{P}(-1)^P \delta_{\mu P(\lambda)} 
\delta_{\sigma P(\xi)} \cdots \delta_{\psi P(\eta)}, \label{Eq:formula1} \\ 
\sum_{\nu} {\bm \eta}_{\nu\nu} \!\ 
\varepsilon_{\mu\nu\sigma\phi\cdots\psi} \varepsilon_{\lambda\nu\xi\rho\cdots\eta} &= - \left(\begin{array}{ccccc}
\mu & \sigma & \phi & \cdots & \psi \\
\lambda& \xi & \rho & \cdots & \eta \\
\end{array}\right)_{\bm \eta} \equiv -\sum_{P}(-1)^P 
\Big\{ {\bm \eta}_{\mu P(\lambda)} 
\delta_{\sigma P(\xi)} \cdots \delta_{\psi P(\eta)} \nonumber \\
&\hspace{2cm} + \delta_{\mu P(\lambda)} 
{\bm \eta}_{\sigma P(\xi)} \cdots \delta_{\psi P(\eta)}
+ \cdots + \delta_{\mu P(\lambda)} 
\delta_{\sigma P(\xi)} \cdots {\bm \eta}_{\psi P(\eta)} \Big\}. \label{Eq:formula2}
\end{align}
Here $P$ stands for a permutation of the $(D-1)$ elements, a permutation from $(\lambda,\xi,\rho,\cdots,\eta)$ to $(P(\lambda),P(\xi),P(\rho),\cdots,P(\eta))$, and $\sum_{P}$ is a sum over all the elements of the permutation group. $(-1)^P$ is $+1$ and $-1$, respectively, when a permutation $P$ consists of an even or odd number of transpositions. We also used a divergence-free condition for the vortex field~\cite{goldbart2009}, 
\begin{align}
\sum_{\lambda=0,1,\cdots,D-1}\frac{\partial}{\partial x_{\lambda}} J_{\lambda\rho\cdots \sigma}({\bm x}) = \sum_{\rho}\frac{\partial}{\partial x_{\rho}} J_{\lambda\rho\cdots \sigma}({\bm x}) = \cdots = \sum_{\sigma}\frac{\partial}{\partial x_{\sigma}} J_{\lambda\rho\cdots \sigma}({\bm x}) =0. \label{Eq:divergence-free}
\end{align}
The divergence-free condition is due to the fact that any closed vortex hyper-spheres have no boundaries. In the following, we will briefly explain the deduction of Eq.~(\ref{Eq:fugacity-rg2}) for the memorandum. The $\alpha_0$ term is calculated as follows:
\begin{align}
&\sum_{\nu}\sum_{\mu,\lambda,\sigma,\xi,\cdots,\psi,\eta} k_{\mu}k_{\lambda}\varepsilon_{\mu\nu\sigma\phi\cdots\psi} 
\varepsilon_{\lambda\nu\xi\rho\cdots\eta} J_{\sigma\phi\cdots\psi}({\bm k})
J_{\xi\rho\cdots\eta}(-{\bm k}) \nonumber \\
& = \sum_{\mu,\lambda,\sigma,\xi,\cdots,\psi,\eta}\left(\begin{array}{ccccc}
\mu & \sigma &\phi& \cdots & \psi \\
\lambda &\xi &\rho & \cdots & \eta \\
\end{array}\right)_{\bm 1} k_{\mu}k_{\lambda} J_{\sigma\phi\cdots \psi}({\bm k}) J_{\xi\rho\cdots\eta}(-{\bm k}) \nonumber \\
& = \sum_{\sigma,\xi,\cdots,\psi,\eta} 
\sum^{\mu \ne \sigma,\phi,\cdots,\psi}_{\mu} k^2_{\mu} \left(\begin{array}{cccc}
 \sigma & \phi &\cdots & \psi \\
 \xi & \rho &\cdots & \eta \\
\end{array}\right)_{\bm 1} J_{\sigma\phi\cdots\psi} J_{\xi\rho\cdots\eta} 
- (D-2)^2 \sum_{\phi,\rho,\cdots,\psi,\eta}\sum^{\mu \ne \sigma}_{\mu,\sigma} k_{\mu} k_{\sigma}  
\left(\begin{array}{ccc}
\phi& \cdots & \psi \\
\rho & \cdots & \eta \\
\end{array}\right)_{\bm 1}J_{\sigma\phi\cdots\psi} J_{\mu\rho\cdots\eta} \nonumber \\
& = \big((D-2)!\big)^2 \Big\{ \sum_{\sigma<\phi<\cdots<\psi}  J_{\sigma\phi\cdots\psi} J_{\sigma\phi\cdots\psi} \sum^{\mu \ne \sigma,\phi,\cdots,\psi}_{\mu} k^2_{\mu} 
-  \sum_{\phi<\cdots<\psi}
\sum^{\mu \ne \sigma}_{\mu, \sigma } k_{\mu} k_{\sigma} J_{\sigma\phi\cdots\psi} J_{\mu\phi\cdots\psi}  \Big\} \nonumber \\
& =  \big((D-2)!\big)^2 \Big\{ \sum_{\sigma<\phi<\cdots<\psi}  J_{\sigma\phi\cdots\psi} J_{\sigma\phi\cdots\psi} \sum^{\mu \ne \sigma,\phi,\cdots,\psi}_{\mu} k^2_{\mu} 
-  \sum_{\phi<\cdots<\psi}
\sum_{\mu, \sigma } k_{\mu} k_{\sigma} J_{\sigma\phi\cdots\psi} J_{\mu\phi\cdots\psi}  
+ \sum_{\phi<\cdots<\psi}
\sum_{\mu} k^2_{\mu}  J_{\mu\phi\cdots\psi} J_{\mu\phi\cdots\psi}\Big\} \nonumber \\
& = \big((D-2)!\big)^2 \Big\{\sum_{\sigma<\phi<\cdots<\psi}  J_{\sigma\phi\cdots\psi} J_{\sigma\phi\cdots\psi} \sum^{\mu \ne \sigma,\phi,\cdots,\psi}_{\mu} k^2_{\mu} 
+ \sum_{\mu<\phi<\cdots<\psi} (k^2_{\mu} + k^2_{\phi} + \cdots + k^2_{\psi}) J_{\mu\phi\cdots\psi} J_{\mu\phi\cdots\psi}\Big\} \nonumber \\
& = \big((D-2)!\big)^2 k^2 \sum_{\sigma<\phi<\cdots<\psi}  J_{\sigma\phi\cdots\psi} J_{\sigma\phi\cdots\psi}. \label{Eq:alpha0}
\end{align}
From the fifth line to the sixth line, we used $\sum_{\mu}k_{\mu} J_{\mu\phi\cdots\psi}(-{\bm k}) = 0$ from Eq.~(\ref{Eq:divergence-free}). We also used  
\begin{align} 
\sum_{\mu}\sum_{\phi<\cdots<\psi} k^2_{\mu} J_{\mu\phi\cdots\psi}({\bm k}) J_{\mu\phi\cdots\psi}(-{\bm k}) 
= \sum_{\mu<\phi<\cdots<\psi}  (k^2_{\mu}+k^2_{\phi} + \cdots + k^2_{\psi}) J_{\mu\phi\cdots\psi}({\bm k}) J_{\mu\phi\cdots\psi}(-{\bm k}). \label{Eq:formula3}
\end{align}
Similarly, the $\alpha_1$ term can be calculated, 
\begin{align}
&\sum_{\nu}\sum_{\mu,\lambda,\sigma,\xi,\cdots,\psi,\eta} k_{\mu}k_{\lambda}{\bm \eta}_{\nu\nu} \varepsilon_{\mu\nu\sigma\phi\cdots\psi} 
\varepsilon_{\lambda\nu\xi\rho\cdots\eta} J_{\sigma\phi\cdots\psi}({\bm k})
J_{\xi\rho\cdots\eta}(-{\bm k}) \nonumber \\
& = - \sum_{\mu,\lambda,\sigma,\xi,\cdots,\psi,\eta}\left(\begin{array}{ccccc}
\mu & \sigma &\phi& \cdots & \psi \\
\lambda &\xi &\rho & \cdots & \eta \\
\end{array}\right)_{\bm \eta} k_{\mu}k_{\lambda} J_{\sigma\phi\cdots \psi}({\bm k}) J_{\xi\rho\cdots\eta}(-{\bm k}) \nonumber \\
& = - \sum_{\sigma,\xi,\cdots,\psi,\eta} 
\sum^{\mu \ne \sigma,\phi,\cdots,\psi}_{\mu} k^2_{\mu} \Bigg( {\bm \eta}_{\mu\mu} \left(\begin{array}{cccc}
 \sigma & \phi &\cdots & \psi \\
 \xi & \rho &\cdots & \eta \\
\end{array}\right)_{\bm 1} + \delta_{\mu\mu} \left(\begin{array}{cccc}
 \sigma & \phi &\cdots & \psi \\
 \xi & \rho &\cdots & \eta \\
\end{array}\right)_{\bm \eta} \Bigg)
J_{\sigma\phi\cdots\psi} J_{\xi\rho\cdots\eta} \nonumber \\ 
& \hspace{0.5cm} + (D-2)^2 \sum_{\phi,\xi,\cdots,\psi,\eta}\sum^{\mu \ne \sigma}_{\mu,\sigma} k_{\mu} k_{\sigma}  
\Bigg( {\bm \eta}_{\mu\mu} \delta_{\sigma\sigma} \left(\begin{array}{ccc}
\phi& \cdots & \psi \\
\rho & \cdots & \eta \\
\end{array}\right)_{\bm 1} + \delta_{\mu\mu} {\bm \eta}_{\sigma\sigma} \left(\begin{array}{ccc}
\phi& \cdots & \psi \\
\rho & \cdots & \eta \\
\end{array}\right)_{\bm 1} \nonumber \\
&\hspace{2.5cm} + \delta_{\mu\mu} \delta_{\sigma\sigma} \left(\begin{array}{ccc}
\phi& \cdots & \psi \\
\rho & \cdots & \eta \\
\end{array}\right)_{\bm \eta}\Bigg)
J_{\sigma\phi\cdots\psi} J_{\mu\rho\cdots\eta}\nonumber \\
& = - \big((D-2)!\big)^2 \sum_{\sigma<\phi<\cdots<\psi}  J_{\sigma\phi\cdots\psi} J_{\sigma\phi\cdots\psi} \Big(\sum^{\mu \ne \sigma,\phi,\cdots,\psi}_{\mu} k^2_{\mu} {\bm \eta}_{\mu\mu} + \sum^{\mu \ne \sigma,\phi,\cdots,\psi}_{\mu} k^2_{\mu} 
({\bm \eta}_{\sigma\sigma} + {\bm \eta}_{\phi\phi}+\cdots +{\bm \eta}_{\psi\psi})\Big) \nonumber \\
& \hspace{0.5cm} + \big((D-2)!\big)^2 \sum_{\phi<\cdots<\psi} \sum^{\mu\ne \sigma}_{\mu,\sigma}  k_{\mu} k_{\sigma}   J_{\sigma\phi\cdots\psi} J_{\mu\phi\cdots\psi}  \Big({\bm \eta}_{\mu\mu} + {\bm \eta}_{\sigma\sigma} + {\bm \eta}_{\phi\phi} + \cdots + {\bm \eta}_{\psi\psi}\Big) \nonumber \\
& =  - \big((D-2)!\big)^2 \sum_{\sigma<\phi<\cdots<\psi}  J_{\sigma\phi\cdots\psi} J_{\sigma\phi\cdots\psi} \Big(\sum^{\mu \ne \sigma,\phi,\cdots,\psi}_{\mu} k^2_{\mu} {\bm \eta}_{\mu\mu} + \sum^{\mu \ne \sigma,\phi,\cdots,\psi}_{\mu} k^2_{\mu} 
({\bm \eta}_{\sigma\sigma} + {\bm \eta}_{\phi\phi}+\cdots +{\bm \eta}_{\psi\psi})\Big) \nonumber \\ 
& \hspace{0.5cm} + \big((D-2)!\big)^2 \sum_{\phi<\cdots<\psi}   \sum_{\mu,\sigma} k_{\mu} k_{\sigma} J_{\sigma\phi\cdots\psi} J_{\mu\phi\cdots\psi} \Big({\bm \eta}_{\mu\mu} + {\bm \eta}_{\sigma\sigma} + {\bm \eta}_{\phi\phi} + \cdots + {\bm \eta}_{\psi\psi}\Big) \nonumber \\
& \hspace{0.5cm} -  \big((D-2)!\big)^2 \sum_{\phi<\cdots<\psi} \sum_{\mu} k^2_{\mu}   J_{\mu\phi\cdots\psi} J_{\mu\phi\cdots\psi}  \Big({\bm \eta}_{\mu\mu} + {\bm \eta}_{\mu\mu} + {\bm \eta}_{\phi\phi} + \cdots + {\bm \eta}_{\psi\psi}\Big) \nonumber \\
& = - \big((D-2)!\big)^2 \sum_{\sigma<\phi<\cdots<\psi}  J_{\sigma\phi\cdots\psi} J_{\sigma\phi\cdots\psi} \Big(\sum^{\mu \ne \sigma,\phi,\cdots,\psi}_{\mu} k^2_{\mu} {\bm \eta}_{\mu\mu} + \sum^{\mu \ne \sigma,\phi,\cdots,\psi}_{\mu} k^2_{\mu} 
({\bm \eta}_{\sigma\sigma} + {\bm \eta}_{\phi\phi}+\cdots +{\bm \eta}_{\psi\psi})\Big) \nonumber \\ 
&  \hspace{0.5cm} -   \big((D-2)!\big)^2 \sum_{\mu<\phi<\cdots<\psi}  ( k^2_{\mu} {\bm \eta}_{\mu\mu} + 
k^2_{\phi} {\bm \eta}_{\phi\phi} + \cdots + k^2_{\psi} {\bm \eta}_{\psi\psi})  J_{\mu\phi\cdots\psi} J_{\mu\phi\cdots\psi} \nonumber \\ 
& \hspace{0.5cm} -   \big((D-2)!\big)^2 \sum_{\mu<\phi<\cdots<\psi} (k^2_{\mu} + k^2_{\phi}+\cdots + k^2_{\psi}) J_{\mu\phi\cdots\psi} J_{\mu\phi\cdots\psi}\Big( {\bm \eta}_{\mu\mu} + {\bm \eta}_{\phi\phi} + \cdots + {\bm \eta}_{\psi\psi}\Big) \nonumber \\
& =  - \big((D-2)!\big)^2 \sum_{\sigma<\phi<\cdots<\psi}  J_{\sigma\phi\cdots\psi} J_{\sigma\phi\cdots\psi} \Big(\sum_{\mu} k^2_{\mu} {\bm \eta}_{\mu\mu} \Big) \nonumber \\
& \hspace{0.5cm} - \big((D-2)!\big)^2 \sum_{\sigma<\phi<\cdots<\psi}  J_{\sigma\phi\cdots\psi} J_{\sigma\phi\cdots\psi} 
({\bm \eta}_{\sigma\sigma} + {\bm \eta}_{\phi\phi}+\cdots +{\bm \eta}_{\psi\psi})  
\Big( \sum_{\mu} k^2_{\mu} \Big) \label{Eq:alpha1}
\end{align}
From the fifth line to the sixth line, we used the divergence-free condition for the vortex field. We also used 
\begin{align}
\sum_{\mu}\sum_{\phi<\cdots<\psi} k^2_{\mu} {\bm \eta}_{\mu\mu} J_{\mu\phi\cdots\psi}({\bm k}) J_{\mu\phi\cdots\psi}(-{\bm k}) 
= \sum_{\mu<\phi<\cdots<\psi}  (k^2_{\mu} {\bm \eta}_{\mu\mu}+k^2_{\phi}{\bm \eta}_{\phi\phi} + \cdots + k^2_{\psi}{\bm \eta}_{\psi\psi}) J_{\mu\phi\cdots\psi}({\bm k}) J_{\mu\phi\cdots\psi}(-{\bm k}), \label{Eq:formula4}
\end{align}
as well as  
\begin{align}
& \sum_{\mu}\sum_{\phi<\cdots<\psi} k^2_{\mu} J_{\mu\phi\cdots\psi}({\bm k}) J_{\mu\phi\cdots\psi}(-{\bm k}) ({\bm \eta}_{\mu\mu} + {\bm \eta}_{\phi\phi} + \cdots + {\bm \eta}_{\psi\psi}) \nonumber \\
& = \sum_{\mu<\phi<\cdots<\psi}  (k^2_{\mu}+k^2_{\phi} + \cdots + k^2_{\psi}) J_{\mu\phi\cdots\psi}({\bm k}) J_{\mu\phi\cdots\psi}(-{\bm k}) 
({\bm \eta}_{\mu\mu} + {\bm \eta}_{\phi\phi} + \cdots + {\bm \eta}_{\psi\psi}) . \label{Eq:formula5}
\end{align}
 Eqs.~(\ref{Eq:alpha0},\ref{Eq:alpha1}) give Eq.~(\ref{Eq:fugacity-rg2}). With Eqs.~(\ref{Eq:a0a1eta}), the short-ranged part of the Coulomb interaction between parallelepiped elements take a diagonal form, 
 \begin{align}
 \sum_{j} \delta E_j = - \frac{1}{(D-2)!} \sum_{\sigma,\phi,\cdots,\psi} 
 \frac{e^2}{2} \int_{b^{-1}\Lambda<|{\bm k}|<\Lambda} 
\frac{d^D{\bm k}}{(2\pi)^D} \!\ f_{\sigma\phi\cdots\psi}({\bm k}) J_{\sigma\phi\cdots\psi}({\bm k}) J_{\sigma\phi\cdots\psi}(-{\bm k}). \label{Eq:fugacity-rg3}
 \end{align}
The Fourier-transform of the interaction potential is given by,   
 \begin{align}
 f_{\sigma\phi\cdots\psi}({\bm k}) = \begin{cases} 
  \frac{ {\bm q}^2+(2-\gamma^{-1}_{\tau})\omega^2}{k^4} &  {\rm if} \,\ \tau \in \{\sigma,\phi,\cdots,\psi\}, \\
  \frac{\gamma^{-1}_{\tau} {\bm q} + \omega^2}{k^4} & {\rm if} \,\ \tau \notin \{\sigma,\phi,\cdots,\psi\}, 
 \end{cases} \label{Eq:fugacity-rg4}
 \end{align}
 with ${\bm k}=(\omega,{\bm q})$ and ${\bm k}\cdot {\bm x} \equiv {\bm q}\cdot {\bm r}-{\omega} \tau$.

     The short-ranged part of the interaction gives rise to a fugacity renormalization from the leading-order term in its gradient expansion. To this end, rewrite Eq.~(\ref{Eq:fugacity-rg3}) in the space-time coordinate, 
 \begin{align}
 \sum_{j} \delta E_j =& - \frac{1}{(D-2)!} \sum_{\sigma,\phi,\cdots,\psi} 
\frac{e^2}{2} \int d^D{\bm R}_1 \int d^D{\bm R}_2 \,\ F_{\sigma\phi\cdots\psi}(|{\bm R}_1-{\bm R}_2|) \!\ J_{\sigma\phi\cdots\psi}({\bm R}_1) J_{\sigma\phi\cdots\psi}({\bm R}_2), \nonumber \\
=& - \frac{2\pi^2 e^2}{(D-2)!} \sum_{j}\sum_{\sigma,\phi,\cdots,\psi} \int_{\partial \Gamma_j} d^{D-2}{\bm s}_1 \int_{\partial \Gamma_j} d^{D-2}{\bm s}_2 \,\ F_{\sigma\phi\cdots\psi}(|{\bm R}({\bm s}_1)-{\bm R}({\bm s}_2)|) \!\ {\sf g}_{j,\sigma\phi\cdots\psi}({\bm s}_1) {\sf g}_{j,\sigma\phi\cdots\psi}({\bm s}_2) 
\label{Eq:fugacity-rg5}
 \end{align}
where we used Eq.~(\ref{def-J}) with the oriented volume 
\begin{align}
{\sf g}_{j,\sigma\phi\cdots\psi}({\bm s}) \equiv \sum^{D-2}_{a,b,\cdots,c=1} 
\varepsilon_{ab\cdots c} \frac{\partial R_{j,\sigma}({\bm s})}{\partial s_{a}}\frac{\partial R_{j,\phi}({\bm s})}{\partial s_{b}}\cdots \frac{\partial R_{j,\psi}({\bm s})}{\partial s_{c}}.  \label{Eq:oriented-volume}
\end{align}
$F_{\sigma\phi\cdots\psi}({\bm x})$ stands for the short-ranged part of the anisotropic Coulomb interaction, 
\begin{align}
F_{\sigma\phi\cdots\psi}({\bm x}) =&  \int_{b^{-1}\Lambda<|{\bm k}|<\Lambda} 
\frac{d^D{\bm k}}{(2\pi)^D}  \,\ e^{i{\bm q}{\bm r}-i\omega\tau} f_{\sigma\phi\cdots\psi}({\bm k}) \equiv \begin{cases}
   \ln b\,\ F_\tau ({\bm x}) &  {\rm if} \,\ \tau \in \{\sigma,\phi,\cdots,\psi\}, \\
 \ln b \,\ F_{\bm r}({\bm x}) & {\rm if} \,\ \tau \notin \{\sigma,\phi,\cdots,\psi\}, 
\end{cases} \nonumber \\
\ln b\,\ F_{\tau}({\bm x}) \equiv & \int_{b^{-1}\Lambda<|{\bm k}|<\Lambda} 
\frac{d^D{\bm k}}{(2\pi)^D}  \,\ e^{i{\bm q}{\bm r}-i\omega\tau} 
\frac{{\bm q}^2+(2-\gamma^{-1}_{\tau})\omega^2}{k^4}, \label{Eq:Ft} \\
\ln b \,\ F_{\bm r}({\bm x}) \equiv & \int_{b^{-1}\Lambda<|{\bm k}|<\Lambda} 
\frac{d^D{\bm k}}{(2\pi)^D}  \,\ e^{i{\bm q}{\bm r}-i\omega\tau} 
  \frac{\gamma^{-1}_{\tau} {\bm q^2} + \omega^2}{k^4},  \label{Eq:Fr}
\end{align}
with  ${\bm k}=(\omega,{\bm q})$ and ${\bm x}=(\tau,{\bm r})$.  Since $F_{\sigma\phi\cdots\psi}({\bm x})$ ranges only over the lattice constant length scale $[|{\bm x}|<\Lambda^{-1}]$, we kept only the interaction within the same vortex hyper-surface in Eq.~(\ref{Eq:fugacity-rg5}). Thanks to the short-ranged nature of the interaction, we may also use a gradient expansion on the hyper-surface and keep the leading order in the expansion, 
\begin{align}
  \delta E_j = -\frac{2\pi^2 e^2}{(D-2)!} \sum_{\sigma, \phi, \cdots,\psi} \int_{\partial \Gamma_j} d^{D-2}{\bm s} \,\ {\sf g}^2_{j,\sigma\phi\cdots\psi}({\bm s})  \int d^{D-2}\Delta {\bm s} \,\ F_{\sigma\phi\cdots\psi}\Big(\sum^{D-2}_{a=1} \Delta s_a \frac{\partial {\bm R}_j({\bm s})}{\partial s_a}\Big) + \cdots, 
\label{Eq:fugacity-rg6}
\end{align}
with $\Delta{\bm s}\equiv {\bm s}_1-{\bm s}_2$,  and ${\bm s}\equiv ({\bm s}_1+{\bm s}_2)/2$.  Note that $f_{\sigma\phi\cdots\psi}({\bm k})$ are even functions of their arguments, so are $F_{\sigma\phi\cdots\psi}({\bm x})$ as well. Moreover, $f_{\sigma\phi\cdots\psi}({\bm k}=({\bm q},\omega))$  is symmetric under the $O(D-1)$ rotation of ${\bm q}$ $[({\bm q},\omega)\rightarrow (\tilde{\bm q},\omega)]$ with $\tilde{\bm q}=O(D-1)\cdot {\bm q}$ . Therefore,  $F_{\sigma\phi\cdots\psi}({\bm x})$ must be functions of $\tau^2$ and $|{\bm r}|^2$ as well. The $\Delta {\bm s}$-integration of such functions with ${\bm x} \equiv \sum^{D-2}_{a=1} \Delta s_a \partial_{s_a} {\bm R}_{j}({\bm s})$ must yield functions only of $\hat{\bf g}_{j,\tau}({\bm s})$ and $\hat{\bf g}_{j,\bm r}(\bm s)$. Let us call  the $\Delta {\bm s}$-integral of  $F_{\tau}({\bm x})$ and $F_{\bm r}({\bm x})$  with ${\bm x}=\sum^{D-2}_{a=1}\Delta s_a \partial_{s_a} {\bm R}({\bm s})$ as $f_{\tau}$ and $f_{\bm r}$, respectively, 
\begin{align}
f_{\tau}(\hat{\bf g}_{j,\tau}({\bm s}),\hat{\bf g}_{j,\rm r}({\bm s})) \equiv &
\int d^{D-2}\Delta {\bm s}\,\ F_{\tau} \Big(\sum^{D-2}_{a=1} \Delta s_a \frac{\partial {\bm R}_j({\bm s})}{\partial s_a}\Big), \label{Eq:ft} \\ 
f_{\bm r}(\hat{\bf g}_{j,\tau}({\bm s}),\hat{\bf g}_{j,\rm r}({\bm s})) \equiv &
\int d^{D-2}\Delta {\bm s}\,\ F_{\bm r} \Big(\sum^{D-2}_{a=1} \Delta s_a \frac{\partial {\bm R}_j({\bm s})}{\partial s_a}\Big). \label{Eq:fr} 
\end{align}
Importantly, the function {\it forms} of $f_{\tau}$ and $f_{\bm r}$ do not depend on the shapes of vortex surfaces, and the same function forms apply to all vortex surfaces. This is because Eqs.~(\ref{Eq:Ft},\ref{Eq:Fr}) apply to all vortex surfaces.

       With the use of these functions together with $h_{\tau} \equiv f^{-\frac{D-3}{D-2}}_{\bm r} f_{\tau}$ and $h_{1}=h_2=\cdots=h_{D-1} \equiv f^{\frac{1}{D-2}}_{\bm r}$, 
 the leading-order gradient-expansion term can be expressed as a function of  $\hat{\bf g}_{j,\tau}({\bm s})$ and $\hat{\bf g}_{j,\bm r}(\bm s)$, 
 \begin{align}
 \delta E_j =& -\frac{2\pi^2 e^2}{(D-2)!} \ln b  \int_{\partial \Gamma_j} d^{D-2}{\bm s} \,\ 
\sum^{D-2}_{a,\cdots,c,a^{\prime},\cdots,c^{\prime}=1} \varepsilon_{ab\cdots c}\varepsilon_{a^{\prime}b^{\prime}\cdots c^{\prime}} \nonumber \\
& \sum_{\sigma}\Big(\frac{\partial R_{j,\sigma}({\bm s})}{\partial s_a} h_{\sigma} \frac{\partial R_{j,\sigma}({\bm s})}{\partial s_{a^{\prime}}}\Big) \sum_{\phi} \Big(\frac{\partial R_{j,\phi}({\bm s})}{\partial s_b} h_{\phi} \frac{\partial R_{j,\phi}({\bm s})}{\partial s_{b^{\prime}}}\Big) \cdots \sum_{\psi}\Big(\frac{\partial R_{j,\psi}({\bm s})}{\partial s_c} h_{\psi} \frac{\partial R_{j,\psi}({\bm s})}{\partial s_{c^{\prime}}}\Big) \nonumber \\
 =& -2\pi^2 e^2 \ln b \int_{\partial \Gamma_j} d^{D-2} {\bm s}\,\ \det\Big[f^{-\frac{D-3}{D-2}}_{\bm r} f_{\tau}\!\ \hat{\bf g}_{j,\tau} + f^{\frac{1}{D-2}}_{\bm r} \hat{\bf g}_{j,\bm r}\Big].   \label{Eq:fugacity-rg7}
 \end{align}
 Here $\det$ is a determinant of a $(D-2)\times (D-2)$ matrix $f^{-\frac{D-3}{D-2}}_{\bm r} f_{\tau}\!\ \hat{\bf g}_{j,\tau} + f^{\frac{1}{D-2}}_{\bm r} \hat{\bf g}_{j,\bm r}$, and $f_{\bm r}$ and $f_{\tau}$ are scalar functions of the temporal and spatial components of the metric tensor, $\hat{\bf g}_{j,\tau}$ and $\hat{\bf g}_{j,\bm r}$.  The function forms can be determined by integrals in Eqs.~(\ref{Eq:Ft},\ref{Eq:Fr}, \ref{Eq:ft}, \ref{Eq:fr}).  Note that Eq.~(\ref{Eq:fugacity-rg7}) inductively justifies the assumption given in Eq.~(\ref{Eq:fugacity-2}).

     To put this vortex fugacity renormalization into Eq.~(\ref{Eq:Zv-9}), we finally obtain 
 \begin{align}
 Z_{\rm v} =& \int {\cal D}{\bm u} \exp \bigg[-\frac{1}{2}\int_{|\partial u|<b^{-1}\Lambda|u|} d^D{\bm x} \!\  \Big\{ \overline{\gamma}_{\tau} 
(\partial_{\mu} u_{\mu\tau})(\partial_{\lambda} u_{\lambda\tau})  + \overline{\gamma}_{\bm r}\sum^{D-1}_{\nu=1}
(\partial_{\mu} u_{\mu\nu})(\partial_{\lambda} u_{\lambda\nu})  \Big\}\bigg] \nonumber \\
&\Bigg\{ 1 + \sum^{\infty}_{N=1} \frac{1}{N!} \prod^{N}_{j=1} \bigg(\int {\cal D}^D {\bm X}_j \int_{|{\bm Y}_j|>a_0 b} {\cal D}^D {\bm Y}_j \int {\cal D}{\sf g}_j({\bm s}) \! \  \exp\big[w\big]\!\ \exp\bigg[ \int_{\partial \Gamma_j} d^{D-2}{\bm s}\,\ \overline{t}\Big(\hat{\bf g}_{j,\tau},\hat{\bf g}_{j,{\bm r}}\Big)\bigg]\bigg)  \nonumber \\ 
& \exp\bigg[-\frac{ie}{2} \int_{|\partial u|<b^{-1}\Lambda|u|} d^D{\bm x}\,\ u_{\mu\nu}({\bm x}) v_{\mu\nu}({\bm x}) - i2\pi \overline{\chi} \sum^{N}_{j=1} {\cal A}_{\tau,j} \bigg] \Bigg\}, \label{Eq:Zv-10}
 \end{align}
 with Eqs.~(\ref{Eq:rg3},\ref{Eq:rg4},\ref{Eq:rg5},\ref{Eq:rg6},\ref{Eq:rg7}), and 
 \begin{align}
 \overline{t}\big(\hat{\bf g}_{\tau},\hat{\bf g}_{\bm r}\big) =  t\big(\hat{\bf g}_{\tau},\hat{\bf g}_{\bm r}\big) - 2\pi^2 e^2 \ln b \!\ \det\Big[f^{-\frac{D-3}{D-2}}_{\bm r}(\hat{\bf g}_{\tau},\hat{\bf g}_{\bm r}) \!\ f_{\tau}(\hat{\bf g}_{\tau},\hat{\bf g}_{\bm r})\!\ \hat{\bf g}_{\tau} + f^{\frac{1}{D-2}}_{\bm r} (\hat{\bf g}_{\tau},\hat{\bf g}_{\bm r})\!\ \hat{\bf g}_{\bm r}\Big],   \label{Eq:rg8}
 \end{align}
 with Eqs.~(\ref{Eq:ft},\ref{Eq:Ft},\ref{Eq:fr},\ref{Eq:Fr}).   

\subsubsection{\label{secIIb2}$D=2$}
      The fugacity renormalization for $D=2$ and $\gamma_{\tau} \ne 1$ can be calculated from Eq.~(\ref{Eq:fugacity-rg1}), where $v_{\mu\nu}$ reduces to a scalar quantity;
\begin{align}
v_{1\tau}({\bm x})=-v_{\tau 1}({\bm x}) = 2\pi \sum_{j} 
\Big(\delta^2({\bm x}-{\bm R}_{j,{\rm v}})- \delta^2({\bm x}-{\bm R}_{j,{\rm av}})\Big) \equiv J({\bm x}). 
\end{align}
Since $v_{\mu\nu}$ is scalar for $D=2$,  instead of Eq.~(\ref{Eq:fugacity-rg2}), we have 
\begin{align}
k_{\mu}k_{\lambda} v_{\mu\nu}({\bm k}) v_{\lambda\nu}(-{\bm k}) 
\big(\alpha_0 \delta_{\nu\nu} + \alpha_1 {\bm \eta}_{\nu\nu}\big) 
&=  J({\bm k}) J(-{\bm k}) \Big(\alpha_0 k^2 - \alpha_1 (\sum_{\mu=\tau,1}{\bm \eta}_{\mu\mu} k^2_{\mu})\Big) \nonumber \\ 
& =  J({\bm k}) J(-{\bm k}) \Big( \gamma^{-1}_{\tau} q^2 + \omega^2 \Big), 
\end{align}
with ${\bm k}\equiv (\omega,q)$.  The short-ranged part of the logarithmic Coulomb interaction can be readily calculated, 
\begin{align}
\sum_{j} \delta E_j &= -\frac{e^2}{2} \ln b \int d^2{\bm R}_1 \int d^2{\bm R}_2 \,\  
F_{\rm 2D}({\bm R}_1-{\bm R}_2) J({\bm R}_1) J({\bm R}_2) \nonumber \\
&= - 2e^2 \pi^2 \ln b \sum_{j,m} \int d^2{\bm R}_1 \int d^2 {\bm R}_2  \,\ F_{\rm 2D}({\bm R}_1-{\bm R}_2) \nonumber \\
&\hspace{2cm} \Big(\delta^2({\bm R}_1-{\bm R}_{j,{\rm v}}) - \delta^2({\bm R}_1-{\bm R}_{j,{\rm av}})\Big)\Big(\delta^2({\bm R}_2-{\bm R}_{m,{\rm v}}) - \delta^2({\bm R}_2-{\bm R}_{m,{\rm av}})\Big) \nonumber \\
&\simeq -\sum_{j} 4\pi^2 e^2 F_{\rm 2D}({\bm 0}) \!\ \ln b,  \label{Eq:fugacity-rg2D-1}
\end{align}
with 
\begin{align}
F_{\rm 2D}({\bm R}) &= \int_{\Lambda b^{-1}<|{\bm k}|<\Lambda} \frac{d^2{\bm k}}{(2\pi)^2}  e^{i{\bm k}\cdot {\bm R}} \frac{\gamma^{-1}_{\tau} q^2 + \omega^2}{k^4}. 
\end{align}
In Eq.~(\ref{Eq:fugacity-rg2D-1}), we only kept ``intra-vortex interactions" $({\bm R}_1={\bm R}_2)$ mediated by the scalar potential $u_{\tau 1}$, while neglecting the inter-vortex interactions $({\bm R}_1 \ne {\bm R}_2)$, because $F_{\rm 2D}({\bm R}_1-{\bm R}_2)$ is a short-ranged function. The strength of the ``intra-vortex interaction" $F_{\rm 2D}({\bm 0})$ can be evaluated with a soft UV cutoff $\Lambda$, 
\begin{align}
F_{\rm 2D}({\bm 0}) \simeq  -\Lambda^{-1} \frac{\partial}{\partial \Lambda^{-1}} \bigg(
\int \frac{d^2{\bm k}}{(2\pi)^2} e^{-\Lambda^{-1}|{\bm k}|}  \frac{\gamma^{-1}_{\tau} q^2 + \omega^2}{k^4}\bigg) = \frac{\gamma^{-1}_{\tau}+1}{4\pi}. \label{Eq:fugacity-rg2D-2}  
\end{align}
To put this into Eq.~(\ref{Eq:Zv-9}) for $D=2$, we obtain, 
\begin{align}
 Z_{\rm v} =& \int {\cal D}{\bm u} \exp \bigg[-\frac{1}{2}\int_{|\partial u|<b^{-1}\Lambda|u|} d^D{\bm x} \!\  \Big\{ \overline{\gamma}_{\tau} 
(\partial_{\mu} u_{\mu\tau})(\partial_{\lambda} u_{\lambda\tau})  + \overline{\gamma}_{\bm r}\sum^{D-1}_{\nu=1}
(\partial_{\mu} u_{\mu\nu})(\partial_{\lambda} u_{\lambda\nu})  \Big\}\bigg] \nonumber \\
&\Bigg\{ 1 + \sum^{\infty}_{N=1} \frac{1}{N!} \prod^{N}_{j=1} \bigg(\int {\cal D}^D {\bm X}_j \int_{|{\bm Y}_j|>a_0 b} {\cal D}^D {\bm Y}_j  \! \  \exp\big[\overline{w}\big]\bigg)  \nonumber \\ 
& \exp\bigg[-\frac{ie}{2} \int_{|\partial u|<b^{-1}\Lambda|u|} d^D{\bm x}\,\ u_{\mu\nu}({\bm x}) v_{\mu\nu}({\bm x}) - i2\pi \overline{\chi} \sum^{N}_{j=1} {\cal A}_{\tau,j} \bigg] \Bigg\}, \label{Eq:Zv-10a}
 \end{align}
with 
\begin{align}
\overline{w} = w - \pi e^2 \big(\overline{\gamma}^{-1}_{\tau}+1\big) \ln b. 
\end{align}
 \subsection{\label{secIIc}renormalization group equation}
   After the integrations over the short-distance degrees of freedom, the partition function reduces to the same form as the original partition function, except for the renormalization of the coupling constants and the enlarged UV cutoff length. To put the enlarged UV length $a_0 b$ back into the original UV length $a_0$, we employ a length rescaling in terms of the RG scale factor $b$,
 \begin{align}
 {\bm x} \rightarrow {\bm x}^{\prime} = {\bm x} b^{-1},\!\ {\bm s}\rightarrow {\bm s}^{\prime}={\bm s}b^{-1}, \!\ {\bm Y}_{j} \rightarrow {\bm Y}^{\prime}_{j} = {\bm Y}_j b^{-1}, 
 {\bm X}_{\bm j} \rightarrow {\bm X}^{\prime}_j = {\bm X}_j b^{-1}, \!\ {\cal A}_{j,\tau} \rightarrow {\cal A}^{\prime}_{j,\tau} = {\cal A}_{j,\tau} b^{-(D-1)},  \label{Eq:length-rescale}
 \end{align}
 with $\partial_{\cdots} \rightarrow \partial^{\prime}_{\cdots} =b\partial_{\cdots} $, $\hat{\bf g}_{j,\cdots}({\bm s})\rightarrow \hat{\bf g}^{\prime}_{j,\cdots}({\bm s}^{\prime})=\hat{\bf g}_{j,\cdots}({\bm s})$, ${\sf g}_{j,\lambda\rho\cdots\sigma}({\bm s}) \rightarrow {\sf g}^{\prime}_{j,\lambda\rho\cdots\sigma}({\bm s}^{\prime})={\sf g}_{j,\lambda\rho\cdots\sigma}({\bm s})$, and $v_{\cdots}({\bm x}) \rightarrow v^{\prime}_{\cdots}({\bm x}^{\prime}) = b^2 v_{\cdots}({\bm x})$.  We choose the normalization of the dual vector potential $u_{\mu\nu}({\bm x})$ in such a way that the space component $\overline{\gamma}_{\bm r}$ of the electromagnetic constitutive constant is invariant after the renormalization, i.e. $\overline{\gamma}_{\bm r}=1$,  
 \begin{align}
 u_{\cdots}({\bm x}) &\rightarrow u^{\prime}_{\cdots}({\bm x}^{\prime}) = \bar{\gamma}^{\frac{1}{2}}_{\bm r} b^{\frac{D-2}{2}} u_{\cdots}({\bm x}).  \label{Eq:field-renormalization-1} 
 \end{align}
 This determines the partition function after renormalization as follows,  
 \begin{align}
  Z_{\rm v} =& \int {\cal D}{\bm u} \exp \bigg[-\frac{1}{2}\int_{|\partial^{\prime} u^{\prime}|<\Lambda|u^{\prime}|} d^D{\bm x}^{\prime}\!\ \Big\{ \gamma^{\prime}_{\tau} \!\ (\partial^{\prime}_{\mu} u^{\prime}_{\mu\tau})(\partial^{\prime}_{\lambda} u^{\prime}_{\lambda\tau})  + \sum^{D-1}_{\nu=1}
(\partial^{\prime}_{\mu} u^{\prime}_{\mu\nu})(\partial^{\prime}_{\lambda} u^{\prime}_{\lambda\nu})  \Big\}\bigg] \nonumber \\
&\Bigg\{ 1 + \sum^{\infty}_{N=1} \frac{1}{N!} \prod^{N}_{j=1} \bigg(\int {\cal D}^{D} {\bm X}^{\prime}_j \int_{|{\bm Y^{\prime}_j}|>a_0} {\cal D}^D {\bm Y}^{\prime}_j \int {\cal D}{\sf g}^{\prime}_j({\bm s}^{\prime}) \, \  \exp[w^{\prime}] \exp\bigg[ \int_{\partial \Gamma_j}^{\prime} d^{D-2}{\bm s}^{\prime}\,\ t^{\prime}\Big(\hat{\bf g}^{\prime}_{j,\tau},\hat{\bf g}^{\prime}_{j,{\bm r}}\Big)\bigg] \bigg)  \nonumber \\ 
& \exp\bigg[-\frac{ie^{\prime}}{2} \int_{|\partial^{\prime} u^{\prime}|<\Lambda|u^{\prime}|} d^D{\bm x}^{\prime}\,\ u^{\prime}_{\mu\nu}({\bm x}^{\prime}) v^{\prime}_{\mu\nu}({\bm x}^{\prime}) - i2\pi \chi^{\prime} \sum^{N}_{j=1} {\cal A}^{\prime}_{\tau,j} \bigg] \Bigg\}, \label{Eq:Zv-11}
\end{align}
with 
\begin{align}
&\gamma^{\prime}_{\tau} = \frac{\overline{\gamma}_{\tau}}{\overline{\gamma}_{\bm r}},  \,\ 
(e^{\prime})^2 = e^2 b^{D-2} \frac{1}{\overline{\gamma}_{\bm r}}, \,\  \chi^{\prime} = \overline{\chi} b^{D-1}, \label{Eq:rg9} 
\end{align}
and 
\begin{align}
\begin{cases}
  t^{\prime}\big(\hat{\bf g}^{\prime}_{\tau},\hat{\bf g}^{\prime}_{\bm r}\big)= b^{D-2}\overline{t}\big(\hat{\bf g}_{\tau},\hat{\bf g}_{\bm r}\big) , \,\ w^{\prime} = w + 2D \ln b, & \ \ \ {\rm for} \,\ D>2, \\
 w^{\prime} = \overline{w}+4 \ln b = w + 4\ln b - \pi e^2 (\gamma^{-1}_{\tau}+1) \ln b, 
 & \ \ \  {\rm for} \,\ D=2. \\
\end{cases}
\end{align}
From this, we obtain the coupled RG equations for $D>2$,  
\begin{align}
\frac{dw}{d\ln b} & = 2D, \label{Eq:w} \\
\frac{d t}{d \ln b} &= 
(D-2)\!\ t - 2\pi^2 e^2 {\rm det} \Big[ f^{-\frac{D-3}{D-2}}_{\bm r} f_{\tau} \hat{\bf g}_{\tau} + f^{\frac{1}{D-2}}_{\bm r}\hat{\bf g}_{\bm r} \Big], 
\label{Eq:t} \\
\frac{d \gamma_{\tau}}{d\ln b} &= 
16\pi^2 e^2 \frac{a^{D-2}_{0}}{2^{2D-2}} A^2 \int_{S_{D-1}} 
{\cal D}^{D-1}{\bm e}_0 \!\  e^{E_0(\theta)} \cos\Big[4\pi \chi \big(\frac{a_0}{2}\big)^{D-1} A \cos\theta\Big] \Big(\cos^2\theta-\gamma_{\tau} \frac{\sin^2\theta}{D-1}\Big),
\label{Eq:gammat} \\ 
\frac{d e^2}{d\ln b} &= (D-2)\!\ e^2 - 16\pi^2 e^4 \frac{a^{D-2}_{0}}{2^{2D-2}} \frac{A^2}{D-1} 
\int_{S_{D-1}} {\cal D}^{D-1}{\bm e}_0 \!\ e^{E_0(\theta)} \cos\Big[4\pi \chi \big(\frac{a_0}{2}\big)^{D-1} A \cos\theta\Big] \sin^2\theta, \label{Eq:e2} \\ 
\frac{d\chi}{d \ln b} & = (D-1) \chi - 4\pi e^2 \frac{1}{\gamma_{\tau}} 
\frac{a^{-1}_0}{2^{D-1}} A \int_{S_{D-1}} {\cal D}^{D-1}{\bm e}_0 \!\ e^{E_0(\theta)} \sin\Big[4\pi\chi 
\big(\frac{a_0}{2}\big)^{D-1} A \cos\theta \Big] \cos\theta. \label{Eq:chi}
\end{align}
$f_{\bm r}$ , $f_{\tau}$ and $t$ in Eq.~(\ref{Eq:t}) are scalar functions of the temporal and spatial components of the metric tensor,  $\hat{\bf g}_{\tau}$ and $\hat{\bf g}_{\bm r}$. As in Eq.~(\ref{Eq:fugacity-2}),  the vortex energy for each closed vortex surface is given by an ${\bm s}$-integral of  $t(\hat{\bf g}_{\tau}({\bm s}),\hat{\bf g}_{\bm r}({\bm s}))$ over the surface. The second term of Eqs.~(\ref{Eq:e2}, \ref{Eq:chi}) stands for the screening of the smaller vortex surface on the long-ranged Coulomb interaction. As for the smaller vortex surface, we considered the coplanar $(D-2)$-sphere, whose diameter is the lattice constant $a_0$, and who resides in a $(D-2)$-dimensional plane. ${\bm e}_0$ in Eqs.~(\ref{Eq:gammat},\ref{Eq:e2},\ref{Eq:chi}) is the unit vector perpendicular to the $(D-2)$-dimensional plane. The ${\bm e}_0$-integral over the $(D-1)$-sphere in Eqs.~(\ref{Eq:gammat},\ref{Eq:e2},\ref{Eq:chi}) enumerates all possible coplanar $(D-2)$-spheres. $\theta$ in Eqs.~(\ref{Eq:gammat},\ref{Eq:e2},\ref{Eq:chi}) stands for the angle between ${\bm e}_0$ and the $\tau$-axis. As discussed in a text below Eq.~(\ref{Eq:e0-angles}), the fugacity $E_0(\theta)$ of the coplanar vortex sphere is given by a symmetric function of $\theta$, $E_0(\theta)=E_0(\pi-\theta)$. 

     Eqs. (\ref{Eq:w},\ref{Eq:t}) indicates that vortex core energy density $t$ decreases/increases exponentially in the RG scale factor $\ln b$, while $w$ increases only in the power of $\ln b$. This suggests that whenever the core energy density decreases or increases, it always dominates over $w$ in the fugacity parameter $y_j = \exp[E_j]$.  Thus, for simplicity, we ignore the scaling equation of $w$ for the $D>2$ case.

      For $D=2$, the RG equations for $w$ and $t$ should be unified into an equation for $w$,   
\begin{align}
\frac{dw}{d\ln b} = 4-\pi e^2 (\gamma^{-1}_{\tau}+1), \label{Eq:w-2D}
\end{align}
while $E_0(\theta)$ in Eqs.~(\ref{Eq:e2}, \ref{Eq:gammat}, \ref{Eq:chi}) becomes a scalar quantity $w$.  In the following section, we analyze these RG equations in a few relevant cases. 

\section{\label{secIII}Phases, phase transitions, and universal criticality}
\subsection{\label{secIIIa} An $\epsilon = D-2$ expansion analysis on the phase transition at $\chi = 0$}
        In the absence of the Berry phase term $[\chi=0]$, the theory becomes isotropic, where $\gamma_{\tau}=1$ is invariant under renormalization, the fugacity of a vortex hyper-surface is proportional to the $(D-2)$-dimensional volume of the surface [Eq.~(\ref{Eq:fugacity-1})], and the fugacity $E_0(\theta)$ of the smallest coplanar vortex sphere becomes independent of $\theta$. In fact, Eqs.~(\ref{Eq:Fr}, \ref{Eq:Ft}) with $\gamma_{\tau}=1$ dictate that $F_{\tau}({\bm x})=F_{{\bm r}}({\bm x})\equiv F({\bm x})$ depends only on $|{\bm x}|=\sqrt{\tau^2+|{\bm r}|^2}$, and Eqs.~(\ref{Eq:fr},\ref{Eq:ft}) indicate that $f_{\tau}=f_{\bm r}\equiv f$ depends only on $\hat{\bf g}=\hat{\bf g}_{\tau} + \hat{\bf g}_{\bm r}$ [see Eq.~(\ref{Eq:1st-f})]. By calculating the integrals with a soft UV cutoff [see section.~\ref{secIVa}], one obtains, 
\begin{align}
f= f_{\tau} = f_{\bm r} = \frac{1}{2\pi}\frac{1}{\det[\hat{\bf g}]^{1/2}}. \label{Eq:isotropic-general-D}
\end{align} 
By substituting Eq.~(\ref{Eq:isotropic-general-D}) into Eq.~(\ref{Eq:t}), one can see that the fugacity renormalization is given by $\pi e^2 \det[\hat{\bf g}]^{1/2}$, justifying Eq.~(\ref{Eq:fugacity-1}): the fugacity of a vortex hyper-surface is proportional to the $(D-2)$-dimensional volume of the surface. Therefore, Eq~(\ref{Eq:t}) reduces to a differential equation for a real-valued scalar parameter $t$, 
\begin{align}
\frac{dt}{d\ln b} = (D-2) t - \pi e^2.  \label{Eq:t2} 
\end{align}
According to Eq.~(\ref{Eq:fugacity-1}), the fugacity for the smallest coplanar vortex sphere [$(D-2)$-sphere] in Eq.~(\ref{Eq:E0}) is independent of $\theta$,
\begin{align}
E_0 = t\!\  \big(\frac{a_0}{2}\big)^{D-2} \frac{2 \pi^{\frac{D-1}{2}}}{\Gamma\big(\frac{D-1}{2}\big)}, 
\label{Eq:E0-3}
\end{align}
so that the integral over ${\bm e}_0$ in Eqs.~(\ref{Eq:e2},\ref{Eq:gammat},\ref{Eq:chi}) can be easily calculated at $\chi=0$, 
\begin{align}
\frac{d e^2}{d\ln b} =& (D-2) e^2 - 16\pi^2 \!\ \exp[E_0]\!\ e^4 \frac{a^{D-2}_{0}}{2^{2D-2}}  \!\ \frac{A^2}{D-1}  \int_{S_{D-1}} {\cal D}^{D-1}{\bm e}_0  \sin^2\theta, \nonumber \\
=& (D-2) e^2  - 16\pi^2 \!\ \exp[E_0] \!\ e^4 \frac{a^{D-2}_0}{2^{2D-2}}  \!\ \frac{\pi^{\frac{D}{2}}}{\Gamma\big(\frac{D+2}{2}\big)} A^2, \label{Eq:e2-4}
\end{align}
with $d\gamma_{\tau}/d\ln b=0$, and $d\chi/d\ln b=0$. Here we used 
\begin{align}
\int_{S_{D-1}} {\cal D}^{D-1}{\bm e}_0 \sin^2\theta &= \int^{\pi}_{0} (\sin\theta)^{D} d\theta 
\int^{\pi}_{0} (\sin\theta^{\prime})^{D-3} d\theta^{\prime} \cdots \int^{\pi}_{0} 
\sin\theta^{\prime\prime}  d\theta^{\prime\prime} \int^{2\pi}_{0} d\theta^{\prime\prime\prime}  \nonumber \\
&= \frac{\sqrt{\pi}\Gamma\big(\frac{D+1}{2}\big)}{\Gamma\big(\frac{D+2}{2}\big)}\frac{\sqrt{\pi}\Gamma\big(\frac{D-2}{2}\big)}{\Gamma\big(\frac{D-1}{2}\big)} \cdots \frac{\sqrt{\pi}\Gamma\big(\frac{1+1}{2}\big)}{\Gamma\big(\frac{1+2}{2}\big)} 2\pi = (D-1) \frac{\pi^{\frac{D}{2}}}{\Gamma\big(\frac{D+2}{2}\big)}. 
\end{align}
 $d\gamma_{\tau}/d\ln b=d\chi/d\ln b=0$ justifies that $\gamma_{\tau}=1$ and $\chi=0$ are invariant under renormalization.  

Let us normalize $t$ and $e^2$ into dimensionless coupling constants, 
\begin{align}
t_{\rm new} = t_{\rm old}\!\ \big(\frac{a_0}{2}\big)^{D-2} \frac{2 \pi^{\frac{D-1}{2}}}{\Gamma\big(\frac{D-1}{2}\big)}, \,\ e^2_{\rm new} = \pi e^2_{\rm old} \!\ 
\big(\frac{a_0}{2}\big)^{D-2} \frac{2 \pi^{\frac{D-1}{2}}}{\Gamma\big(\frac{D-1}{2}\big)}. \label{Eq:old-new}
\end{align}
We obtain 
\begin{align}
\frac{dt}{d\ln b} = (D-2) t - e^2, \ \ \frac{de^2}{d\ln b} = (D-2) e^2 
- \frac{4\sqrt{\pi} }{D-1}
\frac{\big(\frac{\pi}{2}\big)^{D}}{\Gamma\big(\frac{D+1}{2}\big) \Gamma\big(\frac{D+2}{2}\big)} e^4 \!\ \exp[t].  \label{Eq:tande2}
\end{align}
With $D=2+\epsilon$ and $e^2 \equiv x$,  
\begin{align}
\frac{dt}{d\ln b} =  \epsilon t -x, \ \ \frac{dx}{d\ln b} = \epsilon x 
- \big(2 \pi^2 -\Gamma \epsilon \big) x^2 \exp[t].  \label{Eq:tande2-2}
\end{align}
where $\Gamma \equiv \pi^2(3-\gamma+\ln[4/\pi^2] +\psi^{(0)}(3/2))=15.36...$ . The differential equation has a fixed-point solution at  $x=x_*=\epsilon t_*$, $t=t_{*}=t_{(0),*}+\epsilon t_{(1),*}+{\cal O}(\epsilon^2)$. Here $t_{(0),*}$ and $t_{(1),*}$ satisfy  
\begin{align}
1 &=2\pi^2\!\  t_{(0),*} e^{t_{(0),*}}, \nonumber \\
t_{(1),*}&= \frac{\Gamma}{2\pi^2} \frac{t_{(0),*}}{1+t_{(0),*}} = 0.0359.., \nonumber  
\end{align}
with $t_{(0),*}=0.0483...$  Around the fixed point, the differential equations can be linearized,  
\begin{align} 
\frac{d}{d\ln b} \left(\begin{array}{c}
\delta t \\
\delta x \\
\end{array}\right) = \left(\begin{array}{cc} 
\epsilon & - 1 \\
-\epsilon^2 t_* & - \epsilon \\
\end{array}\right) \left(\begin{array}{c} 
\delta t \\
\delta x \\
\end{array}\right).  \label{Eq:rg-isotropic-linearized}
\end{align}
The matrix has positive and negative eigenvalues, 
\begin{align}
\lambda_{\pm}&=\pm \epsilon \sqrt{1+t_*}  = \pm \bigg(\epsilon \sqrt{1+t_{(0),*}} + \frac{1}{2} \epsilon^2 \frac{t_{(1),*}}{\sqrt{1+t_{(0),*}}} + {\cal O}(\epsilon^3)\bigg) \nonumber \\
&\simeq \pm (1.024 \epsilon + 0.0175 \epsilon^2), \nonumber 
\end{align}
indicating that the fixed point is a saddle fixed point characterizing the universal criticality of the ordered-disordered transition. Eigenvectors that belong to $\lambda_{\pm}$ are given by $(1\pm \sqrt{1+t_*},-\epsilon t_*)$.  This suggests that a relevant (thermal) scaling variable associated with $\lambda_+$ consists mostly of $\delta t$. A leading order estimate of the critical exponent for the ordered-disordered transition is
\begin{align}
\nu &= \frac{1}{\epsilon} \frac{1}{\sqrt{1+t_{(0),*}}} - \frac{1}{2} \frac{t_{(1),*}}{(1+t_{(0),*})^{\frac{3}{2}}} + {\cal O}(\epsilon)  
\simeq \frac{0.977}{\epsilon}-0.0167. 
\end{align}
When $\epsilon=1$ ($D=3$), $\nu\simeq 0.960$.  Note also that $x_*=\epsilon t_* ={\cal O}(\epsilon)$ at the saddle fixed point for $D=2+\epsilon$ justifies the perturbative treatment of $e\equiv \sqrt{x}$ in Sec.~\ref{secIIa} for small $\epsilon$ $[D\gtrsim 2]$.  

     Apart from the thermal relevant scaling variable, the Berry phase parameter $\chi$ is also a relevant (magnetic) scaling variable whose scaling dimension $y_{\chi}$ determines other critical properties around the ordered-disordered transition point at $\chi=0$.  To determine the scaling dimension $y_{\chi}$ of the Berry phase parameter $\chi$, let us linearize the RG equation for $\chi$ around the saddle-fixed point $(t_*,x_*)$ on the non-magnetic $[\chi=0]$ plane;
\begin{align}
\frac{d\chi}{d\ln b} = \bigg\{(D-1) - 16\pi^2 e^2 \chi A^2 \Big(\frac{a_0}{2}\Big)^{D-2} \frac{1}{2^D}\frac{\pi^{\frac{D}{2}}}{\Gamma\big(\frac{D+2}{2}\big)}\bigg\}\!\ \chi, \label{Eq:chi-scaling-0}
\end{align}
where we used 
\begin{align} 
\int_{S_{D-1}} d^{D-1}{\bm e}_0 \cos\theta^2 = \frac{1}{D-1}
\int_{S_{D-1}} d^{D-1}{\bm e}_0 \sin\theta^2 = \frac{\pi^{\frac{D}{2}}}{\Gamma\big(\frac{D+2}{2}\big)}. 
\end{align}
With Eq.~(\ref{Eq:old-new}), this becomes, 
\begin{align}
\frac{d\chi}{d\ln b} =&\Big\{ (D-1) - \frac{4\sqrt{\pi}}{D-1}\frac{\big(\frac{\pi}{2}\big)^{D}}{\Gamma\big(\frac{D+2}{2}\big)\Gamma\big(\frac{D+1}{2}\big)} \!\ e^2 \!\ \exp[t]  \Big\} \chi. \label{Eq:chi-scaling-2}
\end{align}
Eqs.~(\ref{Eq:chi-scaling-2},\ref{Eq:tande2}) indicates that the scaling dimension $y_{\chi}$ of the Berry phase parameter $\chi$ is always 1, being independent from the spatial dimension $D$,  
\begin{align}
\frac{d\chi}{d\ln b} =\Big\{(D-1)-(D-2)\Big\} \chi = \chi \,\ \,\ @ \,\  (e,t,\chi)=(e_*,t_*, 0).  \label{Eq:chi-scaling-3}
\end{align}

    $y_{\chi}=1$ at the fixed point $(e,t,\chi)=(e_*,t_*,0)$ originates from the absence of the renormalization factor to the electric charge $e$. To see this point, let us first include $\chi$ into $\partial_{\mu}u_{\mu\tau}$, and rewrite it into a form of an external magnetic field $\chi_h$ in the Maxwell action. Namely, with  $\chi_h \equiv \chi/e$ and $\partial_{\mu}u_{\nu\mu} + \chi_{h}\delta_{\nu\tau}\rightarrow \partial_{\mu}u_{\nu\mu}$, the Berry phase term in the partition function can be absorbed into the coupling term between the vector potential and vortex field,  
   \begin{align}
   \frac{ie}{2} \int d^D{\bm x} \Big\{ u_{\mu\nu}({\bm x}) v_{\mu\nu}({\bm x}) +2\chi_h b_{\tau}({\bm x}) \Big\} &= ie \int d^D{\bm x} \Big\{ \big(\partial_{\mu} u_{\nu\mu}({\bm x})\big) b_{\nu}({\bm x}) + \chi_h b_{\tau}({\bm x}) \Big\} \nonumber \\
   &\rightarrow ie \int d^D{\bm x} \big(\partial_{\mu} u_{\nu\mu}({\bm x})\big) b_{\nu}({\bm x}),
   \end{align}
while the transformation gives rise to the magnetic field term in the Maxwell action, 
   \begin{align}
   \frac{\gamma_{\tau}}{2} \int d^D{\bm x}
   \big(\partial_{\lambda} u_{\nu\lambda}\big)
   \big(\partial_{\mu} u_{\nu\mu}\big) \rightarrow   \frac{\gamma_{\tau}}{2} \int d^D{\bm x}
   \Big\{\big(\partial_{\lambda} u_{\nu\lambda}\big)
   \big(\partial_{\mu} u_{\nu\mu}\big) - 2\chi_h \partial_{\lambda} u_{\tau\lambda}\Big\}.
   \end{align}
Thereby, the partition function can be generally rewritten into another form as well:
\begin{align}
Z_{\rm v} =& \int {\cal D}{\bm u} \, \  \exp \bigg[-\frac{1}{2}\int_{|\partial u|<\Lambda|u|} d^D{\bm x} \!\  \Big\{ 
(\partial_{\mu} u_{\mu\nu})(\partial_{\lambda} u_{\lambda\nu}) + 2\chi_{h} (\partial_{\lambda} u_{\lambda\tau})  \Big\}\bigg]\nonumber \\
& \Bigg\{ 1 + \sum^{\infty}_{N=1} \frac{1}{N!} \prod^N_{j=1} \bigg(\int {\cal D}^D {\bm X}_j \int_{|{\bm Y}_j|>a_0} {\cal D}^D {\bm Y}_j \int {\cal D}{\sf g}_j({\bm s}) \, \  \exp\big[ E_{j}\big]\bigg) \nonumber \\
&\exp\Big[-\frac{i}{2}  e \int_{|\partial u|<\Lambda |u|} d^D{\bm x}\,\ u_{\mu\nu}({\bm x}) v_{\mu\nu}({\bm x})\Big]\Bigg\}. \label{Eq:Zv-12}
\end{align}
The partition function undergoes an ordered-disordered phase transition. When $\chi_h=0$ [the space-time isotropic case],  the universal criticality of the phase transition is characterized by the saddle-fixed point at $(e,t)=(e_{*},t_{*})$ with $E_{j}= t \int_{\partial \Gamma_j} d^{D-2}{\bm s} \{\det \hat{\bf g}_{j}({\bm s})\}^{1/2}$.  The form of Eq.~(\ref{Eq:Zv-12}) suggests that the scaling property of $\chi_h$ around the isotropic saddle fixed point is determined by the field renormalization $Z_u$ of the dual vector potential at the fixed point;
\begin{align}
 \chi_h \rightarrow \chi^{\prime}_h = b^{D-1}  Z^{-\frac{1}{2}}_{u} b^{-\frac{D-2}{2}} \chi_h,  \label{Eq:chih-renormalize}
\end{align}
with 
\begin{align}
&u_{\cdots}(\bm x) \rightarrow u^{\prime}_{\cdots}({\bm x}^{\prime}) = Z^{\frac{1}{2}}_{u} b^{\frac{D-2}{2}} u_{\cdots}({\bm x}).  \label{Eq:u-renormalize} 
\end{align}
     Notice that $Z_{u}=b^{D-2}$ at the saddle fixed point. This is because the integration of the short-distance degrees of freedom at $\chi_h=0$ yields only the renormalization to the electromagnetic constitutive constant and vortex fugacity, while keeping intact the electric charge $e$,  
\begin{align}
Z_{\rm v} =&  \int {\cal D}{\bm u} \, \  \exp \bigg[-\frac{1}{2}\int_{|\partial u|<\Lambda |u|} d^D{\bm x} \!\  
(\partial_{\mu} u_{\mu\nu})(\partial_{\lambda} u_{\lambda\nu})  \bigg] 
\,\ \Bigg\{ 1 + \sum^{\infty}_{N=1} \frac{1}{N!}\nonumber \\
&  \prod^N_{j=1} \bigg(\int {\cal D}^D {\bm X}_j \int_{|{\bm Y}_j|>a_0} {\cal D}^D {\bm Y}_j \int {\cal D}{\sf g}_j({\bm s}) \, \  \exp\big[ t \int_{\partial \Gamma_j} d^{D-2}{\bm s} \sqrt{\det \hat{\bf g}({\bm s})} \big]\bigg) \nonumber \\
&\exp\Big[-\frac{i}{2}  e \int_{|\partial u|<\Lambda |u|} d^D{\bm x}\,\ u_{\mu\nu}({\bm x}) v_{\mu\nu}({\bm x})\Big]\Bigg\} \nonumber \\
=& \int {\cal D}{\bm u} \, \  \exp \bigg[-\frac{1}{2}\int_{|\partial u|<\Lambda b^{-1}|u|} d^D{\bm x} \!\  (1+ \delta \gamma)
(\partial_{\mu} u_{\mu\nu})(\partial_{\lambda} u_{\lambda\nu})  \bigg] \,\ 
\Bigg\{ 1 + \sum^{\infty}_{N=1} \frac{1}{N!}  \nonumber \\
&  \prod^N_{j=1} \bigg(\int {\cal D}^D {\bm X}_j \int_{|{\bm Y}_j|>a_0 b} {\cal D}^D {\bm Y}_j \int {\cal D}{\sf g}_j({\bm s}) \, \  \exp\big[ (t+\delta t)  \int_{\partial \Gamma_j} d^{D-2}{\bm s} \sqrt{\det \hat{\bf g}({\bm s})}  \big]\bigg) \nonumber \\
&\exp\Big[-\frac{i}{2}  e \int_{|\partial u|<\Lambda b^{-1}|u|} d^D{\bm x}\,\ u_{\mu\nu}({\bm x}) v_{\mu\nu}({\bm x})\Big]\Bigg\}. \label{Eq:Zv-13}
\end{align}
As a result, the scaling property of the electric charge $e$ is entirely determined by $Z_u$, 
 \begin{align} 
 e \rightarrow e^{\prime} =  \frac{b^{\frac{D-2}{2}}}{\sqrt{Z_u}} e, \label{Eq:scaling-e1}
 \end{align}
with $Z_{u}=1+\delta \gamma$. This yields the following identity at the fixed point with $e^{\prime}=e=e_*$,     
\begin{align}
{Z_u}_{|e^2=e^2_*, t=t_{*}} = b^{D-2}. \label{Eq:SFP}
\end{align}
Then, from Eq.~(\ref{Eq:chih-renormalize}), the scaling dimension of $\chi_h$ around the saddle fixed point must always be one,   
\begin{align}
\chi^{\prime}_h  = b \chi_h.  
\end{align}
The scaling dimension $y_{\chi}$ of the Berry phase parameter $\chi$  is also 1 as well, being independent of the space-time dimension $D$,
\begin{align}
\chi^{\prime} \equiv \chi^{\prime}_h e^{\prime} =b \chi_h e =  b\chi.
\end{align} 

\subsection{\label{secIIIb}$D=2$ in the presence of the Berry phase term $(\chi\ne 0)$} 
   Let us next analyze the $\chi\ne 0$ RG equation in $D=2$. At $D=2$, $A=1$, and $E_0=w$, and $\int D {\bm e}_0=\int^{2\pi}_{0} d\theta$. With a change of variables, $y \equiv e^w$ and $\nu \equiv 2\pi a_0 \chi$, this leads to  
\begin{align}
\frac{dy}{d\ln b}&= \big(4-\pi (\gamma^{-1}_{\tau}+1) e^2\big)y, \label{Eq:y-2D}  \\
\frac{d\gamma_{\tau}}{d\ln b} &= 4\pi^2 e^2 y \int^{2\pi}_{0} d\theta \cos\big(
\nu\cos\theta\big) \Big(\frac{1-\gamma_{\tau}}{2}  + \frac{1+\gamma_{\tau}}{2} \cos2\theta\Big), \nonumber \\
& = 8\pi^3 e^2 y \Big(\frac{1-\gamma_{\tau}}{2} J_{0}(\nu) + \frac{1+\gamma_{\tau}}{2} J_{2}(\nu)\Big), 
\label{Eq:gammat-2D}  \\
\frac{de^2}{d\ln b} &= - 4\pi^2 e^4 y \int^{2\pi}_{0} d\theta \cos\big(\nu\cos\theta\big) 
\frac{1-\cos2\theta}{2} = -8\pi^3 e^4 y \frac{J_0(\nu)+J_2(\nu)}{2}, \label{Eq:e2-2D}  \\ 
\frac{d\nu}{d\ln b} & = \nu - 4\pi^2 e^2 \frac{y}{\gamma_{\tau}} \int^{2\pi}_{0} 
\sin\big(\nu\cos\theta\big) \cos\theta = \nu - 8\pi^3 e^2 \frac{y}{\gamma_\tau} J_1(\nu), \label{Eq:nu-2D}  
\end{align}
with the Bessel function of the first kind $J_{n}(x)$. On the non-magnetic plane [$\gamma_{\tau}=1$ and $\nu=0$], the RG equation shows the celebrated phase diagram of the Berezinskii-Kosterlitz-Thouless (BKT) phase transition:
\begin{align}
\frac{dy}{d\ln b} = (4-2\pi e^2)y, \,\ \frac{de^2}{d\ln b} = -4\pi^3 e^4 y, 
\end{align}
with the universal KT transition `temperature'; $1/e^2_{\rm KT}=\pi/2$.  A phase boundary between  KT and disordered phases is given by the following separatrix,  
\begin{align}
y = \frac{1}{2\pi^2} \ln \Big(\frac{\pi}{2} e^2 \Big) - \frac{1}{2\pi^2} + \frac{1}{\pi^3} \frac{1}{e^2}, \label{Eq:separatrix} 
\end{align}
with $e^2>e^2_{\rm KT}$.

    The $e^4$ term in the right hand side of Eq.~(\ref{Eq:e2-2D}) represents the screening onto the 2D Coulomb interaction, which shows an oscillatory behavior as a function of the Berry phase parameter $\nu$. The unphysical oscillation of the screening is an artifact of the sharp UV cut-off in Eqs.~(\ref{Eq:Zv-3},\ref{Eq:Zv-4}). In fact, the oscillation disappears when one uses a soft UV cut-off in these equations~\cite{giamarchi2003}, 
\begin{align}
\int D^2 {\bm X}_0 \int_{a_0<|{\bm Y}_0|<a_0 b} D^2 {\bm Y}_0 =& a^{-4}_0 \int d^{2}{\bm X}_0 \int^{a_0 b}_{a_0} Y_0 d Y_0  
\int^{2\pi}_0 d\theta \nonumber \\
\rightarrow& \frac{1}{2} a^{-4}_0 \int d^{2}{\bm X}_0 \,\  a_0 \ln b \frac{\partial}{\partial a_0} \Big(\int^{\infty}_{0} dY_0 \!\ Y_0 e^{-Y_0/a_0} \Big)  
\int^{2\pi}_0 d\theta \nonumber \\
=& \frac{1}{2} a^{-2}_0 \ln b \int d^{2}{\bm X}_0 \int^{\infty}_{0} dy_0 \!\ y^2_0 e^{-y_0}  
\int^{2\pi}_0 d\theta, 
\end{align}
with $y_0 \equiv Y_0/a_0$. Thereby, the soft UV cut-off modifies the renormalization factors to the constitutive constants and the 1D Berry phase term in Eqs.~(\ref{Eq:rg3},\ref{Eq:rg4},\ref{Eq:rg5},\ref{Eq:rg6},\ref{Eq:rg7}),
\begin{align}
\overline{\gamma}_{\tau} &= 
\gamma_{\tau} + 2\pi^3 \ln b \!\ e^w  e^2 \int^{\infty}_0 dy_0 \!\ y^4_0 e^{-y_0} \big(J_0(\nu y_0) - J_{2}(\nu y_0)\big), \nonumber \\
\overline{\gamma}_{r} & = 1 + 2\pi^3 \ln b \!\ e^w e^2  \int^{\infty}_0 dy_0 \!\ y^4_0 e^{-y_0} \big(J_0(\nu y_0) + J_{2}(\nu y_0)\big), \nonumber \\
\overline{\chi} &=\chi - 2\pi^2 \ln b \!\ \frac{e^2}{\gamma_{\tau}}  e^w  a^{-1}_0  \int^{\infty}_0 dy_0 \!\ y^3_0 e^{-y_0} J_{1}(\nu y_0).  
\end{align}
With $y \equiv e^w$, $\nu \equiv 2\pi a_0 \chi$ and $e^2_{\rm new}= 4\pi^2 e^2_{\rm old}$, we finally obtain $\chi\ne 0$ RG equations for $D=2$ as follows,  
\begin{align}
\frac{d\gamma_{\tau}}{d\ln b} &= \pi y e^2 \frac{12-81 \nu^2 + 12\nu^4 - \gamma_{\tau}\big(12+9\nu^2-3\nu^4\big)}{(\nu^2+1)^{\frac{9}{2}}} 
\equiv \pi y e^2  \big(z_{2,\tau}(\nu)-\gamma_{\tau} z_{2,r}(\nu)\big), \label{Eq:gammat-2D-rev} \\
\frac{de^2}{d\ln b} &= - \pi y e^4 \frac{12 + 9 \nu^2 - 3\nu^4}{(\nu^2+1)^{\frac{9}{2}}} \equiv - \pi y e^4 z_{2,r}(\nu)
, \label{Eq:e2-2D-rev} \\
\frac{d\nu}{d\ln b} &= \nu - \frac{\pi ye^2}{\gamma_{\tau}} \frac{3\nu(4-\nu^2)}{(\nu^2+1)^{\frac{7}{2}}} \equiv \nu - \frac{\pi ye^2}{\gamma_{\tau}} z_{1}(\nu), \label{Eq:nu-2D-rev} \\
\frac{dy}{d\ln b} & = \Big(4-\frac{\gamma^{-1}_{\tau}+1}{4\pi} e^2\Big) y, \label{Eq:y-2D-rev} 
\end{align}
with 
\begin{align}
z_{1}(\nu)&\equiv \int^{\infty}_{0} dy \!\ y^3 e^{-y} J_1(\nu y) = \frac{3\nu (4-\nu^2)}{(\nu^2+1)^{\frac{7}{2}}}, \nonumber \\
z_{2,\tau}(\nu) &\equiv \frac{1}{2}\int^{\infty}_0 dy \!\ y^4 e^{-y} \big(J_0(\nu y) - J_2(\nu y)\big) = \frac{12-81 \nu^2 + 12 \nu^4}{(\nu^2+1)^{\frac{9}{2}}}, \nonumber \\
z_{2,r}(\nu) &\equiv \frac{1}{2}\int^{\infty}_0 dy \!\ y^4 e^{-y} \big(J_0(\nu y) + J_2(\nu y)\big) = \frac{12 + 9 \nu^2 -3 \nu^4}{(\nu^2+1)^{\frac{9}{2}}}= \frac{z_{1}(\nu)}{\nu}. \nonumber 
\end{align}
$z_{2,\tau}(\nu)$, and $z_{2,r}(\nu)$ are even functions of $\nu$, while $z_{1}(\nu)$ is an odd  function of $\nu$ [Fig.~\ref{Fg:zzz}]. 
\begin{figure}
\includegraphics[width=0.9\textwidth]{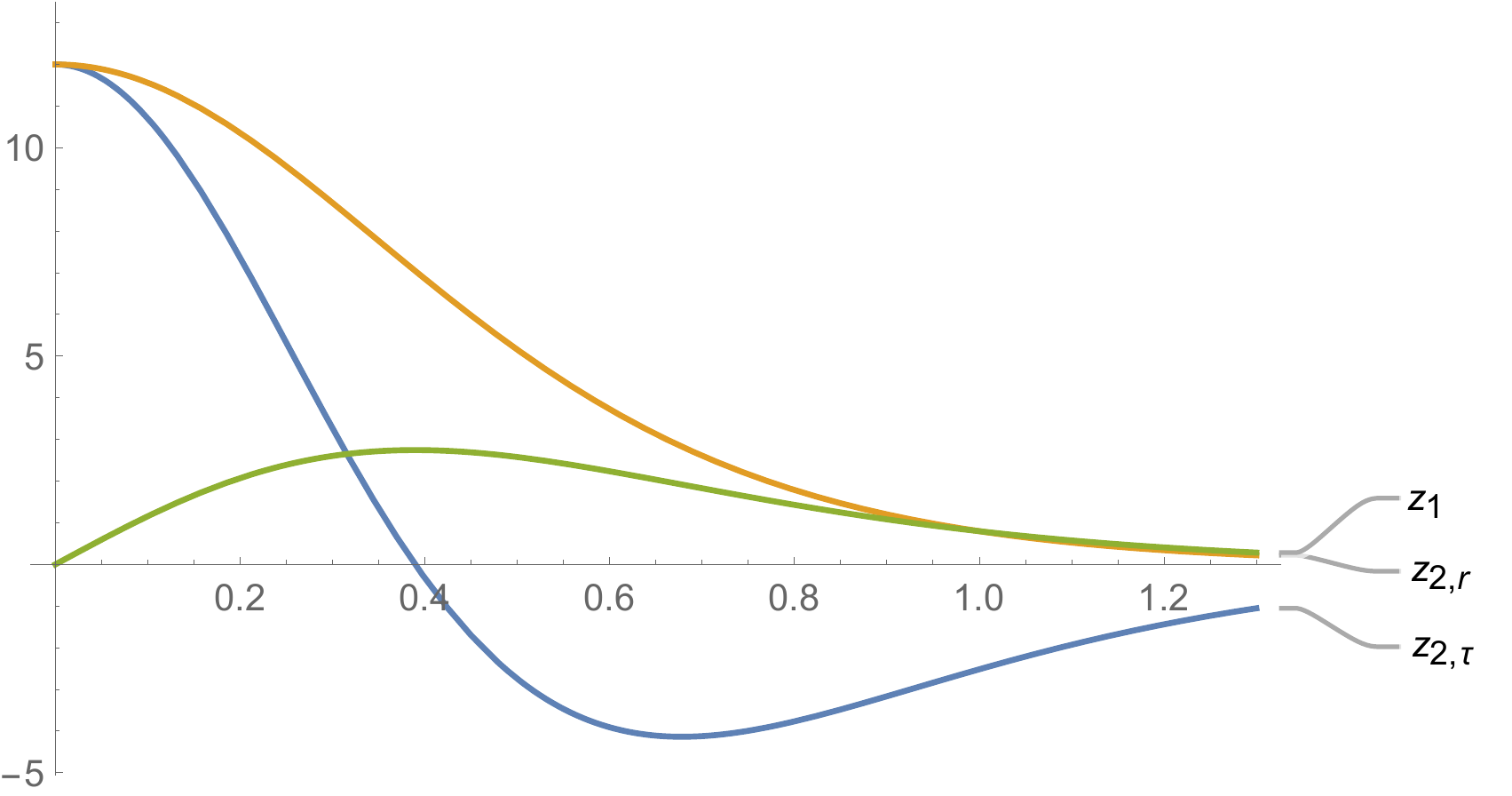}
\caption{$z_1(\nu)$ (green), $z_{2,\tau}(\nu)$(blue) and $z_{2,r}(\nu)$(orange) as functions of $\nu$}
\label{Fg:zzz}
\end{figure}

\begin{figure}[h]
\subfigure[$\nu=0.2$]{
\includegraphics[width=0.45\textwidth]{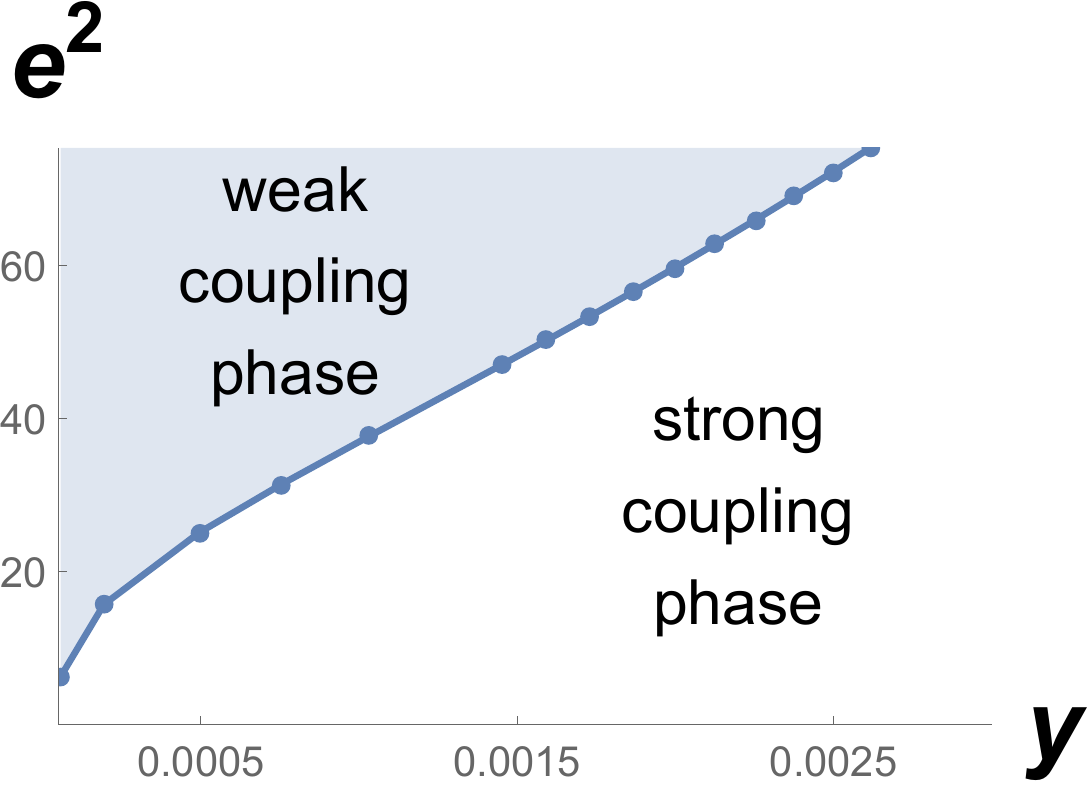}
}
\hfill
\subfigure[$\nu=0.6$]{
\includegraphics[width=0.45\textwidth]{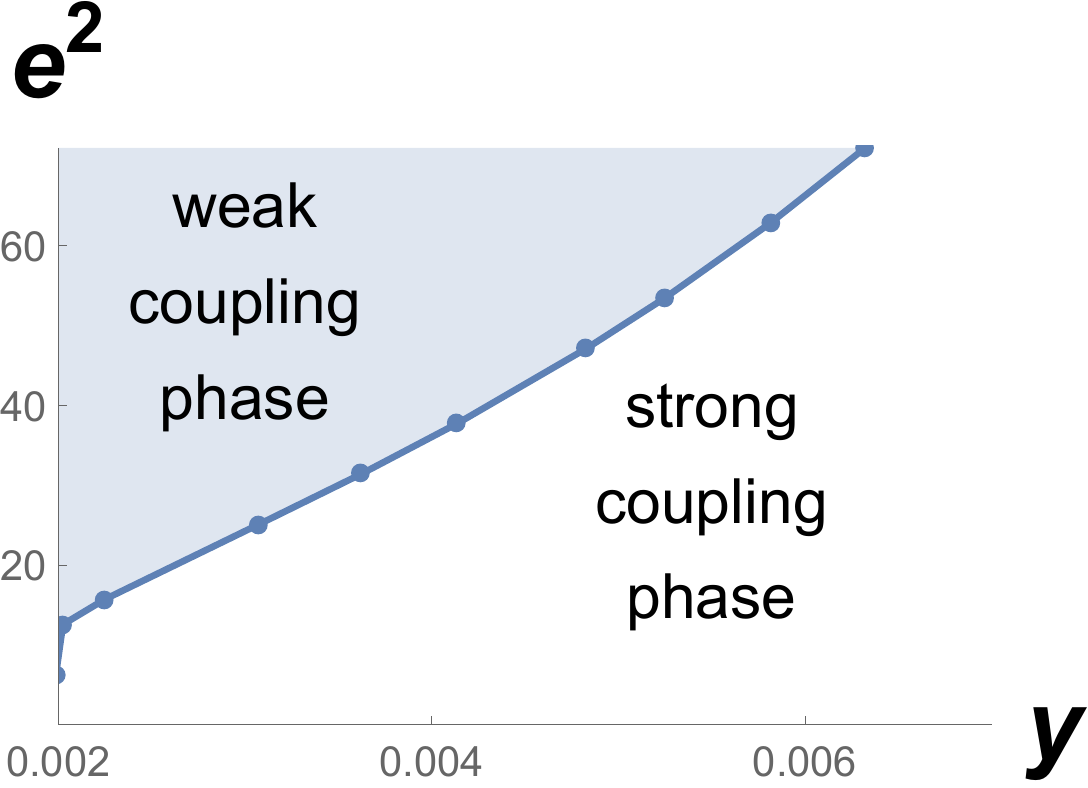}
}
\caption{numerical RG phase diagram with initial values of $\gamma_{\tau}=1$ and $\nu=0.2$ (left) and $\nu=0.6$ (right). The horizontal and vertical axes are initial values of $y$, and $e^2$, respectively. When the initial parameters are in the blue and white-colored regions, the RG equations drive them into the weak-coupling fixed line and strong-coupling fixed point, respectively.}
\label{Fg:2D-pd}
\end{figure} 

         The RG equation has an anisotropic weak-coupling fixed line with finite $e^2$ and $\gamma_{\tau}<1$, around which the screening onto $e^2$, $\gamma^{-1}_{\tau} e^2$ and $\nu$ are vanishingly small. The weak coupling fixed line characterizes a similar critical phase as the Kosterlitz-Thouless phase. The strong coupling fixed point of Eqs.~(\ref{Eq:gammat-2D-rev},\ref{Eq:e2-2D-rev},\ref{Eq:nu-2D-rev},\ref{Eq:y-2D-rev}) turns out to be identical to the isotropic strong coupling fixed point on the non-magnetic plane $[\nu=0]$ with $\gamma_\tau=1$, divergent fugacity parameter $y$, and vanishing $e^2$ and $\nu$, around which the screening effects reduce both $e^2$ and $\nu$ to zero. The strong coupling fixed point represents the conventional 2D disordered phase. An ordered-disordered phase transition from the weak-coupling KT-like phase to the strong-coupling conventional disordered phase is characterized by a saddle fixed point with an {\it infinite} space-time anisotropy $[\gamma_{\tau}=0]$ together with $e^2=0$, and finite universal values of $y$, $\nu$, and $e^2/\gamma_{\tau}$. In the following, we will explain them in details. 
         
         The weak-coupling fixed line is characterized by the absence of the screening, divergent Berry phase parameter $\nu$, and finite (non-universal) values of $e^2$ and $\gamma^{-1}_{\tau} e^2$.  When the value of $(\gamma^{-1}_{\tau}+1) e^2$ is greater than $16\pi$, the fugacity parameter $y$ vanishes exponentially in the large $\ln b$ limit.  Otherwise, $y$ diverges exponentially in the large $\ln b$ limit, $y \propto b^{\alpha}$ with $\alpha<4$.  In either case, the quenching of screening in Eqs.~(\ref{Eq:gammat-2D-rev},\ref{Eq:e2-2D-rev},\ref{Eq:nu-2D-rev}) is induced by the divergent $\nu \propto b$ , where a vanishingly small $ z_{2,\tau(r)}(\nu)  \propto \nu^{-5} \propto b^{-5}$ in the large $\ln b$ limit suppresses the second terms in Eqs.~(\ref{Eq:e2-2D-rev},\ref{Eq:nu-2D-rev}) irrespective of whether $y$ is divergent or vanishing. Thereby, around the weak-coupling fixed line, the RG equation reduces to $d\gamma_{\tau}/d\ln b = de^2/d\ln b =0$, $d\nu/d\ln b= \nu$, and $dy/d\ln  b = (4-(\gamma^{-1}_{\tau}+1)e^2/4\pi) y$. Note that $z_{2,\tau}(\nu)$ is mostly smaller than $z_{2,r}(\nu)$, so that $\gamma_{\tau}$ is mostly decreasing and the convergent value of $\gamma_{\tau}$ is usually less than 1.

\begin{figure}[t]
\subfigure[$e^2$ and $e^2/\gamma_{\tau}$ as functions of $\ln b$]{
\includegraphics[width=0.3\textwidth]{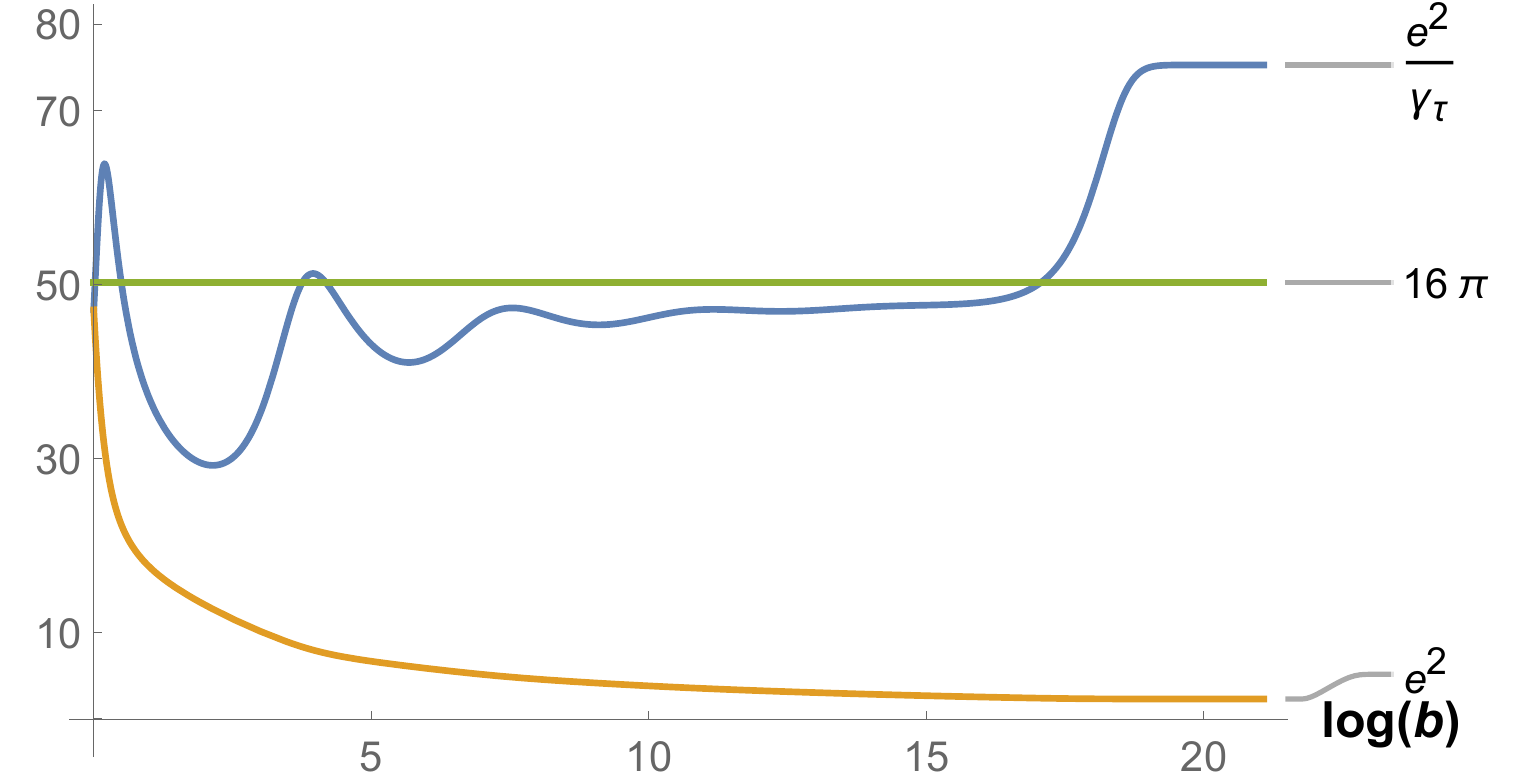}
}
\hfill
\subfigure[$\nu$ as a function of $\ln b$ where $\nu_{-}=0.389...$]{
\includegraphics[width=0.3\textwidth]{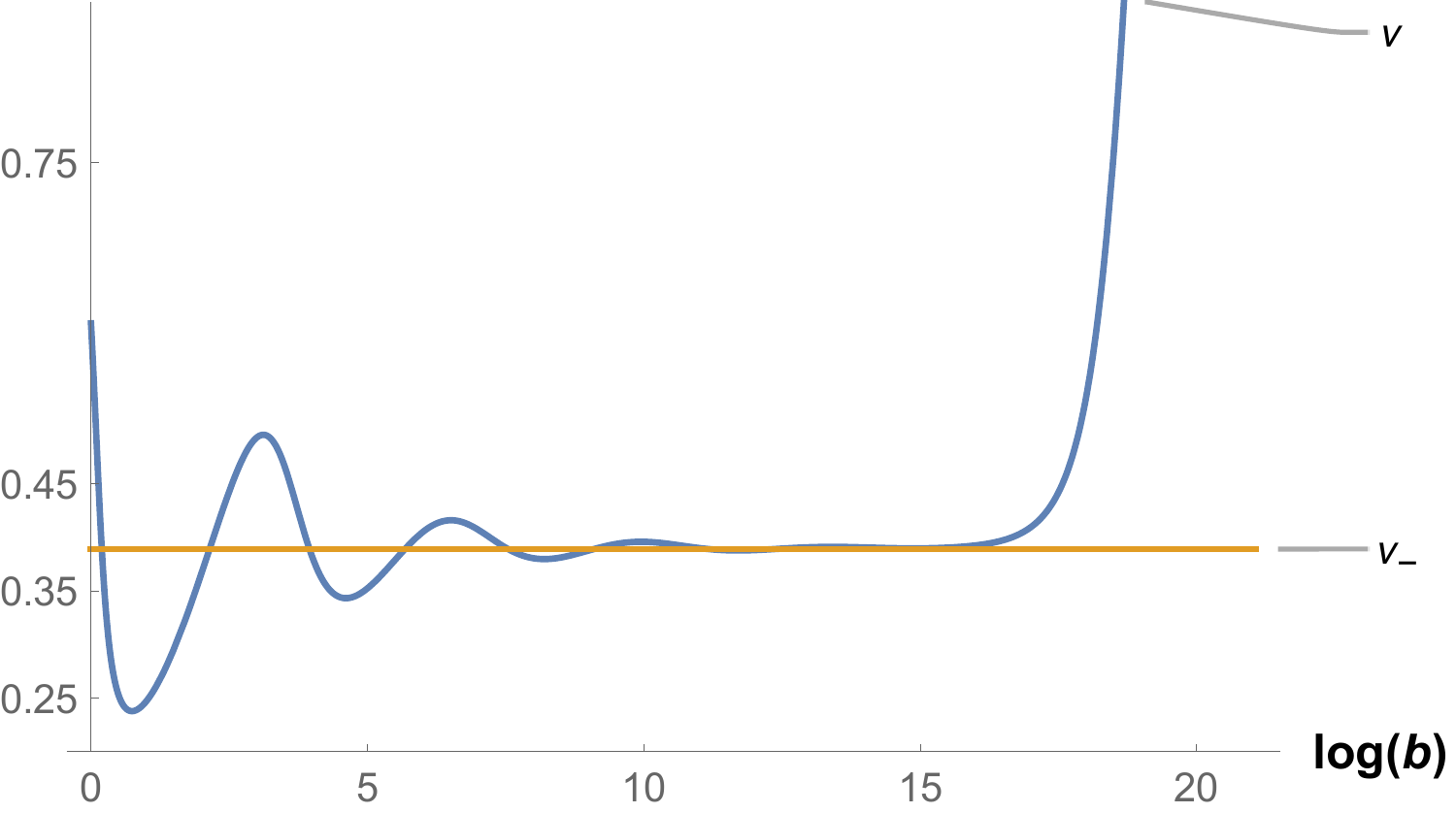}
}
\hfill
\subfigure[$y$ as a function of $\ln b$ where $y_{-}=0.000899...$]{
\includegraphics[width=0.3\textwidth]{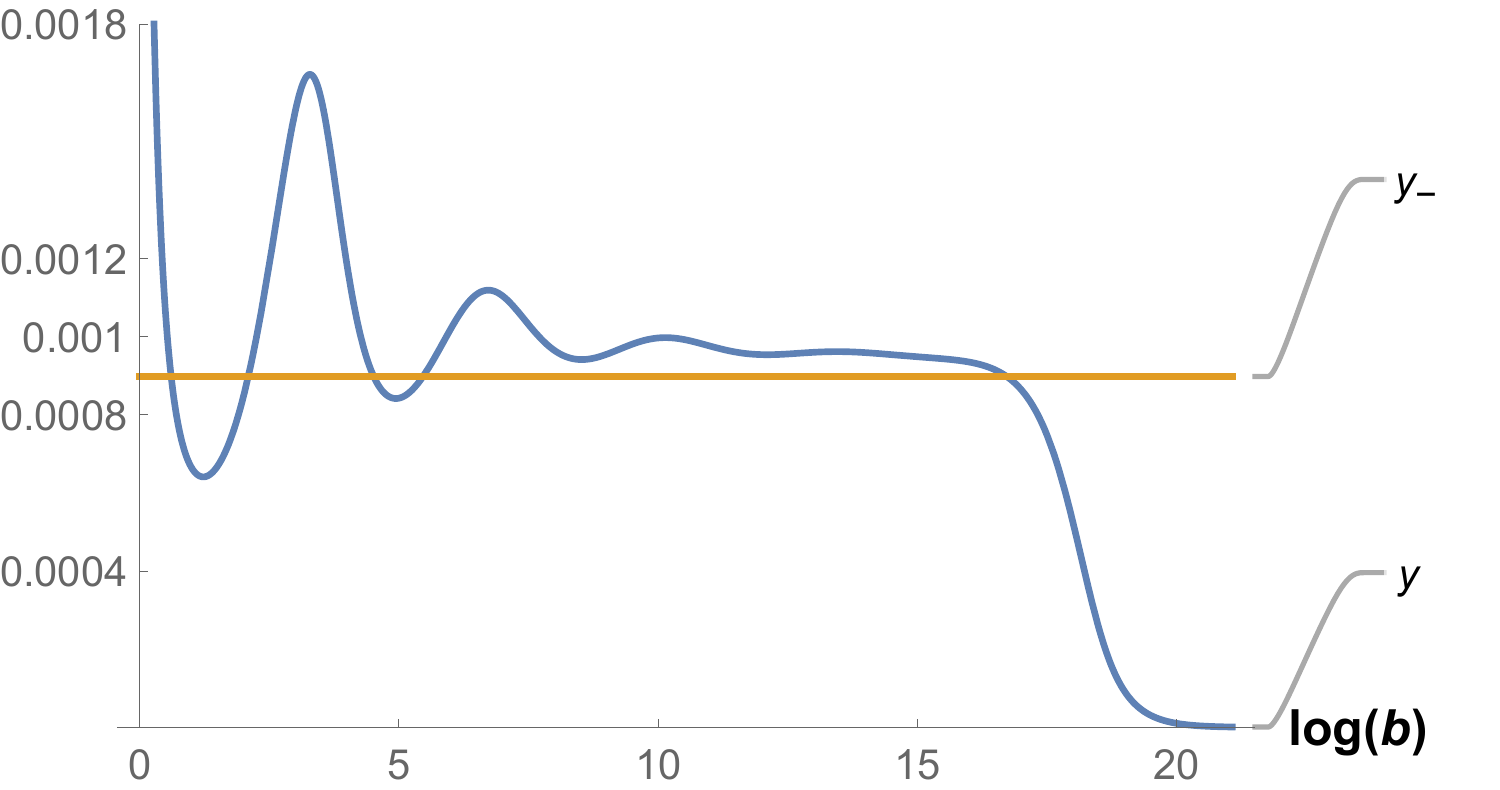}
}
\caption{Numerical solutions of $e^2/\gamma_{\tau}$, $e^2$, $\nu$ and $y$ as functions of the RG scale $\ln b$ for a set of initial parameters closed to the phase boundary between ordered and disordered phases. The initial parameters are chosen in the ordered phase side of the phase boundary: $e^2/\gamma_{\tau}=e^2=15\pi$ and $\nu=0.6$ and $y\lesssim y_c$ where $y_{c}=0.004824...$ (weak-coupling ordered phase).}
\label{Fg:2D-ordered}
\end{figure}
\begin{figure}[t]
\subfigure[$e^2/\gamma_{\tau}$ and $e^2$ as functions of $\ln b$]{
\includegraphics[width=0.3\textwidth]{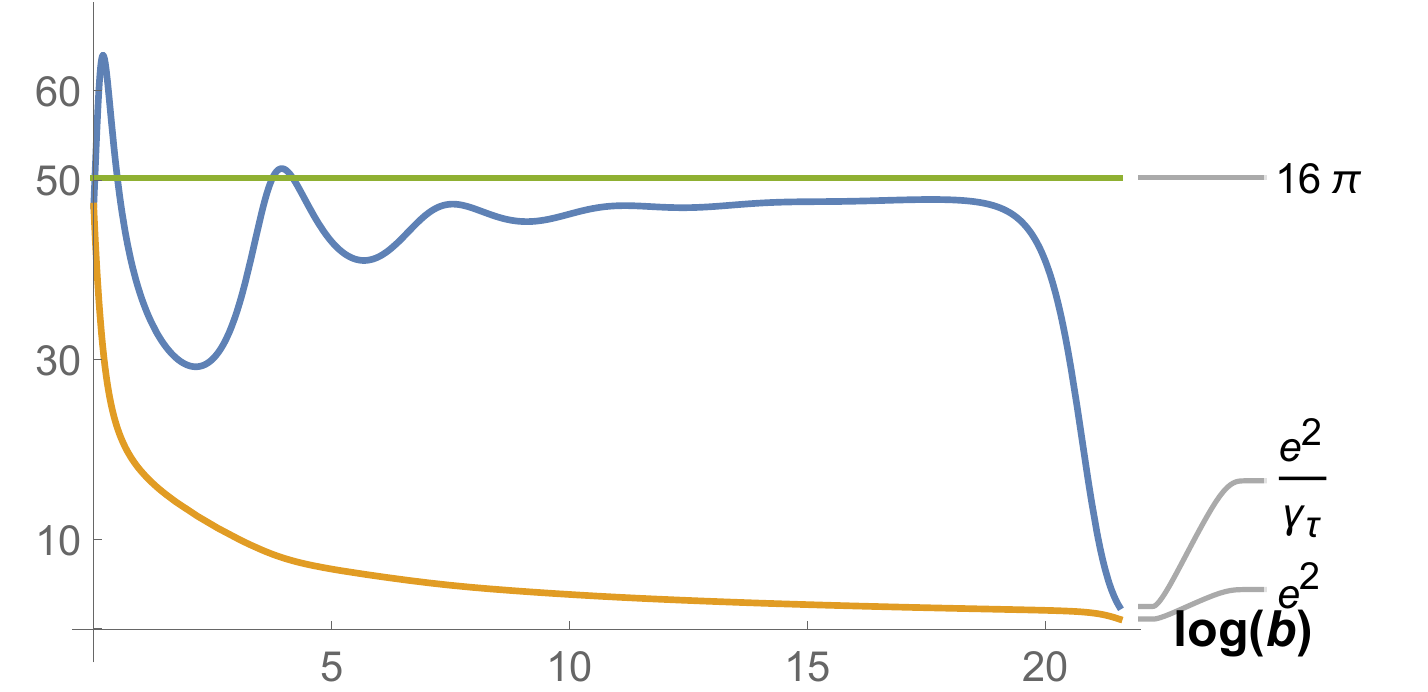}
}
\hfill
\subfigure[$\nu$ as a function of $\ln b$, where $\nu_{-}=0.389...$]{
\includegraphics[width=0.3\textwidth]{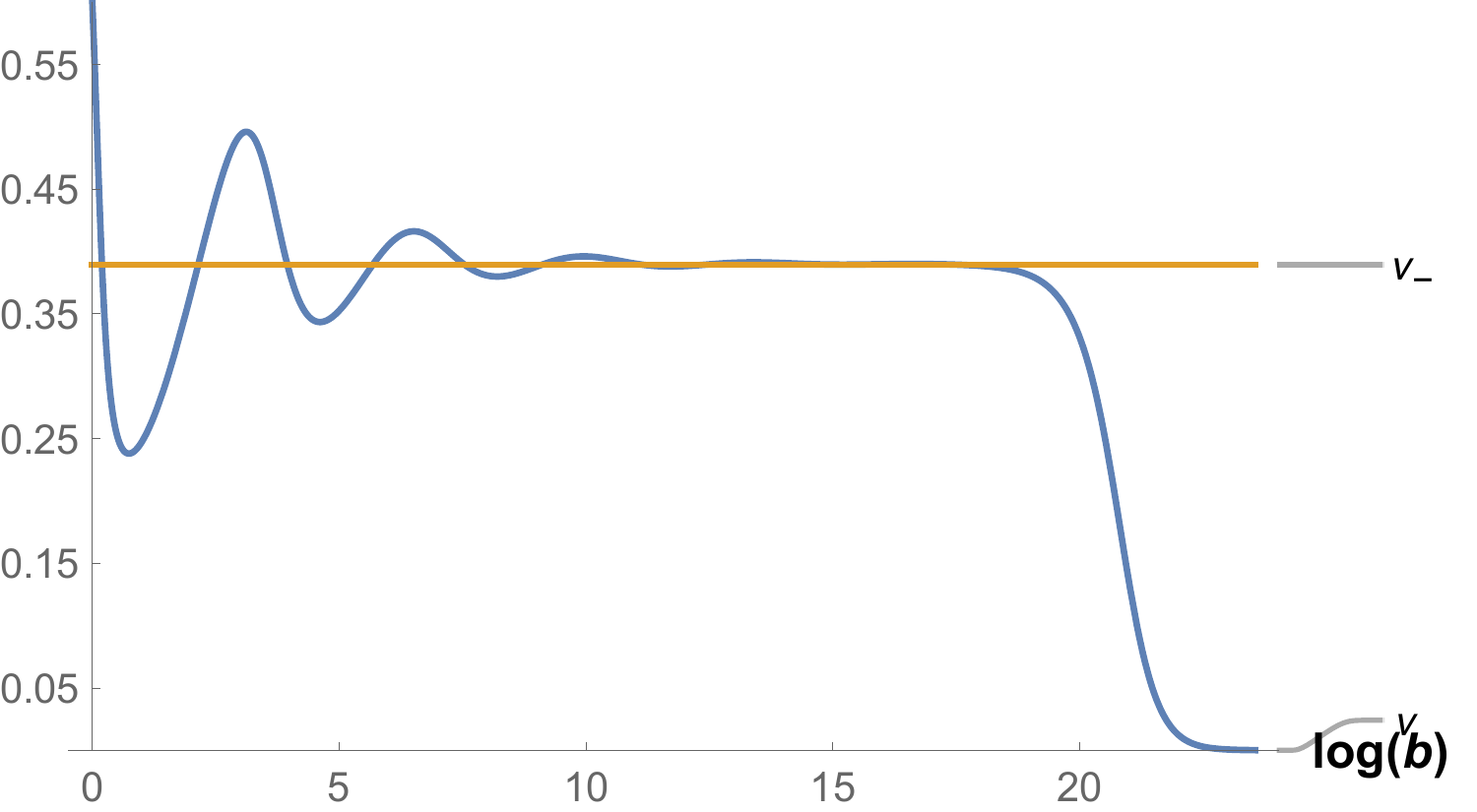}
}
\hfill
\subfigure[$y$ as a function of $\ln b$, where $y_{-}=0.000899...$]{
\includegraphics[width=0.3\textwidth]{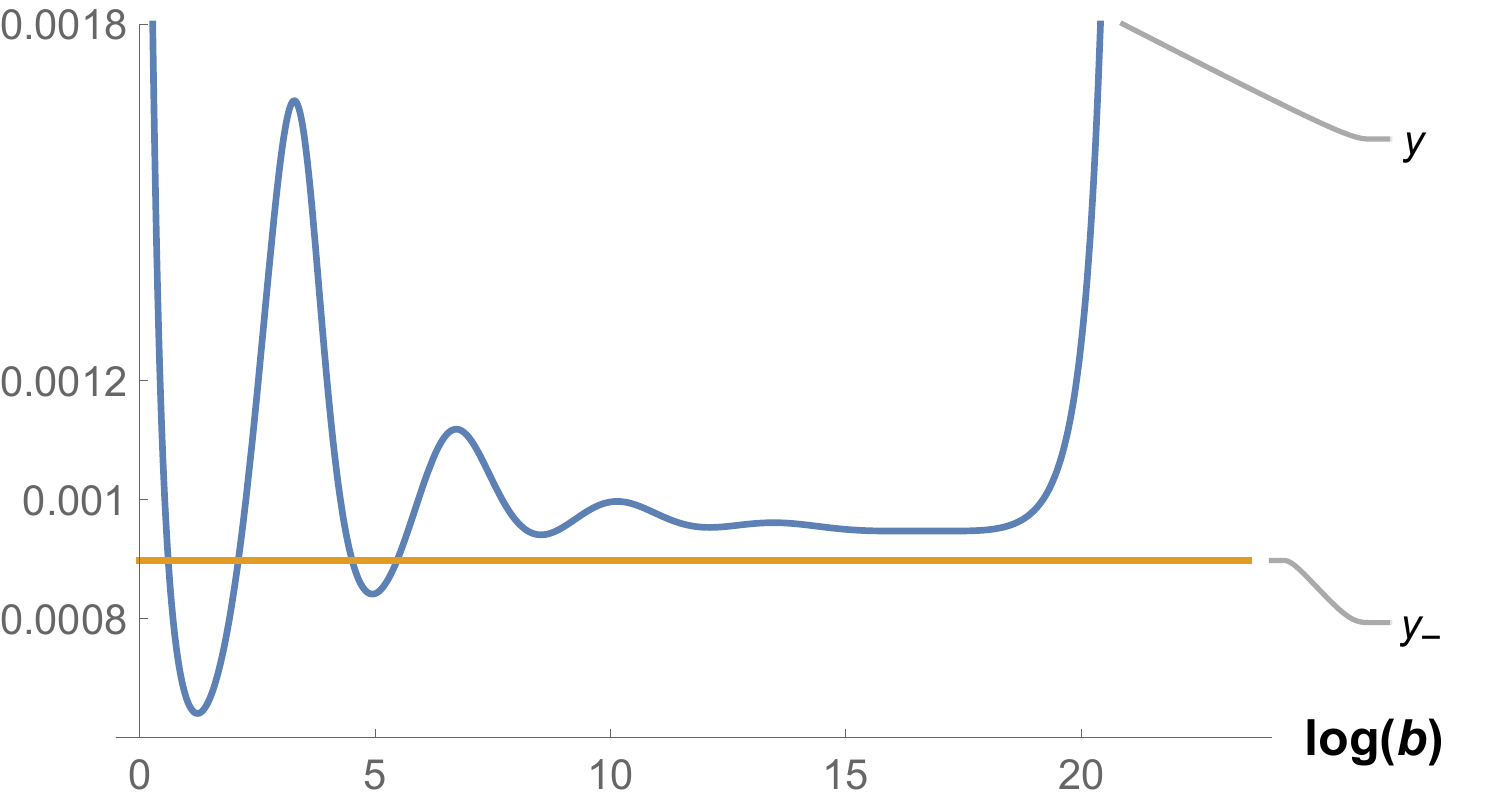}
}
\caption{Numerical solutions of $e^2/\gamma_{\tau}$, $e^2$, $\nu$ and $y$ as functions of the RG scale $\ln b$ for a set of initial parameters closed to the phase boundary between ordered and disordered phases. The initial parameters are chosen in the disordered phase side of the phase boundary:  with $e^2/\gamma_{\tau}=e^2=15\pi$ and $\nu=0.6$ and $y\gtrsim y_c$ (strong-coupling disordered phase).}
\label{Fg:2D-disordered}
\end{figure}

\begin{figure}[t]
\includegraphics[width=0.35\linewidth]{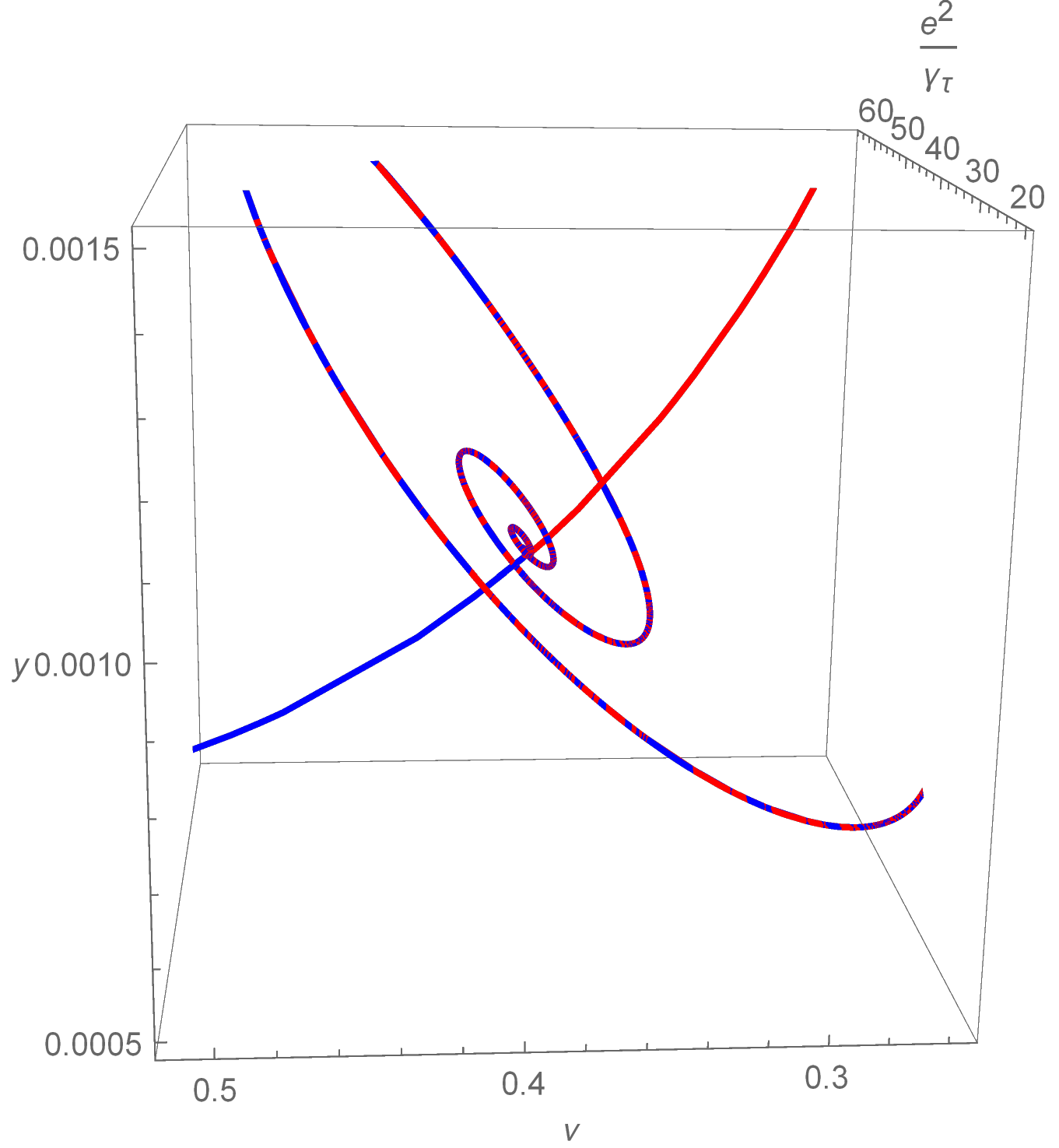}
\caption{Two RG flow trajectories near the saddle-fixed point in the three-dimensional parameter space. The saddle fixed point is at $(e^2/\gamma_{\tau},\nu,y)=(16\pi,\nu_{-},y_{-})=(50.2...,0.389...,0.00899...)$. The blue (red) curve denotes the RG flow trajectory when the initial parameters are chosen in the ordered (disordered) phase side of the phase boundary: $e^2/\gamma_{\tau}=e^2=15\pi$ and $\nu=0.6$ and $y\lesssim  y_c$ $(y\gtrsim y_c)$ respectively. The RG flow trajectories show swirling curves, when it is attracted toward the saddle fixed point. The RG flow trajectories show a rather straight curves, when they are repelled from the saddle fixed point. These behaviors are consistent with the linearization analysis of the RG equation around the fixed point: see Eq.~(\ref{Eq:2D-saddle-fixed-point-linear}) and texts below the equation.}
\label{Fg:15pinearCP}
\end{figure}

        The strong coupling fixed point on the non-magnetic plane is a stable fixed point with $ye^2=1/(3\pi)$, $\nu=0$ $\gamma_{\tau}=1$, $e^2=0$ and divergent $y$. To see this stability, note first that when linearized around the fixed point, the RG equation for $ye^2$ has an IR stable fixed point at $ye^2=1/(3\pi)$ in the limit of $e^2\rightarrow 0$, 
  \begin{align}
  \frac{d(ye^2)}{d\ln b} = 4 ye^2 - 12 \pi (ye^2)^2 - \frac{e^2}{2\pi} ye^2 \rightarrow 4 ye^2 - 12 \pi (ye^2)^2. 
  \end{align}
 The stability of the fixed point in the four-dimensional parameter space can be confirmed by a full set of linearized RG equations, 
 \begin{align}
\frac{d \delta (e^2)}{d\ln b} & = - \pi \Big(\frac{1}{3\pi} + \delta(ye^2)\Big) \delta (e^2) \Big(z_{2,r}(0) + {\cal O}(\nu^2) \Big) = - 4 \delta (e^2), \nonumber  \\ 
 \frac{d\delta (ye^2)}{d\ln b} & = -4 \delta (ye^2)-\frac{1}{6\pi^2} \delta (e^2), \nonumber \\
 \frac{d\delta \nu}{d\ln b} &= \delta \nu - \frac{\pi}{1+\delta \gamma_{\tau}} \Big(\frac{1}{3\pi} 
 + \delta(ye^2)\Big) \Big(\frac{\partial z_1}{\partial\nu}\Big)_{|\nu=0} \delta \nu = \delta \nu - 4 \delta \nu = - 3 \delta \nu, \nonumber \\
 \frac{d\delta \gamma_{\tau}}{d\ln b} &= \pi \big(\frac{1}{3\pi}+ \delta (ye^2)\big) \Big(12 - (1+\delta \gamma_{\tau}) 12 + {\cal O}(\delta\nu^2) \Big) = - 4\delta \gamma_{\tau}. \nonumber 
 \end{align}
with $\delta (ye^2) \equiv ye^2 - 1/(3\pi)$, $\delta \gamma_{\tau} \equiv \gamma_{\tau}-1$, $\delta \nu \equiv \nu$, and $\delta (e^2) \equiv e^2$. The equations show an attractive nature of the fixed point in the four-dimensional parameter space.  

    The ordered-disordered phase transition between the weak-coupling phase and strong coupling phases is controlled by a saddle fixed point with an infinite space-time anisotropy $\gamma_{\tau}=0$.  To elaborate the nature of this fixed point, let us rewrite the RG equations in terms of $\gamma_{\tau}/e^2$, $y$, $\nu$ and $e^2$;
         \begin{align}
         \frac{d(\gamma_{\tau}/e^2)}{d\ln b} &= \pi y z_{2,\tau}(\nu), \label{Eq:2D-RG-rev1a} \\
\frac{de^2}{d\ln b} &= - \pi y e^4 z_{2,r}(\nu), \label{Eq:2D-RG-rev1b} \\
\frac{d\nu}{d\ln b} &= \nu \Big( 1 - \frac{\pi ye^2}{\gamma_{\tau}} z_{2,{r}}(\nu)\Big), \label{Eq:2D-RG-rev1c} \\
\frac{dy}{d\ln b} & = \Big(4-\frac{\gamma^{-1}_{\tau}+1}{4\pi} e^2\Big) y, \label{Eq:2D-RG-rev1d} 
         \end{align}
Note that $z_{2,\tau}(\nu)$ has a pair of zeros at $\nu^2_{\pm} = (27\pm \sqrt{665})/(2\sqrt{2})$  [Fig.~\ref{Fg:zzz}], where $e^2=0$  $e^2/\gamma_{\tau}=16\pi$, $y=y_{\pm}$ with  
\begin{align}
y_{\pm}  = \frac{\gamma_{\tau}}{\pi e^2 z_{2,r}(\nu_{\pm})} = \frac{1}{16 \pi^2} \frac{1}{z_{2,r}(\nu_{\pm})} 
\end{align}
could be possible fixed points of the RG equations. A negative value of $z_{2,r}(\nu_{+})$ denies this possibility for the $\nu=\nu_{+}$ case.  

    The fixed point with $e^2=0$ , $e^2/\gamma_{\tau}=16\pi$, $y=y_{-} = 0.000898...$, and $\nu=\nu_{-}=0.389...$ is a saddle fixed point with a relevant scaling variable, a marginally irrelevant scaling variable, and two irrelevant scaling variables.  $e^2$ plays role of the marginally irrelevant scaling variable around the fixed point: 
\begin{align}
\frac{de^2}{d\ln b} = -\frac{1}{16\pi} e^4 + {\cal O}(e^4 \delta y, e^4 \delta \nu), \label{Eq:marginallye2}
\end{align}
with $\delta y \equiv y - y_{-}$, $\delta \nu\equiv \nu - \nu_{-}$ and $\delta (\gamma_{\tau}/e^2) \equiv \gamma_{\tau}/e^2 - 1/(16\pi)$. The relevant scaling variable and two other irrelevant scaling variables consist of the other three variables. The RG equations for the others are linearized around the fixed point,
\begin{align}
\frac{d}{d\ln b} \left(\begin{array}{c} 
\delta\big(\frac{\gamma_{\tau}}{e^2}\big) \\
\delta \nu \\
\delta y \\
\end{array}\right) = \left(\begin{array}{ccc}  
0 &  \pi y_{-} \big(\frac{dz_{2,\tau}(\nu)}{d\nu}\big)_{|\nu=\nu_{-}} & 0\\
16\pi \nu_{-} & 1 & -\frac{\nu_{-}}{y_{-}}  \\
64\pi y_{-} & 0 & 0 \\
\end{array}\right)  \left(\begin{array}{c} 
\delta\big(\frac{\gamma_{\tau}}{e^2}\big) \\
\delta \nu \\
\delta y \\
\end{array}\right). \label{Eq:2D-saddle-fixed-point-linear}
\end{align}
The 3 by 3 matrix has eigenvalues of $1.95$, $-0.475\pm 1.84 i$, where an eigenvector for the real positive eigenvalue $y_t$ is given by 
\begin{align}
\left(\begin{array}{ccc} 
\delta\big(\frac{\gamma_{\tau}}{e^2} \big) & \delta \nu & \delta y 
\end{array}\right) = \left(\begin{array}{ccc} 
-0.04616...  & 0.999.. & -0.00428...
\end{array}\right). \label{Eq:eigenvector2D} 
\end{align} 
This indicates that the RG flow to the strong/weak-coupling side of the fixed point increases/decreases $y$, while decreasing/increasing $\nu$ and increasing/decreasing $\delta(\gamma_{\tau}/e^2)$. A pair of two complex conjugate eigenvalues with a negative real part suggests that the RG flow streams into the saddle fixed point with a swirling trajectory [Fig.~\ref{Fg:15pinearCP}].   In the following, we name as ${\tt t}$ a scaling variable for the real positive eigenvalue $y_{\tt t}$. 

     The numerical RG phase diagrams are shown in a two-dimensional parameter space subtended by initial values of $y$ and $e^2$ for given initial values of $\nu$ and $\gamma_{\tau}=1$ [Fig.~\ref{Fg:2D-pd}]. The numerical phase diagrams generally show that larger (smaller) initial values of $y$ and smaller (larger) initial values of $e^2$ lead to the strong-coupling disordered phase (weak-coupling ordered phase).  Numerical solutions of $e^2/\gamma_{\tau}$, $e^2$, $\nu$ and $y$ as functions of the RG scale $\ln b$ are shown in Figs.~\ref{Fg:2D-ordered}, \ref{Fg:2D-disordered}. When the initial parameters are on the weak-coupling side of the phase boundary, both $e^2/\gamma_{\tau}$ and $e^2$ approach finite non-universal values in the IR limit $[\ln b\rightarrow \infty]$, while $\nu$ diverges and $y$ vanishes [Fig.~\ref{Fg:2D-ordered}]. When the initial parameters are on the strong-coupling side of the boundary,   both $e^2/\gamma_{\tau}$ and $e^2$ approach zero, while $\gamma_{\tau}$ converges to $1$, $\nu$ approaches zero, and $y$ diverges [Fig.~\ref{Fg:2D-disordered}]. In either case, the parameters make a detour around the saddle fixed point: they first flow into the saddle fixed point before being repelled from the saddle fixed point to strong/weak coupling regions [Figs.~\ref{Fg:15pinearCP}, \ref{Fg:2D-ordered},\ref{Fg:2D-disordered}]. This clearly suggests that the ordered-disordered phase transition is of the second-order, and its universal criticality is controlled by the saddle fixed point at $(y,\nu,e^2/\gamma_{\tau},e^2)=(y_{-},\nu_{-},16\pi,0)$ . For example, the correlation-length exponent is estimated by the scaling dimension of the relevant scaling variable $\tt t$: $1/y_{\tt t} = 1/1.95... = 0.514...$ 

      The divergence of $\gamma_{\tau}$ at the saddle fixed point suggests that the correlation time near the critical point acquires an additional logarithmic divergence compared to the divergent correlation length. To see this point, let us start from a scaling relation between the relevant scaling variable $\tt t$ before and after the renormalization,  ${\tt t}^{\prime} = {\tt t} b^{y_{\tt t}}$. ${\tt t}=0$ corresponds to the critical point, while ${\tt t}$ is either negative (ordered phase) or positive (disordered phase), and $|{\tt t}|$ can be regarded as a `distance' from the critical point. When an initial ${\tt t}$ is infinitesimally small [the system is proximate to the critical point], one needs large $b$ to bring the renormalized $\tt t^{\prime}$ away from a critical point : $b=({\tt t}/{\tt t}_0)^{-1/y_{\tt t}} \gg 1$ for ${\tt t}^{\prime}={\tt t}_0$. For such large $b$, the renormalized $e^2$ must be tiny: In fact, one solves Eq.~(\ref{Eq:marginallye2}) for the renormalized $e^2$,  
\begin{align}
\frac{1}{e^2_0} \simeq \frac{\ln b}{16\pi}. \label{Eq:e2-asym} 
\end{align}
Since the renormalized $e^2/\gamma_{\tau}$ is around $16\pi$, the renormalized space-time anisotropy parameter $\gamma_{\tau}$ must be tiny as well 
\begin{align}
\gamma_{\tau,0}\simeq  \frac{y_{\tt t}}{|\ln {\tt t}|}. \label{Eq:gammat-asym}
\end{align}
The space-time anisotropy appears in a ratio between time and length scales. The form of the Maxwell action suggests that after the renormalization, the time scale $\xi_{\tau,0}$ becomes larger than the length scale $\xi_{r,0}$ by a factor $1/\sqrt{\gamma_{\tau,0}}$: 
\begin{align}
\xi_{\tau,0} = \xi_{0} \frac{1}{\sqrt{\gamma_{\tau,0}}}, \,\ \xi_{r,0} = \xi_0. 
\end{align}
 Since these scales after the renormalization are related to the respective scales before the renormalization by a factor $b$, we finally obtain the correlation time $\xi_{\tau}$ and length $\xi_{r}$ near the critical point as follows, 
 \begin{align}
 \xi_{\tau} &= b \xi_{\tau,0} \sim |{\tt t}|^{-\frac{1}{y_{\tt t}}} \sqrt{|\ln {\tt t}|},  \label{Eq:xitau} \\
 \xi_{r} & = b\xi_{r,0} \sim |{\tt t}|^{-\frac{1}{y_{\tt t}}}. \label{Eq:xir}  
 \end{align}

\subsection{\label{secIIIc}$D=3$ in the presence of Berry phase term $(\chi\ne 0)$}
      In the presence of the Berry phase term, the fugacity renormalization in general $D$ case takes complicated functions of space and time components of the metric tensor, $\hat{\bf g}_{\tau}$ and $\hat{\bf g}_{\bm r}$. In $D=3$, however, these components reduce to scalar quantities, 
 \begin{align}
& \hat{\bf g}_{j,\tau}({\bm s}) \rightarrow {\rm g}_{j,\tau}(s) \equiv \Big(\frac{d R_{j,\tau}(s)}{ds}\Big)^2, \nonumber \\ 
& \hat{\bf g}_{j,{\bm r}}({\bm s}) \rightarrow {\rm g}_{j,{\bm r}}(s) \equiv \sum_{\sigma=1,2}
\Big(\frac{d R_{j,\sigma}(s)}{ds}\Big)^2, \nonumber 
 \end{align}
with ${\rm g}_{j,\tau}(s)+{\rm g}_{j,{\bm r}}(s)=1$, where the fugacity renormalization in Eq.~(\ref{Eq:fugacity-rg7}) becomes  simpler, 
\begin{align}
\delta E_j &= -2\pi^2 e^2 \ln b \int_{\partial \Gamma_j} ds \Big\{ f_{\tau}\big({\rm g}_{j,\tau}(s),{\rm g}_{j,{\bm r}}(s)\big) \!\ {\rm g}_{j,\tau}(s)  + 
f_{\bm r}\big({\rm g}_{j,\tau}(s),{\rm g}_{j,{\bm r}}(s)\big)\!\ {\rm g}_{j,{\bm r}}(s) \Big\}.  
 \label{Eq:deltaEj-3D}
\end{align}
Thereby, the fugacity for each closed vortex loop is given by the $s$-integral of the vortex core energy density $t({\rm g}_{\tau}(s))$ defined in the finite region $0\le {\rm g}_{\tau}(s) \le 1$, $E_j=\int_{\partial \Gamma_j} ds \,\ t({\rm g}_{j,\tau}(s))$, and the renormalization group equation for the vortex core energy density is given by,
\begin{align}
\frac{dt({\rm g}_{\tau})}{d\ln b} = t({\rm g}_{\tau}) - 2\pi^2 e^2 \Big(f_{\tau}\big({\rm g}_{\tau},1-{\rm g}_{\tau}\big) \!\ {\rm g}_{\tau} + f_{\bm r}\big({\rm g}_{\tau},1-{\rm g}_{\tau}\big) \!\ (1-{\rm g}_{\tau})\Big). \label{Eq:RG-t(g)} 
\end{align}
For simplicity, we characterize the vortex core energy density function by its two endpoint values,  $t_{\tau} \equiv t({\rm g}_{\tau}=1)$ and $t_{\bm r} \equiv t({\rm g}_{\tau}=0)$, while interpolating its middle values by $t({\rm g}_{\tau})=t_{\tau} {\rm g}_{\tau} + t_{\bm r} {\rm g}_{\bm r}$, i.e. 
\begin{align}
E_j = \int_{\partial \Gamma_j} ds \big( t_{\tau}{\rm g}_{j,\tau}(s) + t_{\bm r}{\rm g}_{j,{\bm r}}(s) \big) = \int_{\partial \Gamma_j} ds \Big( t_{\tau} \Big(\frac{d{\bm R}_{j,\tau}(s)}{ds}\Big)^2 + t_{\bm r}  \sum_{\sigma=1,2}\Big(\frac{d{\bm R}_{j,\sigma}(s)}{ds}\Big)^2 \Big). \label{Eq:Ej-3d}
\end{align}
From Eq.~(\ref{Eq:RG-t(g)}), the renormalization group equation for the two endpoint values is evaluated by 
\begin{align}
\frac{dt_{\tau}}{d\ln b} &= t_{\tau} - 2\pi^2 e^2 f_{\tau} ({\rm g}_{\tau}=1,{\rm g}_{\bm r}=0\big)=t_{\tau} - \pi e^2, \label{Eq:ttau} \\
\frac{dt_{\bm r}}{d\ln b} &= t_{\bm r} - 2\pi^2 e^2 f_{\bm r} ({\rm g}_{\tau}=0,{\rm g}_{\bm r}=1\big)=t_{\bm r} - \pi e^2 \frac{\gamma^{-1}_{\tau}+1}{2}, \label{Eq:tr} 
\end{align}
[see sec.~\ref{secIVb} for the evaluation]. Eqs.~(\ref{Eq:ttau},\ref{Eq:tr},\ref{Eq:Ej-3d}) with $t_{\tau}=t_{\bm r}=t$ and $\gamma_{\tau}=1$ become consistent with Eqs.~(\ref{Eq:t2},\ref{Eq:fugacity-1}) in the isotropic limit, upholding the validity of the simplification.  In $D=3$, the smallest vortex hyper-sphere becomes a coplanar vortex circle whose diameter is $a_0$.  From Eq.~(\ref{Eq:Ej-3d}), the fugacity  $E_0(\theta)$ for the vortex circle is given by 
\begin{align}
E_0(\theta) &= t_{\tau} \frac{a_0}{2}\int^{2\pi}_{0} d\alpha \sin^2 \theta \sin^2 \alpha + t_{\bm r} \frac{a_0}{2} \int^{2\pi}_{0} d\alpha \big(\cos^2 \theta \sin^2 \alpha + \cos^2 \alpha \big) \nonumber \\
& = \frac{a_0}{2} \pi \big( t_{\tau}  (1-\cos^2\theta) + t_{\bm r} (1+\cos^2\theta)\big), \label{Eq:E0-3D}
\end{align}
where $\theta$ is the angle between a plane with the vortex circle and the time ($\tau$) axis. From Eq.~(\ref{Eq:e_0-integral},\ref{Eq:gammat},\ref{Eq:e2},\ref{Eq:chi},\ref{Eq:E0-3D}), the $D=3$ RG equations for the other coupling constants are given by:
\begin{align}
\frac{d \gamma_{\tau}}{d\ln b} &= 
\frac{\pi^5}{2} e^2 a_{0}  \int^{1}_{-1} ds \,\  e^{\frac{a_0}{2} \pi\big(t_{\tau}(1-s^2)+t_{\bm r}(1+s^2)\big)} e^{-i\frac{\pi^2 a^2_0}{2} \chi  s} \Big(s^2-\gamma_{\tau} \frac{1-s^2}{2}\Big),
\label{Eq:gammat-3D} \\ 
\frac{d e^2}{d\ln b} &= \!\ e^2 - \frac{\pi^5}{2} e^4 a_{0}    
\int^{1}_{-1} ds \,\ e^{\frac{a_0}{2} \pi\big(t_{\tau}(1-s^2)+t_{\bm r}(1+s^2)\big)} e^{-i\frac{\pi^2 a^2_0}{2} \chi  s}
\frac{1-s^2}{2}, \label{Eq:e2-3D} \\ 
\frac{d\chi}{d \ln b} & = 2 \chi - \pi^3 e^2 \frac{1}{\gamma_{\tau}}  
a^{-1}_0  \int^{1}_{-1} ds \,\ e^{\frac{a_0}{2} \pi\big(t_{\tau}(1-s^2)+t_{\bm r}(1+s^2)\big)}  
e^{-i\frac{\pi^2 a^2_0}{2} \chi  s} \!\ i s, \label{Eq:chi-3D}
\end{align}
where $s =\cos\theta$ and $A=\frac{\pi}{2}$ at $D=3$ in Eqs.~(\ref{Eq:gammat},\ref{Eq:e2},\ref{Eq:chi}) . With proper normalization of the constants, 
\begin{align}
e^2_{\rm old} a_0 \equiv e^2_{\rm new}, \,\ \frac{\pi^2 a^2_0}{2} \chi \equiv \nu, \,\ t_{\tau(\bm r),{\rm old}} a_0 \equiv t_{\tau({\bm r}),{\rm new}}, 
\end{align}
the equations are simplified, 
\begin{align}
\frac{de^2}{d\ln  b} &= e^2 - \frac{\pi^5}{2} e^4 Z_{2,{\bm r}} ({\bm t},\nu), \label{Eq:RGe2D_3} \\
\frac{d\nu}{d\ln b} &= 2\nu  - \frac{\pi^5}{2} e^2 \gamma^{-1}_{\tau} Z_{1}({\bm t},\nu), \label{Eq:RGchiD_3}, \\
\frac{d\gamma_{\tau}}{d\ln b} &= \frac{\pi^5}{2} e^2 \Big( Z_{2,{\tau}}({\bm t},\nu) - \gamma_{\tau} Z_{2,{\bm r}}({\bm t},\nu)\Big), \label{Eq:RGgammatauD_3} \\
\frac{dt_{\tau}}{d\ln b} &= t_{\tau} - \pi e^2, \label{Eq:RGttau} \\
\frac{dt_{\bm r}}{d\ln b}&= t_{\bm r} - \pi e^2 \frac{\gamma^{-1}_{\tau}+1}{2}, \label{Eq:RGtr}
\end{align}
with ${\bm t}\equiv (t_{\tau},t_{\bm r})$ and 
\begin{align}
Z_{2,\tau}({\bm t},\nu) &\equiv \int^{1}_{-1} ds \,\ s^2 e^{\frac{\pi}{2}\big(t_{\tau}(1-s^2)+t_{\bm r}(1+s^2)\big)} e^{-i\nu s}, \label{Eq:Z2t} \\
Z_{2,{\bm r}}({\bm t},\nu) &\equiv \int^{1}_{-1} ds \,\ \frac{1-s^2}{2} e^{\frac{\pi}{2}\big(t_{\tau}(1-s^2)+t_{\bm r}(1+s^2)\big)} e^{-i\nu s}, \label{Eq:Z2r} \\ 
Z_{1}({\bm t},\nu) &\equiv \int^{1}_{-1} ds \,\ is \!\ e^{\frac{\pi}{2}\big(t_{\tau}(1-s^2)+t_{\bm r}(1+s^2)\big)} e^{-i\nu s}. \label{Eq:Z1}
\end{align}
On the non-magnetic isotropic plane [$\nu=0$, $\gamma_\tau=1$ and $t_{\tau}=t_{\bm r}=t$], the RG equation describes the ordered-disordered phase transition:
\begin{align}
\frac{de^2}{d\ln b} = e^2 - \frac{\pi^5}{3} e^{\pi t} e^4, \,\ \frac{dt}{d\ln b} = t-\pi e^2 , 
\end{align}
with a saddle fixed point at $(e^2_*,t_*)$ [$\pi e^2_*=t_*$ and $1=\frac{\pi^4}{3} e^{\pi t_*} \pi t_*$]. In the large $b$ limit, parameters in the ordered phase flow into a weak coupling fixed point with divergent $e^2 \sim b$ and negatively divergent $t \sim - b \ln b$, while those in the disordered phase go to a strong coupling fixed point with vanishing $e^2 \sim b \exp [- A\!\ b]$ $(A>0)$, and positively divergent $t \sim b$.   

\begin{figure}[t]
\subfigure[$Z_{2,{\bm r}}({\bm t},\nu)$ with hard UV cutoff (Eq.~(\ref{Eq:Z2r}))]{
\includegraphics[width=0.4\textwidth]{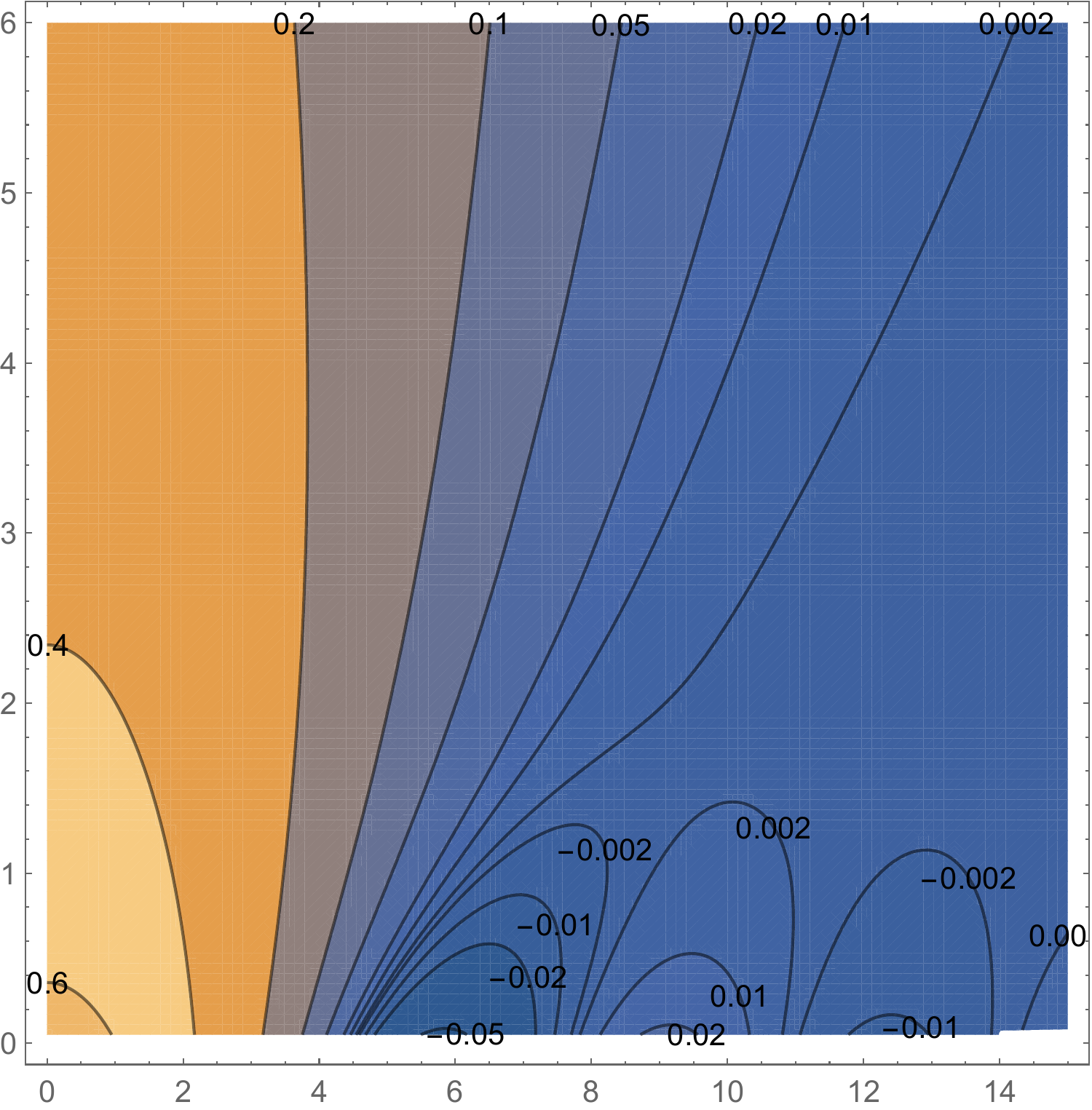}
\label{Fig:hard}
}
\hfill
\subfigure[$Z_{2,{\bm r}}({\bm t},\nu)$ with soft UV cutoff (Eq.~(\ref{Eq:Z2tr-soft-approx}))]{
\includegraphics[width=0.4\textwidth]{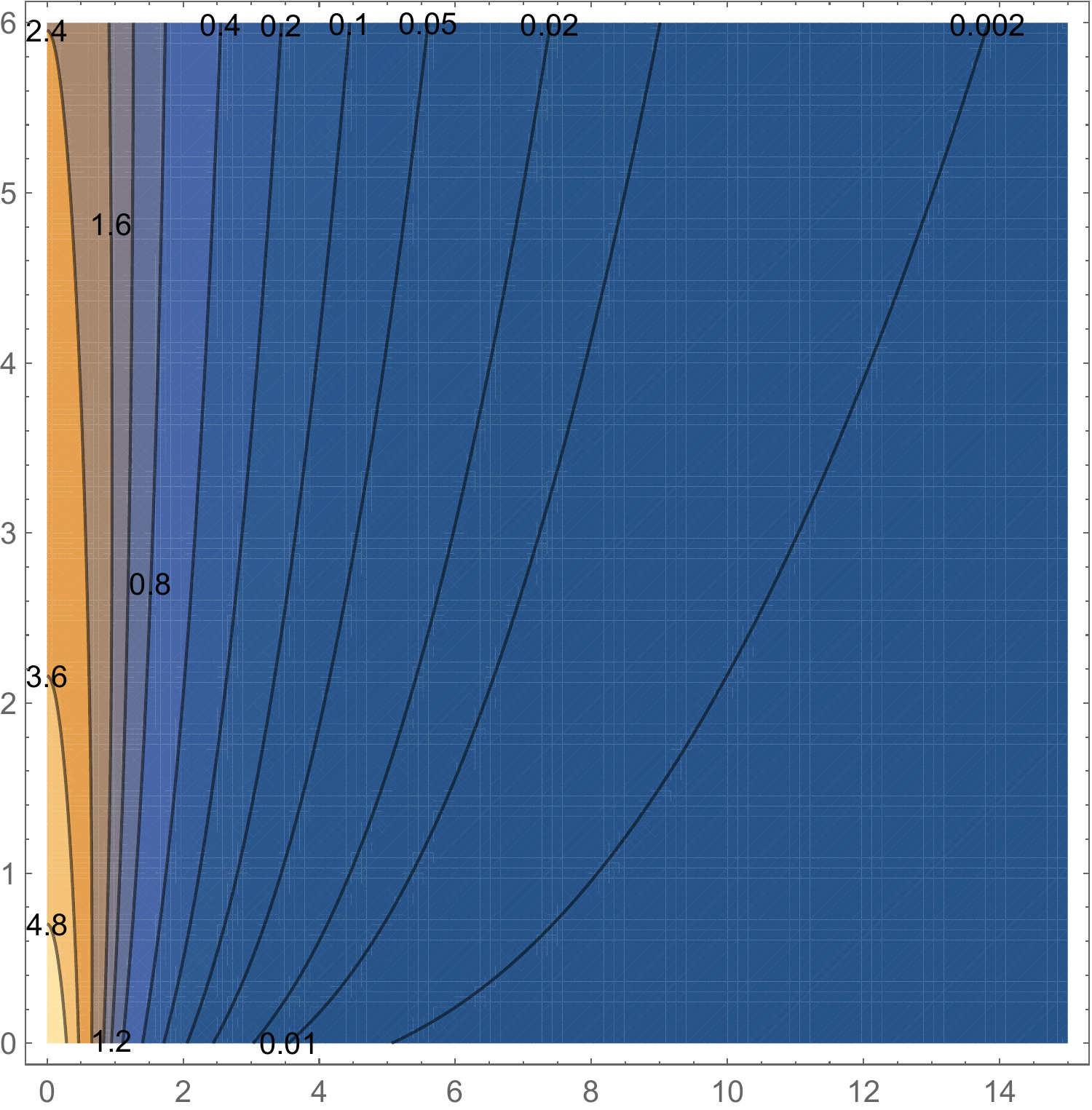}
\label{Fig:soft}
}
\caption{Contour plots of $Z_{2,{\bm r}}({\bm t},\nu)$ as a function of $t_{\tau}=-t_{\bm r} \equiv t_{-}/2$ and $\nu$ with the hard UV cutoff [Eq.~(\ref{Eq:Z2r})] and with the soft UV cutoff [Eq.~(\ref{Eq:Z2tr-soft-approx})]. The horizontal and vertical axes are $\nu$ and $t_{-}$, respectively. Eq.~(\ref{Eq:Z2r}) shows an oscillatory behavior as a function of the Berry phase parameter $\nu$, while Eq.~(\ref{Eq:Z2tr-soft-approx}) does not.}
\end{figure}

     Note that the $e^4$ term on the right hand side of Eq.~(\ref{Eq:RGe2D_3}) represents the screening effect, which shows an unphysical oscillatory behavior as a function of the 1D Berry phase parameter $\nu$ [Fig.~\ref{Fig:hard}]. As in the 2D case, the oscillatory behavior is an artifact of a sharp UV cutoff in Eqs.~(\ref{Eq:Zv-3},\ref{Eq:Zv-4}), and it can be removed by using the soft UV cutoff [see Fig.~\ref{Fig:soft}], 
\begin{align}
\int D^3 {\bm X}_0 \int_{a_0<|{\bm Y}_0|<a_0 b} D^3 {\bm Y}_0 \int D{\bf g}_0(\bm s) 
=& a^{-6}_0 \int d^{3}{\bm X}_0 \int^{a_0 b}_{a_0} Y^2_0 d Y_0  
\int_{S_2} d^2\Omega \nonumber \\
\rightarrow& \frac{1}{\cal N} a^{-6}_0 \int d^{3}{\bm X}_0 \,\  a_0 \ln b \frac{\partial}{\partial a_0} \Big(\int^{\infty}_{0} dY_0 \!\ Y^2_0 e^{-Y^2_0/a^2_0} \Big)  
\int_{S_2} d^2\Omega \nonumber \\
=& \frac{2}{\cal N} a^{-3}_0 \ln b \int d^{3}{\bm X}_0 \int^{\infty}_{0} dy_0 \!\ y^4_0 e^{-y^2_0}  
\int_{S_{2}} d^2\Omega 
\end{align}
with $y_0 \equiv Y_0/a_0$ and a normalization factor ${\cal N}=\int^{\infty}_{0} dy_0 2y^4_0 e^{-y^2_0}= 3\sqrt{\pi}/4$. With the soft UV cutoff, the renormalization of the consecutive constants and the Berry phase term in Eqs.~(\ref{Eq:rg3},\ref{Eq:rg4},\ref{Eq:rg5})  are modified,
\begin{align}
\delta \chi &= \frac{2\pi^3}{\cal N} e a^{-1}_0 \int^{\infty}_{0} dy_0 y^6_0 \!\ e^{-y^2_0} \int^{1}_{-1} ds \!\ e^{\frac{\pi y_0}{2} \big(t_{\tau}(1-s^2) + t_{\bm r}(1+s^2)\big)} \!\ s\sin\big(\nu y^2_0 s\big), \label{Eq:rg3-3D-soft} \\
\left(\begin{array}{c}
\delta \gamma_{\tau}  \\
\delta \gamma_{\bm r} \\
\end{array}\right) &= \frac{\pi^5}{2{\cal N}} e^2 a_0 \int^{\infty}_0 dy_0 y^8_0 e^{-y^2_0} 
\int^{1}_{-1} ds e^{\frac{\pi y_0}{2} \big(t_{\tau}(1-s^2) + t_{\bm r}(1+s^2)\big)} \!\  \left(\begin{array}{c}
s^2  \\
\frac{1-s^2}{2} \\
\end{array}\right) \cos\big(\nu y^2_0 s\big). \label{Eq:rg45-3D-soft} 
\end{align}
 Accordingly, $Z_{1}({\bm t},\nu)$, $Z_{2,{\bm r}}({\bm t},\nu)$ and $Z_{2,\tau}({\bm t},\nu)$ in Eqs.~(\ref{Eq:RGe2D_3},\ref{Eq:RGchiD_3},\ref{Eq:RGgammatauD_3}) are modified into, 
\begin{align}
Z_{1}({\bm t},\nu) &\rightarrow \frac{2}{\cal N}\int^{\infty}_0 dy \!\ y^6 e^{-y^2} \int^{1}_{-1} ds \,\ is \!\ e^{\frac{\pi y}{2}\big(t_{\tau}(1-s^2)+t_{\bm r}(1+s^2)\big)} e^{-i\nu  y^2 s}, \label{Eq:Z1-soft} \\
\left(\begin{array}{c} 
Z_{2,\tau}({\bm t},\nu) \\
Z_{2,{\bm r}}({\bm t},\nu) \\
\end{array}\right)
&\rightarrow \frac{2}{\cal N}\int^{\infty}_0 dy \!\ y^8 e^{-y^2} \int^{1}_{-1} ds \,\ 
\left(\begin{array}{c} 
s^2 \\ 
\frac{1-s^2}{2} \\
\end{array}\right)  e^{\frac{\pi y}{2}\big(t_{\tau}(1-s^2)+t_{\bm r}(1+s^2)\big)} e^{-i\nu y^2 s}. \label{Eq:Z2tr-soft} 
\end{align}
To evaluate these double integrals approximately, we replace $y$ in the fugacity term $E_0(\theta)$ by $1$;
\begin{align}
Z_{1}({\bm t},\nu) &\simeq \frac{2}{\cal N} \int^{1}_{-1} ds \,\ is \!\ e^{\frac{\pi}{2}\big(t_{\tau}(1-s^2)+t_{\bm r}(1+s^2)\big)} 
\int^{\infty}_0 dy \!\ y^6 e^{-y^2}  e^{-i\nu  y^2 s} \nonumber \\ 
&= \frac{5}{2} 
\int^{1}_{-1}ds \!\ is \!\ e^{\frac{\pi}{2}\big(t_{\tau}(1-s^2)+t_{\bm r}(1+s^2)\big)} 
\frac{1}{(1+i\nu s)^{\frac{7}{2}}}, \label{Eq:Z1-soft-approx-a} \\ 
\left(\begin{array}{c} 
Z_{2,\tau}({\bm t},\nu) \\
Z_{2,{\bm r}}({\bm t},\nu) \\
\end{array}\right)
&\simeq  \frac{2}{\cal N} \int^{1}_{-1} ds \,\ 
\left(\begin{array}{c} 
s^2 \\ 
\frac{1-s^2}{2} \\
\end{array}\right)  e^{\frac{\pi}{2}\big(t_{\tau}(1-s^2)+t_{\bm r}(1+s^2)\big)} \int^{\infty}_0 dy \!\ y^8 e^{-y^2} e^{-i\nu y^2 s} \nonumber \\
&= \frac{35}{4} \int^{1}_{-1} ds \!\  
\left(\begin{array}{c} 
s^2 \\ 
\frac{1-s^2}{2} \\
\end{array}\right)  e^{\frac{\pi}{2}\big(t_{\tau}(1-s^2)+t_{\bm r}(1+s^2)\big)} 
\frac{1}{(1+i\nu s)^{\frac{9}{2}}}, \label{Eq:Z2tr-soft-approx-a} 
\end{align}
where the following are used, 
\begin{align}
\int^{\infty}_0 dy \!\ y^6 e^{- \alpha y^2} = \frac{5{\cal N}}{4}\frac{1}{\alpha^{\frac{7}{2}}}, \!\ \int^{\infty}_0 dy \!\ y^8 e^{- \alpha y^2} = \frac{35 {\cal N}}{8}\frac{1}{\alpha^{\frac{9}{2}}}.
\end{align}
$Z_{1}$ and $Z_{2,\tau({\bm r})}$ can be numerically evaluated by the integral over $s \in [-1,1]$, 
\begin{align}
Z_{1}({\bm t},\nu) &= \frac{5}{2} e^{\frac{\pi}{2}(t_{\tau}+t_{\bm r})} \int^{1}_{-1} ds \!\ e^{-\frac{\pi}{2}(t_{\tau}-t_{\bm r})s^2} \frac{s \sin\big(\frac{7}{2}\tan^{-1}(\nu s)\big)}{(1+\nu^2 s^2)^{\frac{7}{4}}}, \label{Eq:Z1-soft-approx} \\
\left(\begin{array}{c}
Z_{2,\tau}({\bm t},\nu) \\
Z_{2,{\bm r}}({\bm t},\nu) \\
\end{array}\right)
&= \frac{35}{4} e^{\frac{\pi}{2}(t_{\tau}+t_{\bm r})} \int^{1}_{-1} ds \!\  
e^{-\frac{\pi}{2}(t_{\tau}-t_{\bm r})s^2}
\left(\begin{array}{c} 
s^2 \\ 
\frac{1-s^2}{2} \\
\end{array}\right)   \frac{\cos\big(\frac{9}{2}\tan^{-1}(\nu s)\big)}{(1+\nu^2 s^2)^{\frac{9}{4}}}. \label{Eq:Z2tr-soft-approx} 
\end{align}
These functions show similar behaviors to $z_1(\nu)$, $z_{2,\tau}(\nu)$, and $z_{2,{\bm r}}(\nu)$ in 2D, respectively. They do not have the oscillatory behaviors as a function of $\nu$ [Fig.~\ref{Fig:soft}].  In the following, we study the RG phase diagram obtained from Eqs.~(\ref{Eq:RGe2D_3},\ref{Eq:RGchiD_3},\ref{Eq:RGgammatauD_3},\ref{Eq:RGtr},\ref{Eq:RGttau}) with Eqs.~(\ref{Eq:Z1-soft-approx},\ref{Eq:Z2tr-soft-approx}).

       In the large $b$ limit, the RG equation drives the coupling constants into a weak coupling region for the ordered phase, an intermediate coupling region for the quasi-disordered phase or a strong coupling region for the disordered phase. The weak-coupling region is characterized by the vanishing screening effect, where $e^2\sim b$, $\nu \sim b^2$, while $\gamma_{\tau}$ converges to a finite positive value less than 1. From Eqs.~(\ref{Eq:RGttau},\ref{Eq:RGtr}) with $e^2\sim b$ and a constant $\gamma_{\tau}$, both $t_{\tau}$ and $t_{\bm r}$ in the large $b$ limit diverge negatively: $t_{\tau}, t_{\bm r} \sim -  b\ln b$. Due to the negatively divergent $t_{\tau}$ and $t_{\bm r}$, $Z_{2,{\bm r}}$, $Z_{2,\tau}$ and $Z_1$ in the large $b$ limit vanish exponentially in $b$, resulting in the vanishing screening on the right hand sides of Eqs.~(\ref{Eq:RGe2D_3}, \ref{Eq:RGchiD_3},\ref{Eq:RGgammatauD_3}).  

       The strong coupling region is characterized by the divergent screening effect, where $t_{\tau}+t_{\bm r}$ diverges positively with $t_{\tau}+t_{\bm r} \sim A_{+} b$ $(A_+>0)$,  $e^2$ vanishes with $e^2 \sim b^{\frac{3}{2}} \exp [-\frac{\pi}{2}A_+ b]$,  $\nu$ vanishes with $\nu \sim b^{2} \exp [-\frac{\pi}{2} A_+ b]$, and the space-time anisotropy parameter diverges with $\gamma^{-1}_{\tau}\sim b$. Thereby, the temporal component $t_{\tau}$ of the vortex-loop fugacity always diverges positively with $t_{\tau}\sim A_{\tau} \!\ b$ $(A_{\tau}>0)$, while the spatial component $t_{\bm r}$ diverges either positively or negatively, depending on its initial value. The distinction between $t_{\tau}$ and $t_{\bm r}$ comes from the enhanced space-time anisotropy parameter $(\gamma^{-1}_{\tau}\sim b)$, which selectively suppresses $t_{\bm r}$ through the fugacity renormalization: the second term of Eq.~(\ref{Eq:RGtr}).  The suppression of $t_{\bm r}$ by the enhanced space-time anisotropy always results in the positively divergent $t_{\tau} - t_{\bm r} \sim A_-\!\ b$  $(A_->0)$. With the divergent $t_{\tau}-t_{\bm r}$, the $s$-integrals in Eqs.~(\ref{Eq:Z1-soft-approx},\ref{Eq:Z2tr-soft-approx}) are dominated by the contributions around $s=0$, allowing asymptotic estimates of  $Z_{2,\tau}({\bm t},\nu)$, $Z_{2,{\bm r}}({\bm t},\nu)$, and $Z_{1}({\bm t},\nu)$:   
          \begin{align}
          \left(\begin{array}{c} 
          Z_{2,\tau} \\
          Z_{2,{\bm r}} \\
          \end{array}\right) 
          &\simeq \frac{35\sqrt{\pi}}{8} \exp[\frac{\pi}{2}(t_{\tau}+t_{\bm r})] \frac{1}{\sqrt{\frac{\pi}{2}(t_{\tau}-t_{\bm r})}}  \left(\begin{array}{c} 
          \frac{1}{\frac{\pi}{2}(t_{\tau}-t_{\bm r})} + {\cal O}(1/(t_{\tau}-t_{\bm r})^2) \\
          1 + {\cal O}(1/(t_{\tau}-t_{\bm r})) \\
          \end{array}\right), \label{Eq:Z2-asym} \\
          Z_{1} &\simeq  \frac{35\sqrt{\pi}}{8} \nu \exp[\frac{\pi}{2}(t_{\tau}+t_{\bm r})] 
          \bigg(\frac{1}{(\frac{\pi}{2}(t_{\tau}-t_{\bm r}))^{3/2}}  + {\cal O}(1/(t_{\tau}-t_{\bm r})^{5/2})\bigg). \label{Eq:Z1-asym}
          \end{align}
Eqs.~(\ref{Eq:RGe2D_3},\ref{Eq:RGgammatauD_3},\ref{Eq:RGchiD_3},\ref{Eq:Z2-asym},\ref{Eq:Z1-asym}) together with $t_{\tau}\pm t_{\bm r}\sim A_{\pm} b$ lead to the strong-coupling asymptotics of $e^2$, $\gamma_{\tau}$ and $\nu$; 
\begin{align}
e^2 \simeq \frac{A_+}{C_+} b^{\frac{3}{2}} \exp [-\frac{\pi}{2}A_{+} b], \,\ \gamma_{\tau} \simeq \frac{2}{\pi A_{-} b},\,\ \nu \simeq {\rm const}\cdot b^2 \exp[-\frac{\pi}{2}A_{+} b],   \label{Eq:3D_Dis_asym}  
\end{align}
with $C_+=\frac{35\pi^4}{32}\sqrt{2/A_-}$.  

\begin{figure}[t]
\subfigure[$\nu=1.2$]{
\includegraphics[width=0.3\textwidth]{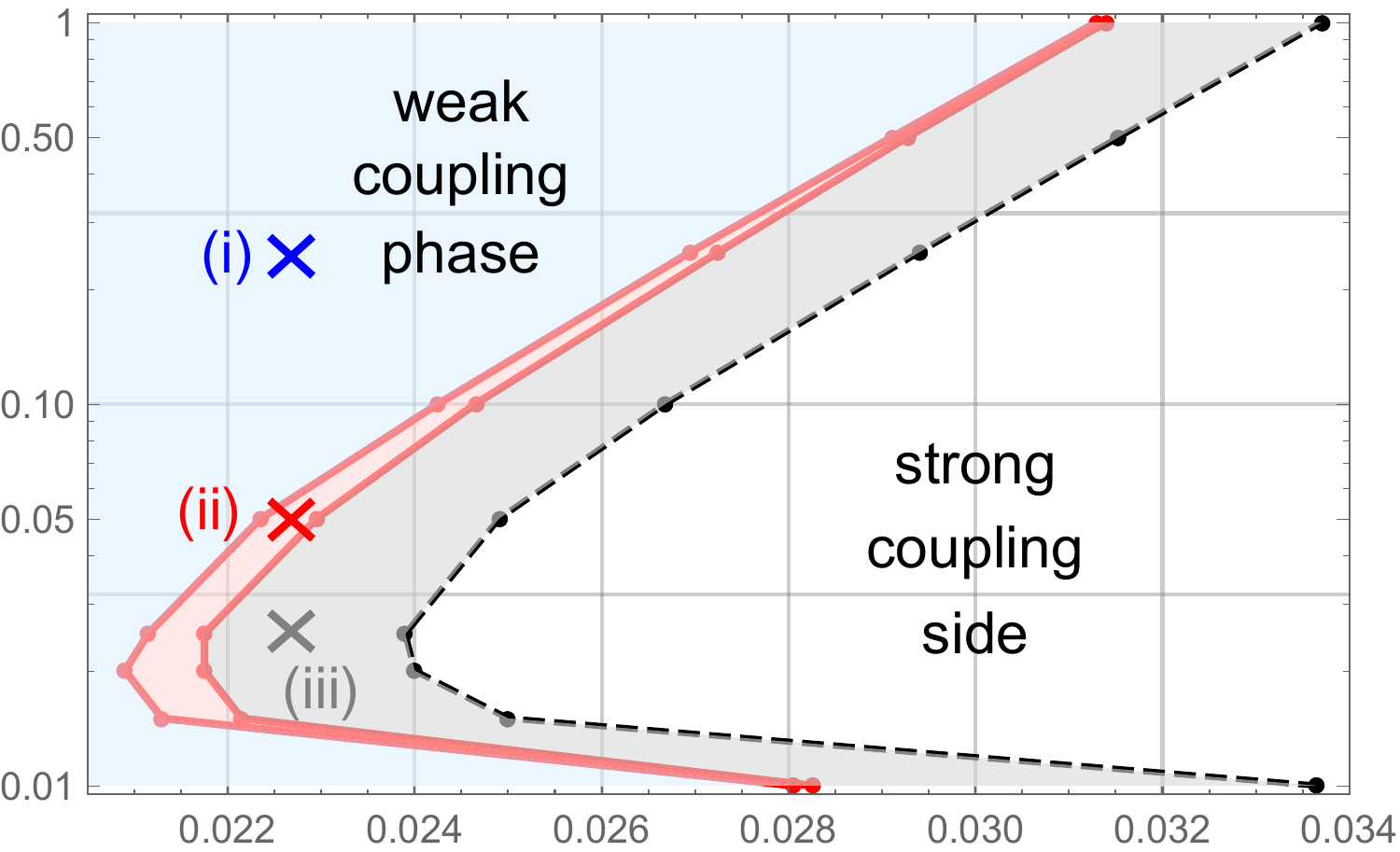}
\label{Fig:nu12_pd}
}
\hfill
\subfigure[$\nu=0.6$]{ 
\includegraphics[width=0.3\textwidth]{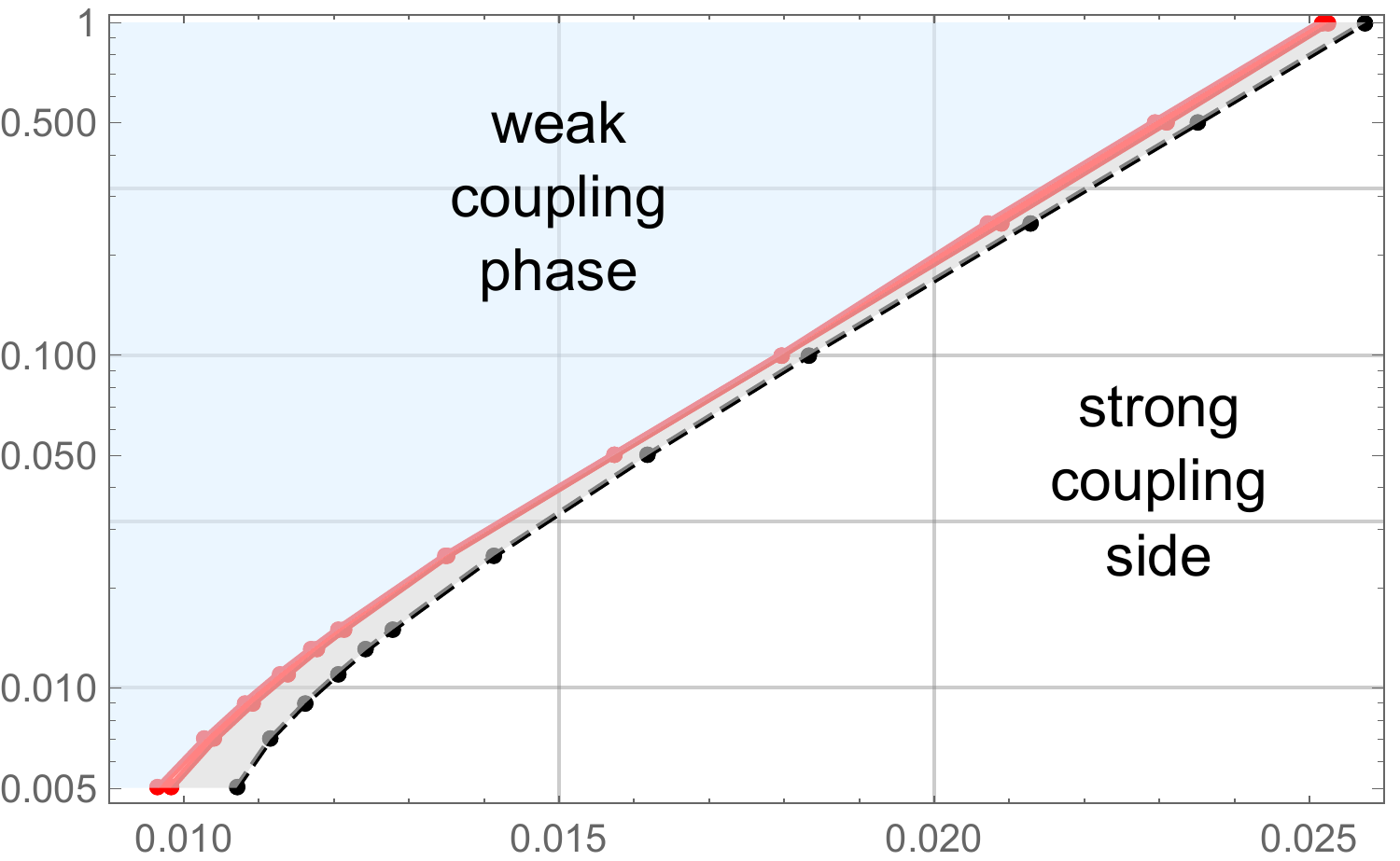}
\label{Fig:nu06_pd1}
}
\hfill
\subfigure[$\nu=0.6$ (enlarged)]{
\includegraphics[width=0.3\textwidth]{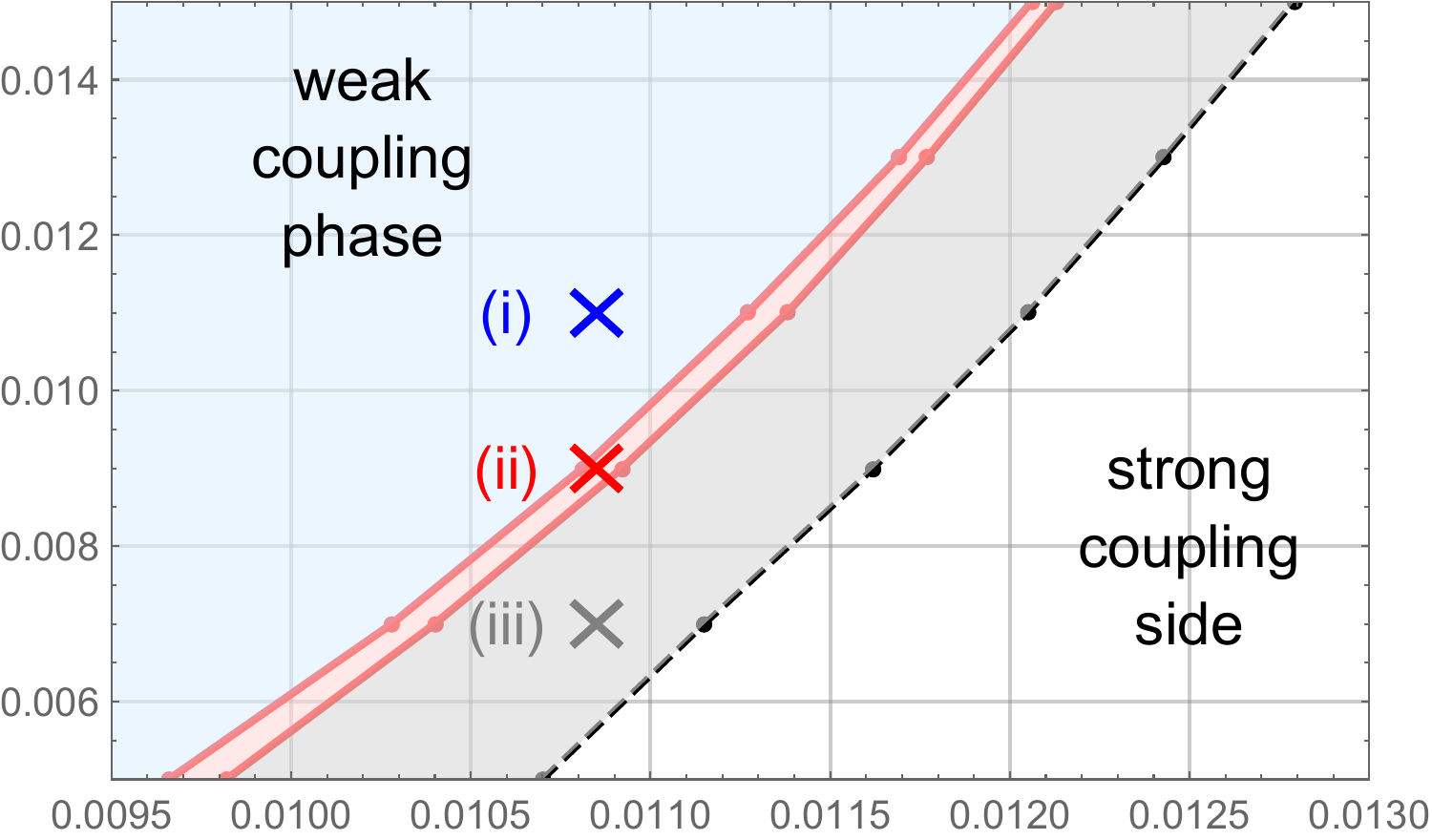}
\label{Fig:nu06_pd2}
}
\caption{Phase diagrams obtained from numerical solutions of the RG equations [Eqs.~(\ref{Eq:RGe2D_3},\ref{Eq:RGchiD_3},\ref{Eq:RGgammatauD_3},\ref{Eq:RGtr},\ref{Eq:RGttau}) together with Eqs.~(\ref{Eq:Z1-soft-approx},\ref{Eq:Z2tr-soft-approx})] with initial values of $e^2$, $t_{\tau}=t_{\bm r} \equiv t$, $\gamma_{\tau}=1$ and $\nu=1.2$ (left) and $0.6$ (middle, and right). The vertical and horizontal axes axis are initial values of $e^2$ and $t$, respectively. The light blue, light red, and light gray (white) regions stand for ordered, quasi-disordered, and disordered phases, respectively.  A set of initial parameters in the light blue region goes to the weak-coupling region with divergent $e^2$ and $\nu$, and negatively divergent $t_{\tau}$ and $t_{\bm r}$, and constant $\gamma_{\tau}$ [see Figs.~\ref{Fig:nu12_Ord_e2nu_gau1}. \ref{Fig:nu12_Ord_tttr_gau1} , \ref{Fig:nu12_Ord_igt_gau1} for how the parameters evolve from the blue cross point ``(i)" in Fig.~\ref{Fig:nu12_pd}; see Figs.\ref{Fig:nu06_Ord_e2nu_gau1}. \ref{Fig:nu06_Ord_tttr_gau1} , \ref{Fig:nu06_Ord_igt_gau1} for how the parameters evolves from the blue cross point ``(i)" in Fig.~\ref{Fig:nu06_pd2}]. Initial parameters in the light gray (white) region go to the strong-coupling regions with vanishing $e^2$ and $\nu$, positively divergent $t_{\tau}$, and negatively (positively) divergent $t_{\bm r}$, and divergent $\gamma^{-1}_{\tau}$ [see Figs.~\ref{Fig:nu12_Dis_e2nu_gau1}. \ref{Fig:nu12_Dis_tttr_gau1} , \ref{Fig:nu12_Dis_igt_gau1} for how the parameters evolve from the gray cross point ``(iii)" in Fig.~\ref{Fig:nu12_pd}; see Figs.\ref{Fig:nu06_Dis_e2nu_gau1}. \ref{Fig:nu06_Dis_tttr_gau1} , \ref{Fig:nu06_Dis_igt_gau1} for how the parameters evolves from the gray cross point ``(iii)" in Fig.~\ref{Fig:nu06_pd2}]. Initial parameters in the light red region go to the intermediate coupling region, where $\gamma^{-1}_{\tau}$ diverges at a finite value of the RG scale $\ln b$ [see Fig.~\ref{Fig:nu12_Bou_igt_gau1} for how $\gamma^{-1}_{\tau}$ evolve from the red cross point ``(ii)" in Fig.~\ref{Fig:nu12_pd}; see Fig.\ref{Fig:nu06_Bou_igt_gau1} for how $\gamma^{-1}_{\tau}$ evolves from the red cross point ``(ii)" in Fig.~\ref{Fig:nu06_pd2}].}
\end{figure}
   
\begin{figure}[t]
\subfigure[$e^2$ and $\nu$ as functions of $\ln b$]{
\includegraphics[width=0.3\textwidth]{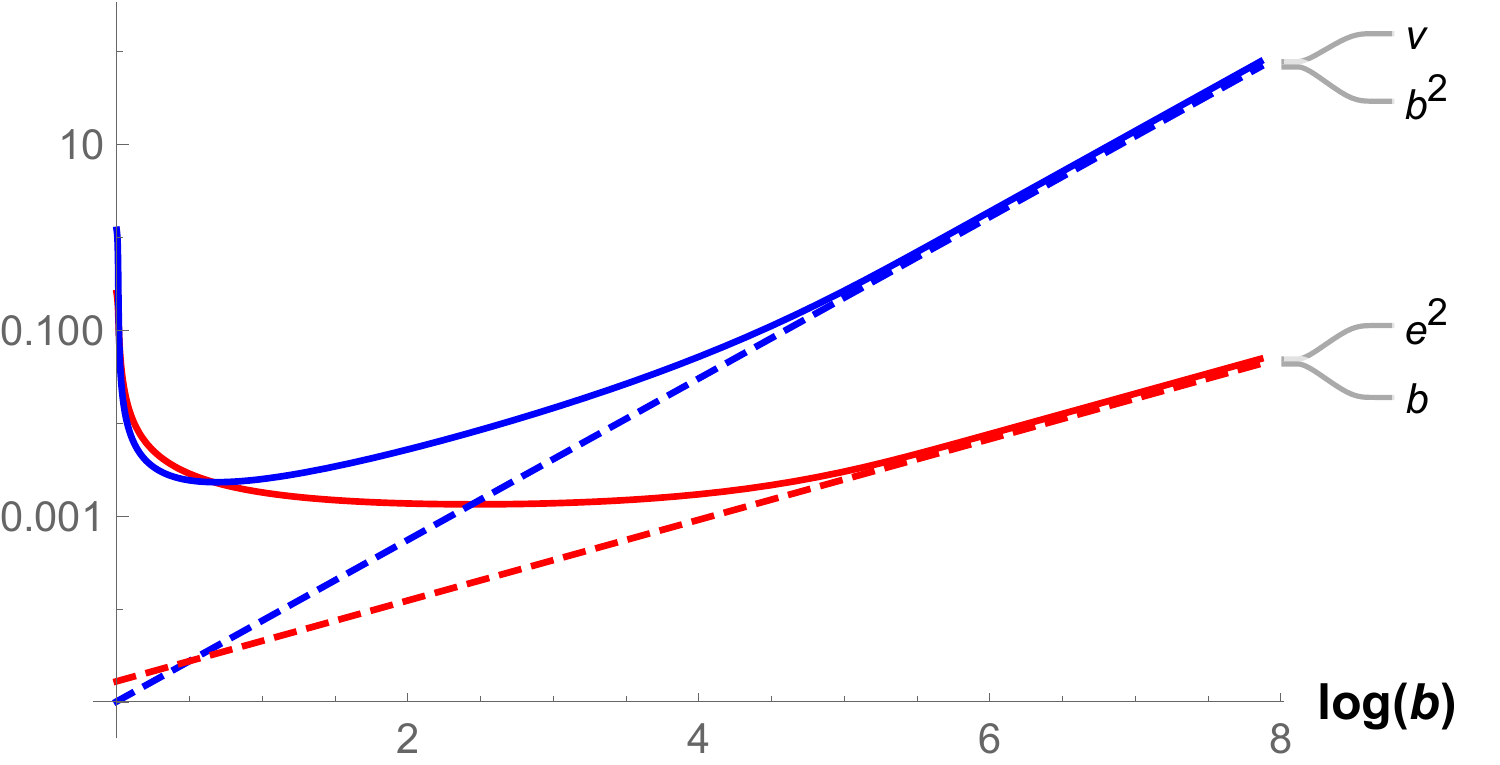}
\label{Fig:nu12_Ord_e2nu_gau1}
}
\hfill
\subfigure[$t_{\tau}$ and $t_{\bm r}$ as functions of $\ln b$]{
\includegraphics[width=0.3\textwidth]{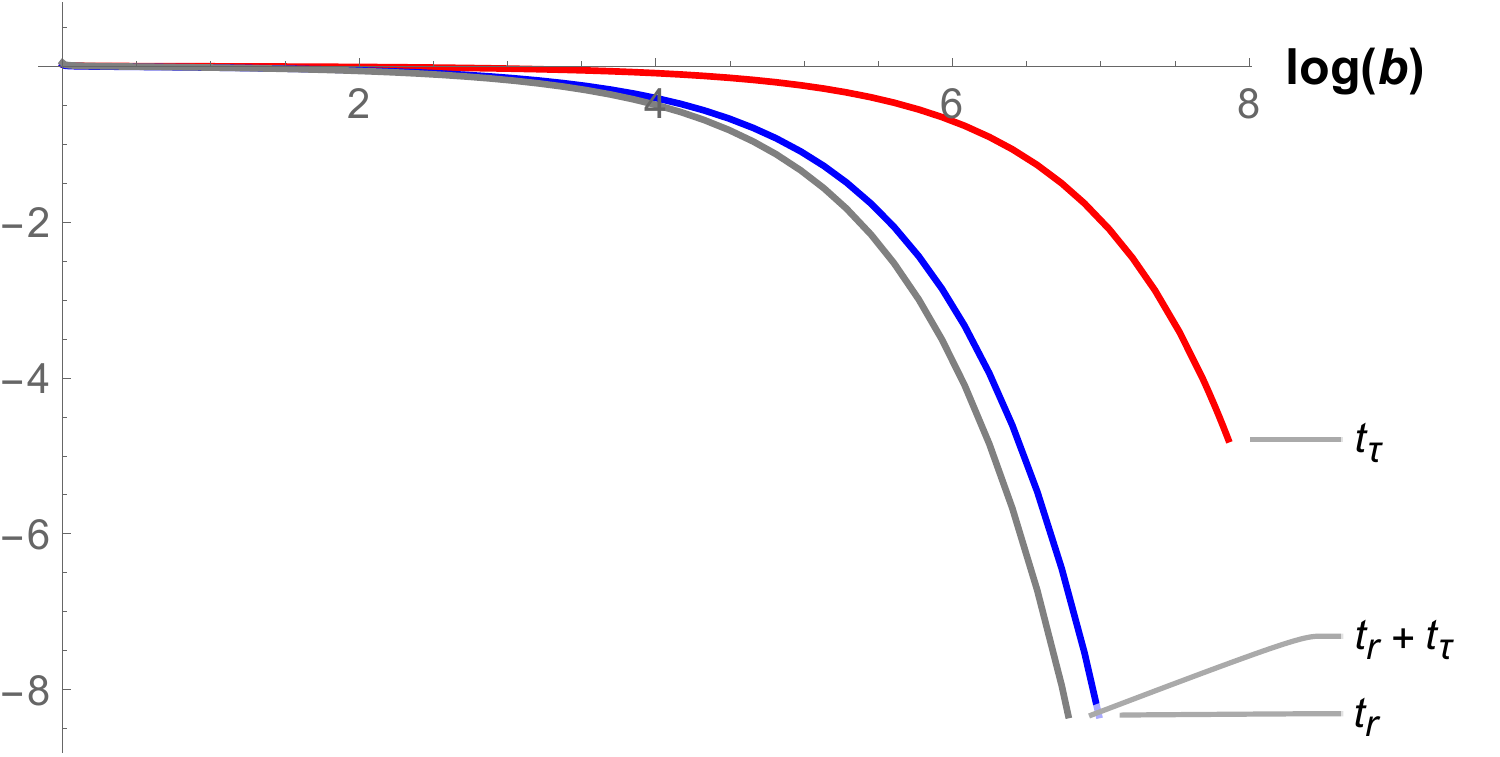}
\label{Fig:nu12_Ord_tttr_gau1}
}
\hfill
\subfigure[$\gamma^{-1}_{\tau}$ as a function of $\ln b$]{
\includegraphics[width=0.3\textwidth]{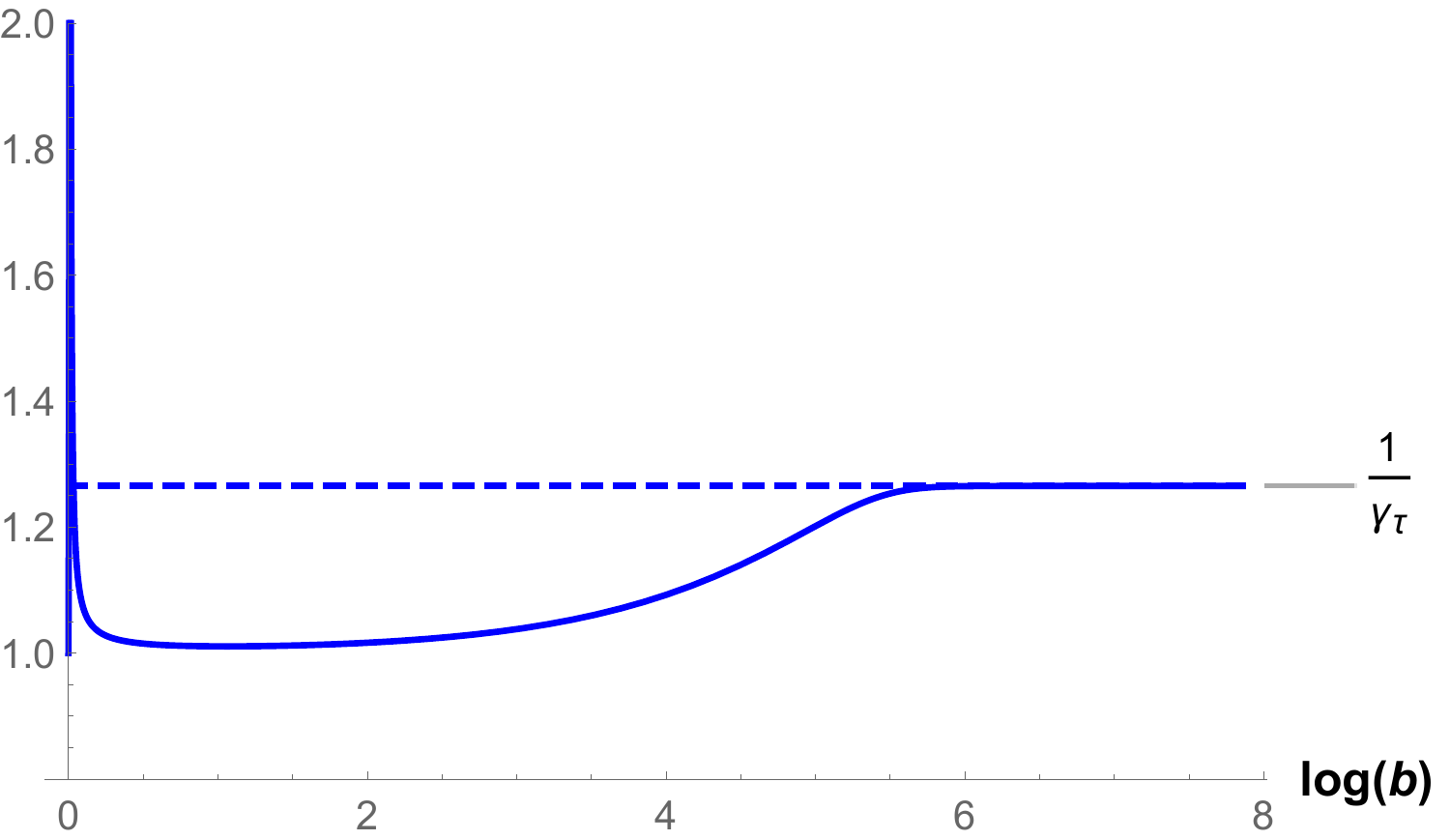}
\label{Fig:nu12_Ord_igt_gau1}
}
\vspace{0.5cm} 

\subfigure[$e^2$ and $\nu$ as functions of $\ln b$]{
\includegraphics[width=0.3\textwidth]{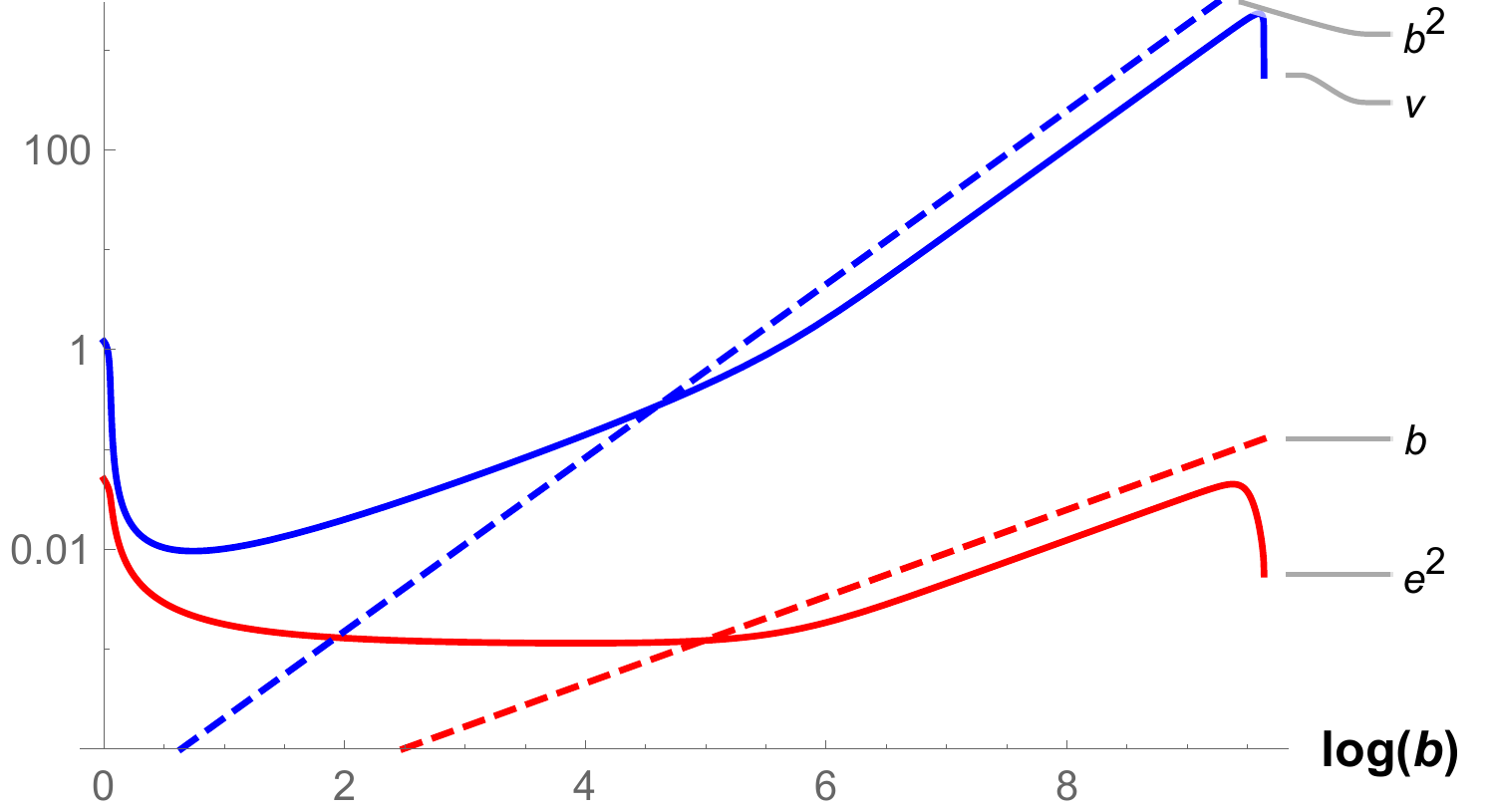}
\label{Fig:nu12_Bou_e2nu_gau1}
}
\hfill
\subfigure[$t_{\tau}$ and $t_{\bm r}$ as functions of $\ln b$]{
\includegraphics[width=0.3\textwidth]{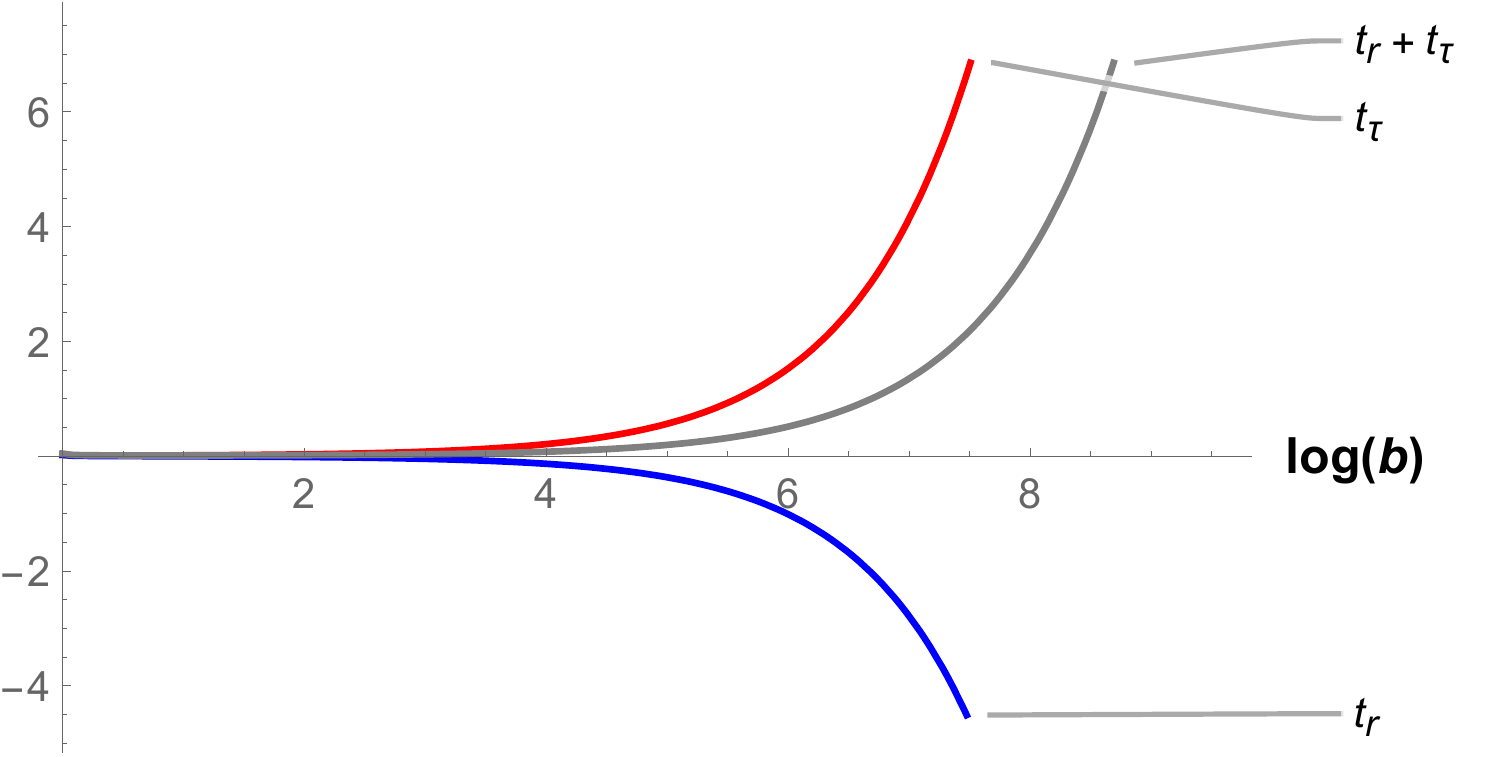}
\label{Fig:nu12_Bou_tttr_gau1}
}
\hfill
\subfigure[$\gamma^{-1}_{\tau}$ as a function of $\ln b$]{
\includegraphics[width=0.3\textwidth]{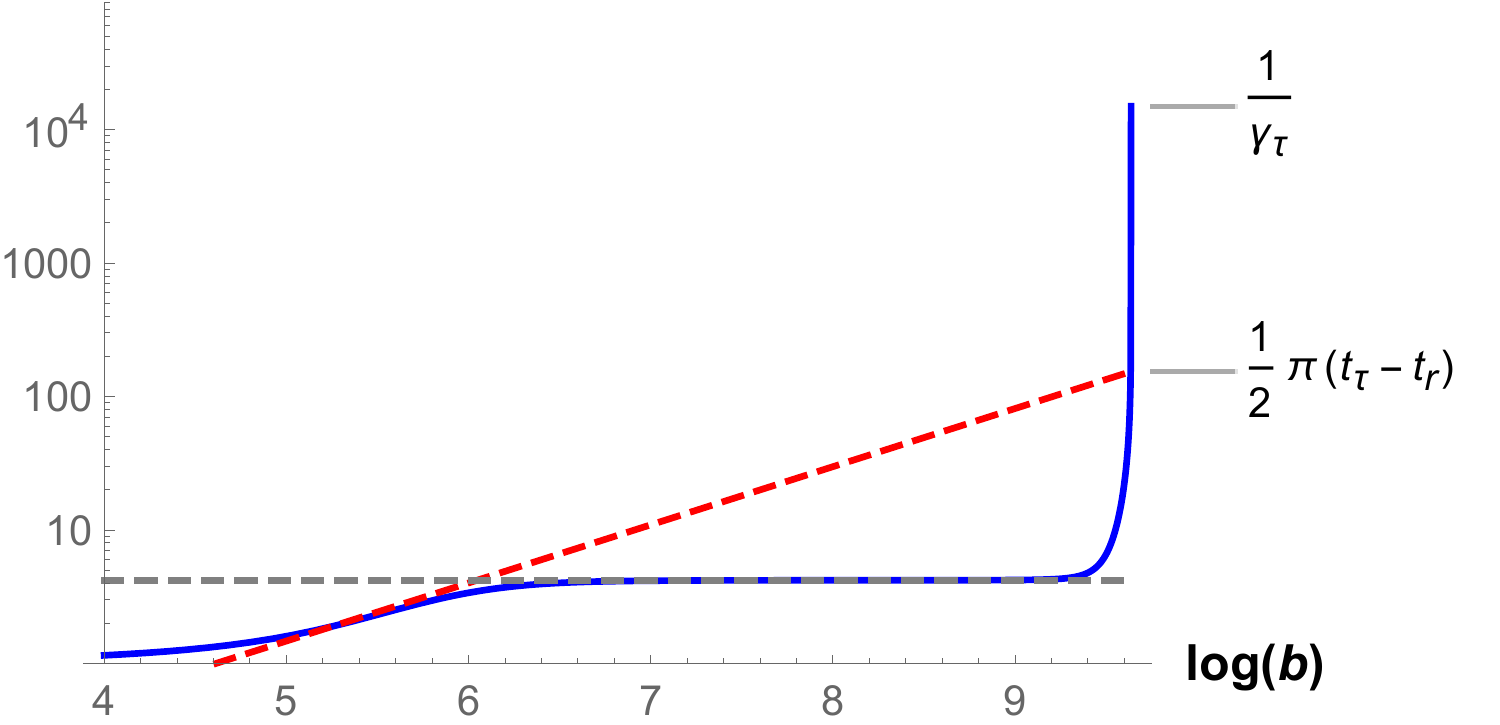}
\label{Fig:nu12_Bou_igt_gau1}
}
\vspace{0.5cm}
\subfigure[$e^2$ and $\nu$ as functions of $\ln b$]{
\includegraphics[width=0.3\textwidth]{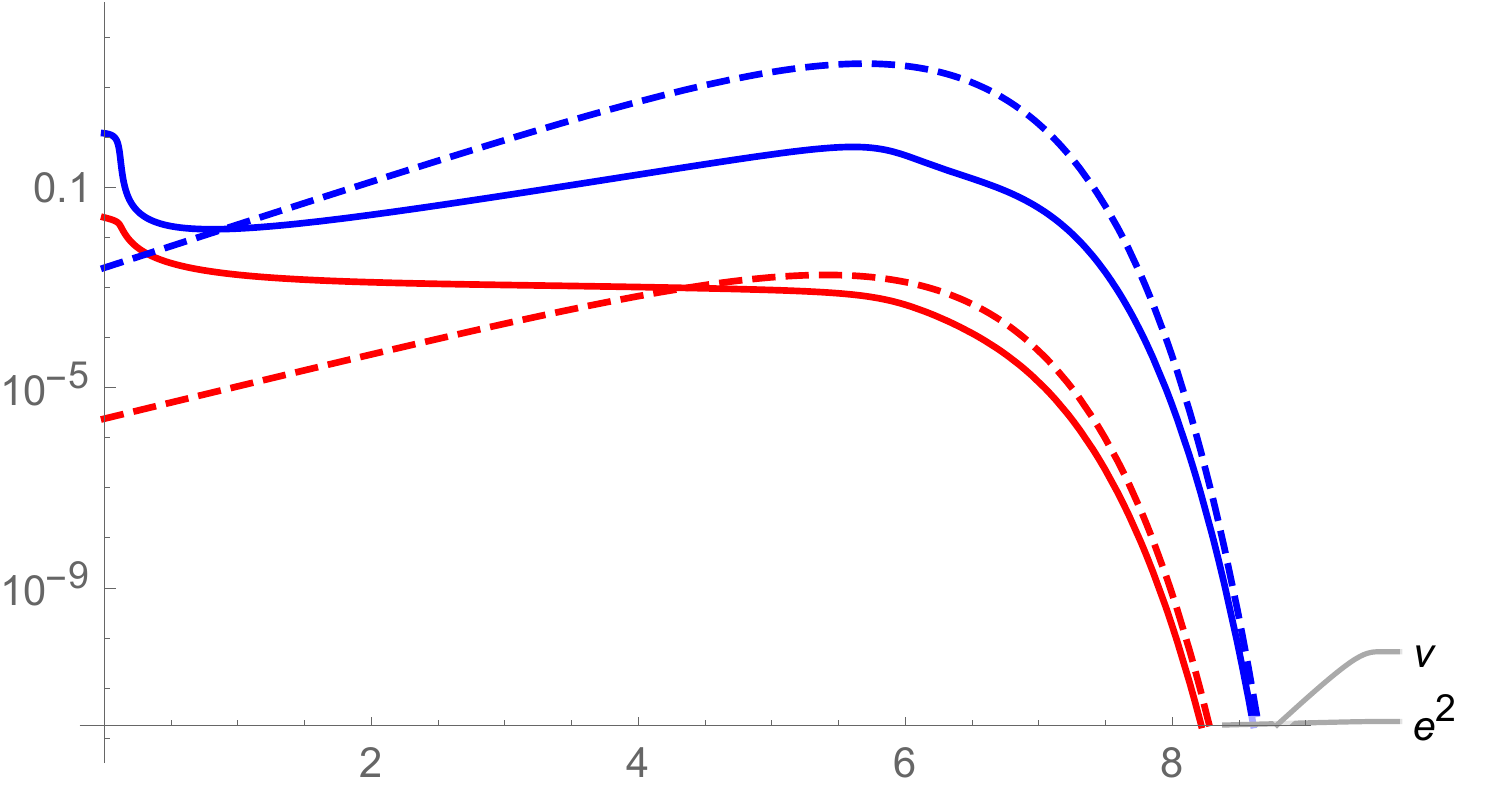}
\label{Fig:nu12_Dis_e2nu_gau1}
}
\hfill
\subfigure[$t_{\tau}$ and $t_{\bm r}$ as functions of $\ln b$]{
\includegraphics[width=0.3\textwidth]{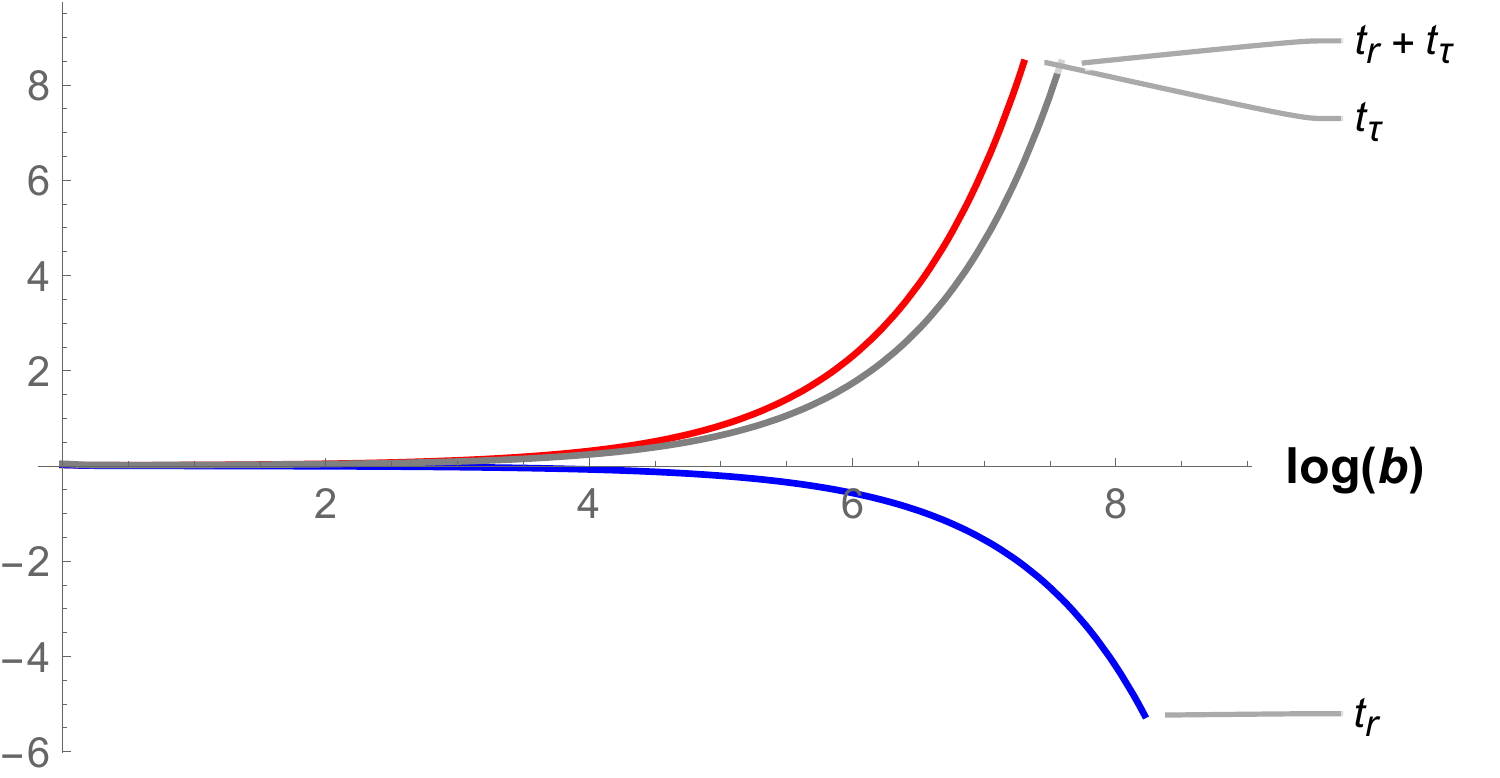}
\label{Fig:nu12_Dis_tttr_gau1}
}
\hfill
\subfigure[$\gamma^{-1}_{\tau}$ as a function of $\ln b$]{
\includegraphics[width=0.3\textwidth]{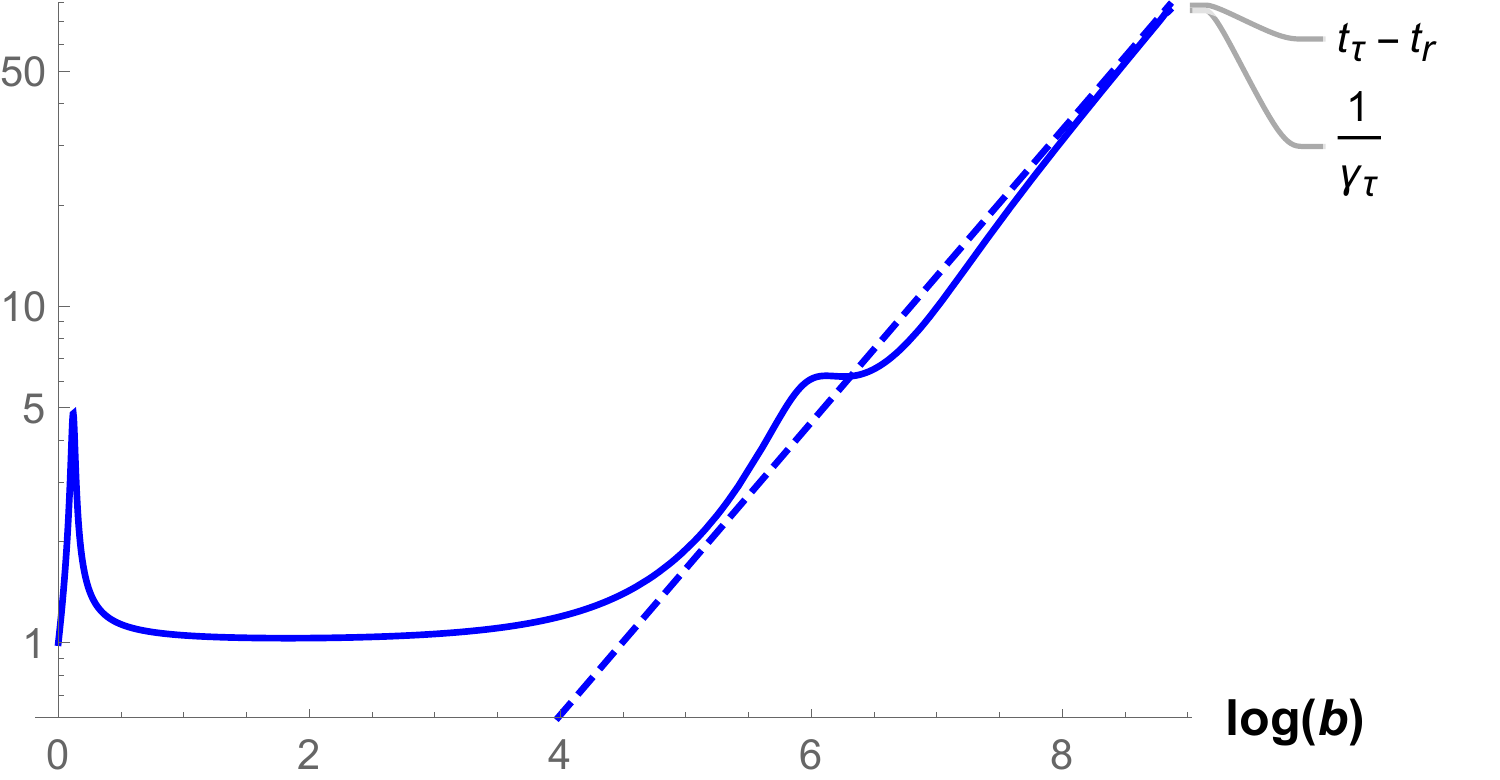}
\label{Fig:nu12_Dis_igt_gau1}
}
\caption{Plots of $e^2$, $\nu$, $t_{\tau}$, $t_{\bm r}$ and ${\gamma^{-1}_{\tau}}$ as functions of the RG scale $\ln b$. The plots are obtained from numerical solutions of the RG equations [Eqs.~(\ref{Eq:RGe2D_3},\ref{Eq:RGchiD_3},\ref{Eq:RGgammatauD_3},\ref{Eq:RGtr},\ref{Eq:RGttau}) together with Eqs.~(\ref{Eq:Z1-soft-approx},\ref{Eq:Z2tr-soft-approx})]. The initial value of $\nu$ and $\gamma^{-1}_{\tau}$ are $1.2$ and $1$, respectively. As for the initial values of $e^2$ and $t_{\tau}=t_{\bm r}=t$, we chose $(e^2,t)_{|\ln b=0}=(0.25,0.02265)$ [denoted by blue cross mark in Fig.~\ref{Fig:nu12_pd}] for Figs.~\ref{Fig:nu12_Ord_e2nu_gau1},\ref{Fig:nu12_Ord_tttr_gau1},\ref{Fig:nu12_Ord_igt_gau1}, $(e^2,t)_{|\ln b=0}=(0.05,0.02265)$ [denoted by red cross mark in Fig.~\ref{Fig:nu12_pd}] for Figs.~\ref{Fig:nu12_Bou_e2nu_gau1},\ref{Fig:nu12_Bou_tttr_gau1},\ref{Fig:nu12_Bou_igt_gau1}, and $(e^2,t)_{|\ln b=0}=(0.025,0.02265)$ [denoted by gray cross mark in Fig.~\ref{Fig:nu12_pd}] for Figs.~\ref{Fig:nu12_Dis_e2nu_gau1},\ref{Fig:nu12_Dis_tttr_gau1},\ref{Fig:nu12_Dis_igt_gau1}. Dashed lines in Figs.~\ref{Fig:nu12_Dis_e2nu_gau1}, \ref{Fig:nu12_Dis_igt_gau1} are from Eq.~(\ref{Eq:3D_Dis_asym}).}
\label{Fig:3D_nu12}
\end{figure}
\begin{figure}[b]
\subfigure[$e^2$ and $\nu$ as functions of $\ln b$]{
\includegraphics[width=0.3\textwidth]{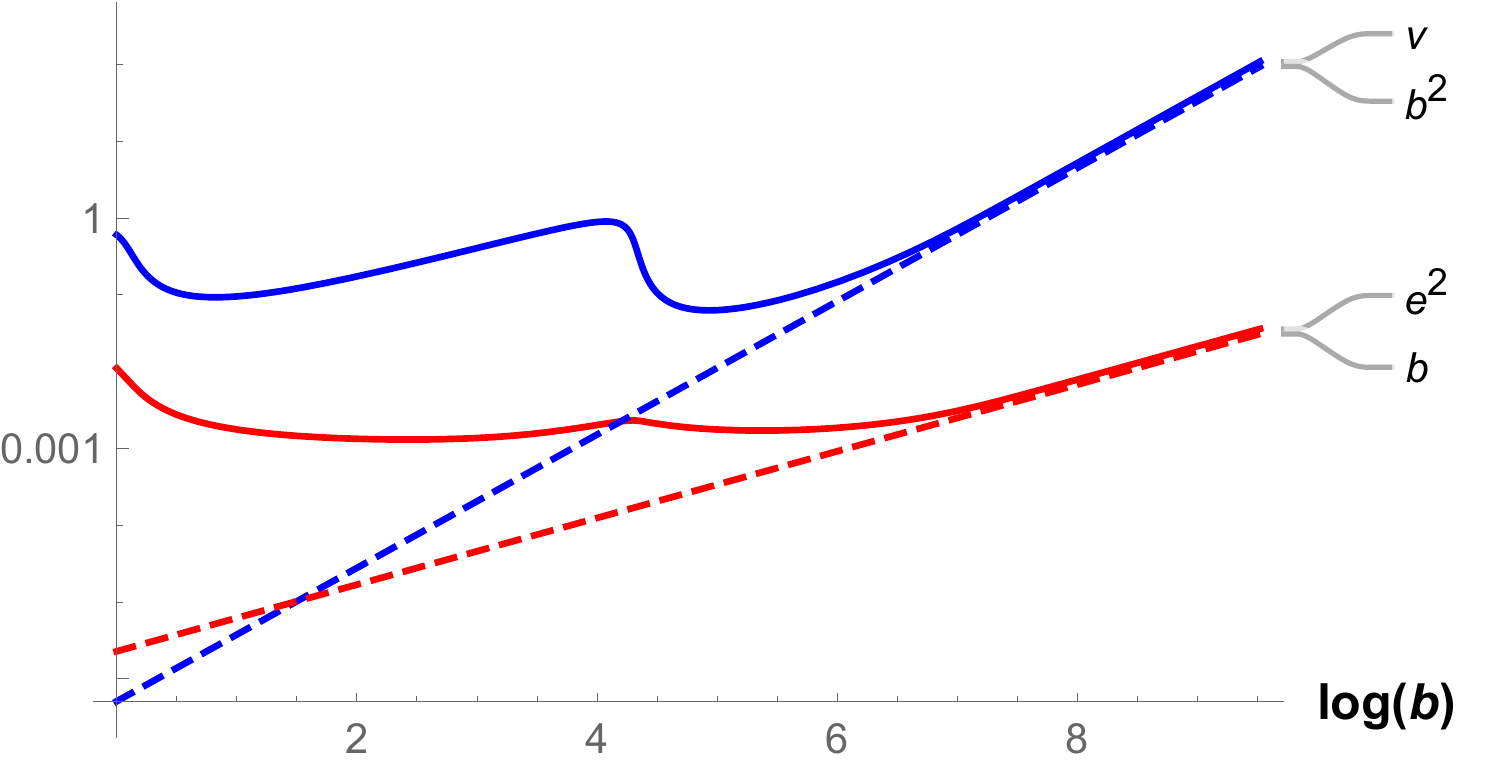}
\label{Fig:nu06_Ord_e2nu_gau1}
}
\hfill
\subfigure[$t_{\tau}$ and $t_{\bm r}$ as functions of $\ln b$]{
\includegraphics[width=0.3\textwidth]{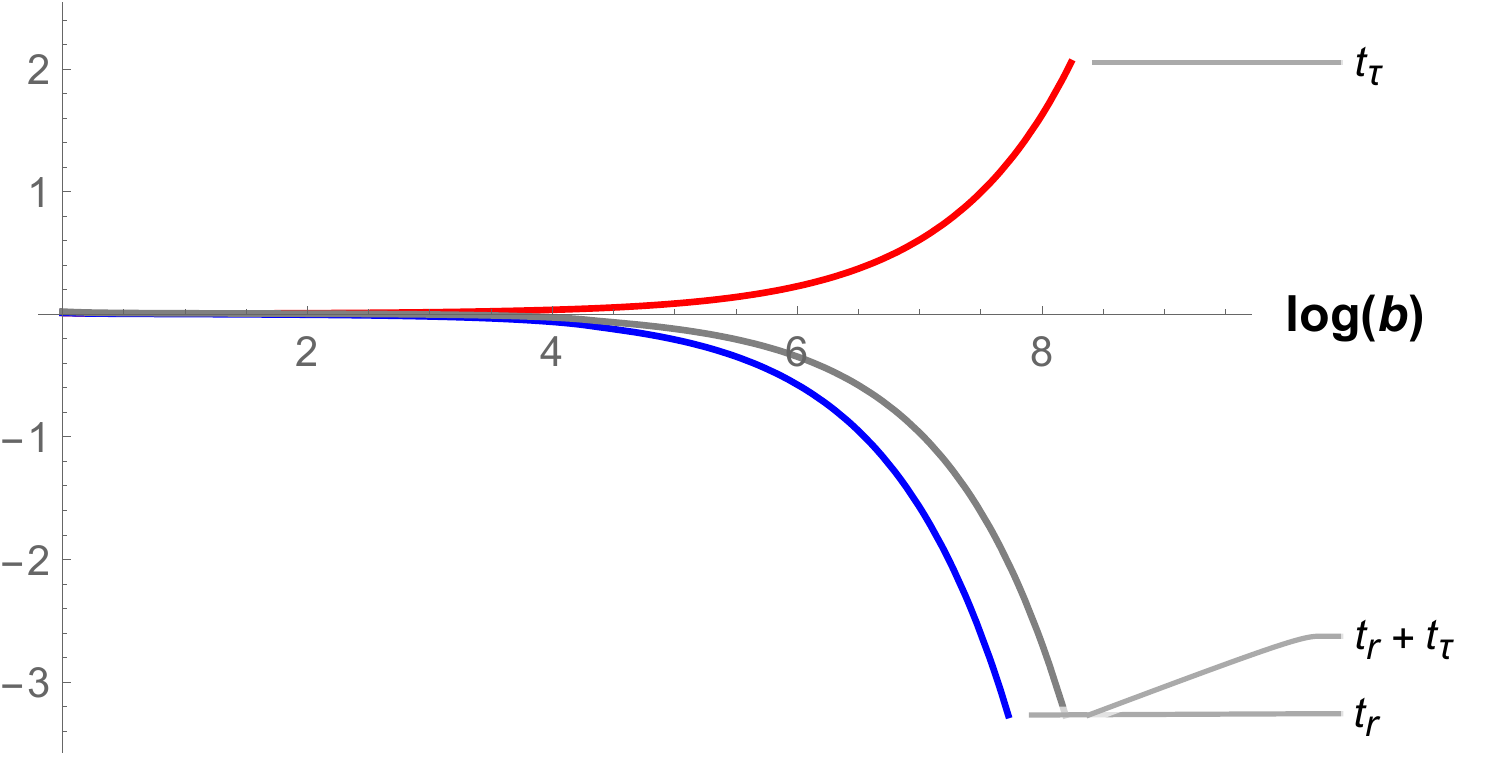}
\label{Fig:nu06_Ord_tttr_gau1}
}
\hfill
\subfigure[$\gamma^{-1}_{\tau}$ as a function of $\ln b$]{
\includegraphics[width=0.3\textwidth]{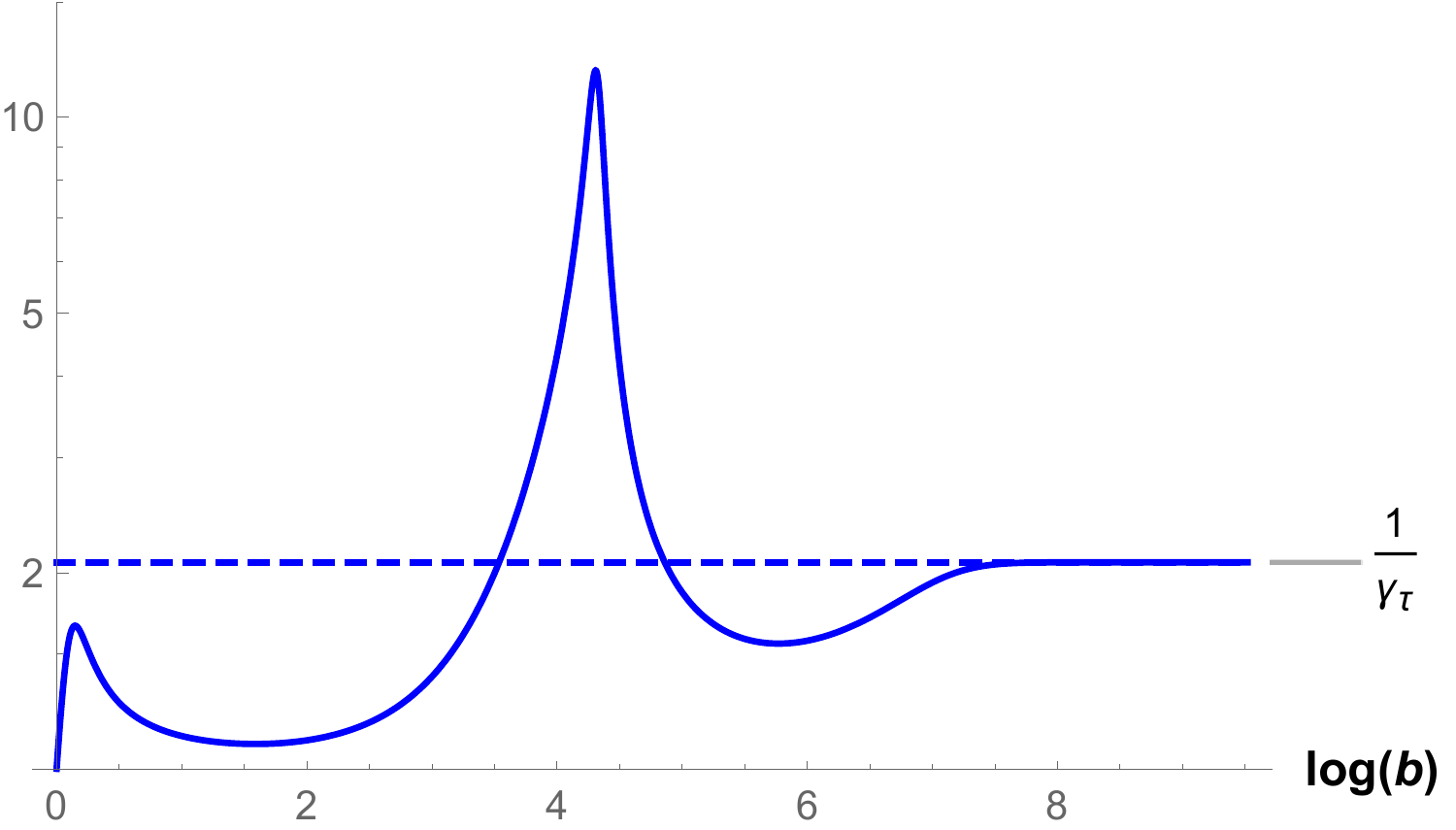}
\label{Fig:nu06_Ord_igt_gau1}
}
\vspace{0.5cm}
\subfigure[$e^2$ and $\nu$ as functions of $\ln b$]{
\includegraphics[width=0.3\textwidth]{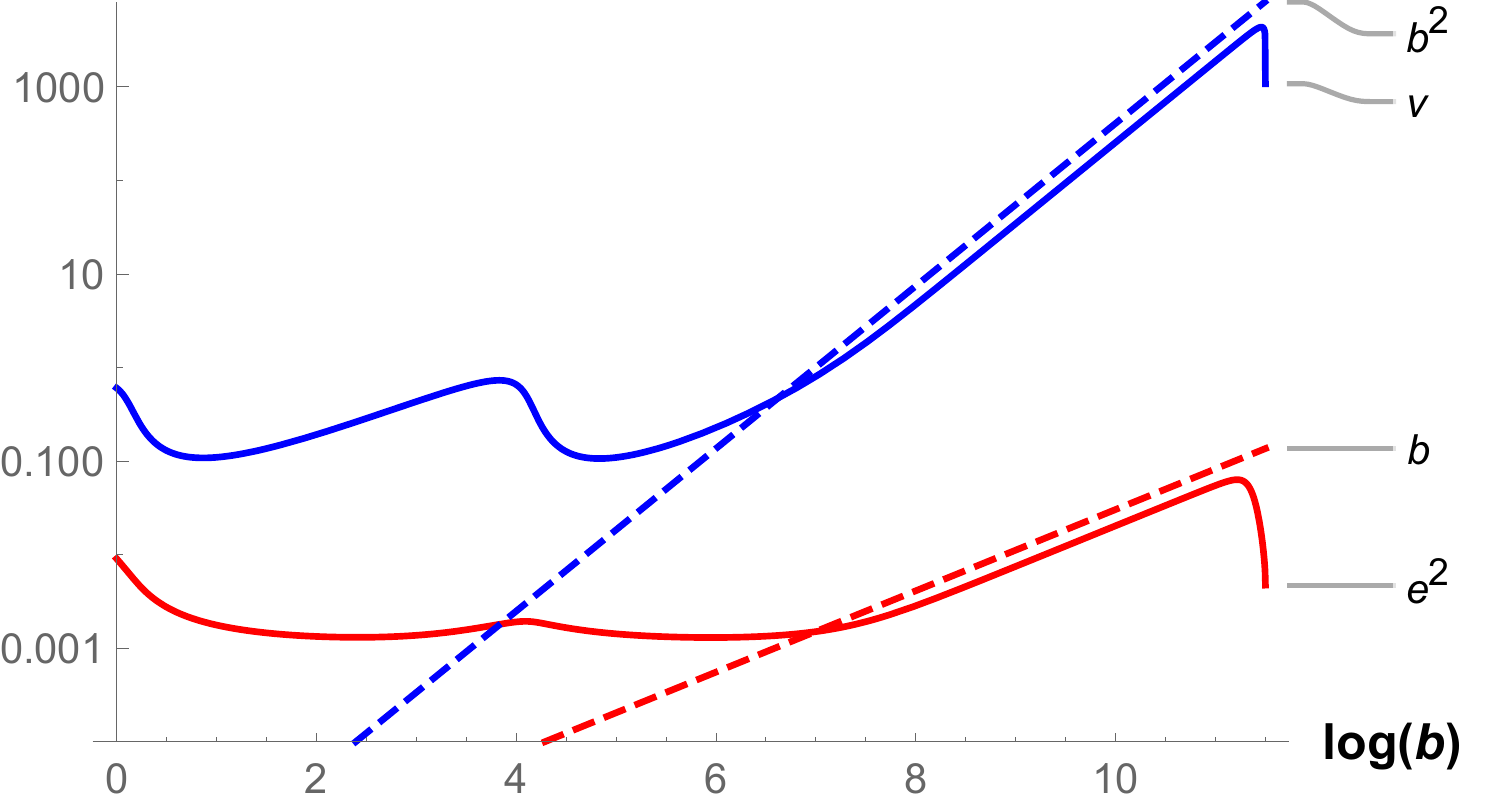}
\label{Fig:nu06_Bou_e2nu_gau1}
}
\hfill
\subfigure[$t_{\tau}$ and $t_{\bm r}$ as functions of $\ln b$]{
\includegraphics[width=0.3\textwidth]{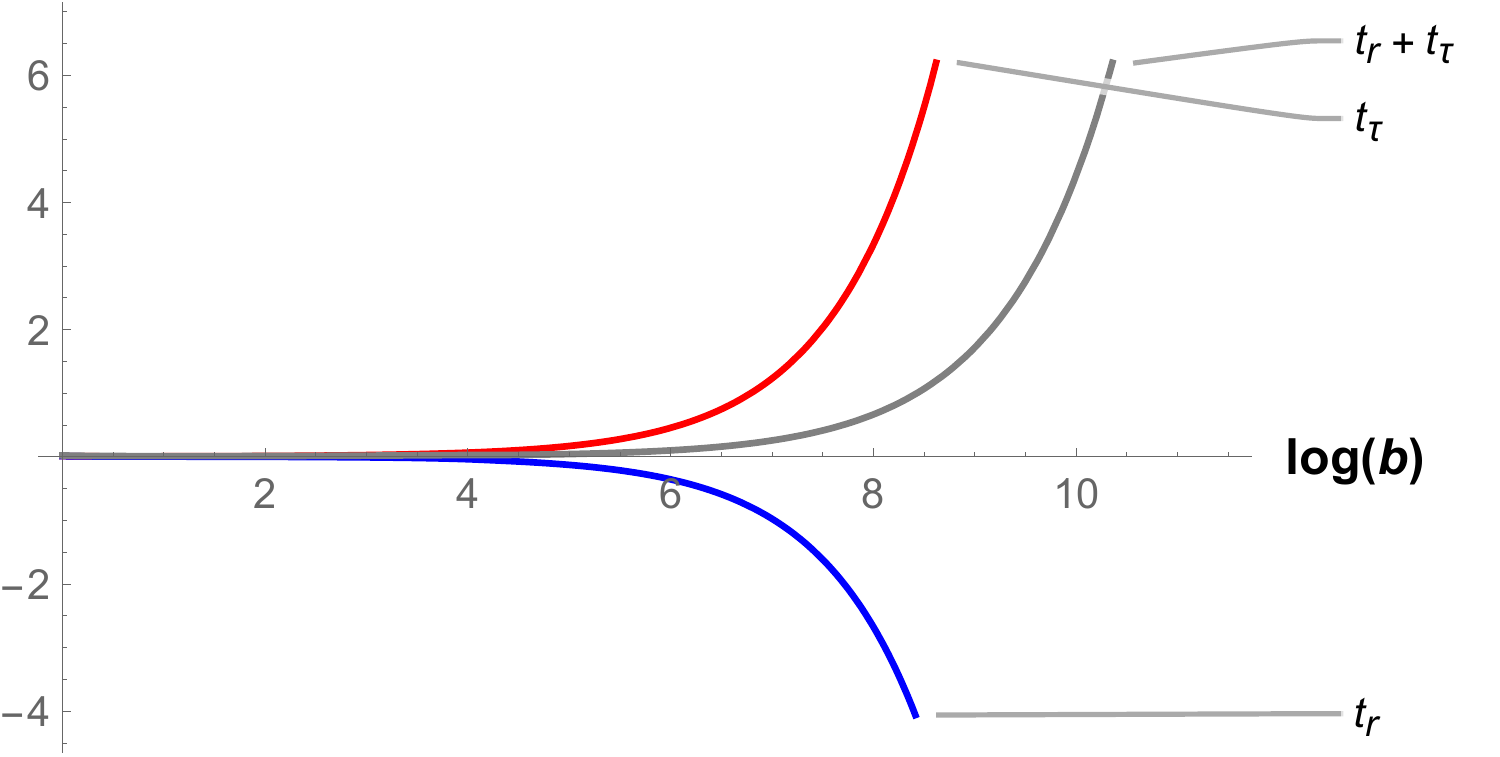}
\label{Fig:nu06_Bou_tttr_gau1}
}
\hfill
\subfigure[$\gamma^{-1}_{\tau}$ as a function of $\ln b$]{
\includegraphics[width=0.3\textwidth]{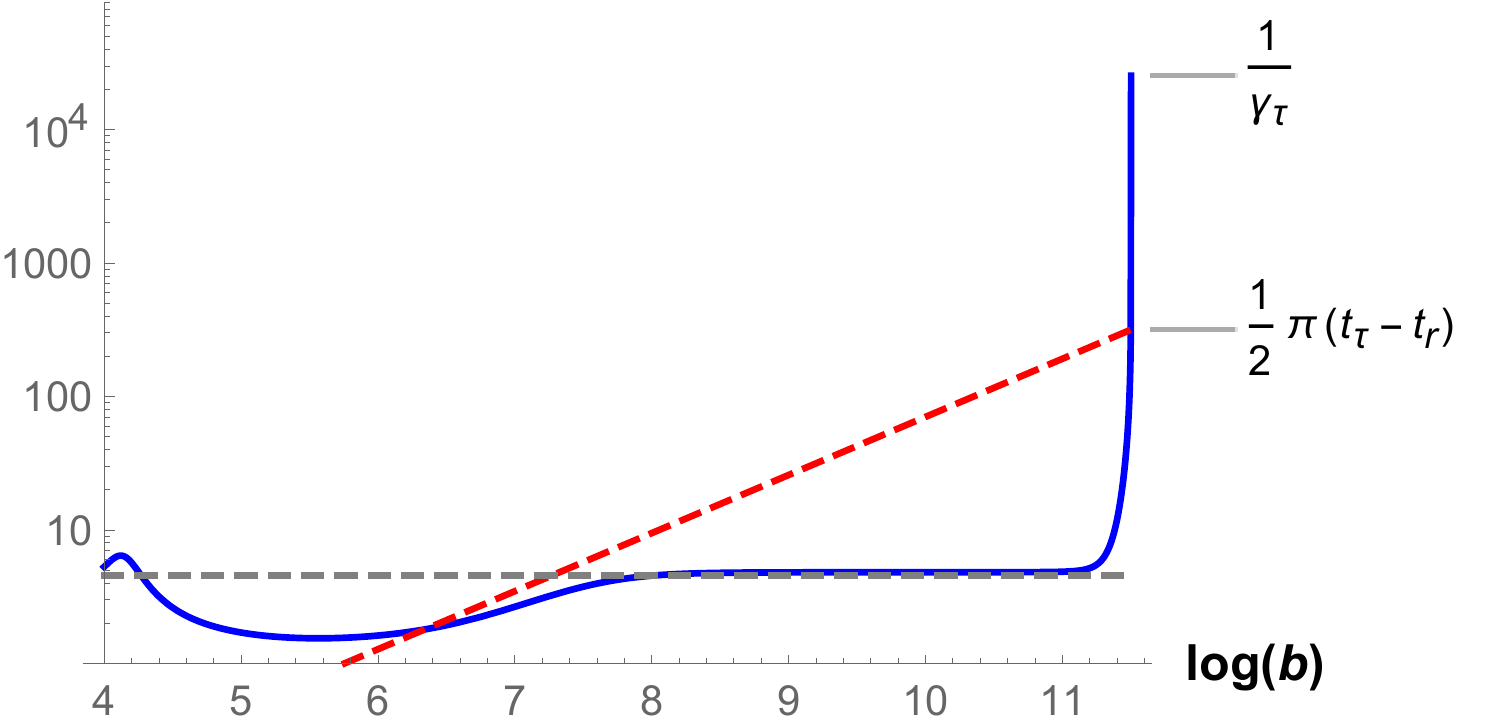}
\label{Fig:nu06_Bou_igt_gau1}
}
\vspace{0.5cm}
\subfigure[$e^2$ and $\nu$ as functions of $\ln b$]{
\includegraphics[width=0.3\textwidth]{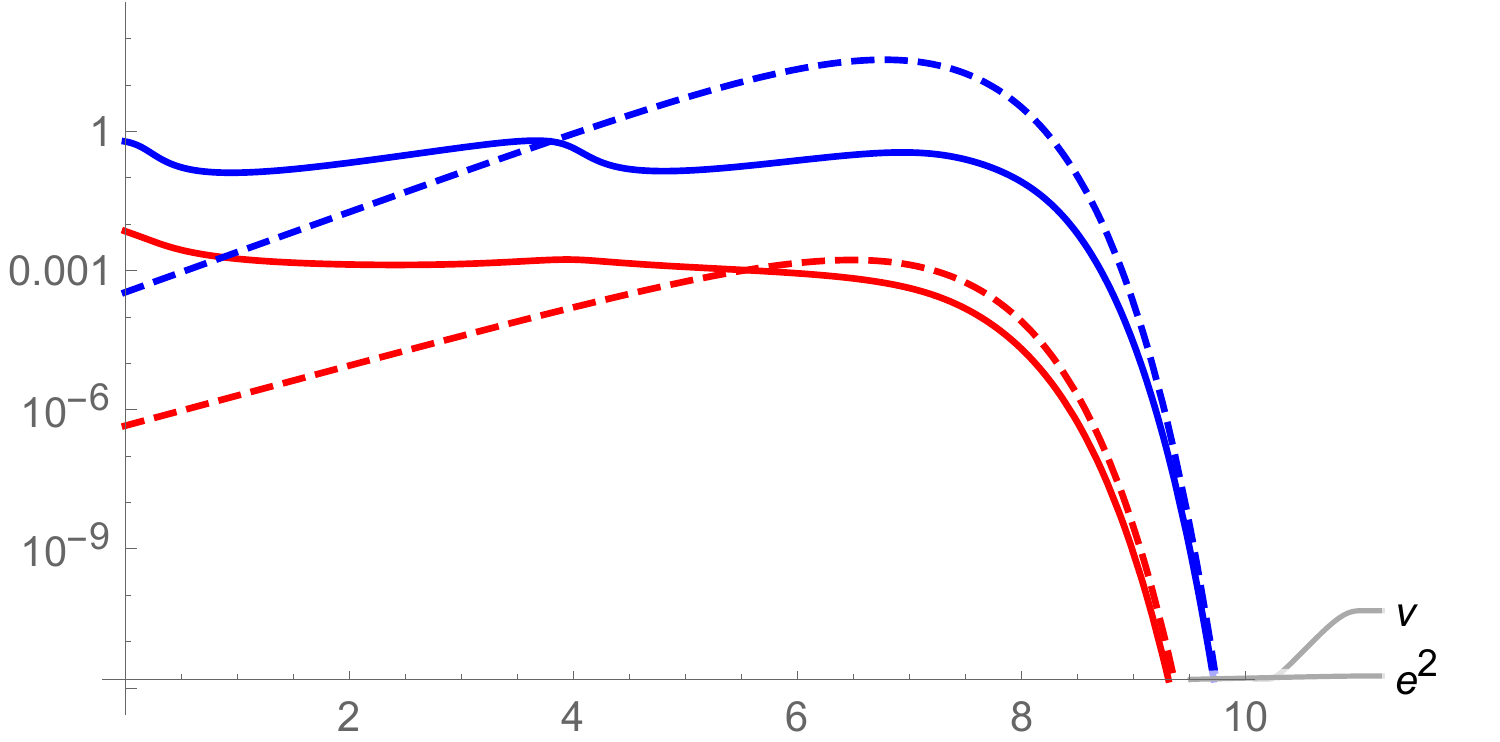}
\label{Fig:nu06_Dis_e2nu_gau1}
}
\hfill
\subfigure[$t_{\tau}$ and $t_{\bm r}$ as functions of $\ln b$]{
\includegraphics[width=0.3\textwidth]{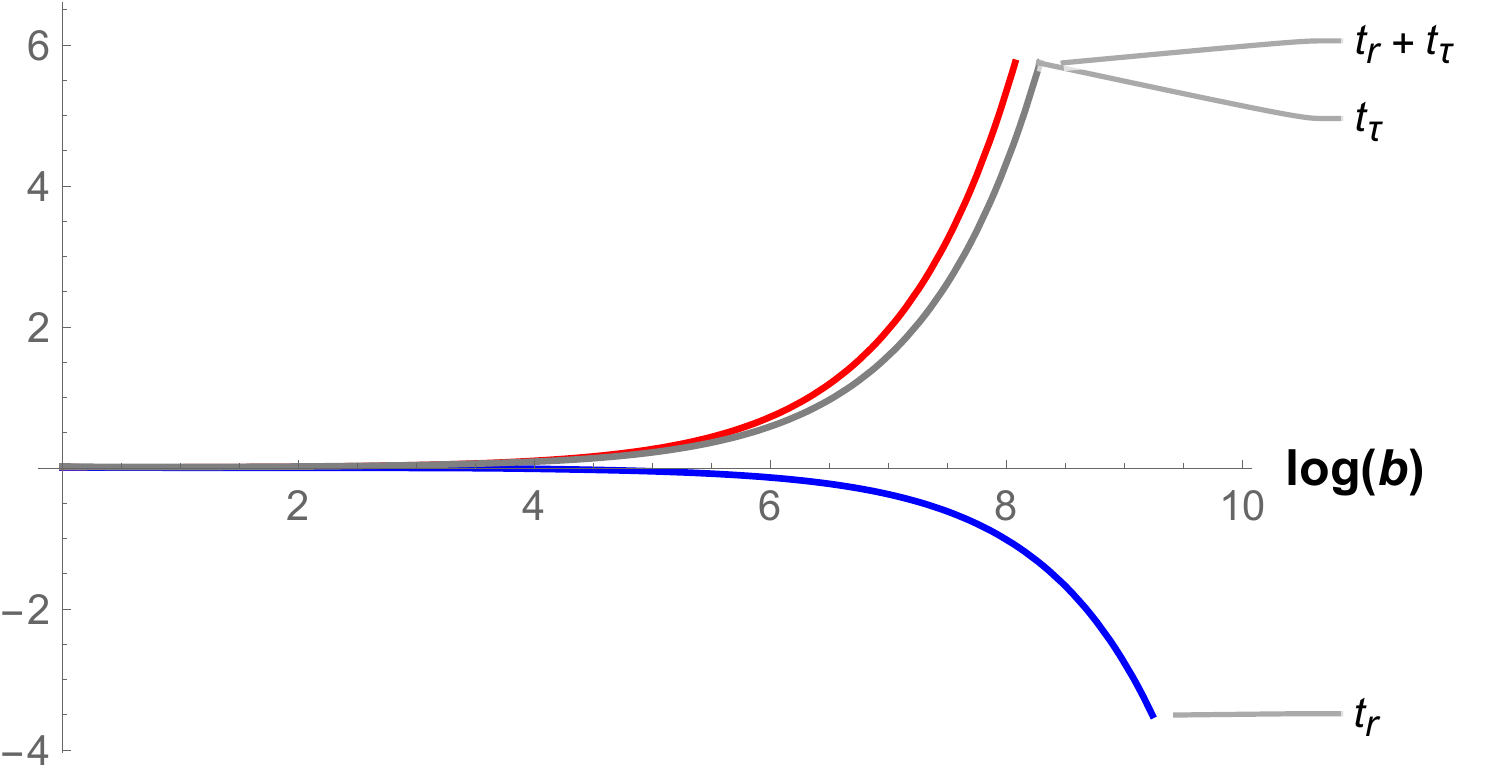}
\label{Fig:nu06_Dis_tttr_gau1}
}
\hfill
\subfigure[$\gamma^{-1}_{\tau}$ as a function of $\ln b$]{
\includegraphics[width=0.3\textwidth]{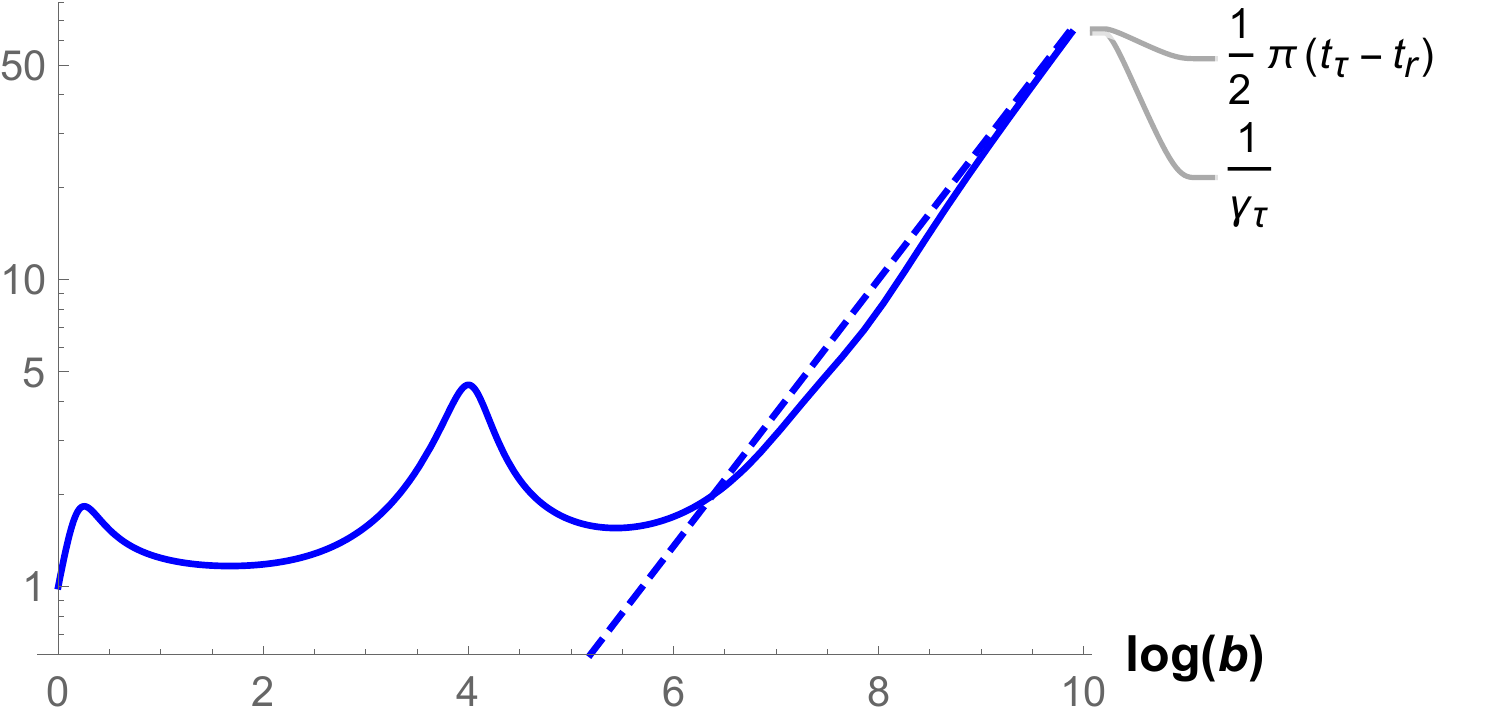}
\label{Fig:nu06_Dis_igt_gau1}
}
\caption{Plots of $e^2$, $\nu$, $t_{\tau}$, $t_{\bm r}$ and ${\gamma^{-1}_{\tau}}$ as functions of the RG scale $\ln b$. The plots are obtained from numerical solutions of the RG equations [Eqs.~(\ref{Eq:RGe2D_3},\ref{Eq:RGchiD_3},\ref{Eq:RGgammatauD_3},\ref{Eq:RGtr},\ref{Eq:RGttau}) together with Eqs.~(\ref{Eq:Z1-soft-approx},\ref{Eq:Z2tr-soft-approx})]. The initial value of $\nu$ and $\gamma^{-1}_{\tau}$ are $0.6$ and $1$, respectively. As for the initial values of $e^2$ and $t_{\tau}=t_{\bm r}=t$, we chose $(e^2,t)_{|\ln b=0}=(0.011,0.01085)$ [denoted by blue cross mark in Fig.~\ref{Fig:nu06_pd2}] for Figs.~\ref{Fig:nu06_Ord_e2nu_gau1},\ref{Fig:nu06_Ord_tttr_gau1},\ref{Fig:nu06_Ord_igt_gau1}, $(e^2,t)_{|\ln b=0}=(0.009,0.01085)$ [denoted by red cross mark in Fig.~\ref{Fig:nu06_pd2}] for Figs.~\ref{Fig:nu06_Bou_e2nu_gau1},\ref{Fig:nu06_Bou_tttr_gau1},\ref{Fig:nu06_Bou_igt_gau1}, and $(e^2,t)_{|\ln b=0}=(0.007,0.01085)$ [denoted by gray cross mark in Fig.~\ref{Fig:nu06_pd2}] for Figs.~\ref{Fig:nu06_Dis_e2nu_gau1},\ref{Fig:nu06_Dis_tttr_gau1},\ref{Fig:nu06_Dis_igt_gau1}. Dashed lines in Figs.~\ref{Fig:nu06_Dis_e2nu_gau1}, \ref{Fig:nu06_Dis_igt_gau1} are from Eq.~(\ref{Eq:3D_Dis_asym}).}
\label{Fig:3D_nu06}
\end{figure}

         
             The phase diagrams of the RG equation are numerically obtained in a two-dimensional parameter space subtended by initial values of $t=t_{\tau}=t_{\bm r}$, and $e^2$ [Figs.~\ref{Fig:nu12_pd}, \ref{Fig:nu06_pd1}]. We choose several initial values of $\nu$ and $\gamma_{\tau}$ to be $(1.2,1)$ and $(0.6,1)$ for Fig.~\ref{Fig:nu12_pd} and 
             Fig.~\ref{Fig:nu06_pd1}, respectively. The phase diagrams generally show that larger (smaller) initial values of $t$ and smaller (larger)  $e^2$ tend to be in the strong-coupling  (weak-coupling) regions. For these two regions, the 
             large-$b$ behaviors of $e^2$, $\nu$, $\gamma_{\tau}$, and $t_{\tau}+t_{\bm r}$ are consistent with the previous arguments [Fig.~\ref{Fig:3D_nu12}]. Namely, when the initial parameters are in the weak-coupling region [light blue region], both $e^2$ and $\nu$ diverge with $e^2\sim b$ and $\nu\sim b^2$  in the large $b$ limit [Figs.~\ref{Fig:nu12_Ord_e2nu_gau1},\ref{Fig:nu06_Ord_e2nu_gau1}], while $t_{\tau}+t_{\bm r}$ diverges negatively [Figs.~\ref{Fig:nu12_Ord_tttr_gau1},\ref{Fig:nu06_Ord_tttr_gau1}], and $\gamma_{\tau}$ converges to a finite non-universal constant [Figs.~\ref{Fig:nu12_Ord_igt_gau1},\ref{Fig:nu06_Ord_igt_gau1}]. When the initial parameters are in the strong-coupling region [gray or white regions], both $e^2$ and $\nu$ vanish exponentially in the large $b$ limit [Figs.~\ref{Fig:nu12_Dis_e2nu_gau1},\ref{Fig:nu06_Dis_e2nu_gau1}], while $t_{\tau}+t_{\bm r}$ diverges positively [Figs.~\ref{Fig:nu12_Dis_tttr_gau1},\ref{Fig:nu06_Dis_tttr_gau1}], and $\gamma_{\tau}$ goes to zero as $\gamma_{\tau} \sim 2/(\pi(t_{\tau}-t_{\bm r}))$ [Figs.~\ref{Fig:nu12_Dis_igt_gau1},\ref{Fig:nu06_Dis_igt_gau1}].

           Importantly, when the initial parameters are in an intermediate region between these two regions [light red region], $\gamma^{-1}_{\tau}$ diverges at a finite RG scale $b$ [Figs.~\ref{Fig:nu12_Bou_igt_gau1},\ref{Fig:nu06_Bou_igt_gau1}] where $e^2$ and $t_{\tau}$ remain finite positive values [Figs.~\ref{Fig:nu12_Bou_e2nu_gau1},\ref{Fig:nu06_Bou_e2nu_gau1}, Figs.~\ref{Fig:nu12_Bou_tttr_gau1},\ref{Fig:nu06_Bou_tttr_gau1}]. Such solutions indicate a certain type of ``Landau-pole" instability in the intermediate region. To discuss this point, note first that the large $\gamma^{-1}_{\tau}$ with finite $e^2$ generally polarizes vortex loops along the $\tau$ direction. Namely, with larger $\gamma^{-1}_{\tau}$, the anisotropic Coulomb interaction energy between ``neighboring" vortex loop elements favors vortex loop polarized along $\tau$. From the analogy to 3-dimensional magnetostatics, the Coulomb energy $E_{\rm Coul}(C,\gamma_{\tau})$ of a vortex loop $C$ is given by  
        \begin{align}
        &\exp[-E_{\rm Coul}(C,\gamma_{\tau})]  \equiv \nonumber \\ 
        &\int D{\sf a}({\bm x}) \exp\bigg[-\textcolor{red}{\frac{1}{2}}\int d^3{\bm x} \!\ (\nabla \times {\sf a})^T \cdot \left(\begin{array}{cc} 
        \gamma_{\tau} & 0 \\ 
        0 & {\bf 1}_{2\times 2} \\
       \end{array}\right) \cdot (\nabla \times {\sf a}) - 2\pi i e \oint_C d{\bm l}\cdot {\sf a}({\bm l})\bigg], \nonumber 
        \end{align}
with ${\bm x} \equiv (\tau,{\bm r})$, $\nabla \equiv (\nabla_{\tau},\nabla_{\bm r})$, ${\bm l} \equiv(l_{\tau},{\bm l}_{\bm r})$, and ${\sf a}\equiv ({\sf a}_{\tau},{\sf a}_{\bm r})$.  To see how $E_{\rm Coul}(C,\gamma_{\tau})$ depends on the shape of the loop $C$, consider the following scale transformation, 
\begin{align}
&\tau^{\prime}=\tau, \,\ {\bm r}^{\prime} = {\bm r} \frac{1}{\sqrt{\gamma_{\tau}}},\,\ 
{\sf a}^{\prime}_{\tau} = {\sf a}_{\tau},\,\ {\sf a}^{\prime}_{\bm r} = \sqrt{\gamma_{\tau}} {\sf a}_{\bm r}, \nonumber \\
&\nabla^{\prime}_{\tau}=\nabla_{\tau}, \,\ 
\nabla^{\prime}_{\bm r}=\sqrt{\gamma_{\tau}}\nabla_{\bm r}, \,\ l^{\prime}_{\tau}=l_{\tau}, \,\ {\bm l}^{\prime}_{\bm r} = {\bm l}_{\bm r} \frac{1}{\sqrt{\gamma_{\tau}}}, 
\end{align}
which brings the Maxwell action with $\gamma_{\tau} \ne 1$ into the action with $\gamma_{\tau}=1$, 
\begin{align} 
&  \int D{\sf a}({\bm x}) \exp\bigg[-\textcolor{red}{\frac{1}{2}}\int d^3{\bm x} (\nabla \times {\sf a})^T \cdot \left(\begin{array}{cc} 
        \gamma_{\tau} & 0 \\ 
        0 & {\bf 1}_{2\times 2} \\
       \end{array}\right) \cdot (\nabla \times {\sf a}) - 2\pi i e \oint_C d{\bm l}\cdot {\sf a}({\bm l})\bigg]. \nonumber   \nonumber \\ 
        & = {\rm const}.\int D{\sf a}^{\prime}({\bm x}^{\prime}) \exp\bigg[-\textcolor{red}{\frac{1}{2}}\int d^3{\bm x}^{\prime} (\nabla^{\prime} \times {\sf a}^{\prime})^T \cdot \left(\begin{array}{cc} 
        1 & 0 \\ 
        0 & {\bf 1}_{2\times 2} \\
       \end{array}\right) \cdot (\nabla^{\prime} \times {\sf a}^{\prime}) - 2\pi i e \oint_{C^{\prime}} d{\bm l}^{\prime}\cdot {\sf a}^{\prime}({\bm l}^{\prime})\bigg], \nonumber \\ 
       &= {\rm const.}\exp[-E_{\rm Coul}(C^{\prime},\gamma_{\tau}=1)].  
\end{align} 
The scale transformation deforms $C$ into $C^{\prime}$ by stretching the loop in the ${\bm r}$ direction by a factor of $1/\sqrt{\gamma_{\tau}}$. The transformation relates the anisotropic Coulomb interaction energy $E_{\rm Coul}(C,\gamma_{\tau})$ of $C$ to the isotropic Coulomb interaction energy $E_{\rm Coul}(C^{\prime},\gamma_{\tau}=1)$ of the stretched loop $C^{\prime}$.  
\begin{align}
E_{\rm Coul}(C,\gamma_{\tau}) = E_{\rm Coul}(C^{\prime},\gamma_{\tau}=1) + {\rm const}. 
\end{align}
Here `const.' on the right-hand side is independent of the loops. Note that the Coulomb interaction energy of a vortex loop is generally dominated by its short-ranged part with the UV logarithmic divergence, $|\log a_0|$. In the isotropic limit, the coefficient of the UV logarithmic divergence is given by a vortex loop length,   
\begin{align}
E_{\rm Coul}(C^{\prime},\gamma_{\tau}=1)  = \frac{\pi e^2}{\textcolor{red}{2}} L(C^{\prime})|\log a_0| + \cdots.   
\end{align}
Thus, when $\gamma^{-1}_{\tau} \gg 1$, the dominant part of the anisotropic Coulomb energy $E_{\rm Coul}(C,\gamma_{\tau})$ disfavors vortex loops with finite projections in the ${\bm r}$ direction. Namely, by the stretch, a loop $C_1$ with a finite projection in ${\bm r}$ results in a longer loop than a loop $C_2$ without any projection in ${\bm r}$: in the limit of  $\gamma^{-1}_{\tau}\rightarrow \infty$, $L(C^{\prime}_1)\gg L(C^{\prime}_2)$.  The vortex loops without any projection to ${\bm r}$ are nothing but vortex lines parallel to the $\tau$ axis.

       Upon the divergence of $\gamma^{-1}_{\tau}$ with finite $e^2$, the vortex excitations take the form of vortex lines fully polarized along $\tau$, where the 3D partition function reduces to the 2D isotropic partition function for the BKT transition; 
\begin{align}
\lim_{\gamma_{\tau}\rightarrow 0}Z 
&= \int {\cal D} {\sf a}  
\exp \Big[-\frac{1}{2} \int d^3{\bm x} \sum_{\mu=1,2}\big(\partial_{\tau} {\sf a}_{\mu} 
- \partial_{\mu} {\sf a}_{\tau}\big)^2 \Big] \nonumber \\
& \hspace{1cm} \bigg\{
1 + \sum^{\infty}_{N=1} \frac{1}{N!} 
\prod^{N}_{j=1} \Big(\int {\cal D}^2{\bm r}_{2j-1} \!\ e^{\beta_{\tau} t_{\tau}} 
\int {\cal D}^2{\bm r}_{2j} \!\ e^{\beta_{\tau} t_{\tau}}\Big)  \nonumber \\
& \hspace{1cm} \exp \Big[- i 2\pi e \sum^{2N}_{j=1} (-1)^j 
\int^{\beta_{\tau}}_{0} d \tau \!\ {\sf a}_{\tau}({\bm r}_j,\tau)\Big]\bigg\}.
\end{align}
Here we take the inverse temperature $\beta_{\tau}$ [a `system size' along the time direction] to be finite for bookkeeping purposes. ${\bm r}_{j}$ $(j=1,\cdots,2N)$ denotes the two-dimensional spatial coordinate of the $j$th vortex line within the $1$-$2$ plane, $e^{\beta_{\tau} t_{\tau}}$ is the fugacity of each vortex line,  and $(-1)^j=\pm 1$ denotes the vorticity of 
the $j$th vortex line. As the vortex line is polarized along $\tau$, only the temporal component ${\sf a}_{\tau}$ of the vector potential couples with the vortex excitations. Namely, the partition function in the momentum-frequency representation is given by 
\begin{align}
\lim_{\gamma_{\tau}\rightarrow 0} Z
&= \int {\cal D} {\sf a}  
\exp \Big[-\frac{1}{2} \frac{1}{\beta_{\tau}} \sum_{\omega} 
\int \frac{d^2{\bm q}}{(2\pi)^2} \sum_{\mu=1,2}\big|\omega {\sf a}_{\mu}(\omega,{\bm q}) 
-q_{\mu} {\sf a}_{\tau}(\omega,{\bm q})\big|^2 \Big] \nonumber \\
& \hspace{1cm} \bigg\{
1 + \sum^{\infty}_{N=1} \frac{1}{N!} 
\prod^{N}_{j=1} \Big(\int {\cal D}^2{\bm r}_{2j-1} \!\ e^{\beta_{\tau}t_{\tau}} 
\int {\cal D}^2{\bm r}_{2j} \!\ e^{\beta_{\tau}t_{\tau}}\Big)  \nonumber \\
& \hspace{1cm} \exp \Big[- i 2\pi e \sum^{2N}_{j=1} \sigma_j 
\int\frac{d^2{\bm q}}{(2\pi)^2} \!\ e^{i{\bm q}\cdot {\bm r}_{j}}
{\sf a}_{\tau}({\omega}=0,{\bm q})\Big]\bigg\},
\end{align}
with the Fourier transform, 
\begin{align}
{\sf a}_{\mu}(\tau,{\bm r}) = \frac{1}{\beta_{\tau}} \sum_{\omega} 
\int \frac{d^2{\bm q}}{(2\pi)^2} e^{i{\bm q}{\bm r}-i\omega \tau} 
{\sf a}_{\mu}(\omega,{\bm q}), 
\end{align}
and $\omega = \frac{2\pi}{\beta_{\tau}} {n}$ $(n=1,2,\cdots)$.  In the right hand side, only ${\sf a}_{\tau}(\omega=0,{\bm q})$ couples with the vortex excitations, 
while ${\sf a}_{\tau}(0,{\bm q})$ is decoupled from the spatial components ${\sf a}_{\mu}(0,{\bm q})$ of the gauge field ($\mu=1,2$) in the 
Maxwell action. Thus, by integrating out ${\sf a}_{\tau}(\omega\ne 0,{\bm q})$ and ${\sf a}_{\mu}(\omega\ne 0,{\bm q})$ ($\mu=1,2$),  we reach  
the 2D isotropic partition function, 
\begin{align}
\lim_{\gamma_{\tau} \rightarrow 0} Z  
&= {\rm const.} \int {\cal D} {\sf a}(0,{\bm q})  
\exp \Big[-\frac{1}{2} \frac{1}{\beta_{\tau}}  
\int \frac{d^2{\bm q}}{(2\pi)^2}  
q^2 |{\sf a}_{\tau}(0,{\bm q})\big|^2 \Big] \nonumber \\
& \hspace{1cm} \bigg\{
1 + \sum^{\infty}_{N=1} \frac{1}{N!} 
\prod^{N}_{j=1} \Big(\int {\cal D}^2{\bm r}_{2j-1} \!\ e^{\beta_{\tau}t_{\tau}} 
\int {\cal D}^2{\bm r}_{2j} \!\ e^{\beta_{\tau}t_{\tau}}\Big)  \nonumber \\
& \hspace{1cm} \exp \Big[- i 2\pi e \sum^{2N}_{j=1} \sigma_j 
\int\frac{d^2{\bm q}}{(2\pi)^2} \!\ e^{i{\bm q}\cdot {\bm r}_{j}}
{\sf a}_{\tau}(0,{\bm q})\Big]\bigg\}. \label{Eq:Zv-2x}
\end{align}
Eq.~(\ref{Eq:Zv-2x}) is identical to Eq.~(\ref{Eq:Zv-2}) with $D=2$, $\gamma_{\tau}=1$, and $\chi=0$, where ${\sf a}_{\tau}(\omega=0,{\bm r})/\sqrt{\beta_{\tau}}$ plays the role of the 2D electrostatic potential $u_{\tau 1}({\bm r})$, and the relevant parameters in the two equations are related by,  
\begin{align}
 y \equiv e^{w} \!\ \!\ ({\rm 2D} \,\ \gamma_{\tau}=1, \,\ \chi=0) &\Leftrightarrow e^{2\beta_{\tau} t_{\tau}}  \!\ \!\ ({\rm 3D}  \,\ \gamma_{\tau}=0), \label{Eq:corres1} \\
 e^2 \!\ \!\ ({\rm 2D}\,\ \gamma_{\tau}=1,  \,\ \chi=0) &\Leftrightarrow e^2 \beta_{\tau} \!\ \!\ ({\rm 3D} \,\ \gamma_{\tau}=0). \label{Eq:corres2}
\end{align}
By substituting Eqs.~(\ref{Eq:corres1}, \ref{Eq:corres2}) into Eq.~(\ref{Eq:separatrix}), we can readily see that $\lim_{\beta_{\tau}\rightarrow \infty}\lim_{\gamma_{\tau}\rightarrow 0} Z$ with positive $t_{\tau}$ and a finite $e^2$ is in a quasi-disordered phase, while $\lim_{\beta_{\tau}\rightarrow \infty}\lim_{\gamma_{\tau}\rightarrow 0} Z$ with negative $t_{\tau}$ and a finite $e^2$ is in the ordered phase. In the quasi-disordered phase, the SF phase correlation becomes short-ranged in the spatial direction due to the proliferation of vortex lines, while the SF phase correlation remains long-ranged in the time direction. In the ordered phase, the fugacity of vortex lines vanishes, going far below the separatrix of the KT transition, and the SF phase correlation remains long-ranged in any directions. Numerical solutions of the RG equations show that the divergence of $\gamma^{-1}_{\tau}$ at the finite RG scale always comes with positive $t_{\tau}$ [Figs.~\ref{Fig:nu12_Bou_tttr_gau1}, \ref{Fig:nu06_Bou_tttr_gau1}] and finite $e^2$ [Figs.~\ref{Fig:nu12_Bou_e2nu_gau1}, \ref{Fig:nu06_Bou_e2nu_gau1}]. Thus, we conclude that the intermediate coupling region is unstable to a formation of the quasi-disordered phase with short-range spatial correlation but persistent temporal correlation.  

\begin{figure}[t]
\subfigure[$Z_{1}({\bm t},\nu)$ with soft UV cutoff (Eq.~(\ref{Eq:Z1-soft-approx-a}))]{
\includegraphics[width=0.3\textwidth]{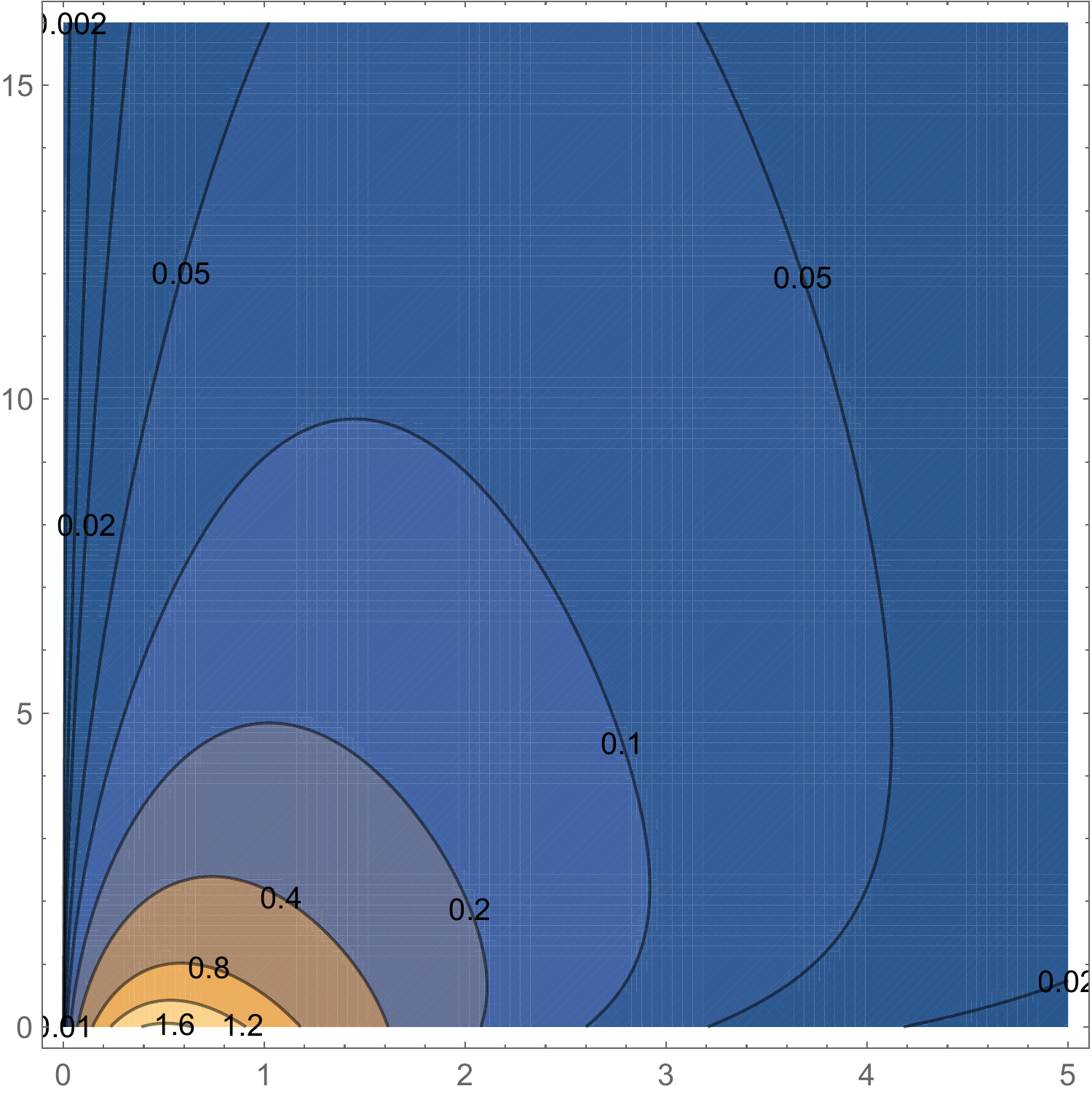}
\label{Fig:soft-z1}
}
\hfill
\subfigure[$Z_{2,\tau}({\bm t},\nu)$ with soft UV cutoff (Eq.~(\ref{Eq:Z2tr-soft-approx}))]{
\includegraphics[width=0.3\textwidth]{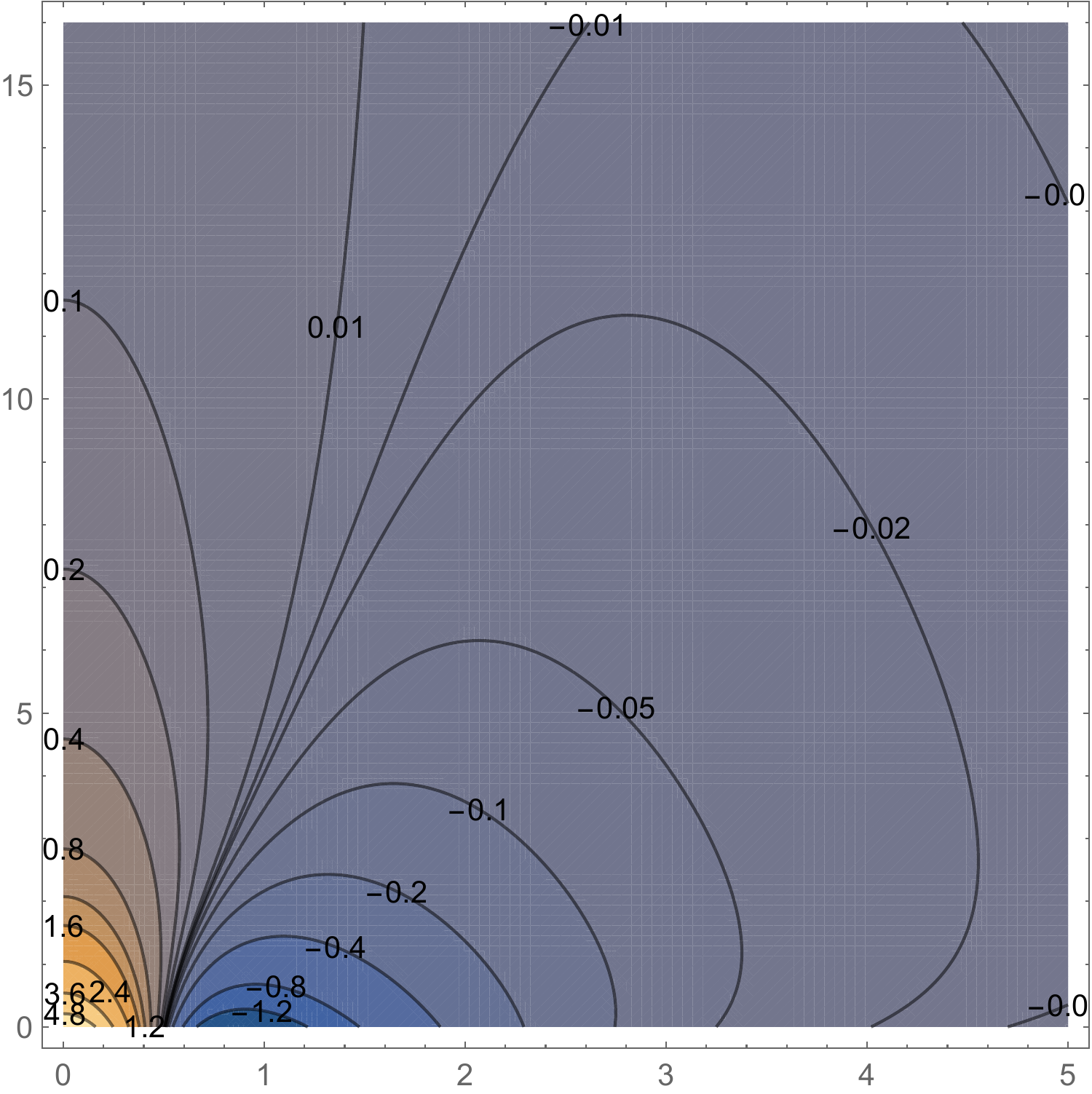}
\label{Fig:soft-z2t}
}
\hfill
\subfigure[$Z_{2,\tau}({\bm t},\nu)$ with soft UV cutoff (larger scale)]{
\includegraphics[width=0.3\textwidth]{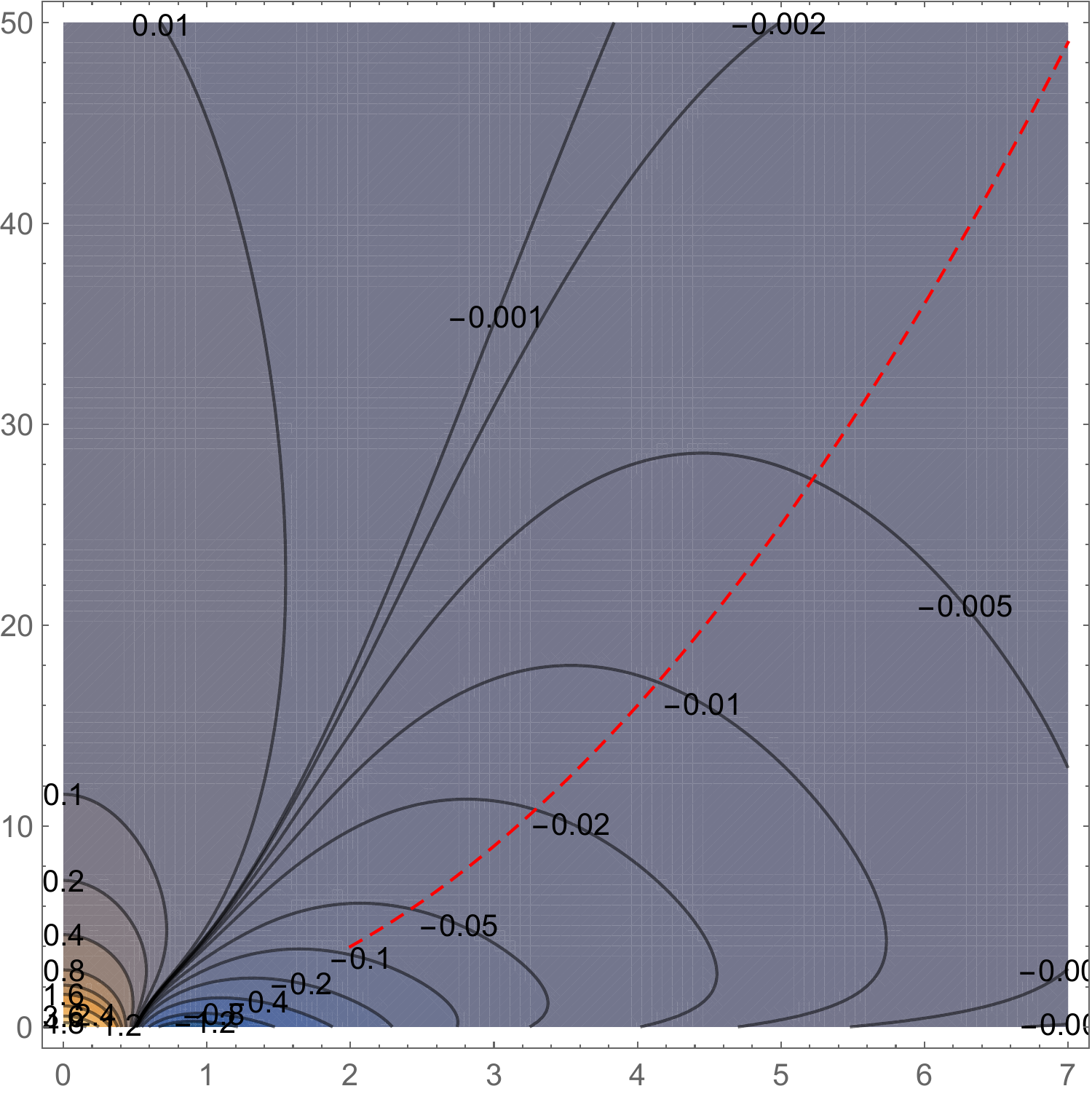}
\label{Fig:soft-z2t_a}
}
\caption{Contour plots of $Z_{2,\tau}({\bm t},\nu) \equiv \partial_{\nu} Z_{1}({\bm t},\nu)$ and $Z_{1}({\bm t},\nu)$ as functions of $t_{\tau}=-t_{\bm r} \equiv t_{-}/2$ and $\nu$ with the soft UV cutoff [Eq.~(\ref{Eq:Z2tr-soft-approx})]. The horizontal and vertical axes are $\nu$ and $t_{-}$, respectively.  $Z_{1}({\bm t},\nu)$ for $\nu>0$ has a positive `ridge' along $t_{-}\sim \nu^2$, while $Z_{2,\tau}({\bm t},\nu)$ has a negative `gorge' along $t_{-}\sim \nu^2$ [dashed red line in Fig.~\ref{Fig:soft-z2t_a}]}
\end{figure}

    The intermediate coupling region appears universally in the RG phase diagram upon a transformation from the weak coupling region to the strong coupling region. In the weak coupling side, $t_{\tau}+t_{\bm r}$ diverges negatively, and $\nu$ increases with $\nu\sim b^2$, where $Z_{2,t}({\bm t}, \nu)$ is vanishingly small, taking a negative value for the larger $\nu$ [Figs.~\ref{Fig:soft-z2t},\ref{Fig:soft-z2t_a}]. In the strong coupling side, $t_{\tau}+t_{\bm r}$ diverges positively, and $\nu$ becomes vanishingly small, where $Z_{2,t}({\bm t},\nu)$ gets exponentially larger, taking a positive value for the smaller $\nu$ [Fig.\ref{Fig:soft-z2t}]. The transformation of parameters between these two regions necessitates passage through an intermediate region where $Z_{2,t}({\bm t},\nu)$ attains large negative values. This arises from a causal constraint: the parameter $\nu$ begins to decrease only after $t_{\tau}+t_{\bm r}$ diverges to large positive values. Consequently, by the time $\nu$ enters the negative `gorge' of  $Z_{2,t}({\bm t},\nu)$ (Fig.~\ref{Fig:soft-z2t_a}), $t_{\tau}+t_{\bm r}$ has already grown significantly positive. From the first term on the right hand side of Eq.~(\ref{Eq:RGgammatauD_3}), the large negative $Z_{2,t}({\bm t},\nu)$ reduces $\gamma_{\tau}$ to zero at a finite RG scale $\ln b$.

       The ordered-quasi-disordered transition must be the first order, as the correlation length $\xi_{\bm r}$ on the strong-coupling side of the boundary remains finite, not showing any critical divergence. Namely, in the large $\beta_{\tau}$ limit, the substitution of Eqs.~(\ref{Eq:corres1}, \ref{Eq:corres2}) into Eq.~(\ref{Eq:separatrix}) always brings the relevant parameters into a 2D disordered phase region far above the separatrix of the KT transition, where the correlation length $\xi_{\bm r}$ must be fairly small.

\section{\label{secIV} fugacity renormalization : Further Considerations} 
\subsection{\label{secIVa}Derivation of Eq.~(\ref{Eq:isotropic-general-D})}
     In the isotropic and general $D$ case, we take $\gamma_{\tau}=1$ in Eqs.~(\ref{Eq:Ft},\ref{Eq:Fr}) and calculate $F_{\tau}({\bm x})=F_{\bm r}({\bm x})$ with a soft ultraviolet (UV) cutoff for the momentum integral, 
\begin{align}
F({\bm x}) \equiv &F_{\tau}({\bm x})=F_{\bm r}({\bm x})
= \Lambda \frac{\partial}{\partial \Lambda} \bigg[\int \frac{d^D{\bm k}}{(2\pi)^D}\frac{e^{i{\bm k}\cdot{\bm x}}}{k^2} e^{-\Lambda^{-1}k}\bigg] \nonumber \\
 =& \frac{2\pi^{\frac{D}{2}}}{\Lambda} \bigg[\int^{\infty}_0 \frac{dk}{(2\pi)^D} k^{D-2} e^{-\Lambda^{-1}k}  \!\ _0\tilde{F}_1\big(\frac{D}{2},-\frac{k^2 x^2}{4}\big) \bigg],  
\end{align}
where $_0\tilde{F}_1(a,b)$ stands for the regularized hypergeometric function. Following a reference~\cite{goldbart2009}, we further calculate the $\Delta {\bm s}$-integral in Eqs.~(\ref{Eq:ft},\ref{Eq:fr}) for $f\equiv f_{\tau}=f_{\bm r}$,
\begin{align}
f& = \int d^{D-2}\Delta{\bm s} \,\  
F\Big(\sum^{D-2}_{a=1}\Delta s_a\frac{\partial {\bm R}_{j}({\bm s})}{\partial s_a}\Big) 
= \frac{1}{\sqrt{\det \hat{\bf g}_j}} \int d^{D-2}{\bm \xi} \,\  F({\bm \xi}).  \label{Eq:f=1ovsqg}
 \end{align}
In the first line, the $(D-2)$-dimensional vector $\Delta {\bm s}$ is transformed into a $(D-2)$-dimensional vector ${\bm \xi}$ associated with the first fundamental form $\hat{\bf g}_j({\bm s})$, 
\begin{align}
\hat{\bf g}_{j,ab}({\bm s})\equiv  \sum_{\mu=\tau,1,\cdots,D-1}\frac{\partial {\bm R}_{j,\mu}}{\partial s_a} \frac{\partial {\bm R}_{j,\mu}}{\partial s_b} 
\end{align}
$(a,b=1,2,\cdots,D-2)$.  Firstly, this $(D-2)\times (D-2)$ real symmetric matrix $\hat{\bf g}_j$ is diagonalized by an orthogonal matrix with real positive eigenvalues $\eta_a$ $(a=1,\cdots,D-2)$.  Then, $\Delta {\bm s}$ is linearly transformed into a $(D-2)$-dimensional vector ${\bm \zeta}$ by the orthogonal matrix, while  ${\bm \xi}$ is related to ${\bm \zeta}$ by a dilatation transformation with the positive eigenvalues,  $\xi_a \equiv \zeta_a \sqrt{\eta_a}$ $(a=1,\cdots, D-2)$.  ${\bm \xi}$ thus introduced can be substituted for the argument of the integrand of Eq.~(\ref{Eq:f=1ovsqg}), as $F({\bm x})$ depends only on $|{\bm x}|^2$, and 
\begin{align}
\sum_{\mu=\tau,1,\cdots,D-1}\bigg(\sum^{D-2}_{a=1}\Delta s_a\frac{\partial {\bm R}_{j,\mu}({\bm s})}{\partial s_a}\bigg)^2 = \sum^{D-2}_{a,b=1}\Delta s_a \Delta s_b \!\ \hat{\bf g}_{j,ab} = \sum^{D-2}_{a=1} \zeta^2_a \!\ \eta_a = \sum^{D-2}_{a=1} \xi^2_a \equiv \big|{\bm \xi}\big|^2. \label{Eq:Goldbart1}
\end{align}
When the $\Delta {\bm s}$-integral is rewritten by the ${\bm \xi}$-integral, the Jacobian gives $1/\sqrt{\det \hat{\bf g}_j}$, 
\begin{align}
&\int d^{D-2}\Delta {\bm s} = \int d^{D-2} {\bm \zeta} = \Big(\prod^{D-2}_{a=1} \sqrt{\eta_a}\Big)^{-1} \int d^{D-2}{\bm \xi} = \frac{1}{\sqrt{\det \hat{\bf g}_j}}  \int d^{D-2}{\bm \xi}.\label{Eq:Godbart2}
\end{align} 
The ${\bm \xi}$-integral of $F({\bm \xi})$ in the right hand side of Eq.~(\ref{Eq:f=1ovsqg}) yields a universal constant,  
\begin{align}
\int d^{D-2} {\bm \xi} \,\ F({\bm \xi})& = \frac{1}{\Lambda} \frac{2\pi^{\frac{D}{2}}}{(2\pi)^D} 
\int d^{D-2} {\bm \xi} \,\ \int^{\infty}_{0} k^{D-2}dk \!\ e^{-\Lambda^{-1} k} \!\ 
_0\tilde{F}_1\Big(\frac{D}{2},-\frac{k^2 \xi^2}{4}\Big) \nonumber \\
&=\frac{1}{\Lambda} \frac{2^{D-1}\pi^{\frac{D-1}{2}}}{(2\pi)^D} 
\int d^{D-2} {\bm \xi} \,\
\frac{\Gamma\big(\frac{D-1}{2}\big)}{(\Lambda^{-2}+\xi^2)^{\frac{D-1}{2}}} \nonumber \\
&= \frac{1}{\Lambda}  \frac{2^{D-1}\pi^{\frac{D-1}{2}}}{(2\pi)^D} 
\frac{2\pi^{\frac{D-2}{2}}}{\Gamma\big(\frac{D-2}{2}\big)} \int^{\infty}_{0} \xi^{D-3}d\xi \frac{\Gamma\big(\frac{D-1}{2}\big)}{(\Lambda^{-2}+\xi^2)^{\frac{D-1}{2}}} \nonumber \\
&= \frac{1}{\Lambda} \frac{1}{\pi^{\frac{3}{2}}}\frac{\Gamma\big(\frac{D-1}{2}\big)}{\Gamma\big(\frac{D-2}{2}\big)} \frac{\sqrt{\pi}}{2\Lambda^{-1}} \frac{\Gamma\big(\frac{D-2}{2}\big)}{\Gamma\big(\frac{D-1}{2}\big)} = \frac{1}{2\pi}. \label{Eq:f=1ov2pi}
\end{align}

 Eqs.~(\ref{Eq:f=1ovsqg},\ref{Eq:f=1ov2pi}) give Eq.~(\ref{Eq:isotropic-general-D}).   
 
\subsection{\label{secIVb}Derivation of Eqs.~(\ref{Eq:ttau},\ref{Eq:tr})}
    As we discussed in Section~\ref{secIIb}, the short-range part of the Coulomb interaction gives rise to a fugacity renormalization. In the general $D$ case,  the fugacity renormalization is given by a rather complicated function of the temporal and spatial components of the metric tensor, $\hat{\bf g}_{\tau}$ and $\hat{\bf g}_{\bm r}$ [see Eqs.~(\ref{Eq:fugacity-rg7})]. In $D=3$, however, the components reduce to scalar quantities, and the function reduces to a simpler form. To see this,  let us start from the gradient expansion of the short-range part of the 3D Coulomb interaction within the same vortex loop $\partial \Gamma_j$,   
\begin{align}
\sum_{j} \delta E_j &= \ln b \frac{e^2}{2} \sum_{\sigma=\tau,1,2} \int d^3 {\bm R}_1 
\int d^3{\bm R}_2 F_{\sigma}(|{\bm R}_1-{\bm R}_2|) J_{\sigma}({\bm R}_1) J_{\sigma}({\bm R}_2), \nonumber \\ 
&= 2\pi^2 e^2 \ln b \sum_{\sigma=\tau,1,2} \sum_{j} \int_{\partial \Gamma_j} ds_1 \int_{\partial \Gamma_j} ds_2 \,\ F_{\sigma}(|{\bm R}_j(s_1)-{\bm R}_j(s_2)|) \frac{\partial R_{j,\sigma}(s_1)}{ds_1}  \frac{\partial R_{j,\sigma}(s_2)}{ds_2}, \nonumber \\
&= 2\pi^2 e^2 \ln b \sum_{\sigma=\tau,1,2} \sum_j \int_{\partial \Gamma_j} ds  \Big(\frac{\partial R_{j,\sigma}(s)}{ds}\Big)^2 
\int d \Delta s \,\ F_{\sigma}\Big(\frac{\partial R_{j,\sigma}(s)}{ds} \Delta s\Big) + \cdots. \label{Eq:leading-order}
 \end{align}
 Here we used, 
 \begin{align}
 J_{\sigma}({\bm R}) &= 2\pi \sum_{j} \int_{\partial \Gamma_j} ds \frac{\partial R_{j,\sigma}(s)}{ds} \delta^3({\bm R}-{\bm R}_j(s)),  \nonumber \\
 \ln b \!\ F_{\sigma}({\bm x}) &= \int_{b^{-1}\Lambda<|{\bm k}|<\Lambda} \frac{d^3{\bm k}}{(2\pi)^3} \frac{e^{i{\bm q}{\bm r}-i\omega \tau}}{k^4}  
 \begin{cases}
 {\bm q}^2 + (2-\gamma^{-1}_{\tau})\omega^2  & \ \ \ {\rm for} \,\ \sigma=\tau, \\
  \gamma^{-1}_{\tau}{\bm q}^2 + \omega^2   & \ \ \  {\rm for} \,\ \sigma=1,2. \\
 \end{cases}
 \end{align}
 The short-range part of the interaction $F_{\sigma}({\bm x})$ $(\sigma=\tau,1,2)$ may be calculated with a soft UV cutoff with the cylindrical symmetry,   
 \begin{align}
  F_{\tau}({\bm x}) &= -\Lambda^{-1}\frac{\partial}{\partial \Lambda^{-1}} 
 \int \frac{d^3 {\bm k}}{ (2\pi)^3} e^{i{\bm q}{\bm r}-i\omega \tau} e^{-\Lambda^{-1}|{\bm q}|} \,\ 
 \frac{|{\bm q}|^2 + (2-\gamma^{-1}_{\tau}) \omega^2}{k^4}, \label{Eq:Ft3D} \\
  F_{\bm r}({\bm x}) & \equiv  F_{1}({\bm x}) =  F_{2}({\bm x}) = -\Lambda^{-1}\frac{\partial}{\partial \Lambda^{-1}} 
 \int \frac{d^3 {\bm k}}{ (2\pi)^3} e^{i{\bm q}{\bm r}-i\omega \tau} e^{-\Lambda^{-1}|{\bm q}|} \,\ 
 \frac{\gamma^{-1}_{\tau} |{\bm q}|^2 +  \omega^2}{k^4}, \label{Eq:Fr3D}
 \end{align}
 with ${\bm x}=(\tau,{\bm r})$ and ${\bm k}=(\omega,{\bm q})$.  As $F_{\sigma}\big(\frac{\partial {\bm R}}{\partial s} \Delta s\big)$ comes with $(\partial_{s} {\bm R}_{\sigma})^2$ in Eq.~(\ref{Eq:leading-order}), it is natural to estimate the $\Delta s$-integral of $F_{\tau}\big(\frac{\partial {\bm R}}{\partial s} \Delta s\big)$ using a geometry of $\frac{\partial {\bm R}}{\partial s}\Delta s = \Delta s(1,0,0)$, and the $\Delta s$-integral of $F_{\bm r}\big(\frac{\partial {\bm R}}{\partial s} \Delta s\big)$ by the other geometry: $\frac{\partial {\bm R}}{\partial s}\Delta s = \Delta s(0,\cos\theta,\sin\theta)$. With this simplification, the one-dimensional $\Delta s$-integrals in Eqs.~(\ref{Eq:ft}, \ref{Eq:fr}) can be readily calculated, giving reasonable estimates of $f_{\tau}$ and $f_{\bm r}$ in the $D=3$ case, 
 \begin{align}
 f_{\tau} &= \int d\Delta s \,\ F_{\tau}\big(\Delta s(1,0,0)\big) = 
 - \Lambda^{-1} \frac{\partial}{\partial \Lambda^{-1}} \int \frac{d^2{\bm q}}{(2\pi)^2} 
 \frac{1}{|{\bm q}|^2} e^{-\Lambda^{-1}|{\bm q}|}  
\nonumber \\
&= 
 \Lambda^{-1} \frac{1}{2\pi} \int^{\infty}_{0} d|{\bm q}| e^{-\Lambda^{-1}|{\bm q}|} = \frac{1}{2\pi}, \label{Eq:ft3D} \\
 f_{\bm r} &= \int d\Delta s \,\ F_{\bm r} \big(\Delta s (0,1,0)\big) \nonumber \\
 &= - \Lambda^{-1} \frac{\partial}{\partial \Lambda^{-1}} \int d\Delta s 
 \int^{+\infty}_{-\infty} \frac{d\omega}{2\pi} 
 \int^{\infty}_0 \frac{q d q}{(2\pi)^2} \int^{2\pi}_{0} d\theta  \,\ 
 \frac{1}{(\omega^2+q^2)^2} e^{-\Lambda^{-1}q} e^{-iq |\Delta s| \cos\theta}\,\ \big(\gamma^{-1}_{\tau}q^2 + \omega^2\big)  \nonumber \\
 & = - \Lambda^{-1} \frac{\partial}{\partial \Lambda^{-1}} \int d\Delta s 
 \int^{\infty}_0 \frac{q d q}{(2\pi)^2} \int^{2\pi}_{0} d\theta \,\   
 \frac{\gamma^{-1}_{\tau}+1}{4q} e^{-\Lambda^{-1}q} e^{-iq |\Delta s| \cos\theta} \nonumber \\
 &= \Lambda^{-1} \int d\Delta s \int^{2\pi}_{0} d\theta \,\ \frac{1}{(\Lambda^{-1}+i\Delta s \cos\theta)^2} \frac{\gamma^{-1}_{\tau}+1}{4} \frac{1}{(2\pi)^2} \nonumber \\
 &= \Lambda^{-1} \int^{+\infty}_{-\infty} d\Delta s \frac{\Lambda^{-1}}{(\Lambda^{-2}+|\Delta s|^2)^{\frac{3}{2}}} \frac{\gamma^{-1}_{\tau}+1}{4} \frac{1}{2\pi} = \frac{\gamma^{-1}_{\tau}+1}{4\pi}. \label{Eq:fr3D}  
 \end{align}
Eqs.~(\ref{Eq:ft3D}, \ref{Eq:fr3D}) give Eqs.~(\ref{Eq:ttau},\ref{Eq:tr}). 
 
\subsection{\label{secIVc} fugacity renormalization with different gauge : its impact on physics}
      In the derivation of the fugacity renormalization, we expressed the vector potential $u_{\mu\nu}({\bm k})$ in terms of the divergence-free auxiliary field $\varphi_{\nu}({\bm k})$, Eq.~(\ref{Eq:gauge1}), and integrated the auxiliary field $\varphi_{\nu}({\bm k})$. This gauge choice facilitated the calculation for the general space-time dimension $D$, while for $D=2$ and $3$, one may also integrate the vector potential $u_{\mu\nu}({\bm k})$ directly. The direct integration of the vector potentials replaces the arithmetic mean between $e^2/\gamma_{\tau}$ and $e^2$ in Eqs.~(\ref{Eq:y-2D},\ref{Eq:y-2D-rev},\ref{Eq:tr},\ref{Eq:RGtr})  by their geometric mean. For $D=2$,  Eqs.~(\ref{Eq:y-2D},\ref{Eq:y-2D-rev}) are replaced by 
\begin{align}
    \frac{dy}{d\ln b} = \big(4-\frac{\gamma^{-1/2}_{\tau} e^2}{2\pi}\big) y. \label{Eq:gauge2}
\end{align}
For $D=3$, Eqs.~(\ref{Eq:tr},\ref{Eq:RGtr}) are replaced by 
\begin{align}
\begin{cases}
    \frac{d t_{\tau}}{d\ln b} = t_{\tau} - \pi e^2, \\
    \frac{d t_{\bm r}}{d\ln b} = t_{\bm r} - \pi e^2 \gamma^{-1/2}_{\tau}. \\
\end{cases} \label{Eq:gauge3}
\end{align}
 
    In the following, we will argue that these changes do {\it not} alter the qualitative natures of the phase transitions for $D=2$ and $3$.  
\subsubsection{\label{secIVc1}$D=2$} 

\begin{figure}[h]
\centering
\subfigure[numerical RG phase diagram]{
\includegraphics[width=0.5\textwidth]{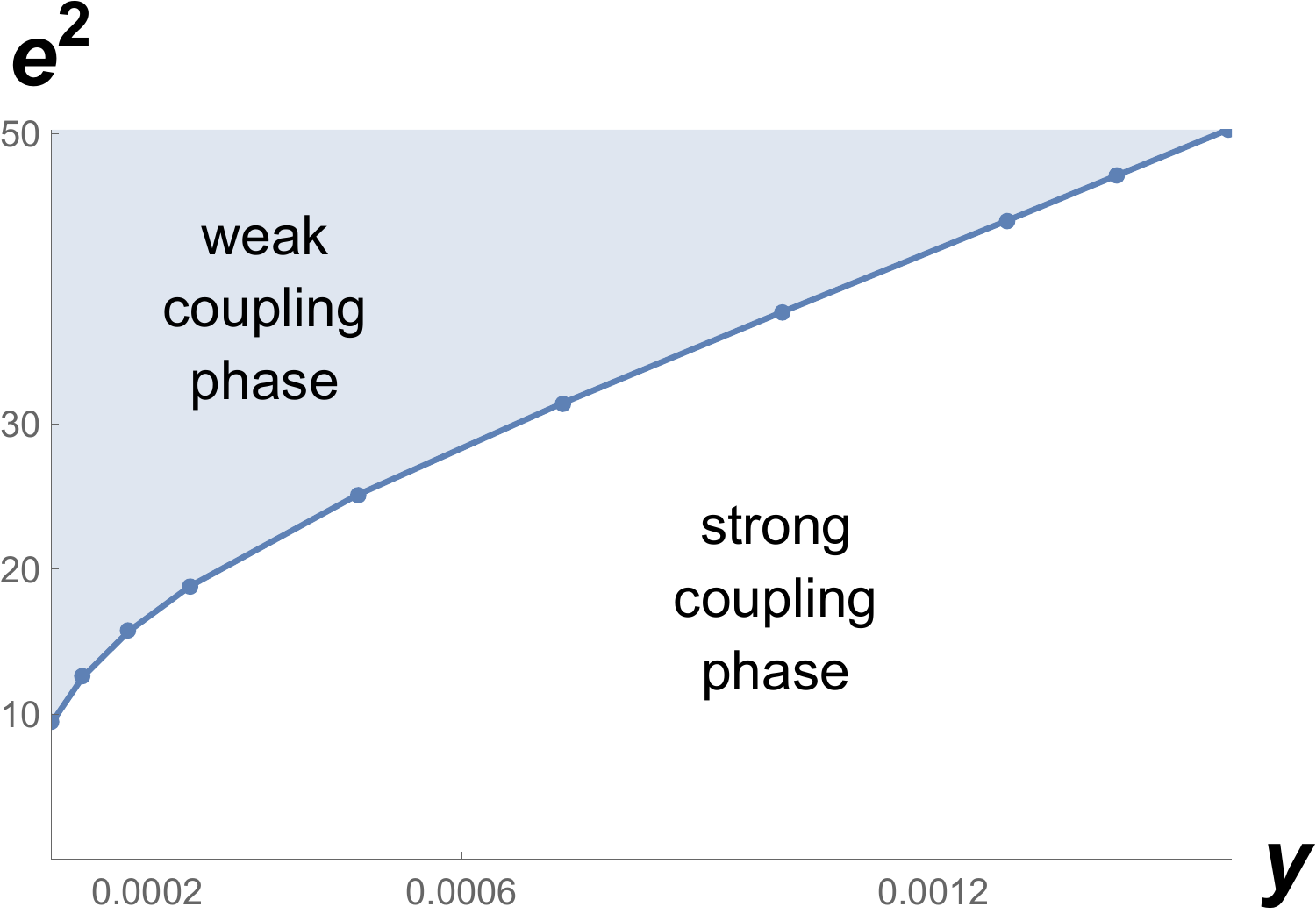}
\label{Fg:2dpd-another}
}
\hfill
\subfigure[RG flow trajectory near the saddle fixed point]{
\includegraphics[width=0.4\textwidth]{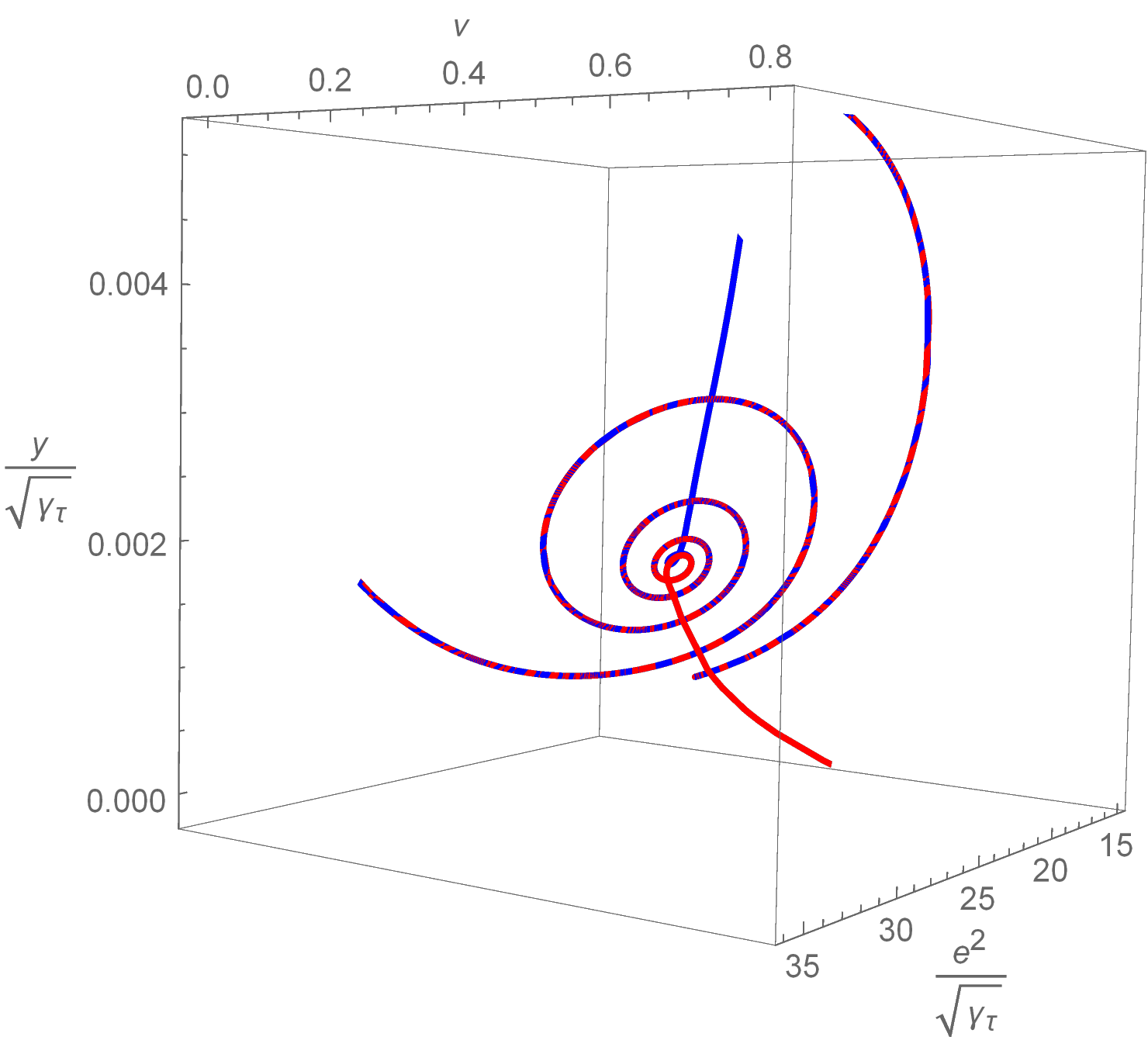}
\label{Fg:2dnearCP-another}
}
\caption{(left) RG phase diagram with initial values of $\gamma_{\tau}=1$ and $\nu=0.2$. The horizontal and vertical axes are initial values of $y$, and $e^2$, respectively. (right) Two RG flow trajectories near the saddle fixed point in the three-dimensional parameter space. The saddle fixed point is at $(e^2/\gamma^{1/2}_{\tau},\nu,y/\gamma^{1/2}_{\tau})=(8\pi,\nu_{-},\rho_{-})=(25.9...,0.389...,0.001798)$. The blue (red) curve denotes the RG flow trajectory when the initial parameters are chosen in the disordered (ordered) phase side of the phase boundary: $e^2/\gamma_{\tau}=e^2=10\pi$ and $\nu=0.2$ and $y\lesssim  y_c$ $(y\gtrsim y_c)$ respectively [$y_c=0.000728...$]. }
\end{figure}

     As in Sec.~\ref{secIIIb}, the modified RG equation has an anisotropic weak-coupling fixed line for the KT-like critical phase, as well as the isotropic strong coupling fixed point for the conventional 2D disordered phase. The phase transition between these two phases is also characterized by a saddle fixed point with an infinite space-time anisotropy as well.  Differently from Sec.~\ref{secIIIb}, the saddle fixed point of the modified RG equation is characterized by finite values of $y/\gamma^{1/2}_{\tau}$, $\nu$ and $e^2/\gamma^{1/2}_{\tau}$ rather than finite values of $y$, $\nu$ and $e^2/\gamma_{\tau}$. To analyze the scaling properties of this saddle fixed point, let us rewrite the RG equations in terms of $e^2/\gamma^{1/2}_{\tau}$, $y/\gamma^{1/2}_{\tau}$, $\nu$ and $e^2$:
         \begin{align}
         \frac{d(e^2/\gamma^{1/2}_{\tau})}{d\ln b} &= -\frac{\pi}{2} \Big(\frac{y}{\gamma^{1/2}_{\tau}}\Big) e^4 z_{2,r}(\nu) - \frac{\pi}{2} \Big(\frac{y}{\gamma^{1/2}_{\tau}}\Big) \Big(\frac{e^2}{\gamma^{1/2}_{\tau}}\Big)^2 z_{2,\tau}(\nu), \label{Eq:2D-RG-rev2a} \\
\frac{de^2}{d\ln b} & =  - \pi y e^4 z_{2,r}(\nu)= - \pi \Big(\frac{y}{\gamma^{1/2}_{\tau}}\Big) 
\Big( \frac{\gamma^{1/2}_{\tau}}{e^2}\Big) e^6 z_{2,r}(\nu), \label{Eq:2D-RG-rev2b} \\
\frac{d\nu}{d\ln b} & = \nu \Big( 1 - \pi \frac{y}{\gamma^{1/2}_{\tau}}\frac{e^2}{\gamma^{1/2}_{\tau}} z_{2,{r}}(\nu)\Big), \label{Eq:2D-RG-rev2c} \\
\frac{d(y/\gamma^{1/2}_{\tau})}{d\ln b} & = \Big(4-\frac{1}{2\pi} 
\frac{e^2}{\gamma^{1/2}_{\tau}} - \frac{\pi}{2} \frac{y}{\gamma^{1/2}_{\tau}}\frac{e^2}{\gamma^{1/2}_{\tau}} z_{2,\tau}(\nu) + \frac{\pi}{2} y e^2 z_{2,r}(\nu) \Big) \frac{y}{\gamma^{1/2}_{\tau}}, \label{Eq:2D-RG-rev2d} 
         \end{align}
As in Sec.~\ref{secIIIb}, let us start from the zero of $z_{2,\tau}(\nu)$: $z=z_{\pm}$. The RG equations could have a fixed point at $e^2=0$,   $\nu=\nu_{\pm}$, $e^2/\gamma^{1/2}_{\tau}=8\pi$ and $y/\gamma^{1/2}_{\tau}=\rho_{\pm}$ with
\begin{align}
\rho_{\pm}  = \frac{\gamma^{1/2}_{\tau}}{\pi e^2 z_{2,r}(\nu_{\pm})} = \frac{1}{8 \pi^2} \frac{1}{z_{2,r}(\nu_{\pm})}. 
\end{align}
 Since $z_{2,r}(\nu_{+})<0$, we only consider the $z=z_{-}$ case, with $\nu_{-}=0.389$, and $\rho_{-}=0.001798$. In fact,  $(e^2,\nu,e^2/\gamma^{1/2}_{\tau},y/\gamma^{1/2}_{\tau})=(0,\nu_{-},8\pi,\rho_{-})$ is a saddle fixed point: it has a relevant scaling variable, a marginally irrelevant scaling variable, and two irrelevant scaling variables. As in sec.~\ref{secIIIb}, $e^2$ becomes the marginally irrelevant scaling variable, 
\begin{align}
\frac{de^2}{d\ln b} = - \frac{1}{64\pi^2} e^6 + {\cal O}\big(\delta(e^2/\gamma^{1/2}_{\tau}) e^6, \delta(y/\gamma^{1/2}_{\tau}) e^6, 
\delta(\nu) e^6\Big).
\end{align}
where $\delta (e^2/\gamma^{1/2}_{\tau}) \equiv e^2/\gamma^{1/2}_{\tau} - 8\pi$, $\delta \nu\equiv \nu - \nu_{-}$ and $\delta (y/\gamma^{1/2}_{\tau}) \equiv y/\gamma^{1/2}_{\tau} - \rho_{-}$. The RG equations for these three quantities are linearized around the fixed point,
\begin{align}
\frac{d}{d\ln b} \left(\begin{array}{c} 
\delta\big(\frac{\gamma^{1/2}_{\tau}}{e^2}\big) \\
\delta \nu \\
\delta \big( \frac{y}{\gamma^{1/2}_{\tau}}\big) \\
\end{array}\right) = \left(\begin{array}{ccc}  
0 &  \frac{\pi}{2} \rho_{-} \big(\frac{dz_{2,\tau}(\nu)}{d\nu}\big)_{|\nu=\nu_{-}} & 0 \\
8\pi \nu_{-} & 1& -\frac{\nu_{-}}{\rho_{-}}  \\
32\pi \rho_{-} &  -4\pi^2 \rho^2_{-} \Big(\frac{\partial z_{2,\tau}(\nu)}{\partial \nu}\Big)_{|\nu=\nu_{-}} & 0 \\
\end{array}\right)  \left(\begin{array}{c} 
\delta\big(\frac{\gamma^{1/2}_{\tau}}{e^2}\big) \\
\delta \nu \\
\delta  \big( \frac{y}{\gamma^{1/2}_{\tau}}\big)\\
\end{array}\right). 
\end{align}
Thereby, the 3 by 3 matrix has eigenvalues of $1.455$, $-0.227\pm 1.541 i$. The positive real eigenvalue is nothing but the scaling dimension of the relevant scaling variable, while a pair of two complex conjugate eigenvalues with a negative real part is for the two irrelevant scaling variables. The relevant scaling variable is given by a linear superposition of  $\delta (e^2/\gamma^{1/2}_{\tau})$, $\delta \nu$, and $\delta (y/\gamma^{1/2}_{\tau})$ with the following coefficients,  
\begin{align}
\left(\begin{array}{ccc}
\delta\big(\frac{\gamma^{1/2}_{\tau}}{e^2}\big) & \delta \nu & \delta\big(\frac{y}{\gamma^{1/2}_{\tau}}\big) \\
\end{array}\right) = \left(\begin{array}{ccc}
-0.0618... & 0.998... & - 0.00489... \\
\end{array}\right).
\end{align}
This indicates that the RG flow streams away from the saddle fixed point with increasing/decreasing $y/\gamma^{1/2}_{\tau}$, 
decreasing/increasing $\nu$ and increasing/decreasing $\gamma^{1/2}_{\tau}/e^2$. The complex conjugate eigenvalues suggest that the RG flow streams into the fixed point with a swirling trajectory [Fig.~\ref{Fg:2dnearCP-another}].  

      The numerical RG phase diagrams are shown in a two-dimensional parameter space subtended by initial values of $y$ and $e^2$ for given initial values of $\nu$ and $\gamma_{\tau}=1$ [Fig.~\ref{Fg:2dpd-another}]. As in Sec.~\ref{secIIIb}, larger (smaller) initial values of $y$ and smaller (larger) initial values of $e^2$ lead to the strong-coupling disordered phase (weak-coupling ordered phase).  Numerical solutions of $e^2/\gamma^{1/2}_{\tau}$, $e^2$, $\nu$ and $y/\gamma^{1/2}_{\tau}$ as functions of the RG scale $\ln b$ are shown in Figs.~\ref{Fg:2D_gau2}. When the initial parameters are on the weak-coupling side of the phase boundary, both $e^2/\gamma^{1/2}_{\tau}$ and $e^2$ approach finite non-universal values in the IR limit $[\ln b\rightarrow \infty]$, while $\nu$ diverges [Figs.~\ref{Fig:nu02_Ord_e2_gau2},\ref{Fig:nu02_Ord_nu_gau2},\ref{Fig:nu02_Ord_y_gau2}]. When the initial parameters are on the strong-coupling side of the boundary, both $e^2/\gamma^{1/2}_{\tau}$ and $e^2$ approach zero, while $\gamma_{\tau}$ converges to $1$, $\nu$ approaches zero, and $y$ diverges [Figs.~\ref{Fig:nu02_Dis_e2_gau2},\ref{Fig:nu02_Dis_nu_gau2},\ref{Fig:nu02_Dis_y_gau2}]. In either case, the parameters make a detour around the saddle fixed point: they first flow into the saddle fixed point with the swirling trajectory, and then they are repelled from the saddle fixed point to strong/weak coupling regions [Figs.~\ref{Fg:2dnearCP-another}, \ref{Fg:2D_gau2}]. As in Sec.~\ref{secIIIb}, the saddle fixed point with infinite space-time anisotropy also endows the correlation time with the additional logarithmic divergence near the critical point. Namely, the correlation time $\xi_{\tau}$ and correlation length $\xi_{r}$ near the critical point show the same space-time anisotropic divergence as in Sec.~\ref{secIIIb},  
 \begin{align}
 \xi_{\tau} &= b \xi_{\tau,0} \sim |{\tt t}|^{-\frac{1}{y_{\tt t}}} \sqrt{|\ln {\tt t}|},  \label{Eq:xitau2} \\
 \xi_{r} & = b\xi_{r,0} \sim |{\tt t}|^{-\frac{1}{y_{\tt t}}}. \label{Eq:xir2}  
 \end{align}
with $y_{\tt t}= 1/1.455...=0.687...$. This is because the renormalized space-time anisotropy parameter shows the same logarithmic divergence as in Eq.~(\ref{Eq:gammat-asym}):   
\begin{align}
e^4_0 \sim \frac{32\pi^2 y_{\tt t}}{|\ln {\tt t}|}, \,\ \,\ \frac{e^2_0}{\gamma^{1/2}_{\tau,0}} 
\sim 8\pi. \nonumber 
\end{align}  

\begin{figure}[t]
\subfigure[$e^2/\gamma^{1/2}_{\tau}$ and $e^2$ as functions of $\ln b$]{
\includegraphics[width=0.3\textwidth]{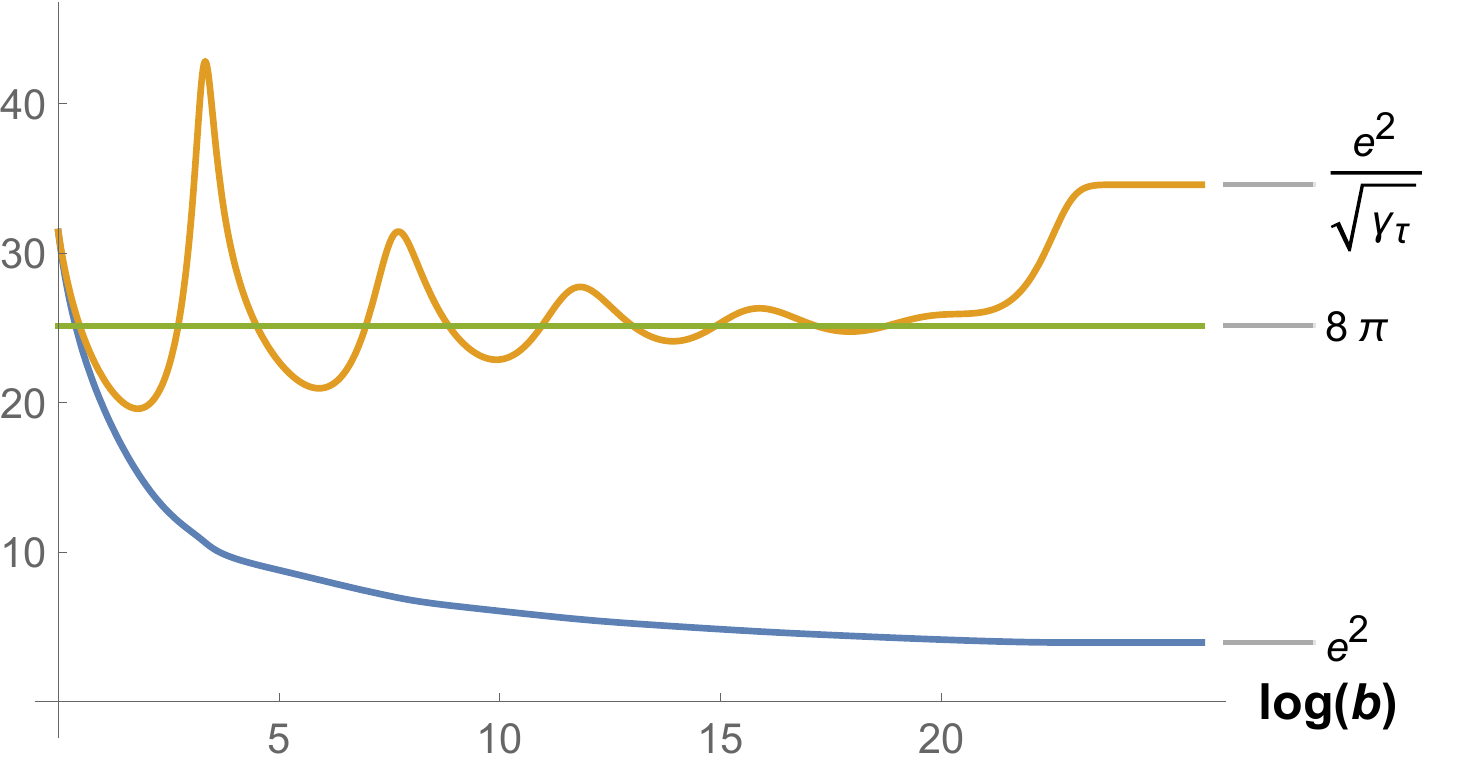}
\label{Fig:nu02_Ord_e2_gau2}
}
\hfill
\subfigure[$\nu$ as a function of $\ln b$]{
\includegraphics[width=0.3\textwidth]{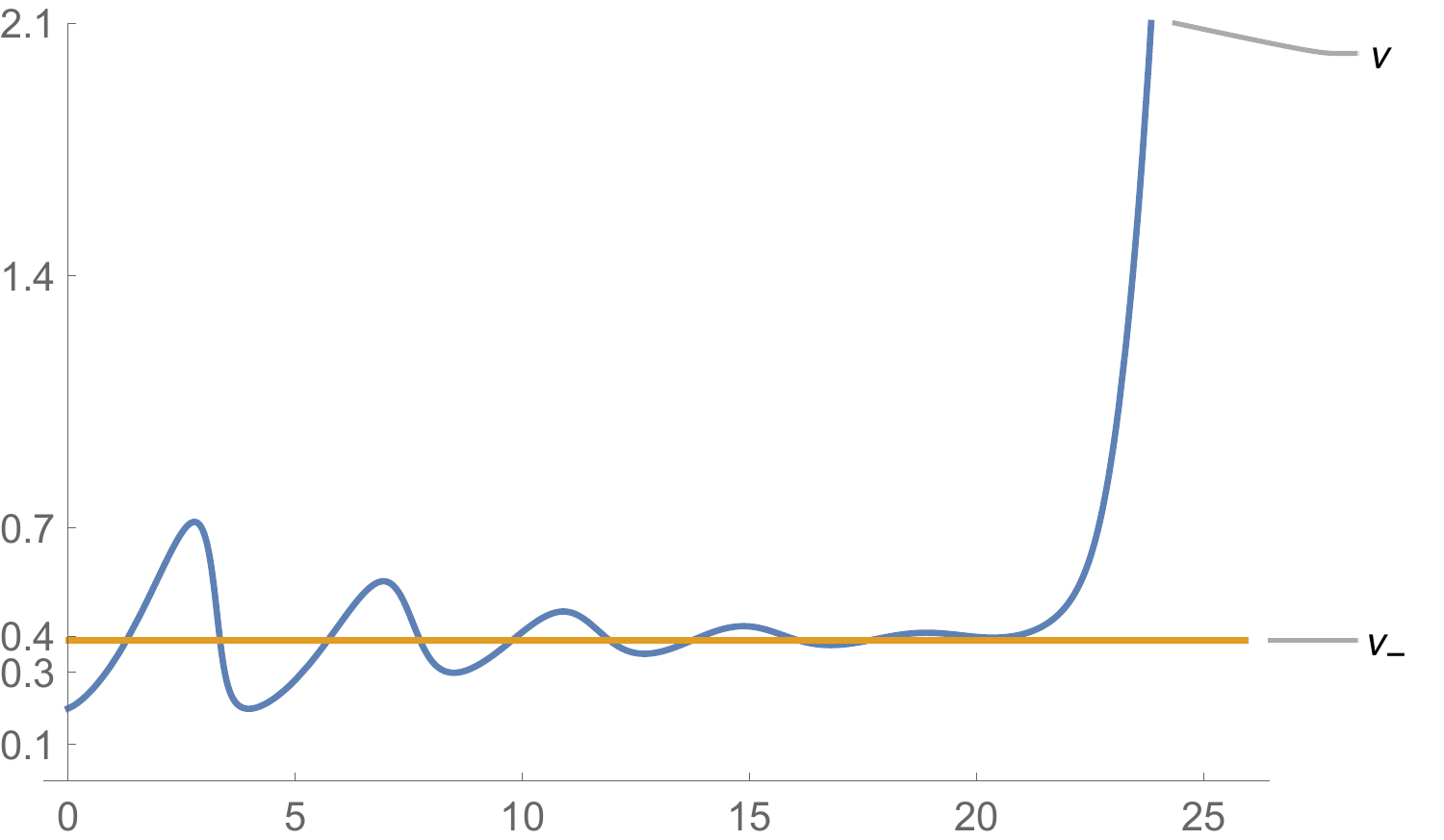}
\label{Fig:nu02_Ord_nu_gau2}
}
\hfill
\subfigure[$y/\gamma^{1/2}_{\tau}$ as a function of $\ln b$]{
\includegraphics[width=0.3\textwidth]{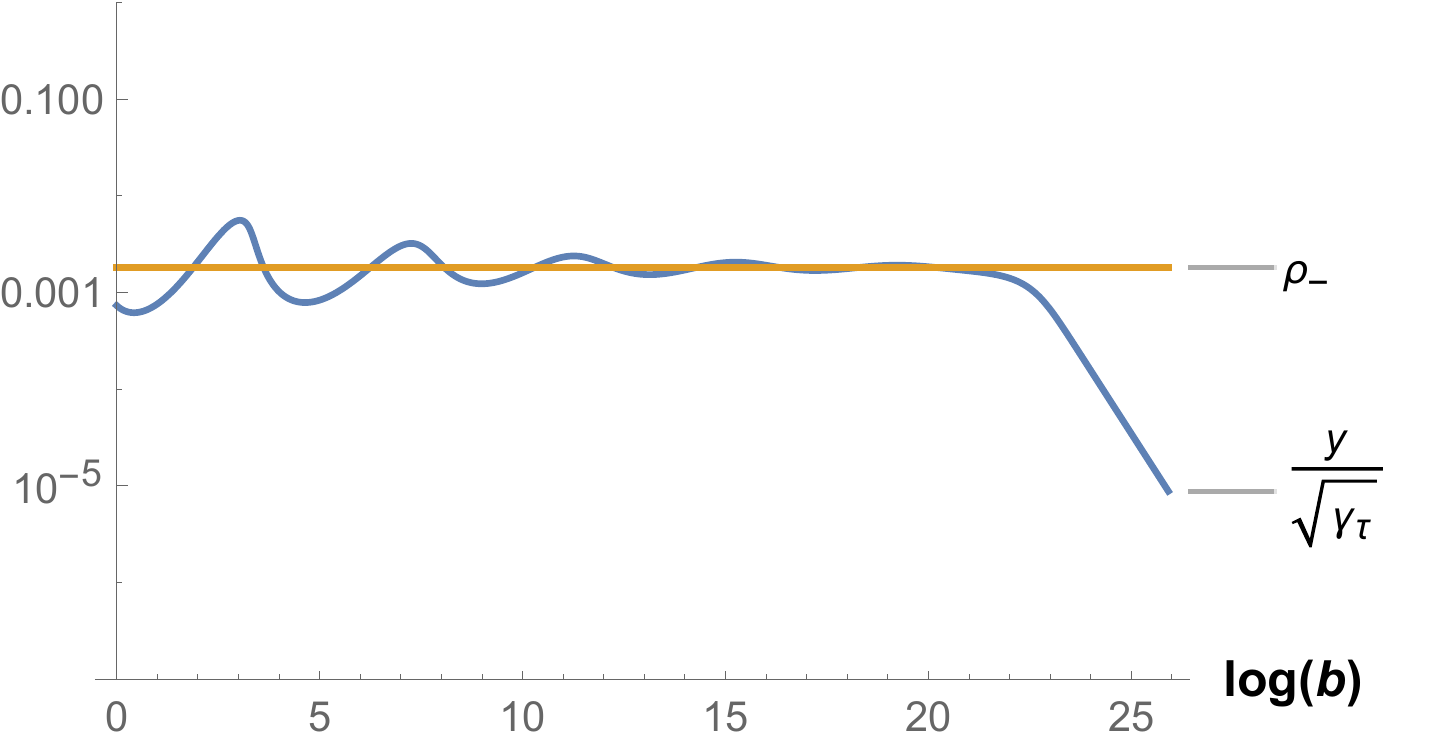}
\label{Fig:nu02_Ord_y_gau2}
}
\vspace{0.5cm}
\subfigure[$e^2/\gamma^{1/2}_{\tau}$ and $e^2$ as functions of $\ln b$]{
\includegraphics[width=0.3\textwidth]{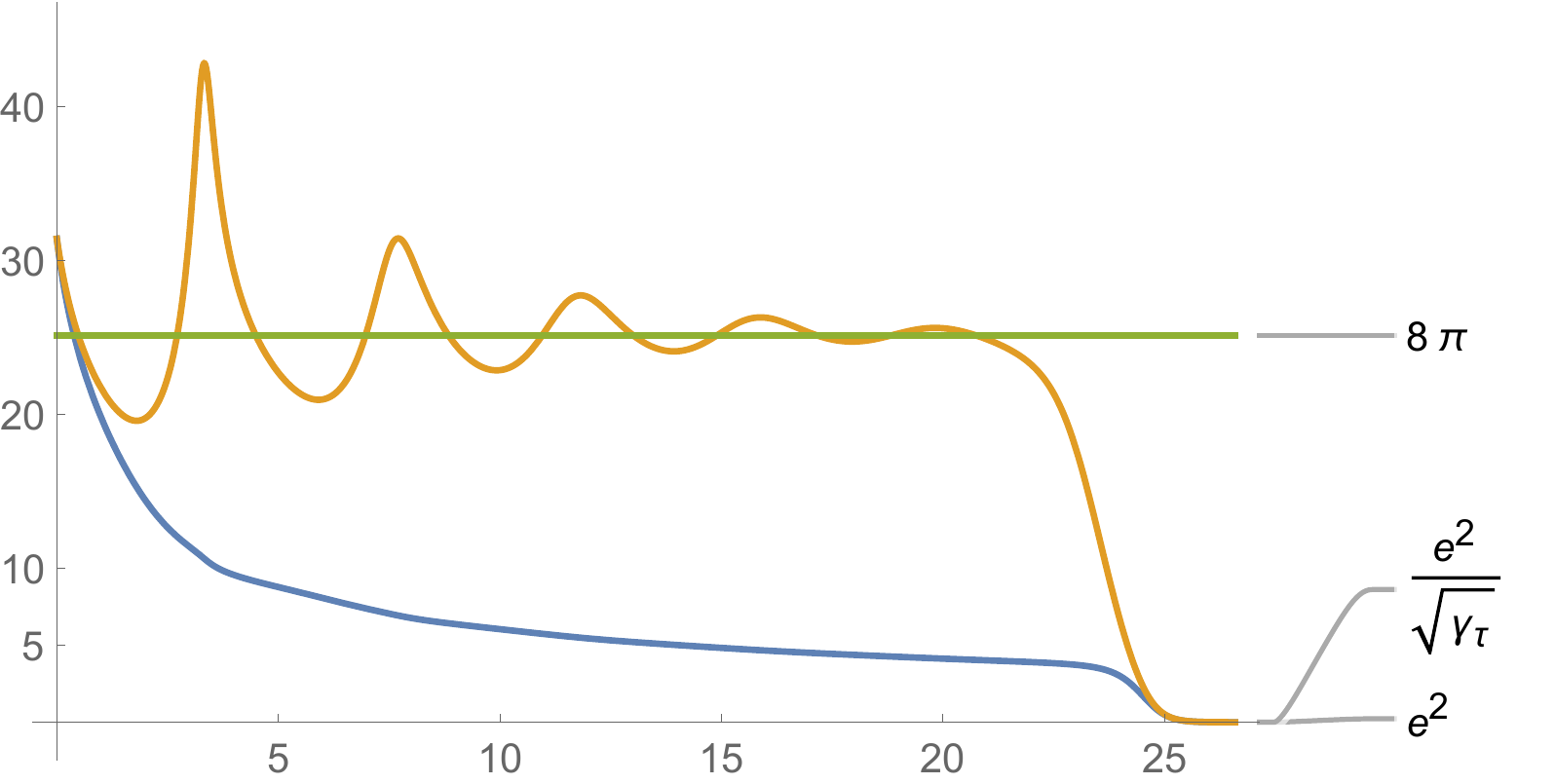}
\label{Fig:nu02_Dis_e2_gau2}
}
\hfill
\subfigure[$\nu$ as a function of $\ln b$]{
\includegraphics[width=0.3\textwidth]{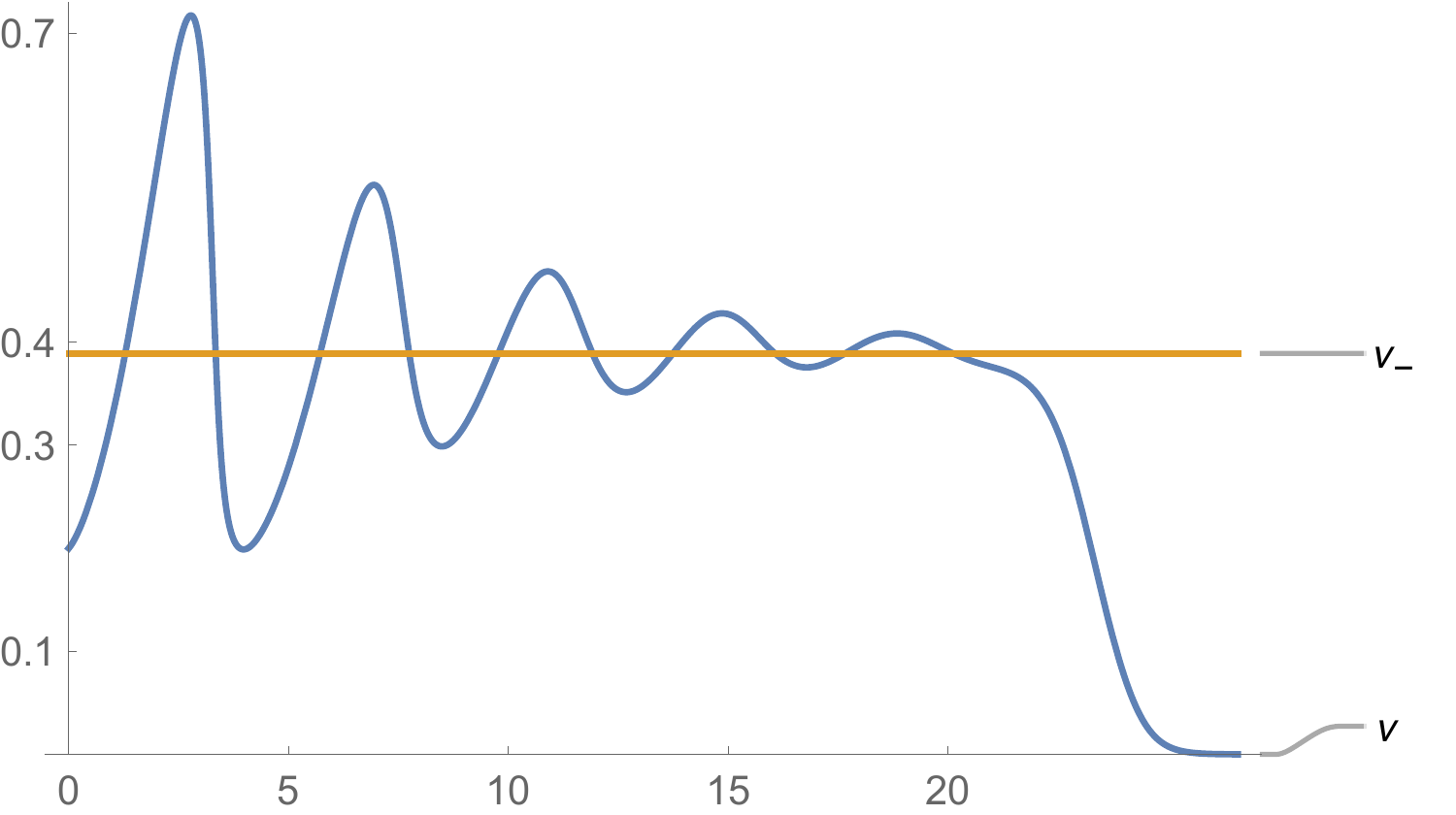}
\label{Fig:nu02_Dis_nu_gau2}
}
\hfill
\subfigure[$y/\gamma^{1/2}_{\tau}$ as a function of $\ln b$]{
\includegraphics[width=0.3\textwidth]{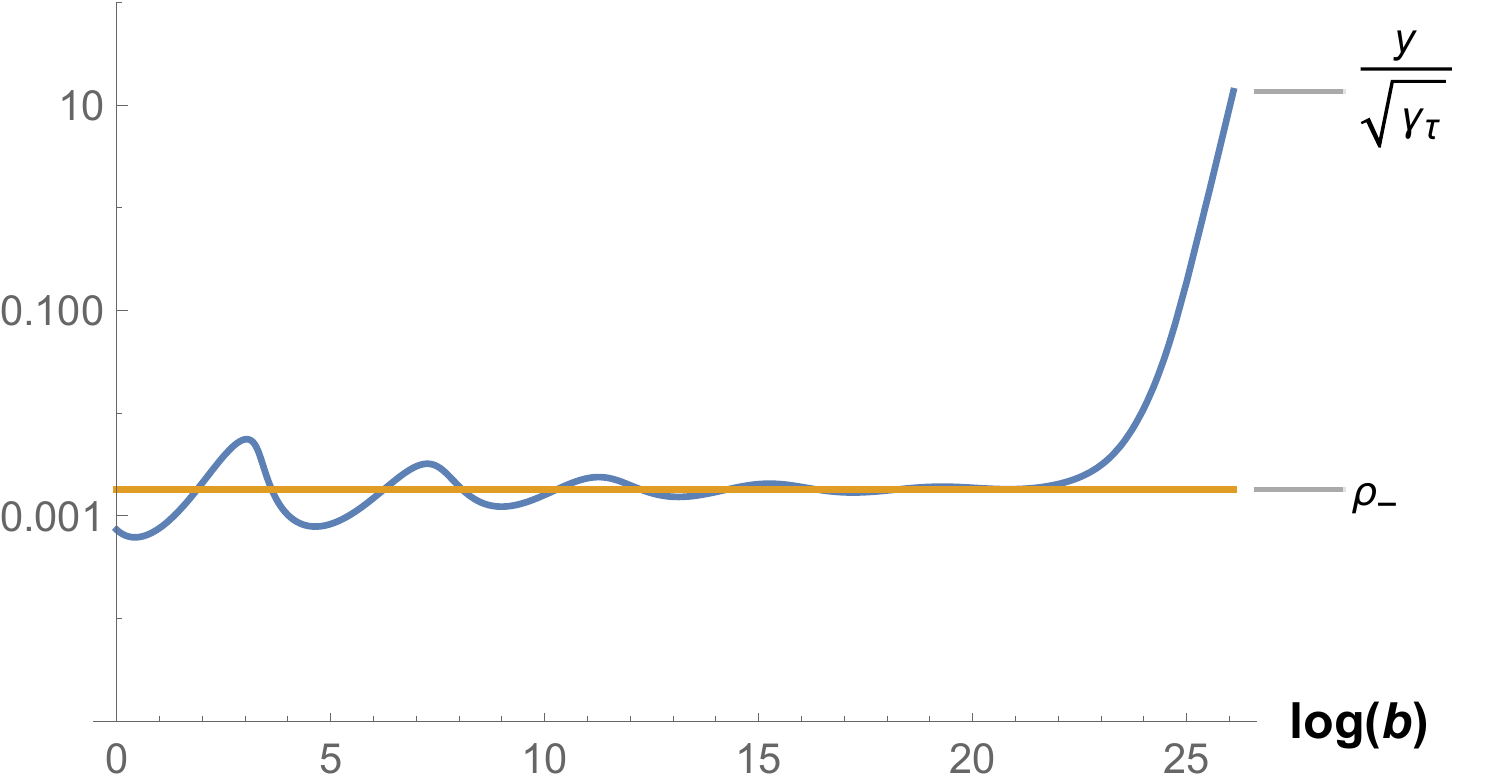}
\label{Fig:nu02_Dis_y_gau2}
}
\caption{Numerical solutions of $e^2/\gamma^{1/2}_{\tau}$, $e^2$, $\nu$ and $y/\gamma^{1/2}_{\tau}$ as functions of the RG scale $\ln b$ for a set of initial parameters closed to the phase boundary between ordered and disordered phases. For Figs.~\ref{Fig:nu02_Ord_e2_gau2}, \ref{Fig:nu02_Ord_nu_gau2}, \ref{Fig:nu02_Ord_y_gau2}, the initial parameters are chosen in the ordered phase side of the phase boundary: $e^2/\gamma_{\tau}=e^2=10\pi$ and $\nu=0.2$ and $y\gtrsim y_c$ where $y_{c}=0.000728..$.. For Figs.~\ref{Fig:nu02_Dis_e2_gau2}, \ref{Fig:nu02_Dis_nu_gau2}, \ref{Fig:nu02_Dis_y_gau2}, the initial parameters are chosen in the disordered phase side of the phase boundary.}
\label{Fg:2D_gau2}
\end{figure}

\subsubsection{\label{secVb}$D=3$}

      The modified RG equation in $D=3$ gives the qualitatively same phase transition as in Sec.~\ref{secIIIc} [see Figs.~\ref{Fig:nu12_pd_gau2}, \ref{Fig:3D_nu12_gauge2}].  The equation in the large $b$ limit has the weak-coupling region with divergent $e^2$ and $\nu$ [Fig.~\ref{Fig:nu12_Ord_e2nu_gau2}], negatively divergent $t_{\tau}$ and $t_{\bm r}$ [Fig.~\ref{Fig:nu12_Ord_tttr_gau2}], and convergent $\gamma_{\tau}$ [Fig.~\ref{Fig:nu12_Ord_igt_gau2}], the intermediate coupling region with divergent $\gamma^{-1}_{\tau}$ at finite RG scale $b$ [Fig.~\ref{Fig:nu12_Bou_igt_gau2}], and the strong-coupling region with vanishing $e^2$ and $\nu$  [Fig.~\ref{Fig:nu12_Dis_e2nu_gau2}], positively divergent $t_{+}=t_{\tau}+t_{\bm r}$ [Fig.~\ref{Fig:nu12_Dis_tttr_gau2}], and divergent $\gamma_{\tau}$ [Fig.~\ref{Fig:nu12_Dis_igt_gau2}]. For larger initial values of $e^2$ and smaller $t\equiv t_{\tau}=t_{\bm r}$ [light blue region in Fig.~\ref{Fig:nu12_pd_gau2}], the coupling constants in the large $b$ limit flow into the weak-coupling region, describing the conventional 3D ordered phase. For smaller values of $e^2$ and larger $t=t_{\tau}=t_{\bm r}$ [light gray and white regions in Fig.~\ref{Fig:nu12_pd_gau2}], they flow into the strong-coupling region for the conventional 3D disordered phase. When initial $e^2$ and $t$ are in the intermediate coupling region [light red region in Fig.~\ref{Fig:nu12_pd_gau2}], the space-time anisotropy parameter $\gamma^{-1}_{\tau}$ always diverges at a finite RG scale $\ln b$ [Fig.~\ref{Fig:nu12_Bou_igt_gau2}] with finite $e^2$ and positive $t_{\tau}$ [\ref{Fig:nu12_Bou_e2nu_gau2},\ref{Fig:nu12_Bou_tttr_gau2}]. This indicates that the intermediate coupling region undergoes the Landau-pole instability to the quasi-disordered phase. In the quasi-disordered phase, the spatial proliferation of vortex lines polarized along $\tau$ renders the U(1) phase correlation along the spatial direction to be short-ranged, while leaving intact the U(1) phase correlation along the temporal direction.  
\begin{figure}[t]
\includegraphics[width=0.5\linewidth]{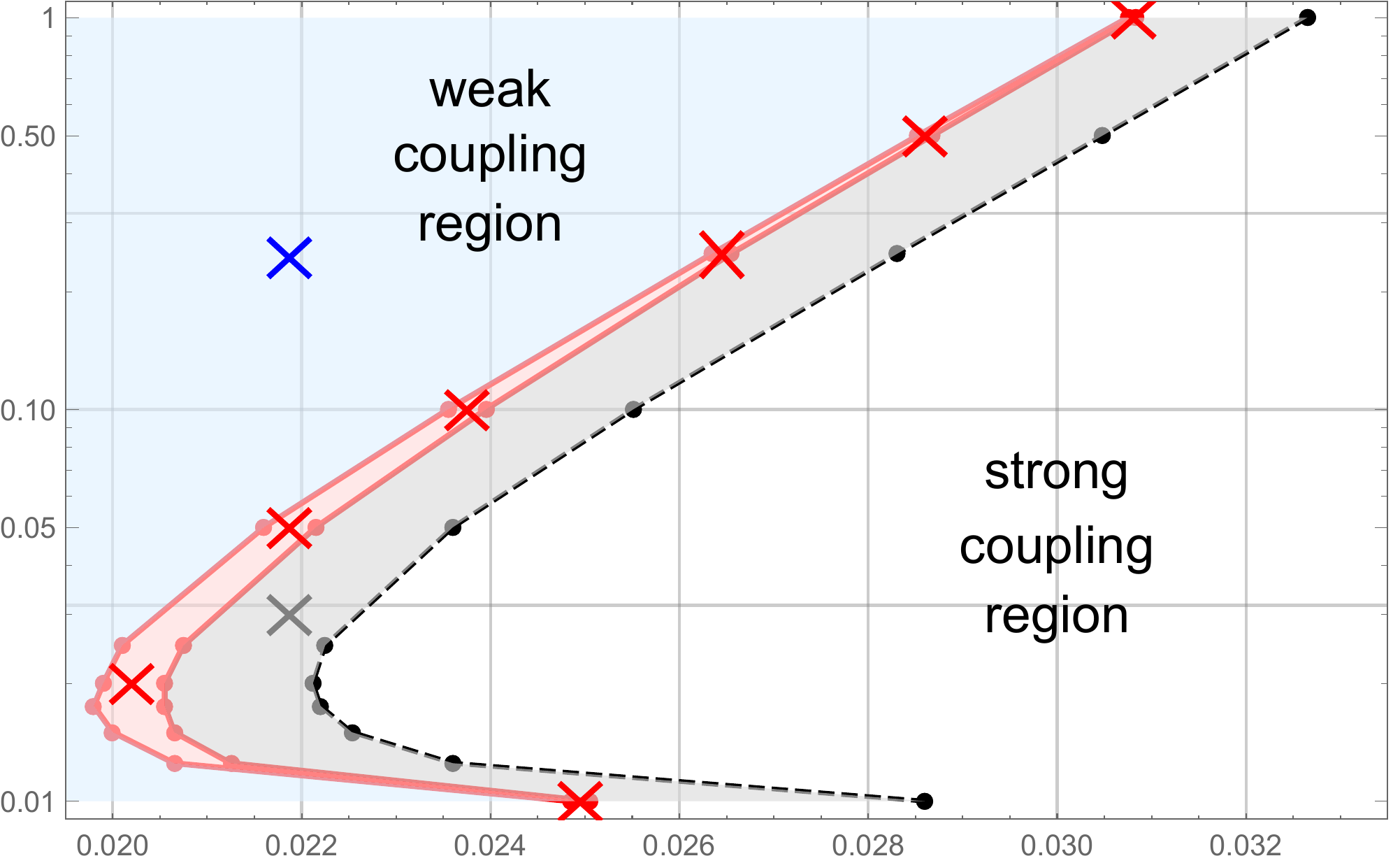}
\caption{Phase diagrams obtained from numerical solutions of the RG equations [Eqs.~(\ref{Eq:RGe2D_3},\ref{Eq:RGchiD_3},\ref{Eq:RGgammatauD_3},\ref{Eq:gauge3}) together with Eqs.~(\ref{Eq:Z1-soft-approx},\ref{Eq:Z2tr-soft-approx})] with initial values of $e^2$, $t_{\tau}=t_{\bm r} \equiv t$, $\gamma_{\tau}=1$ and $\nu=1.2$. The vertical and horizontal axes axis are initial values of $e^2$ and $t$, respectively. The light blue, light red, and light gray (white) regions stand for ordered, quasi-disordered, and disordered phases, respectively. Namely, a set of initial parameters in the light blue region go to the weak-coupling region with divergent $e^2$ and $\nu$, and negatively divergent $t_{\tau}$ and $t_{\bm r}$, and constant $\gamma_{\tau}$ [see Figs.~\ref{Fig:nu12_Ord_e2nu_gau2}. \ref{Fig:nu12_Ord_tttr_gau2} , \ref{Fig:nu12_Ord_igt_gau2} for how the parameters evolve from the blue cross point ``(i)"]. Initial parameters in the light gray (white) region go to the strong-coupling region with vanishing $e^2$ and $\nu$, positively divergent $t_{+}=t_{\tau}+t_{\bm r}$, and negatively (positively) divergent $t_{\bm r}$, and divergent $\gamma^{-1}_{\tau}$ [see Figs.~\ref{Fig:nu12_Dis_e2nu_gau2}. \ref{Fig:nu12_Dis_tttr_gau2} , \ref{Fig:nu12_Dis_igt_gau2} for how the parameters evolve from the gray cross point ``(iii)"]. Initial parameters in the light red region go to the intermediate coupling region, where $\gamma^{-1}_{\tau}$ diverges at a finite value of the RG scale $\ln b$ [see Fig.~\ref{Fig:nu12_Bou_igt_gau2} for how $\gamma^{-1}_{\tau}$ evolve from the red cross point ``(ii)"].}
\label{Fig:nu12_pd_gau2}
\end{figure}

\begin{figure}[t]
\subfigure[$e^2$ and $\nu$ as functions of $\ln b$]{
\includegraphics[width=0.3\textwidth]{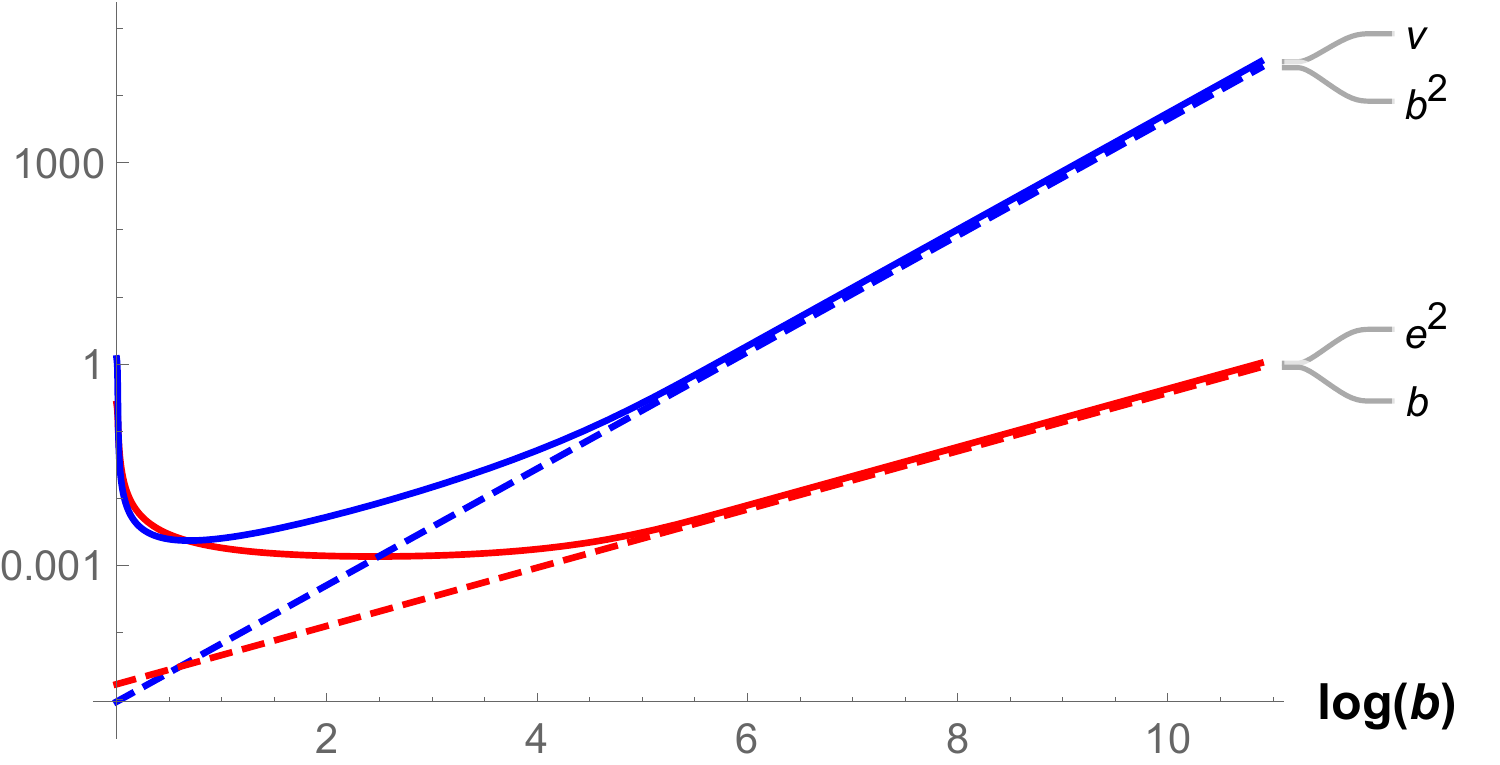}
\label{Fig:nu12_Ord_e2nu_gau2}
}
\hfill
\subfigure[$t_{\tau}$ and $t_{\bm r}$ as functions of $\ln b$]{
\includegraphics[width=0.3\textwidth]{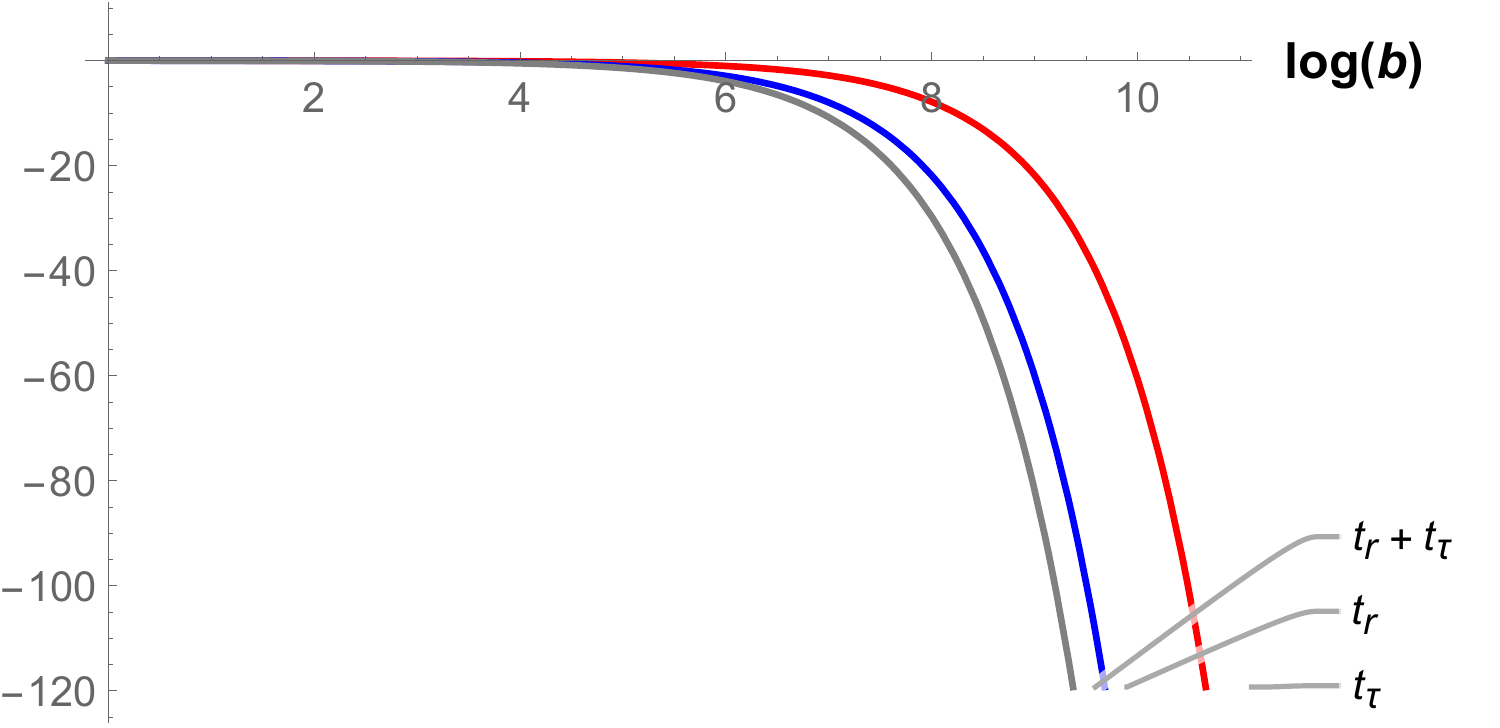}
\label{Fig:nu12_Ord_tttr_gau2}
}
\hfill
\subfigure[$\gamma^{-1}_{\tau}$ as a function of $\ln b$]{
\includegraphics[width=0.3\textwidth]{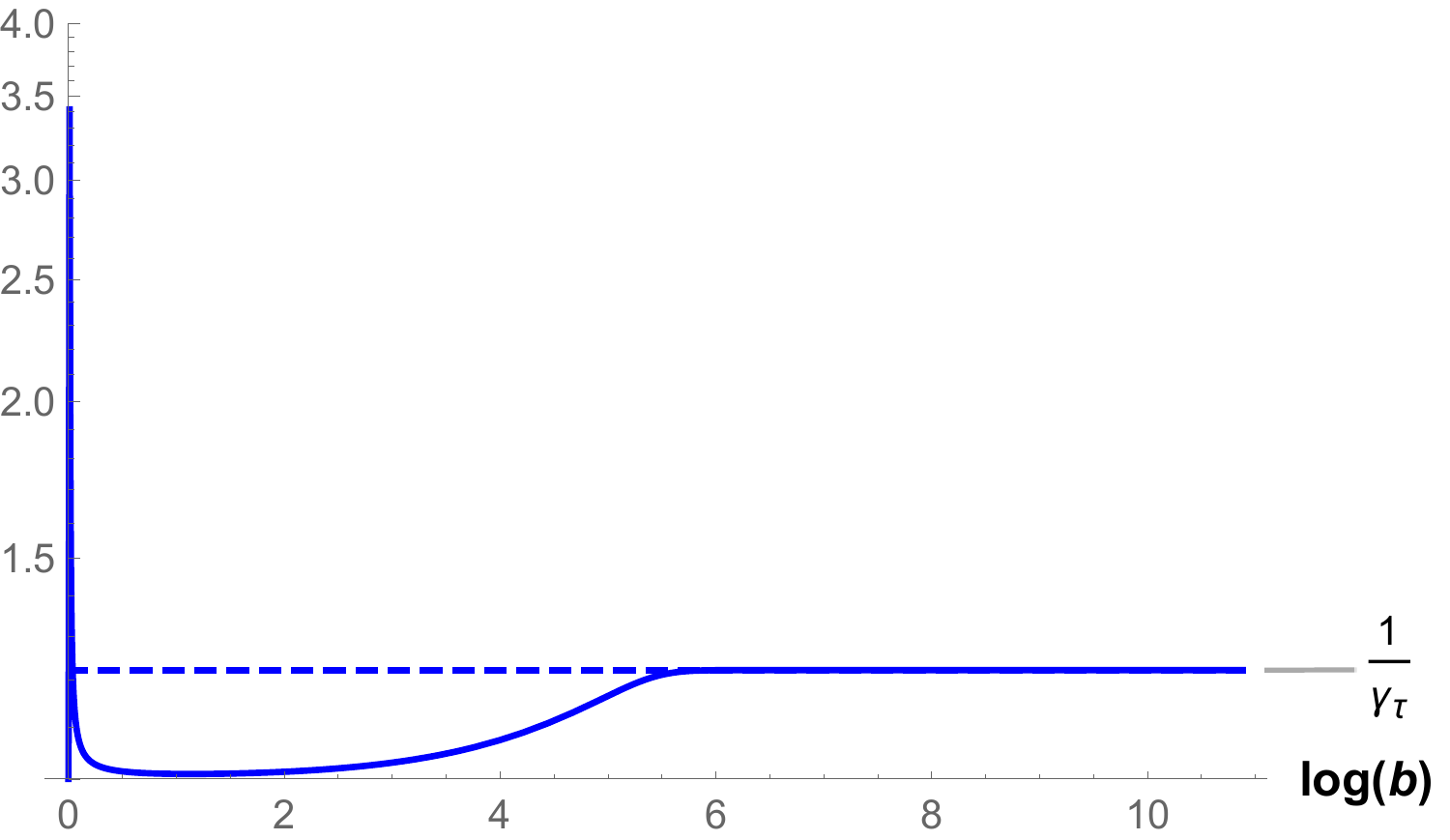}
\label{Fig:nu12_Ord_igt_gau2}
}
\vspace{0.5cm}
\subfigure[$e^2$ and $\nu$ as functions of $\ln b$]{
\includegraphics[width=0.3\textwidth]{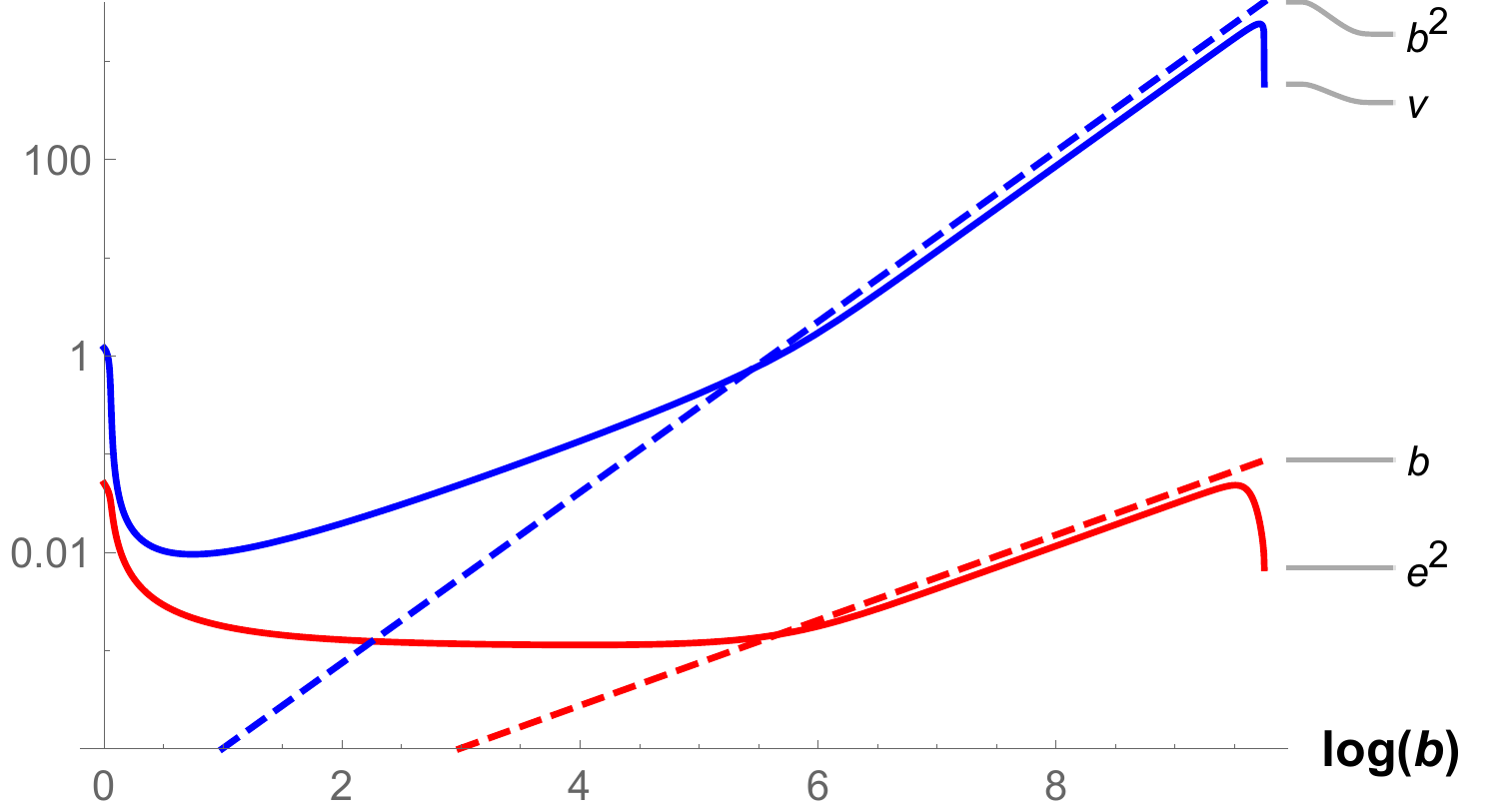}
\label{Fig:nu12_Bou_e2nu_gau2}
}
\hfill
\subfigure[$t_{\tau}$ and $t_{\bm r}$ as functions of $\ln b$]{
\includegraphics[width=0.3\textwidth]{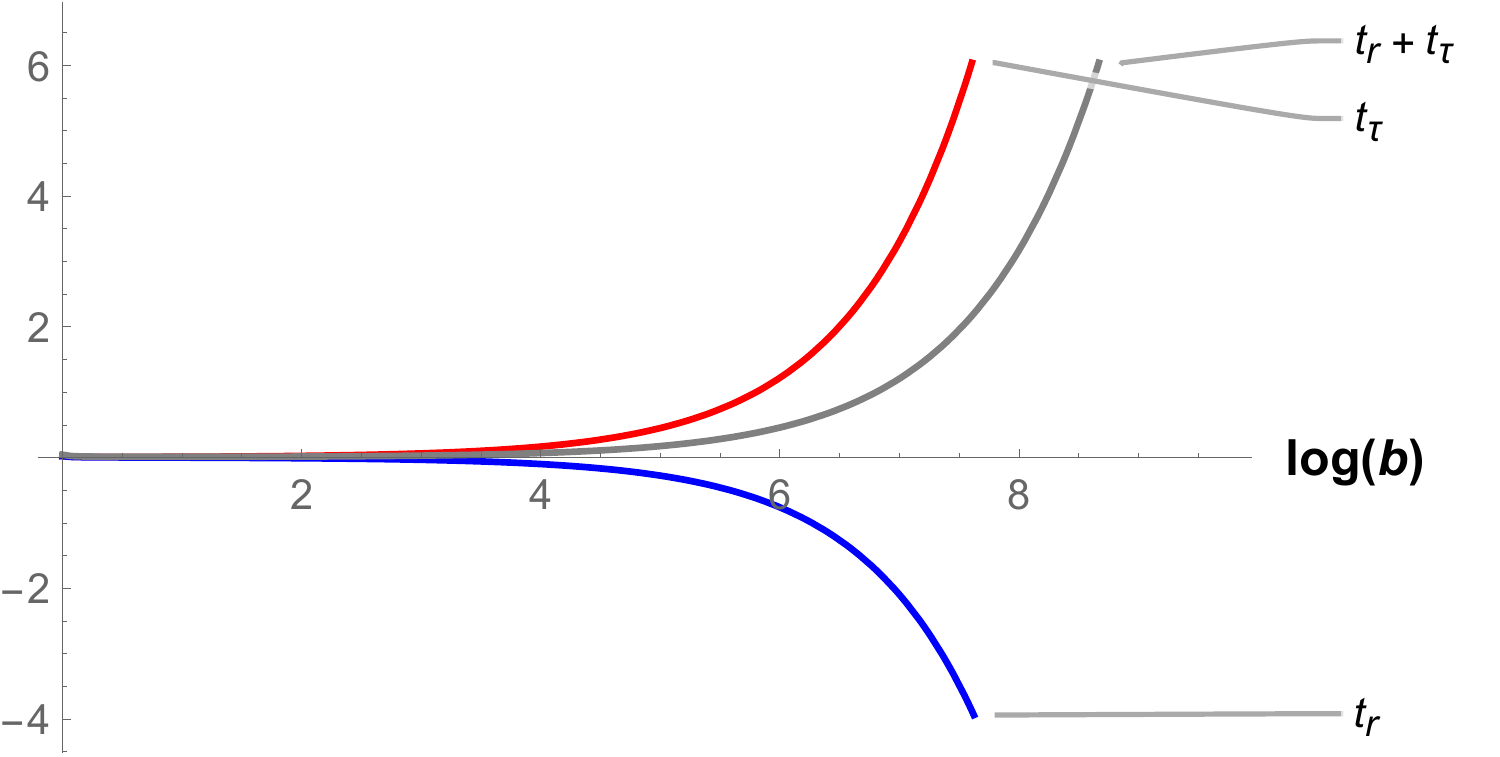}
\label{Fig:nu12_Bou_tttr_gau2}
}
\hfill
\subfigure[$\gamma^{-1}_{\tau}$ as a function of $\ln b$]{
\includegraphics[width=0.3\textwidth]{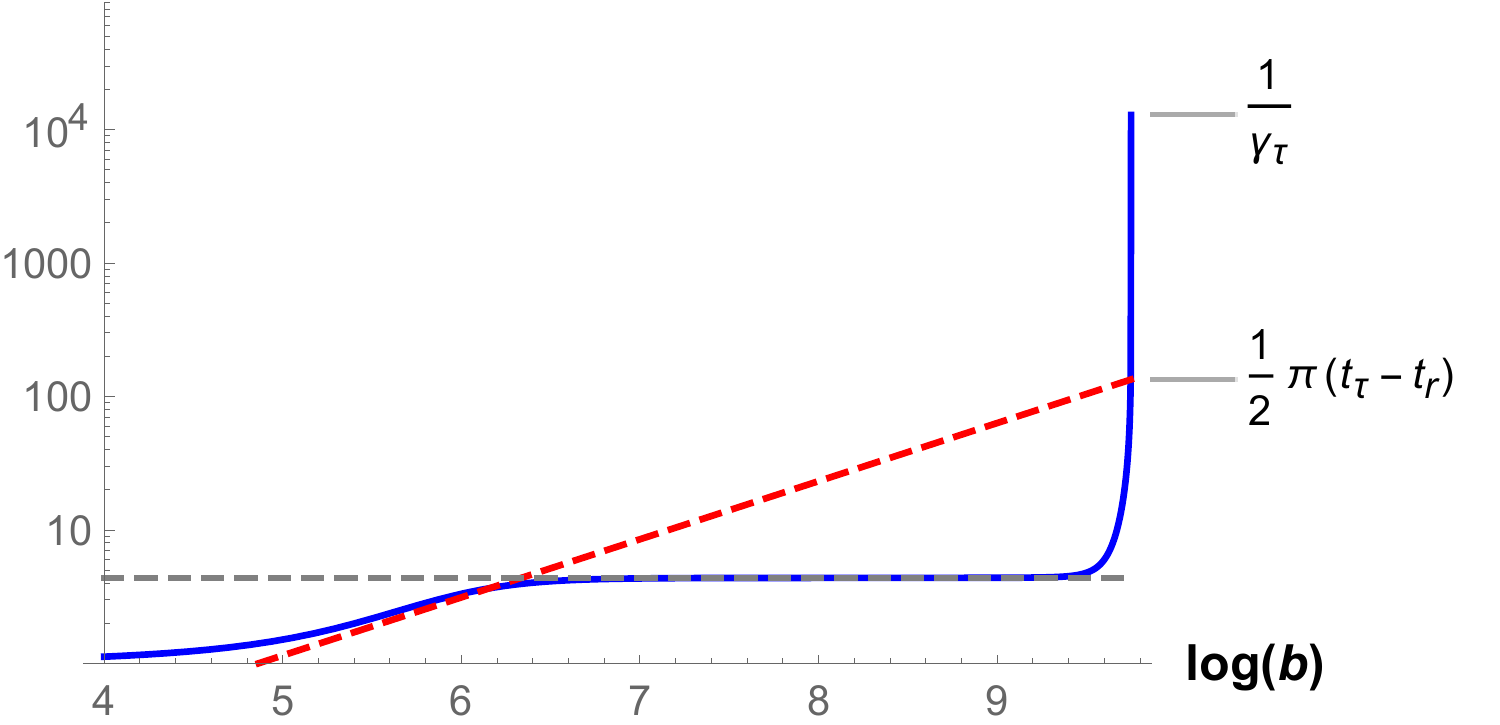}
\label{Fig:nu12_Bou_igt_gau2}
}
\vspace{0.5cm}
\subfigure[$e^2$ and $\nu$ as functions of $\ln b$]{
\includegraphics[width=0.3\textwidth]{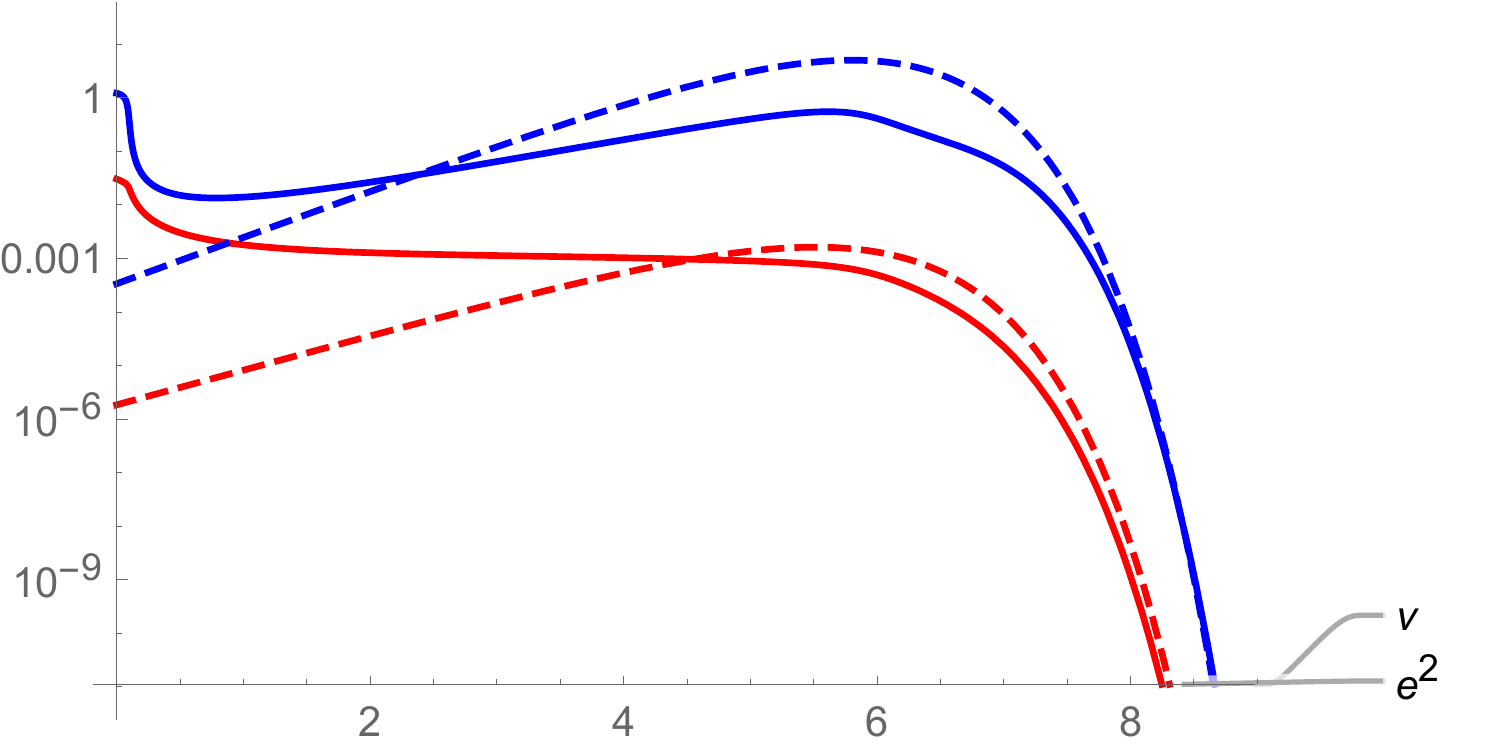}
\label{Fig:nu12_Dis_e2nu_gau2}
}
\hfill
\subfigure[$t_{\tau}$ and $t_{\bm r}$ as functions of $\ln b$]{
\includegraphics[width=0.3\textwidth]{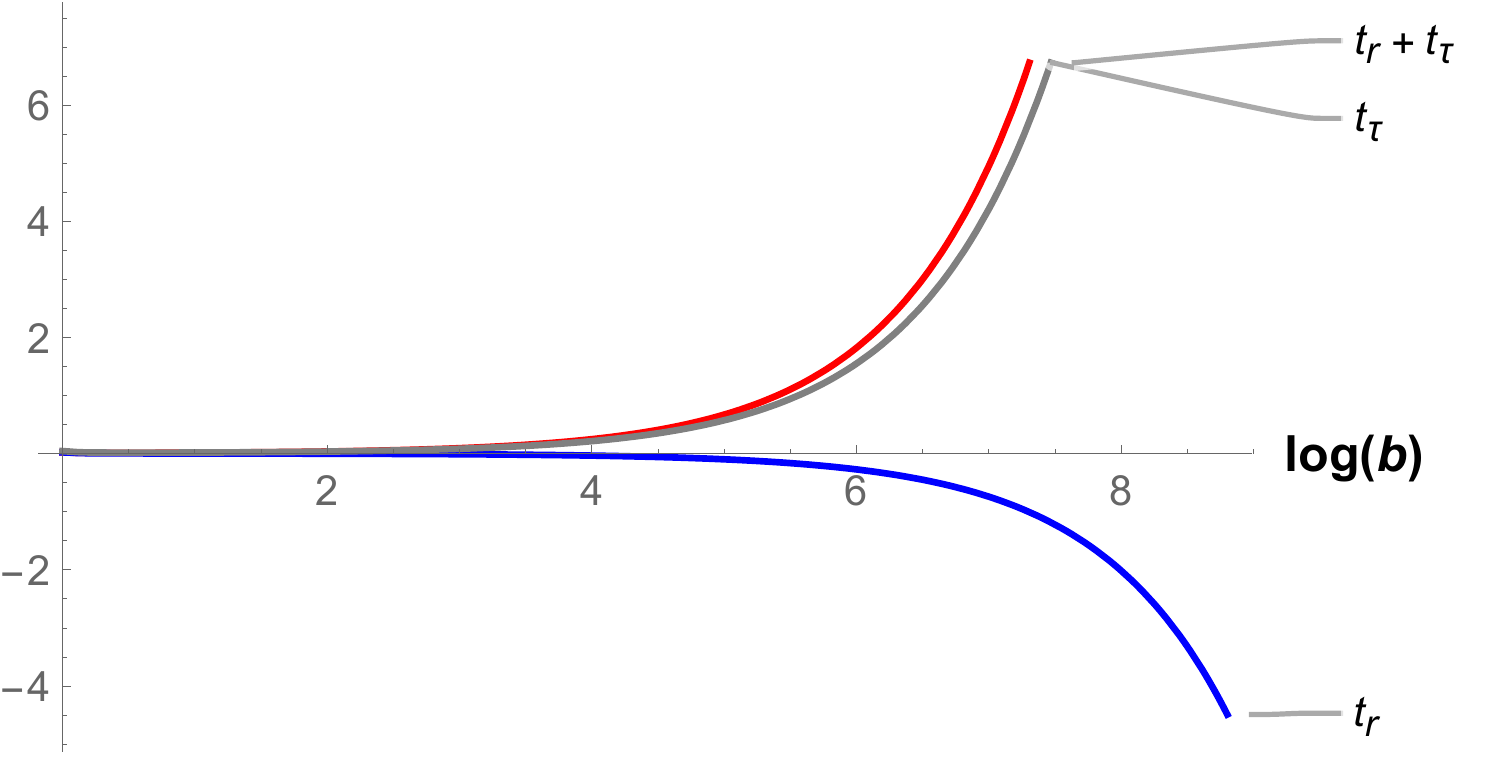}
\label{Fig:nu12_Dis_tttr_gau2}
}
\hfill
\subfigure[$\gamma^{-1}_{\tau}$ as a function of $\ln b$]{
\includegraphics[width=0.3\textwidth]{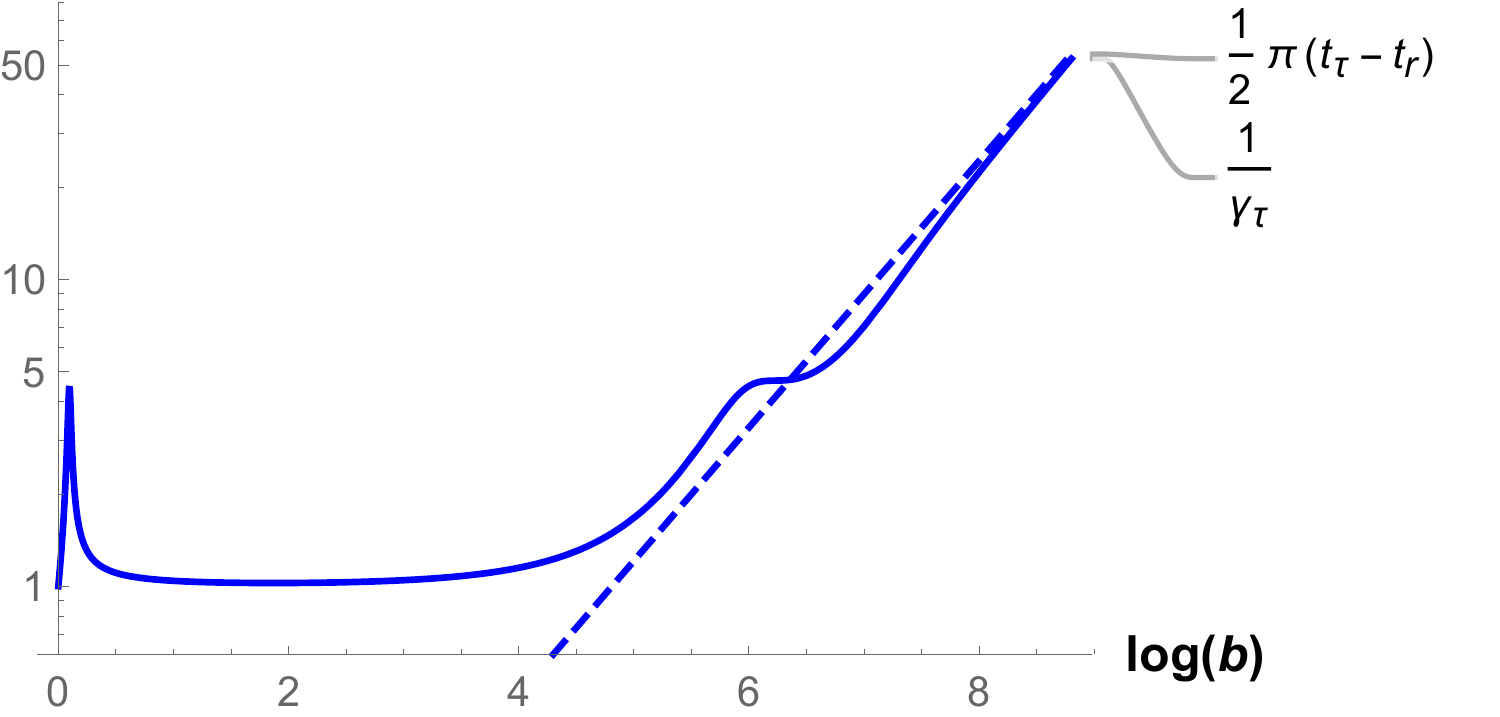}
\label{Fig:nu12_Dis_igt_gau2}
}
\caption{Plots of $e^2$, $\nu$, $t_{\tau}$, $t_{\bm r}$ and ${\gamma^{-1}_{\tau}}$ as functions of the RG scale $\ln b$. The plots are obtained from numerical solutions of the RG equations [Eqs.~(\ref{Eq:RGe2D_3},\ref{Eq:RGchiD_3},\ref{Eq:RGgammatauD_3},\ref{Eq:gauge3}) together with Eqs.~(\ref{Eq:Z1-soft-approx},\ref{Eq:Z2tr-soft-approx})]. The initial value of $\nu$ and $\gamma^{-1}_{\tau}$ are $1.2$ and $1$, respectively. As for the initial values of $e^2$ and $t_{\tau}=t_{\bm r}=t$, we chose $(e^2,t)_{|\ln b=0}=(0.25,0.0219)$ [denoted by blue cross mark in Fig.~\ref{Fig:nu12_pd_gau2}] for Figs.~\ref{Fig:nu12_Ord_e2nu_gau2},\ref{Fig:nu12_Ord_tttr_gau2},\ref{Fig:nu12_Ord_igt_gau2}, $(e^2,t)_{|\ln b=0}=(0.05,0.0219)$  [denoted by red cross mark in Fig.~\ref{Fig:nu12_pd_gau2}] for Figs.~\ref{Fig:nu12_Bou_e2nu_gau2},\ref{Fig:nu12_Bou_tttr_gau2},\ref{Fig:nu12_Bou_igt_gau2}, and $(e^2,t)_{|\ln b=0}=(0.03,0.0219)$  [denoted by gray cross mark in Fig.~\ref{Fig:nu12_pd_gau2}] for Figs.~\ref{Fig:nu12_Dis_e2nu_gau2},\ref{Fig:nu12_Dis_tttr_gau2},\ref{Fig:nu12_Dis_igt_gau2}. Dashed lines in Figs.~\ref{Fig:nu12_Dis_e2nu_gau2}, \ref{Fig:nu12_Dis_igt_gau2} are from Eq.~(\ref{Eq:3D_Dis_asym}).}
\label{Fig:3D_nu12_gauge2}
\end{figure}

\begin{figure}[t]
\subfigure[$e^2$ and $\nu$ as functions of $\ln b$]{
\includegraphics[width=0.3\textwidth]{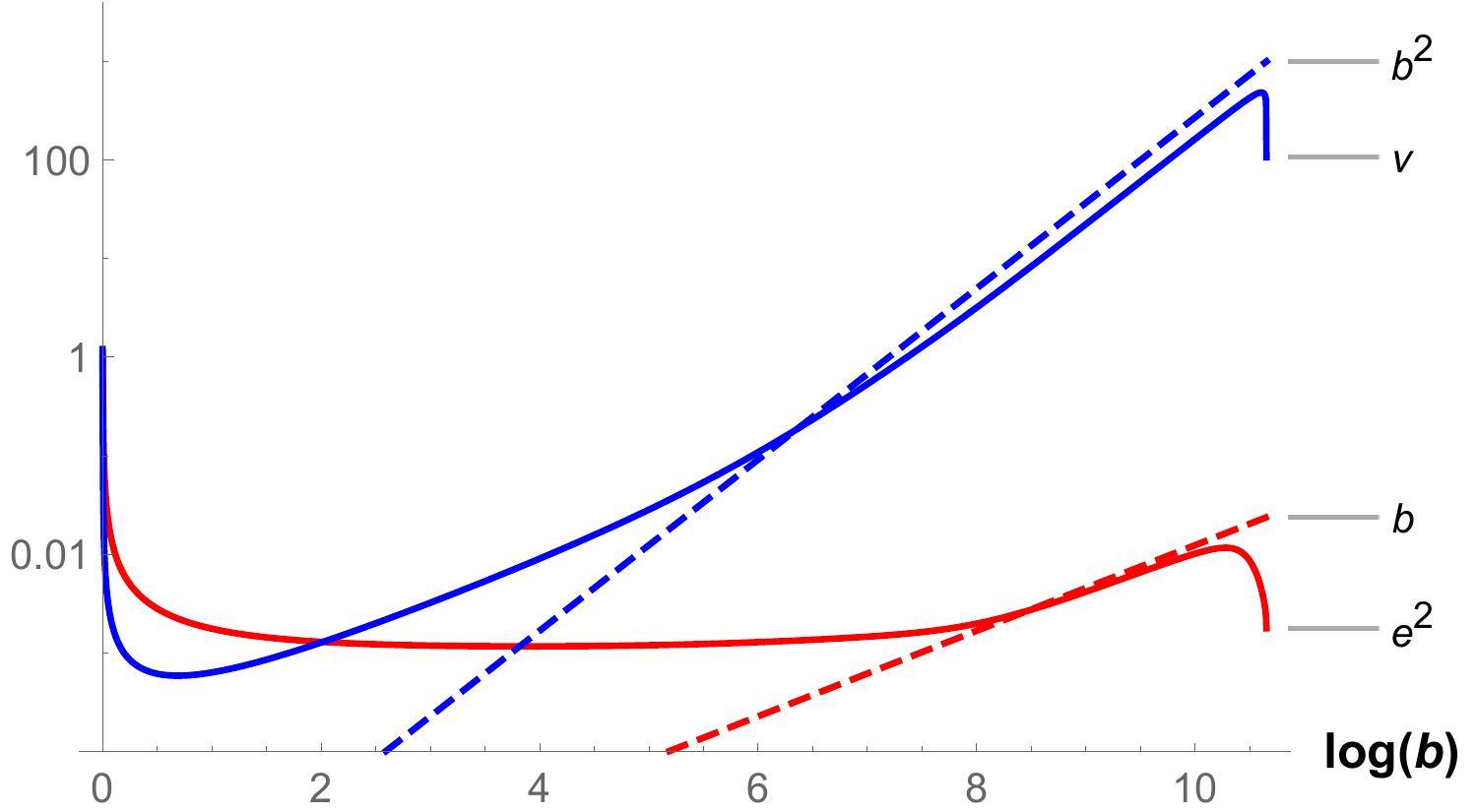}
\label{Fig:nu12_Ord_e2nu_gau2_100}
}
\hfill
\subfigure[$t_{\tau}$ and $t_{\bm r}$ as functions of $\ln b$]{
\includegraphics[width=0.3\textwidth]{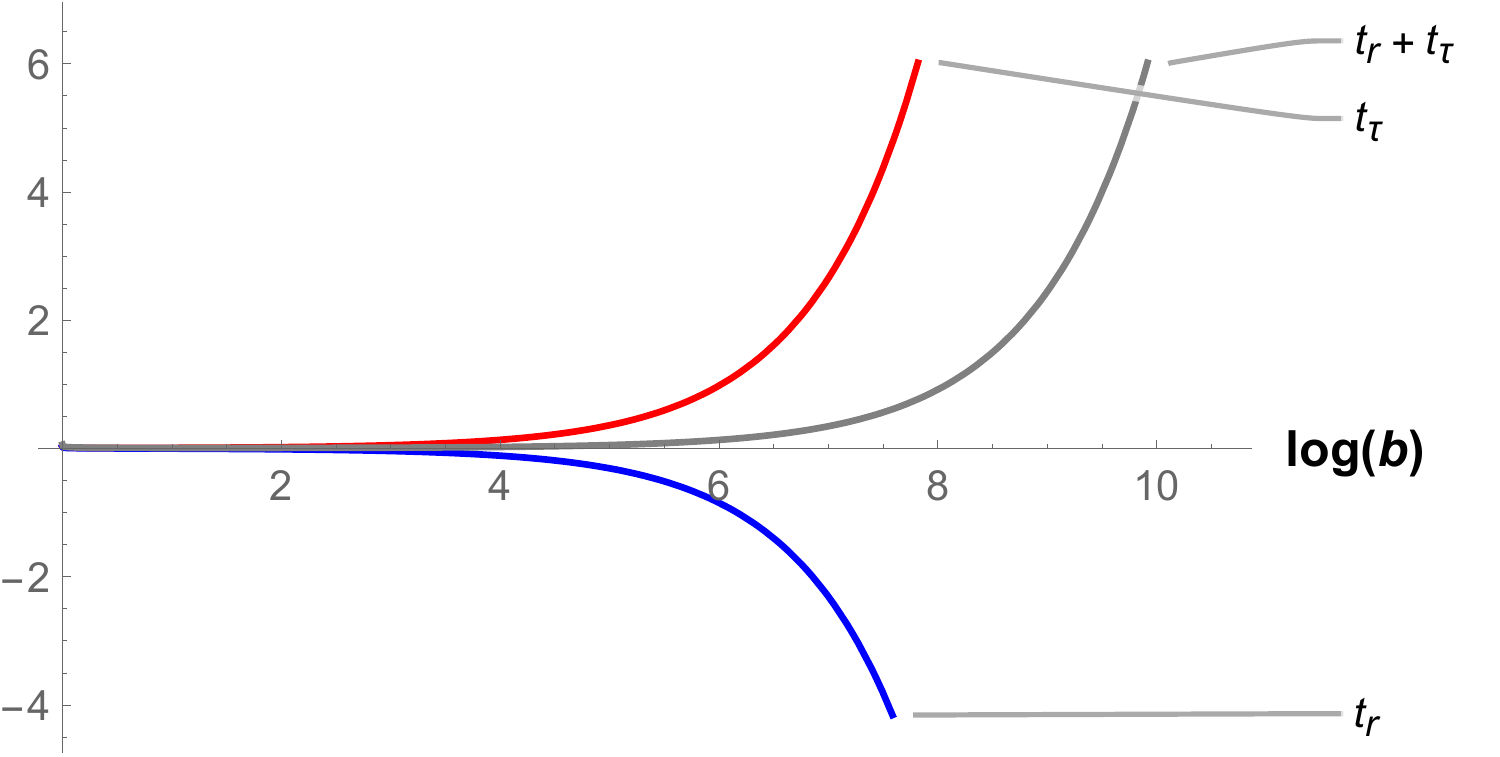}
\label{Fig:nu12_Ord_tttr_gau2_100}
}
\hfill
\subfigure[$\gamma^{-1}_{\tau}$ as a function of $\ln b$]{
\includegraphics[width=0.3\textwidth]{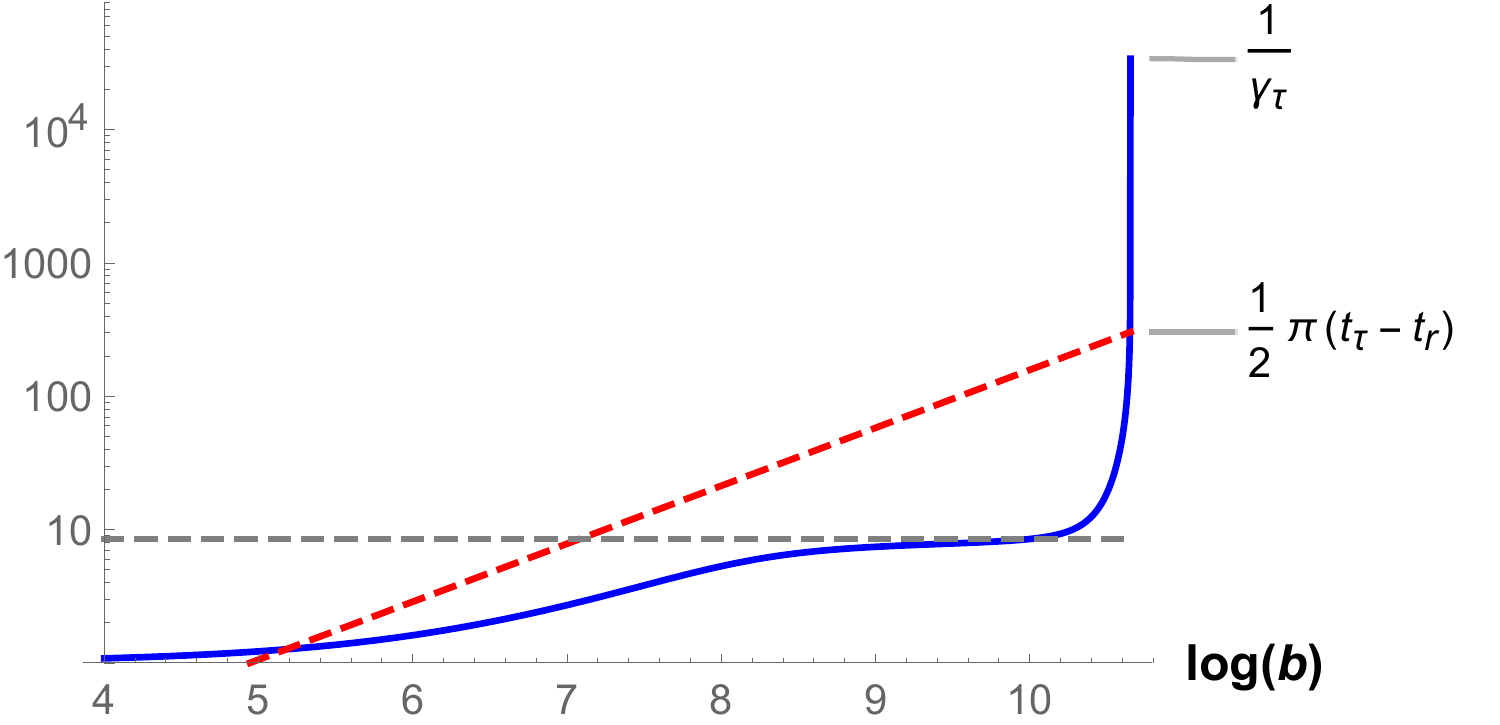}
\label{Fig:nu12_Ord_igt_gau2_100}
}
\vspace{0.5cm}
\subfigure[$e^2$ and $\nu$ as functions of $\ln b$]{
\includegraphics[width=0.3\textwidth]{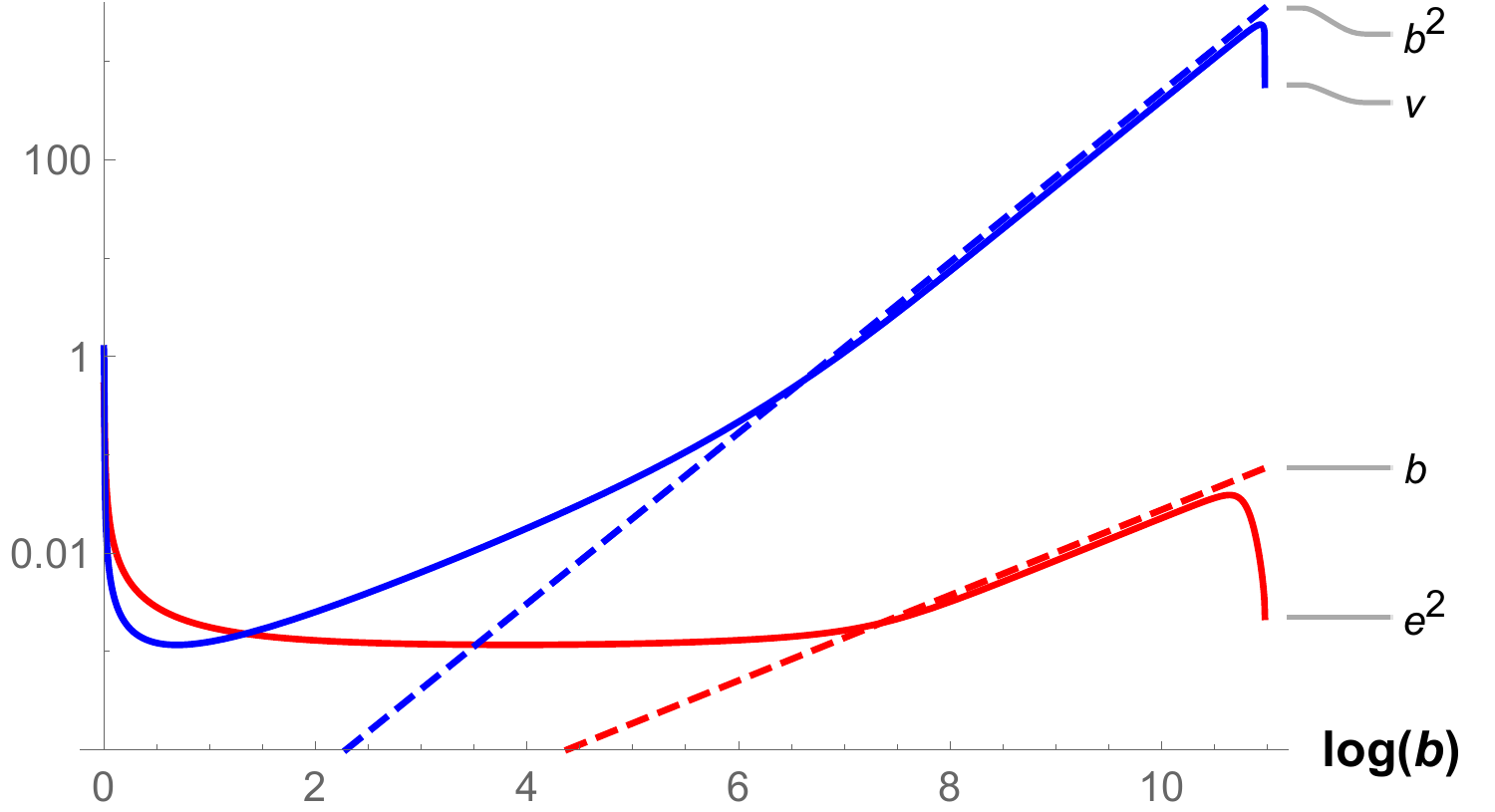}
\label{Fig:nu12_Bou_e2nu_gau2_050}
}
\hfill
\subfigure[$t_{\tau}$ and $t_{\bm r}$ as functions of $\ln b$]{
\includegraphics[width=0.3\textwidth]{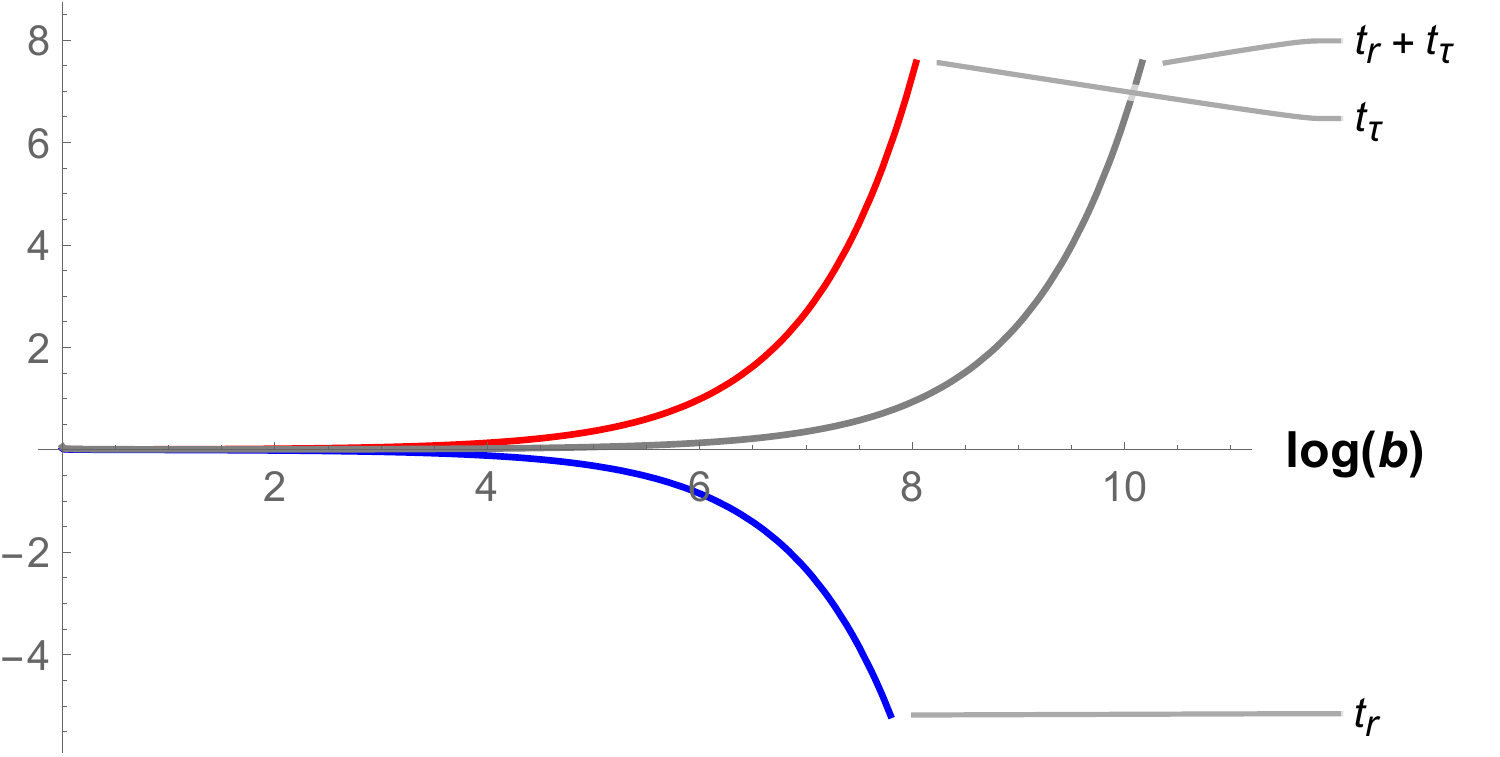}
\label{Fig:nu12_Bou_tttr_gau2_050}
}
\hfill
\subfigure[$\gamma^{-1}_{\tau}$ as a function of $\ln b$]{
\includegraphics[width=0.3\textwidth]{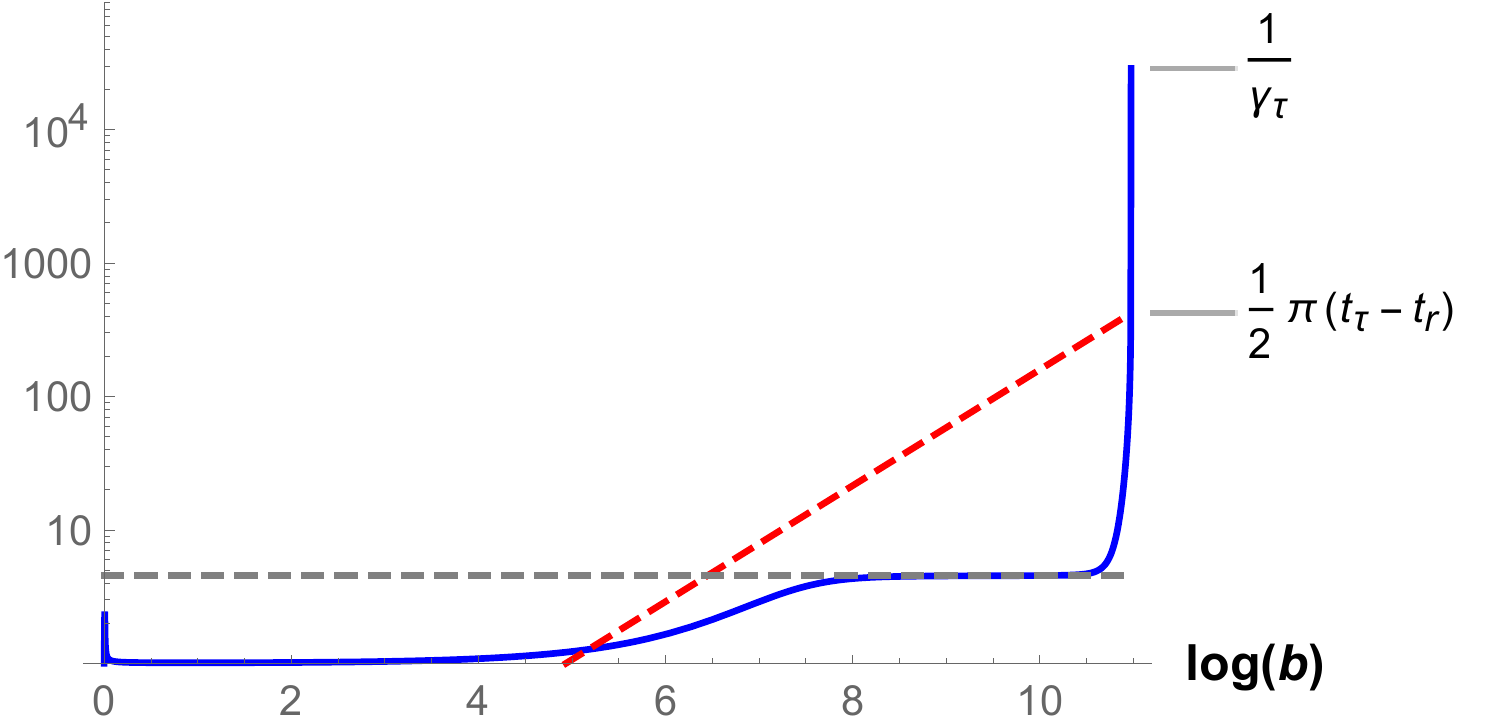}
\label{Fig:nu12_Bou_igt_gau2_050}
}
\vspace{0.5cm}
\subfigure[$e^2$ and $\nu$ as functions of $\ln b$]{
\includegraphics[width=0.3\textwidth]{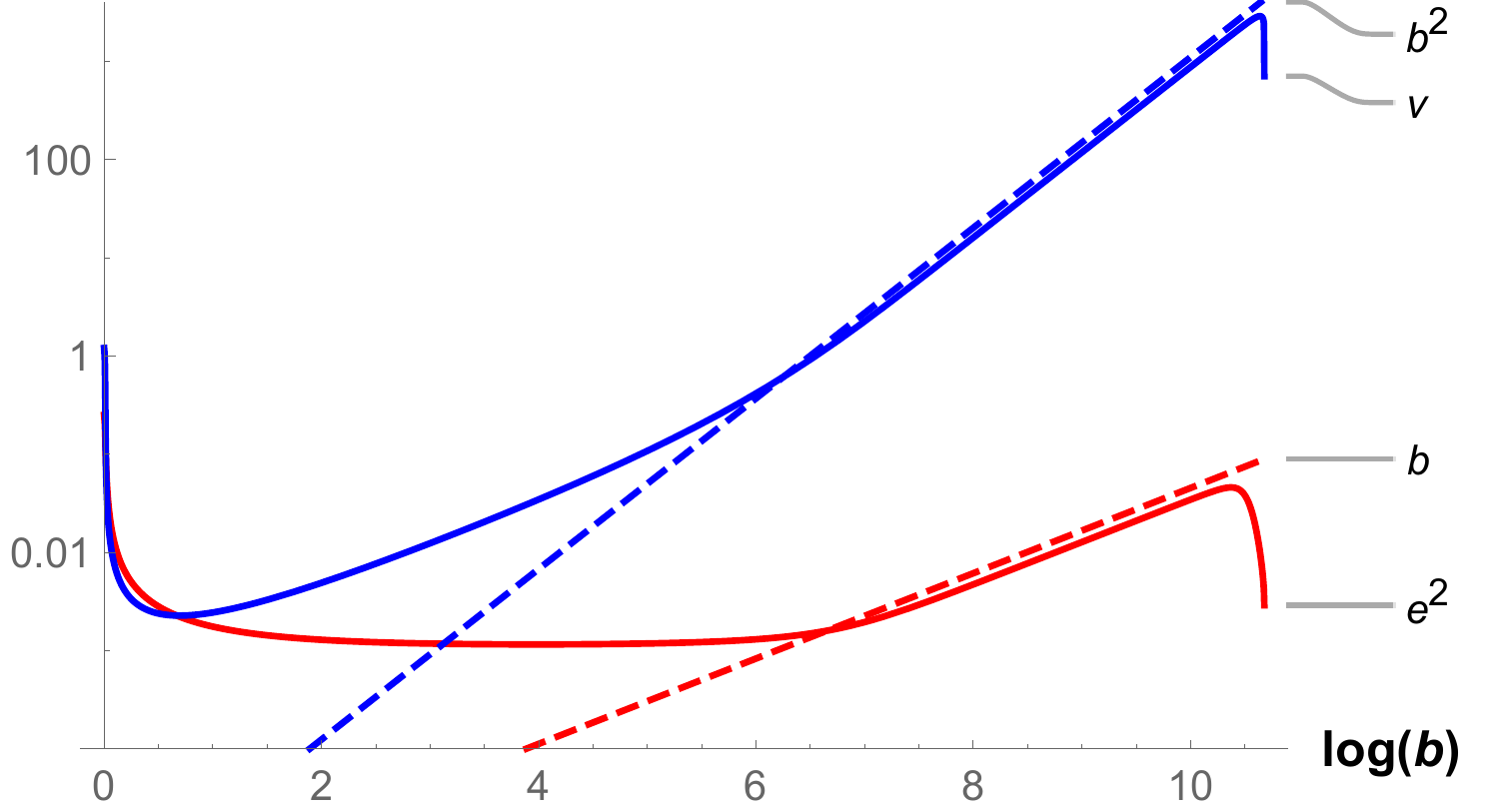}
\label{Fig:nu12_Dis_e2nu_gau2_025}
}
\hfill
\subfigure[$t_{\tau}$ and $t_{\bm r}$ as functions of $\ln b$]{
\includegraphics[width=0.3\textwidth]{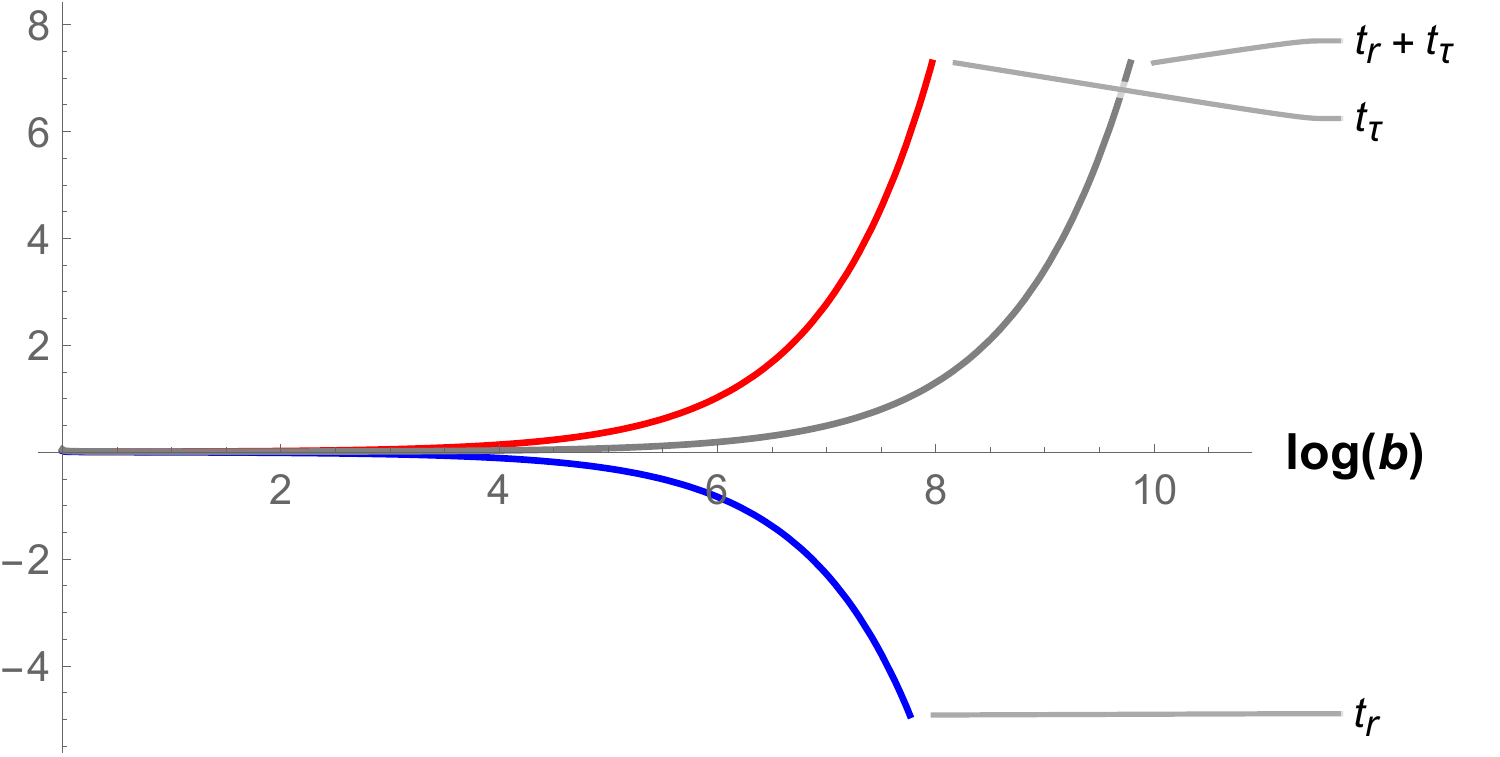}
\label{Fig:nu12_Dis_tttr_gau2_025}
}
\hfill
\subfigure[$\gamma^{-1}_{\tau}$ as a function of $\ln b$]{
\includegraphics[width=0.3\textwidth]{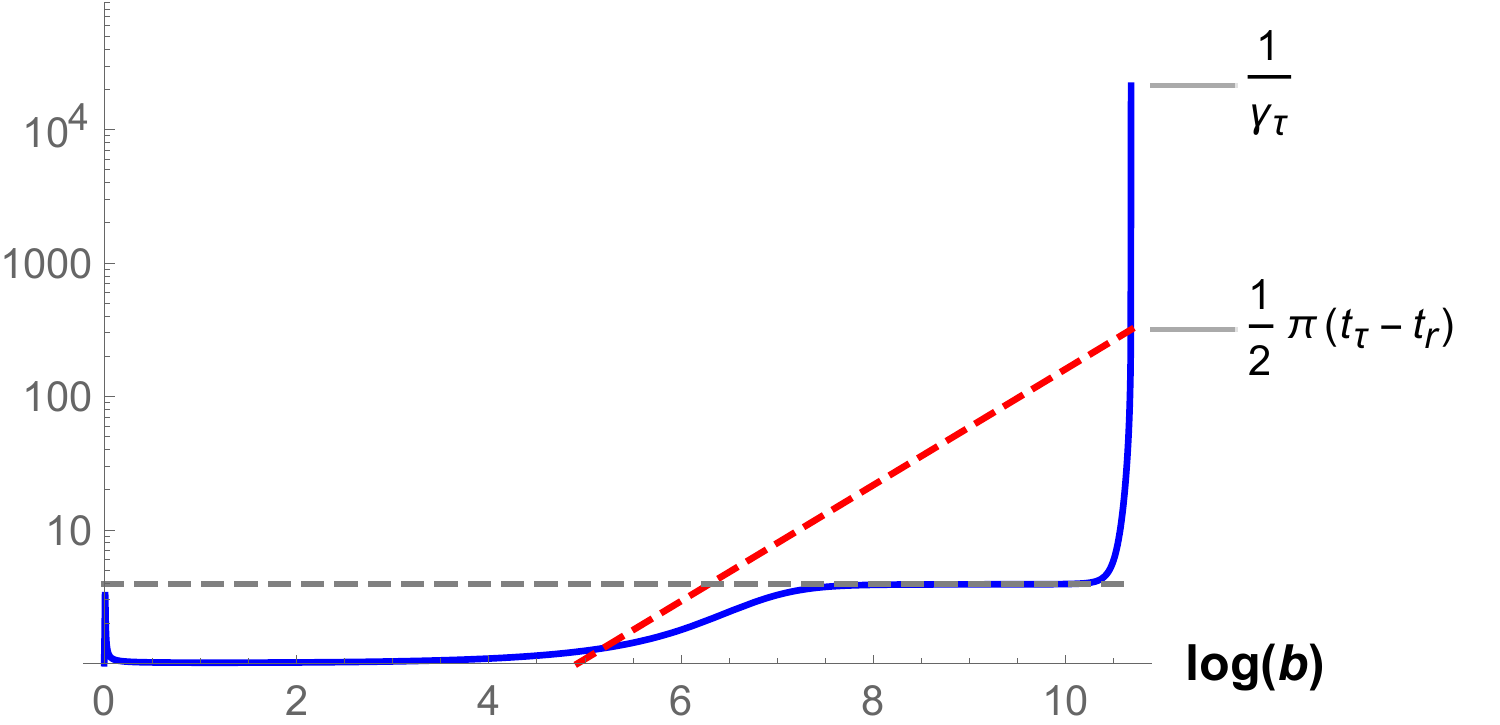}
\label{Fig:nu12_Dis_igt_gau2_025}
}
\caption{Plots of $e^2$, $\nu$, $t_{\tau}$, $t_{\bm r}$ and ${\gamma^{-1}_{\tau}}$ as functions of the RG scale $\ln b$ in the intermediate region [red color region in Fig.~\ref{Fig:nu12_pd_gau2}], where the initial value of $\nu$ and $\gamma^{-1}_{\tau}$ are $1.2$ and $1$, respectively. As for the initial values of $e^2$ and $t_{\tau}=t_{\bm r}=t$, we chose $(e^2,t)_{|\ln b=0}=(1.0,0.03081)$ [denoted by red cross mark in Fig.~\ref{Fig:nu12_pd_gau2}] for Figs.~\ref{Fig:nu12_Ord_e2nu_gau2_100},\ref{Fig:nu12_Ord_tttr_gau2_100},\ref{Fig:nu12_Ord_igt_gau2_100}, $(e^2,t)_{|\ln b=0}=(0.50,0.0286)$  [denoted by red cross mark in Fig.~\ref{Fig:nu12_pd_gau2}] for Figs.~\ref{Fig:nu12_Bou_e2nu_gau2_050},\ref{Fig:nu12_Bou_tttr_gau2_050},\ref{Fig:nu12_Bou_igt_gau2_050}, and $(e^2,t)_{|\ln b=0}=(0.25,0.02645)$  [denoted by red cross mark in Fig.~\ref{Fig:nu12_pd_gau2}] for Figs.~\ref{Fig:nu12_Dis_e2nu_gau2_025},\ref{Fig:nu12_Dis_tttr_gau2_025},\ref{Fig:nu12_Dis_igt_gau2_025}.}
\label{Fig:intermediate_3D_1}
\end{figure}

\begin{figure}[t]
\subfigure[$e^2$ and $\nu$ as functions of $\ln b$]{
\includegraphics[width=0.3\textwidth]{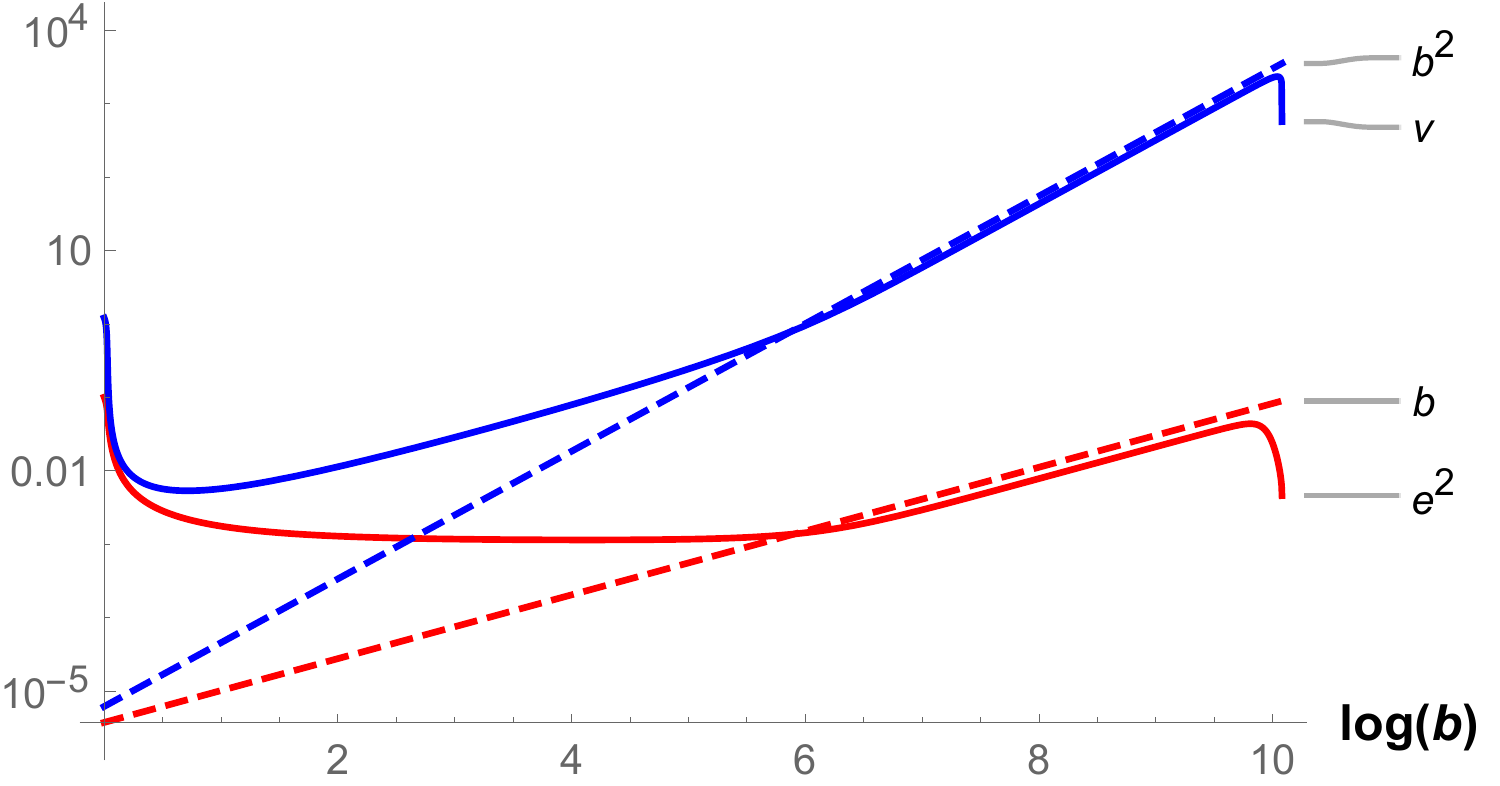}
\label{Fig:nu12_Ord_e2nu_gau2_010}
}
\hfill
\subfigure[$t_{\tau}$ and $t_{\bm r}$ as functions of $\ln b$]{
\includegraphics[width=0.3\textwidth]{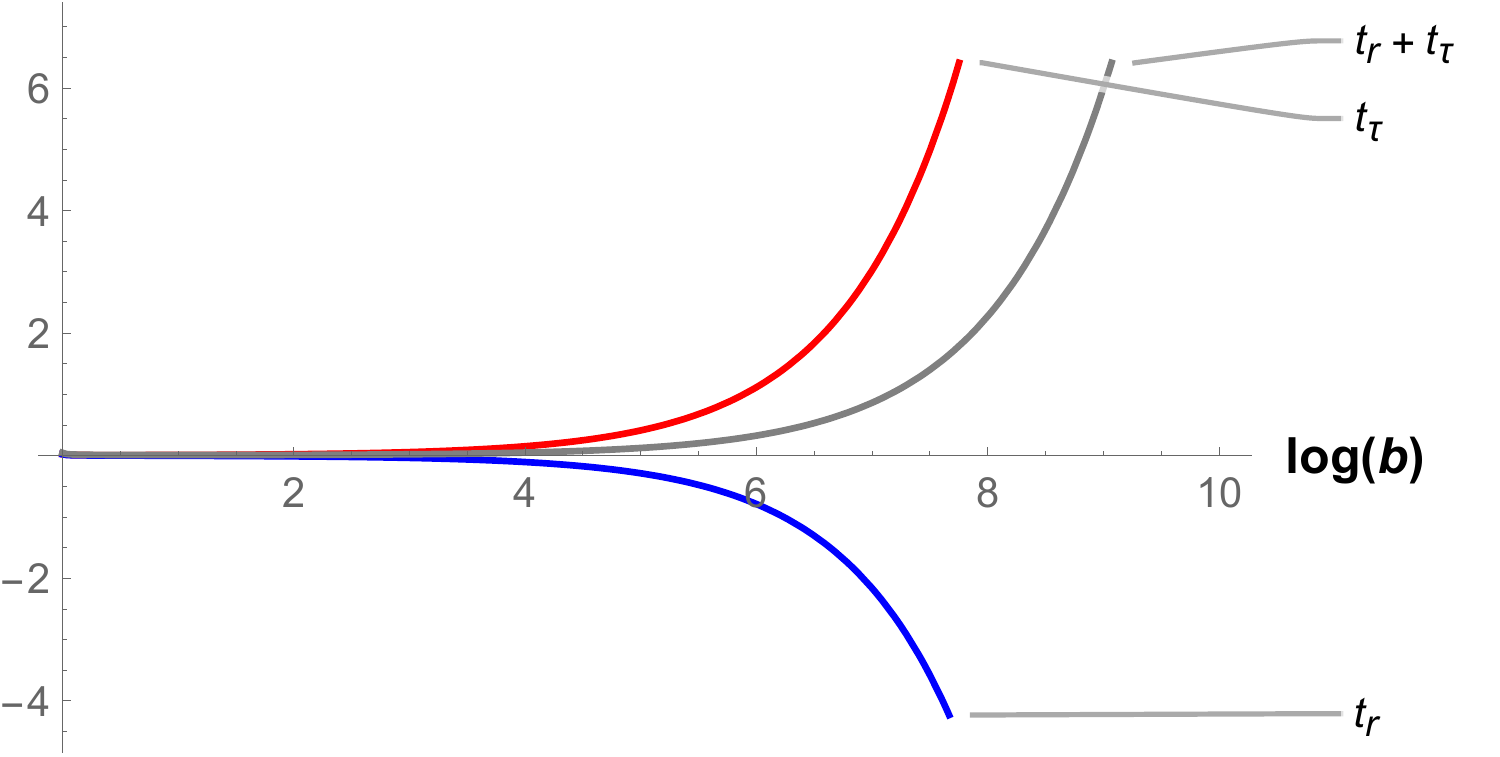}
\label{Fig:nu12_Ord_tttr_gau2_010}
}
\hfill
\subfigure[$\gamma^{-1}_{\tau}$ as a function of $\ln b$]{
\includegraphics[width=0.3\textwidth]{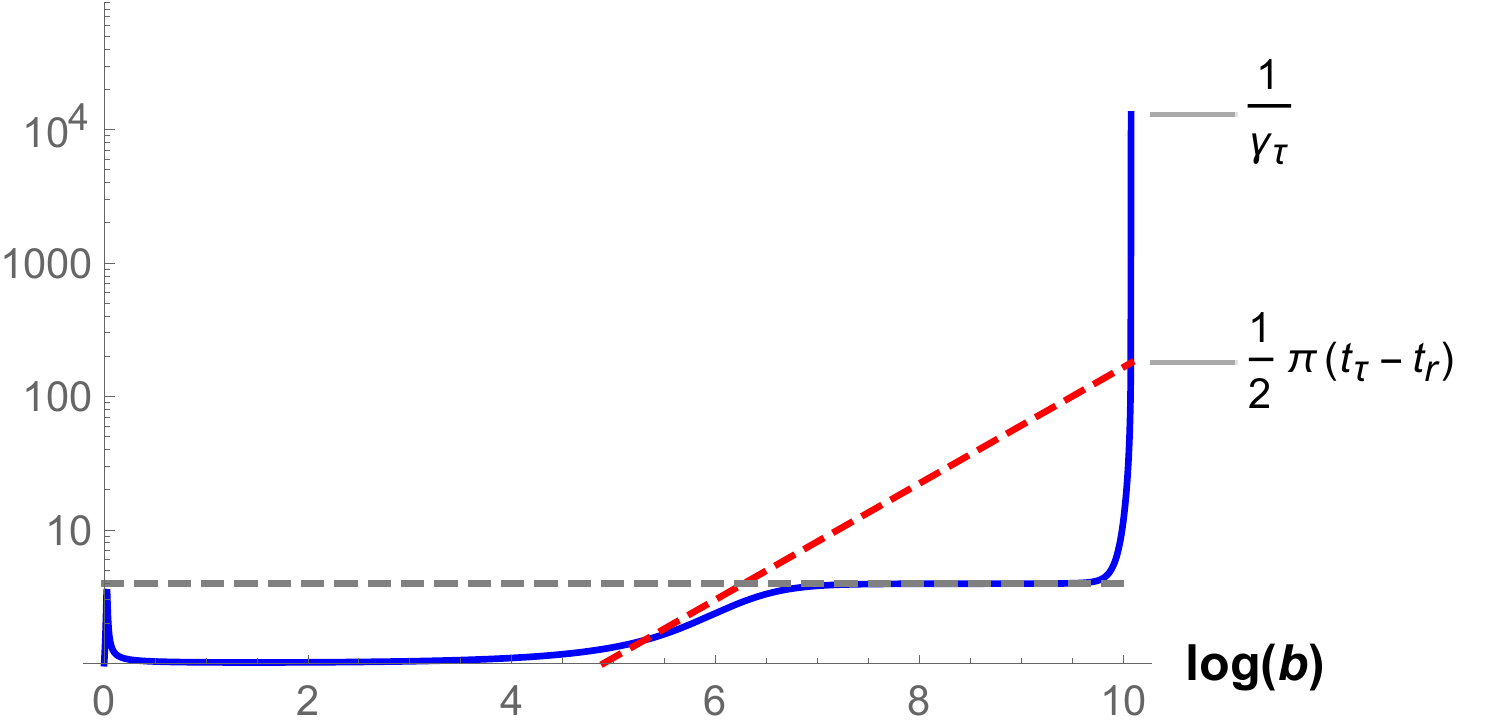}
\label{Fig:nu12_Ord_igt_gau2_010}
}
\vspace{0.5cm}
\subfigure[$e^2$ and $\nu$ as functions of $\ln b$]{
\includegraphics[width=0.3\textwidth]{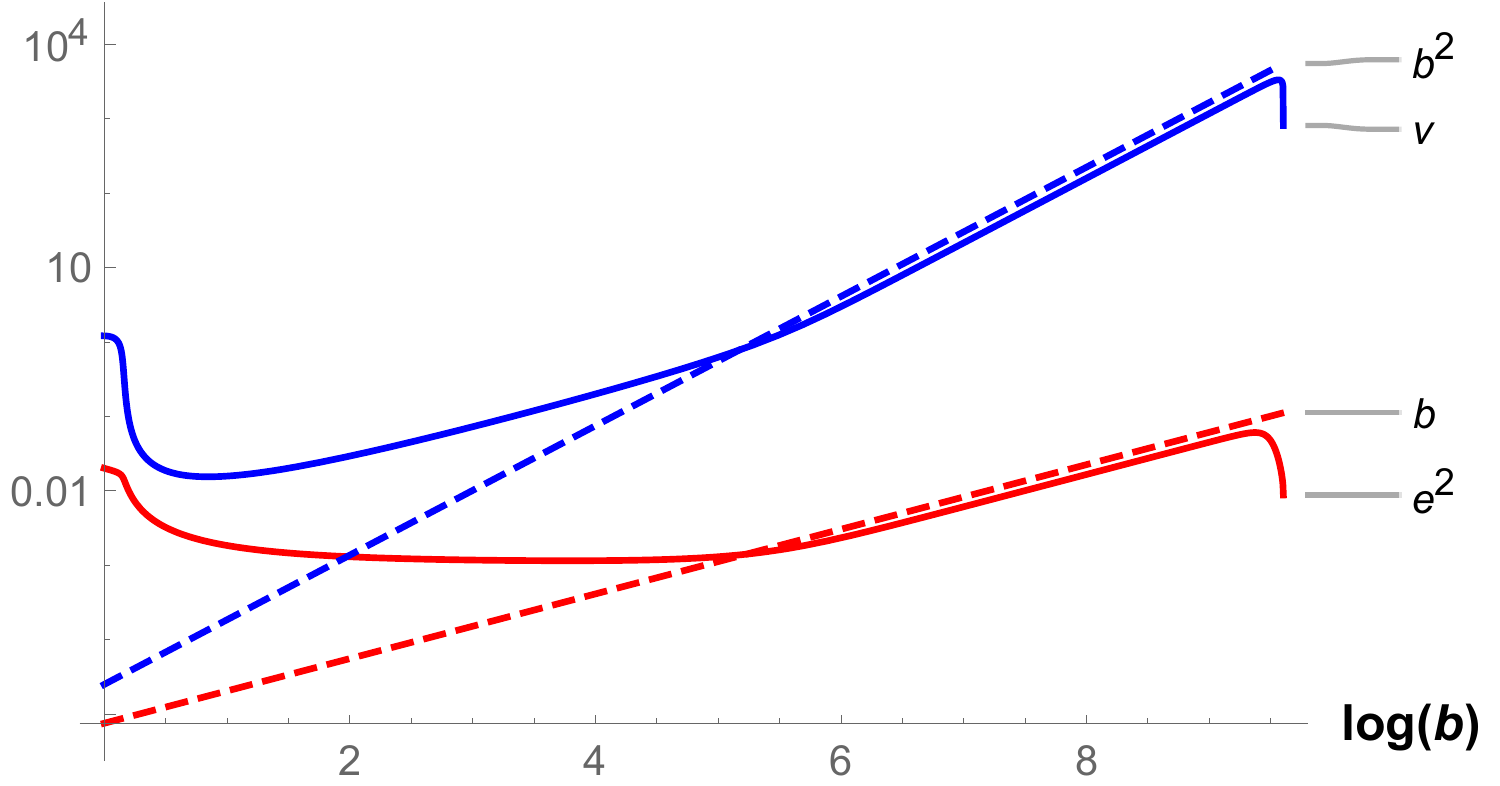}
\label{Fig:nu12_Bou_e2nu_gau2_002}
}
\hfill
\subfigure[$t_{\tau}$ and $t_{\bm r}$ as functions of $\ln b$]{
\includegraphics[width=0.3\textwidth]{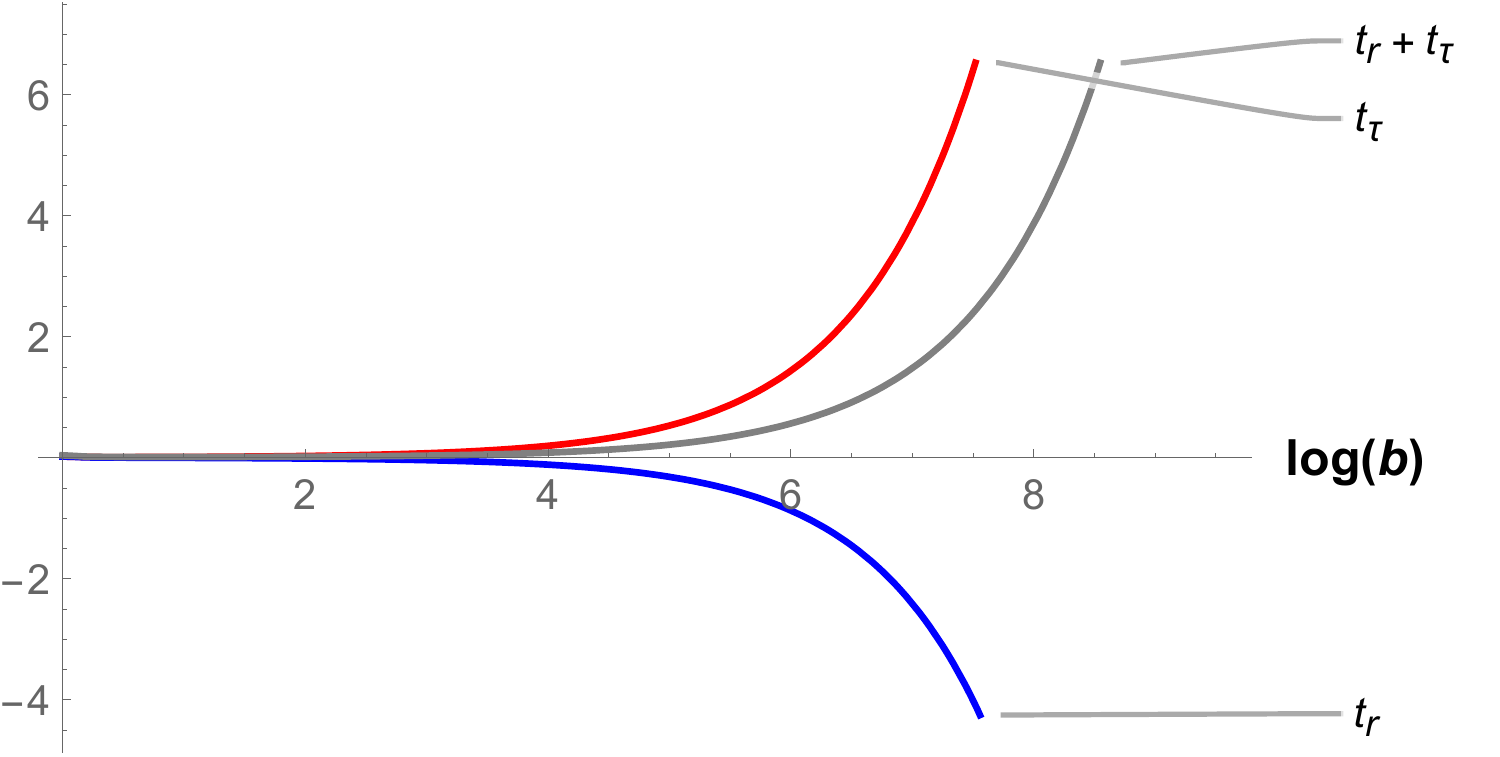}
\label{Fig:nu12_Bou_tttr_gau2_002}
}
\hfill
\subfigure[$\gamma^{-1}_{\tau}$ as a function of $\ln b$]{
\includegraphics[width=0.3\textwidth]{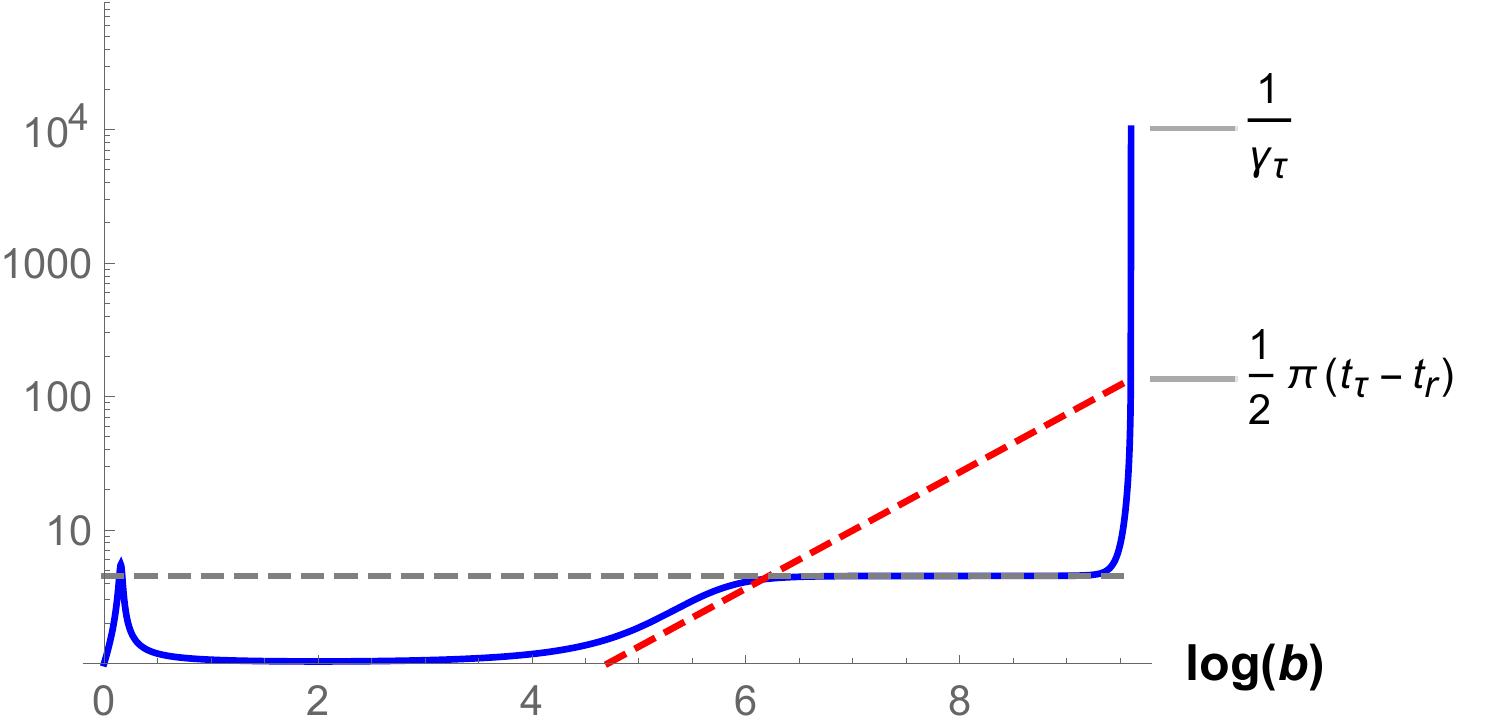}
\label{Fig:nu12_Bou_igt_gau2_002}
}
\vspace{0.5cm}
\subfigure[$e^2$ and $\nu$ as functions of $\ln b$]{
\includegraphics[width=0.3\textwidth]{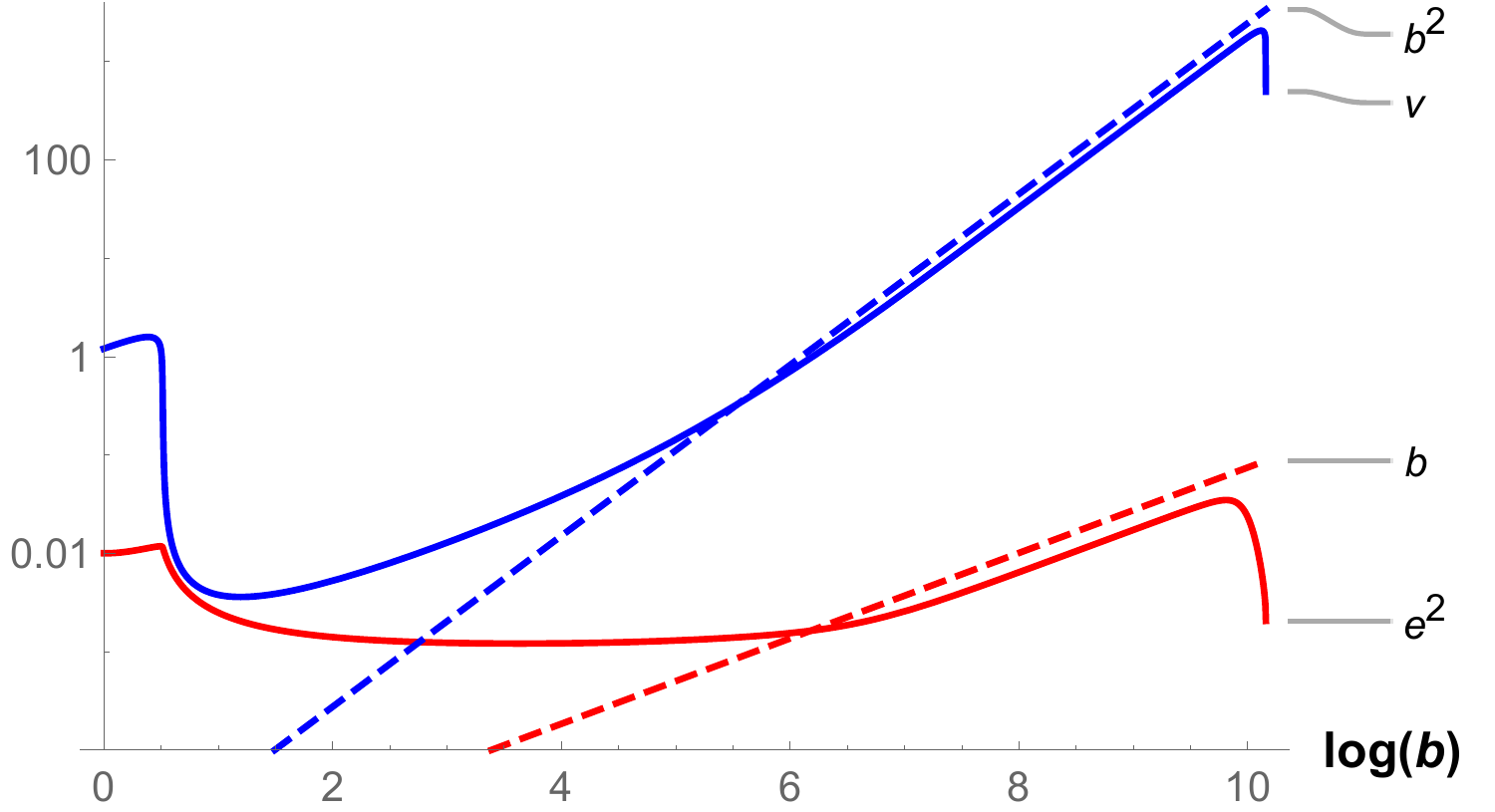}
\label{Fig:nu12_Dis_e2nu_gau2_001}
}
\hfill
\subfigure[$t_{\tau}$ and $t_{\bm r}$ as functions of $\ln b$]{
\includegraphics[width=0.3\textwidth]{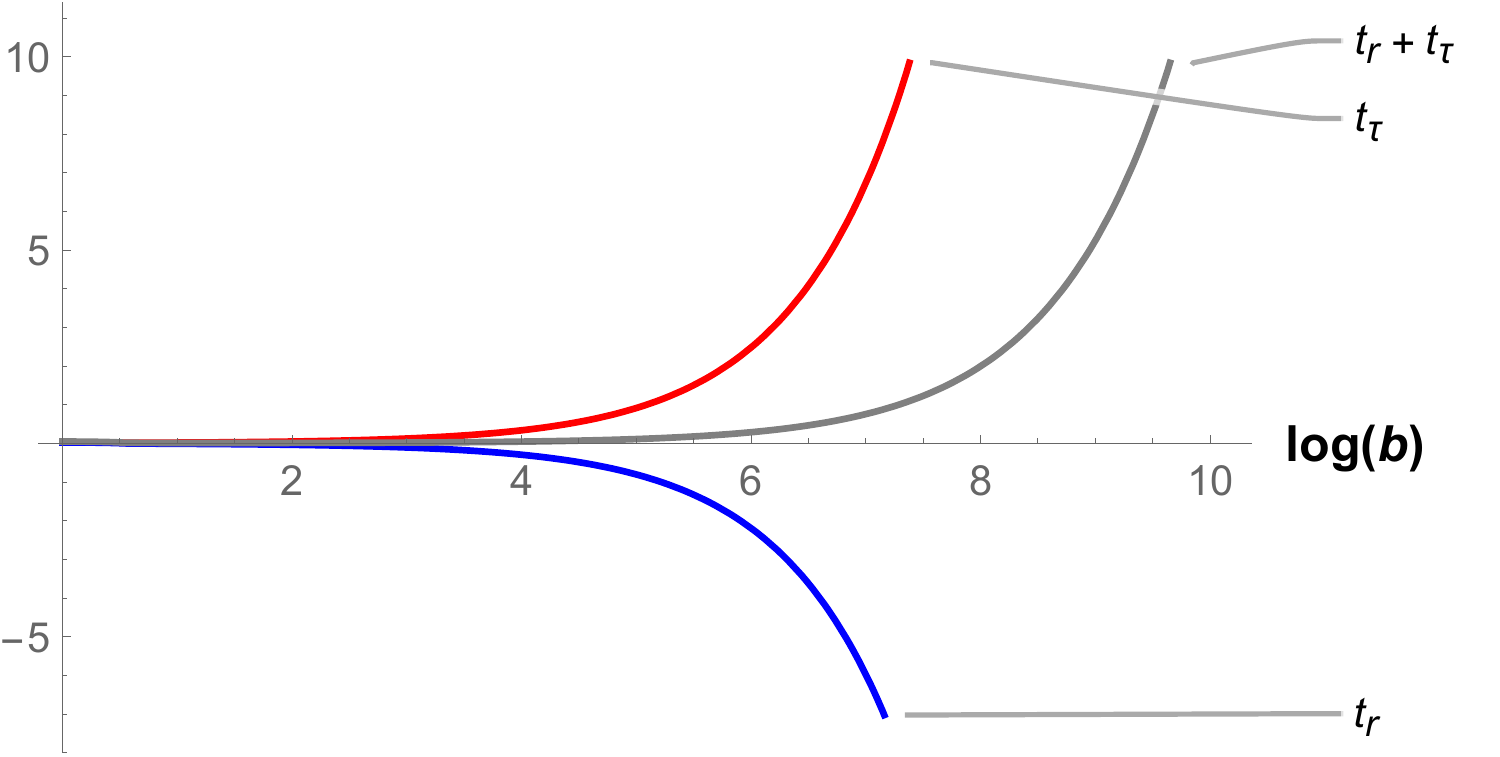}
\label{Fig:nu12_Dis_tttr_gau2_001}
}
\hfill
\subfigure[$\gamma^{-1}_{\tau}$ as a function of $\ln b$]{
\includegraphics[width=0.3\textwidth]{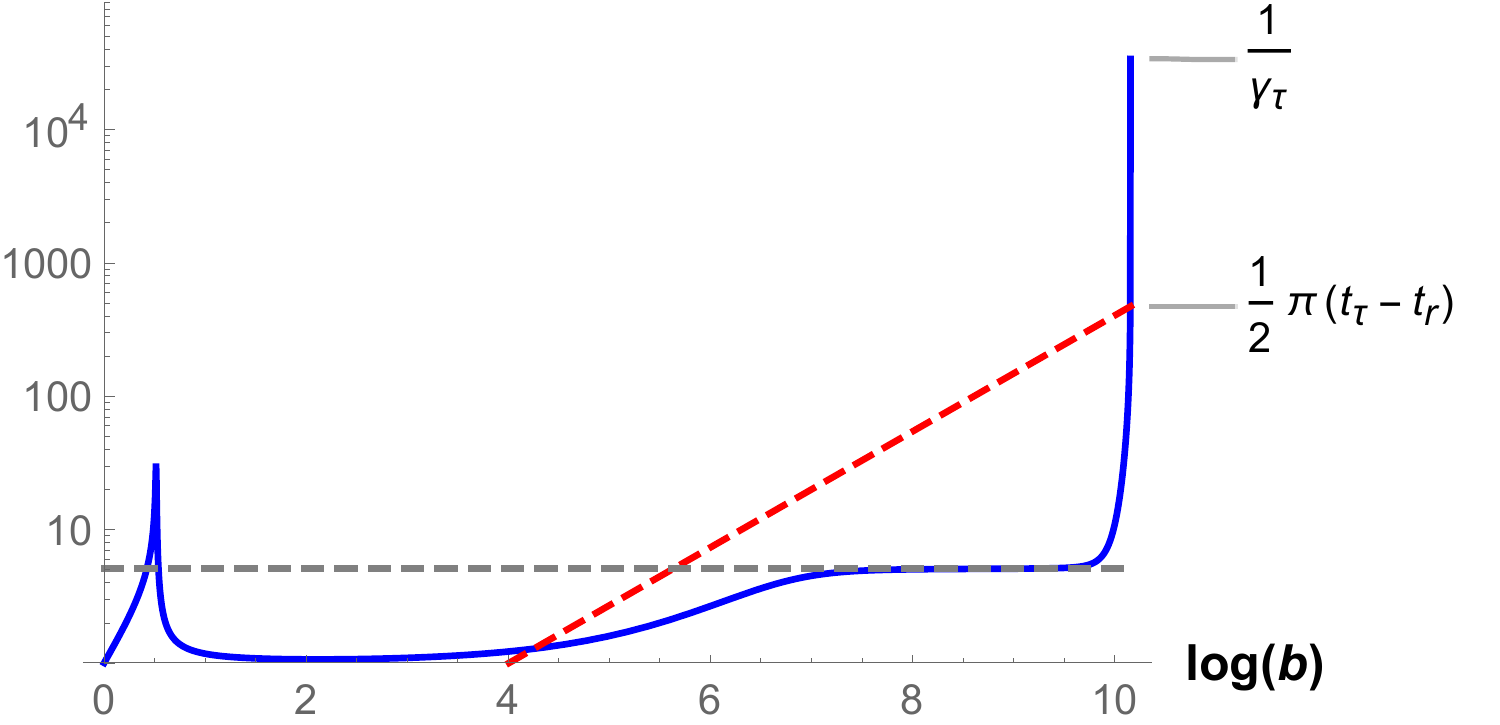}
\label{Fig:nu12_Dis_igt_gau2_001}
}
\caption{Plots of $e^2$, $\nu$, $t_{\tau}$, $t_{\bm r}$ and ${\gamma^{-1}_{\tau}}$ as functions of the RG scale $\ln b$ in the intermediate region [red color region in Fig.~\ref{Fig:nu12_pd_gau2}], where the initial value of $\nu$ and $\gamma^{-1}_{\tau}$ are $1.2$ and $1$, respectively. As for the initial values of $e^2$ and $t_{\tau}=t_{\bm r}=t$, we chose $(e^2,t)_{|\ln b=0}=(0.1,0.02375)$ [denoted by red cross mark in Fig.~\ref{Fig:nu12_pd_gau2}] for Figs.~\ref{Fig:nu12_Ord_e2nu_gau2_010},\ref{Fig:nu12_Ord_tttr_gau2_010},\ref{Fig:nu12_Ord_igt_gau2_010}, $(e^2,t)_{|\ln b=0}=(0.02,0.0202)$  [denoted by red cross mark in Fig.~\ref{Fig:nu12_pd_gau2}] for Figs.~\ref{Fig:nu12_Bou_e2nu_gau2_002},\ref{Fig:nu12_Bou_tttr_gau2_002},\ref{Fig:nu12_Bou_igt_gau2_002}, and $(e^2,t)_{|\ln b=0}=(0.01,0.02495)$  [denoted by red cross mark in Fig.~\ref{Fig:nu12_pd_gau2}] for Figs.~\ref{Fig:nu12_Dis_e2nu_gau2_001},\ref{Fig:nu12_Dis_tttr_gau2_001},\ref{Fig:nu12_Dis_igt_gau2_001}.}
\label{Fig:intermediate_3D_2}
\end{figure}

\section{Nature of the quasi-disordered phase} 
      In the previous section, we argued that the vortex loop gas model in the intermediate coupling regime is unstable toward the spatial proliferation of temporally polarized vortex lines, and that this drives a phase transition from the ordered phase to the quasi-disordered phase. In (2+1)-dimensional spacetime, a closed vortex loop is the geometric representation of a complete quantum-dynamical process: the spontaneous creation of a vortex-antivortex pair, their subsequent quantum-mechanical motion in the 2D spatial plane, and their ultimate mutual annihilation. This is a fundamentally dynamical event. In stark contrast, a vortex line strictly polarized along the imaginary time axis represents a static vortex with frozen dynamics; a boson field $e^{i\psi}$ (where $\psi$ is the spin-wave part of the U(1) phase variable $\theta$) experiences it as a quenched magnetic flux. Consequently, when the time-polarized vortex lines proliferate in space, the partition function of the loop gas model is effectively given by a quantum partition function of a free (or weakly interacting) boson system in a background of dense, randomly-distributed quenched magnetic fluxes. All single-particle states of such a 2D boson system are Anderson localized~\cite{efetov80,efetov1983}. Thereby, utilizing the localized nature of the boson wavefunction~\cite{fisher1989b}, we can argue that the quasi-disordered phase shares all the defining properties of the Bose glass phase~\cite{fisher1989b,giamarchi1987,giamarchi1988}: (i) gapless quasiparticle excitations, (ii) finite compressibility, (iii) algebraic decay of the superfluid correlation in time, reflecting slow dynamics, (iv) exponential decay of the superfluid correlation in space, indicating the absence of superfluid stiffness, and (v) insulating boson transport. In what follows, we explain this in detail. 

\subsection{quasi-disordered phase in correlated boson systems}
    To this end, let us first follow Ref.~\cite{lee1991}  and review how a (2+1)-dimensional correlated boson system with an on-site repulsive interaction $V$ can be mapped to the vortex loop gas model.  The partition function of the boson system is given by, 
\begin{align}
Z_{\rm boson} = \lim_{\beta_{\tau}\rightarrow \infty} \int {\cal D}\phi \exp \bigg[-\int^{\beta_{\tau}}_0 d\tau \int d^2{\bm r} 
\Big(\phi^{\dagger}\partial_{\tau}\phi + \frac{1}{2m} |{\bm \partial}_{\bm r}\phi|^2 + \frac{V}{2} \big(|\phi|^2 - \overline{\rho}\big)^2\Big) \bigg]. \label{Eq:zb}
\end{align}
Here, $\phi=|\phi|e^{i\theta}$ is a complex scalar (boson) field, $m$ is the boson mass, and $\overline{\rho}$ is the mean boson density. As in the sigma model case, the spatial-gradient term of the U(1) phase variables is decoupled by introducing a Hubbard-Stratonovich field ${\bm J}_{\bm r} \equiv (J_{1},J_{2})$, 
\begin{align}
Z_{\rm boson} &= \int {\cal D}\theta {\cal D} \delta\rho {\cal D} {\bm J}_{\bm r} \exp \bigg[-S_{\rm boson} \Big], \label{Eq:zb0} \\
S_{\rm boson} &= \int d\tau \int d^2{\bm r} \bigg\{i(\overline{\rho}+\delta\rho) \partial_{\tau} \theta + i {\bm J}_{\bm r} \cdot {\bm \partial}_{\bm r} \theta + \frac{V}{2} (\delta \rho)^2 
+ \frac{m}{2\rho} |{\bm J}_{\bm r}|^2 + \frac{1}{8m \rho} |{\bm \partial}_{\bm r}(\delta \rho)|^2\bigg\}, \label{Eq:zb1}
\end{align}
with $|\phi|^2\equiv \rho \equiv \overline{\rho}+\delta\rho$. The phase gradient vector ${\bm \nabla} \theta \equiv (\partial_{\tau}\theta,{\bm \partial}_{\bm r}\theta)$ in (2+1)-dimensional spacetime is then decomposed into a rotation-free part ${\bm \nabla} \psi$ and a divergence-free part ${\bm b}$, ${\bm \nabla}\theta = {\bm b}+{\bm \nabla}\psi$ with ${\bm \nabla}\times {\bm b}({\bm x}) = 2\pi \oint_{\Gamma} d{\bm l} \!\ \delta^3({\bm x}-{\bm l})$. Here $\Gamma$ stands for closed vortex loops in the (2+1)-dimensional spacetime. Integrating over the non-singular scalar field $\psi$ imposes a divergence-free condition on $\delta \rho$ and ${\bm J}_{\bm r}$, namely $\partial_{\tau} \delta\rho + {\bm \partial}_{\bm r}\cdot {\bm J}_{\bm r}=0$, which expresses the boson charge conservation. As in the sigma model, this condition is satisfied by introducing a magnetic vector potential ${\sf a}$, such that $\delta \rho = ({\bm \nabla} \times {\sf a})_{\tau}$, and ${\bm J}_{\bm r} = ({\bm \nabla} \times {\sf a})_{\bm r}$. Thus, in this duality transformation, the on-site repulsive interaction $V$ becomes the temporal component $\nu_{\tau}$ of the magnetic reluctivity, the boson mass $m$ normalized by the boson density $\rho$ becomes the spatial component ${\nu}_{\bm r}$ of the reluctivity, and the mean boson density $\overline{\rho}$ becomes the temporal Berry phase parameter $\chi$: 
\begin{align}
Z_{\rm boson} &= \int {\cal D}{\bm b} {\cal D} {\sf a} \exp \big[-S_{\rm boson} \big], \label{Eq:zb2} \\
S_{\rm boson} &= \int d\tau d^2{\bm r} \bigg\{i \!\ \overline{\rho} \!\ b_{\tau}  + i ({\bm \nabla} \times {\sf a})\cdot {\bm b} + \frac{V}{2} ({\bm \nabla}\times {\sf a})^2_{\tau} 
+ \frac{m}{2(\overline{\rho}+({\bm \nabla}\times {\sf a})_{\tau})} |({\bm \nabla}\times {\sf a})_{\bm r}|^2 + \cdots \bigg\}. \label{Eq:zb3}
\end{align}
Here, we omitted the term $|{\bm \partial}_{\bm r}(\delta \rho)|^2$ because it contains higher-order spatial gradients than the Maxwell term: its tree level scaling dimension is negative, rendering it irrelevant in the renormalization-group sense. Similarly, for finite mean boson density $\overline{\rho}$, we may expand the spatial component $\nu_{\bm r}$ of the magnetic reluctivity in powers of $({\bm \nabla}\times {\sf a})_{\tau}$ and retain only the leading order, omitting all higher-order contributions. This leads to the vortex loop gas model, Eq.~(2) in the main text,  
\begin{align}
Z_{\rm boson} =& \int {\cal D} {\sf a} \exp \bigg[- \frac{1}{2} \int d\tau d^2{\bm r} \Big\{V({\bm \nabla}\times {\sf a})^2_{\tau} 
+ \frac{m}{\overline{\rho}} |({\bm \nabla}\times {\sf a})_{\bm r}|^2\Big\} \bigg]  \\
     &\int {\cal D}\Gamma \exp \bigg[ i2\pi  \oint_{\Gamma} d{\bm l}\cdot {\sf a}({\bm l}) -   2\pi i \!\ \overline{\rho} \!\  S_{\bm r}[\Gamma] \bigg].  \label{Eq:zb4}
\end{align}
In the main text, we argued that the finite Berry phase term $2\pi i \!\ \overline{\rho} \!\  S_{\bm r}[\Gamma]$ alters the short-vortex-loop screening physics of the loop gas model, yielding the intermediate coupling regime. In this regime, the temporal component $V$ of the magnetic reluctivity vanishes at a finite RG scale upon renormalization, while the logarithmic fugacity $t_{\tau}$ for the vortex loop segment along the imaginary time $\tau$ direction remains positive. The small $V$ then polarizes the vortex loops $\Gamma$ along the time direction through the fugacity renormalization, Eq.~(10) in the main text. When all vortex loops $\Gamma$ become fully polarized along $\tau$, only the temporally uniform component of ${\sf a}_{\tau}$ couples to the vortex degree of freedom, and $Z_{\rm boson}$ reduces to the classical partition function of a two-dimensional (2D) Coulomb gas~\cite{kosterlitz1973,kosterlitz1974,kosterlitz2017}: 
\begin{align}
Z_{\rm qd} \simeq & \lim_{\beta_{\tau} \rightarrow \infty} \int {\cal D} {\sf a} \exp \bigg[- \frac{1}{2} \int d\tau d^2{\bm r} \Big\{V_{\rm R}({\bm \nabla}\times {\sf a})^2_{\tau} 
+ \frac{m_{\rm R}}{\overline{\rho}_{\rm R}} |({\bm \nabla}\times {\sf a})_{\bm r}|^2\Big\} \bigg] \Bigg(  1  \!\  +   \sum^{\infty}_{N=1} \Big(\frac{1}{N!}\Big)^2 \\
     &  \prod^N_{j=1} \Big(\int {\cal D}^2{\bm R}_{2j-1} \!\ e^{\beta_{\tau} t_{\tau}} \!\ 
     \int {\cal D}^2{\bm R}_{2j} \!\ \!\ e^{\beta_{\tau} t_{\tau}} \Big) \exp \bigg[ i2\pi  \sum^N_{j=1}\int^{\beta_{\tau}}_{0} dl_{\tau} \big( 
     {\sf a}_{\tau}(l_\tau,{\bm R}_{2j-1}) -  {\sf a}_{\tau}(l_\tau,{\bm R}_{2j})\big) \bigg] \Bigg).  \label{Eq:zb5}
\end{align}
Here, ${\bm R}_{2j-1}$ and ${\bm R}_{2j}$ denote the two-dimensional coordinates of the time-polarized vortex line and antivortex line, respectively, and $e^{\beta_{\tau} t_{\tau}}$ is the fugacity of the polarized (anti)vortex line. We retain the renormalized on-site interaction, $V_{\rm R}$, at a small but finite value for bookkeeping purposes. In this reduction, the fugacity $e^{\beta_{\tau}t_{\tau}}$ of the time-polarized vortex line becomes the fugacity of the electric charge in the 2D Coulomb gas. For sufficiently low temperature, $\beta_{\tau}\rightarrow \infty$, the positive logarithmic fugacity $t_{\tau}$ always brings the system into the 2D plasma phase regime, where the spatial proliferation of the time-polarized vortex lines drives a phase transition from the ordered phase to the quasi-disordered phase.

   To address the properties of the quasi-disordered phase, one may transform Eq.~(\ref{Eq:zb5}) back to the boson field while retaining the vortex degrees of freedom as integration variables:
\begin{align}
Z_{\rm qd} = &   \sum_{N} \Big(\frac{1}{N!}\Big)^2 
\prod^N_{j=1} \Big(\int {\cal D}^2{\bm R}_{2j-1} \!\ e^{\beta_{\tau} t_{\tau}} \!\ 
     \int {\cal D}^2{\bm R}_{2j} \!\ \!\ e^{\beta_{\tau} t_{\tau}} \Big)  \int_{\oint_C {\bm \partial}_{\bm r} \theta \cdot d{\bm r} = \pm 2\pi} 
     {\cal D} \phi \nonumber \\
& \exp \bigg[ -\int d\tau \int d^2{\bm r} 
\Big(\phi^{\dagger}\partial_{\tau}\phi + \frac{1}{2m_{\rm R}} |\partial_{\bm r}\phi|^2 + \frac{V_{\rm R}}{2} \big(|\phi|^2 - \overline{\rho}_{\rm R}\big)^2\Big)\bigg] \Bigg). \label{Eq:zb6}
\end{align}
Here, for a given spatial configuration of time-polarized vortex lines $\{{\bm R}_1,{\bm R}_{2},\cdots\}$, the integral over the complex scalar field $\phi=|\phi|e^{i\theta}$ is taken with a boundary condition. The condition requires that the line integral of the spatial gradient of the U(1) phase $\theta$ along any closed loop $C$ around a vortex center ${\bm R}_j$ equals $\pm 2\pi$,
\begin{align}
\oint_{C} {\bm \partial}_{\bm r} \theta(\tau,{\bm r}) \cdot d{\bm r} = \left\{\begin{array}{ll} 
2\pi & {\rm for} \,\ j= {\rm even}, \\
-2\pi & {\rm for}\,\  j= {\rm odd}. \\
\end{array}\right.  \label{Eq:boundary}
\end{align}
One can readily see the equivalence between Eq.~(\ref{Eq:zb5}) and Eq.~(\ref{Eq:zb6}) by noting that the integral over the complex boson field $\phi$ in Eq.~(\ref{Eq:zb}) can be generally decomposed into a sum over possible vortex loop configurations $\int {\cal D}\Gamma$ and an integral over $\phi$ under a boundary condition associated with the vortex loop configuration: 
\begin{align}
\int {\cal D}\phi = \int {\cal D}\Gamma \int_{\oint_{C} {\bm \nabla}\theta \cdot{\bm dl} 
= 2\pi} {\cal D}\phi. 
\end{align}
Here, the generic boundary conditions require that the line integral of the phase gradient ${\bm \nabla}\theta$ in the 
(2+1)-dimensional spacetime along a closed loop $C$ that encircles a vortex loop segment in $\Gamma$ equals $2\pi$.  In the quasi-disordered phase, $\Gamma$ becomes fully polarized along the time direction, for which the generic boundary conditions become time-independent, reducing to Eq.~(\ref{Eq:boundary}). 

     The boundary condition Eq.~(\ref{Eq:boundary}) imposes a phase winding of $\pm 2\pi$ on the boson field $\phi$ at each vortex position. The partition function $Z_{\rm qd}$ in the quasi-disordered phase can therefore be regarded as a weighted average over quantum partition functions of a free (or weakly interacting, since $V_{\rm R}$ is small) boson system in a random, static magnetic flux background. Each magnetic flux locally breaks time-reversal symmetry, placing the free boson system in the 2D unitary class~\cite{evers2008}. As vortex lines and antivortex lines always appear in pairs, the total magnetic flux in the 2D system is always zero. In the quasi-disordered phase, the vortex fugacity $e^{\beta_{\tau}t_{\tau}}$ is large and positive, so the weighted average is dominated by the free-boson partition functions with dense distributions of random magnetic fluxes. Such 2D free boson systems are generally in the Anderson localized phase~\cite{efetov80,evers2008}, in which single-particle boson density of states $\rho(\omega)$ is finite near zero energy, $\omega\equiv {\varepsilon}-V_{\rm R}\overline{\rho}_{\rm R}=0$, and the boson wave functions near zero energy are spatially localized. This manifests gapless quasiparticle excitations and insulating boson transport in the quasi-disordered phase~\cite{fisher1989b}. Due to the localization of the boson quasiparticle wave functions, the superfluid correlation $\langle T_{\tau}\{\phi({\bm r},\tau)\phi^{\dagger}({\bm r}^{\prime},\tau^{\prime})\} \rangle$ decays exponentially in space, evidencing the absence of superfluid stiffness~\cite{fisher1989b}. At the same time, the finite density of states $\rho(0)$ indicates that the equal-position correlation decays algebraically in time, 
\begin{align}
\langle T_{\tau}\{\phi({\bm r},\tau)\phi^{\dagger}({\bm r},0)\} \rangle \simeq 2\pi \Big(\frac{\rho(0)}{|\tau|}+\frac{\rho^{\prime}(0)}{|\tau|^2}+\cdots\Big),
\end{align}
reflecting quasi-persistent temporal U(1) phase coherence~\cite{fisher1989b}. The finite density of states near the zero energy also indicates a finite compressibility in the quasi-disordered phase. Taken together, these properties establish that the quasi-disordered phase is a Bose glass phase, with the crucial distinction that the disorder is self-generated through the Berry-phase-driven proliferation of topological defects rather than imposed externally.

\end{widetext}

\end{document}